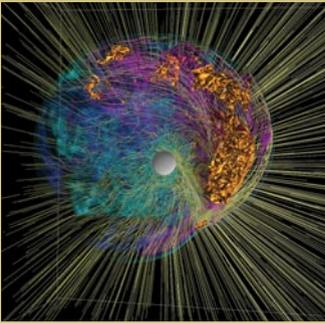
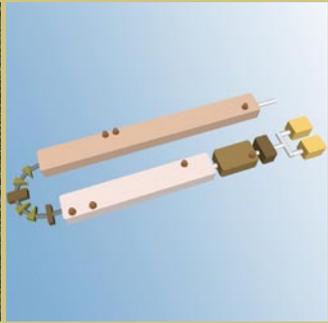
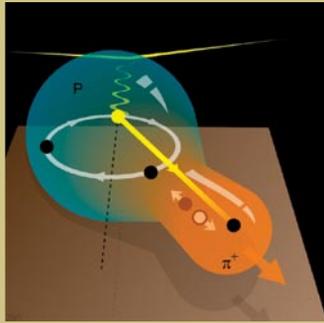
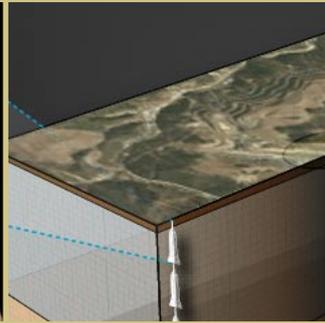
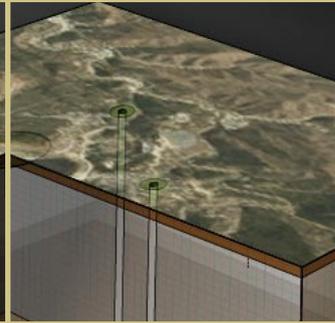
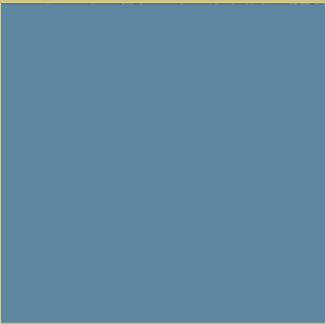
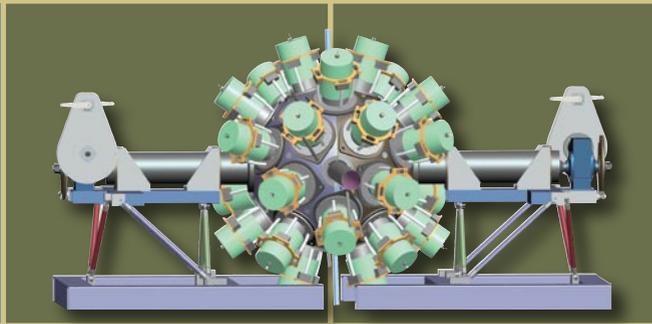
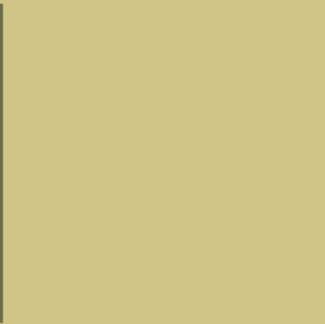
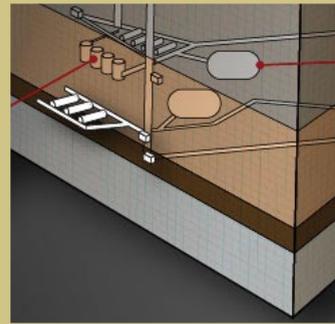
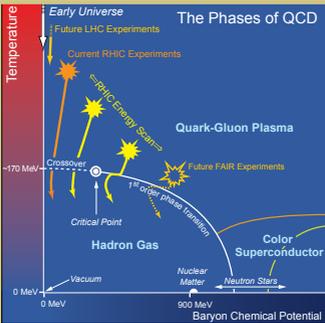
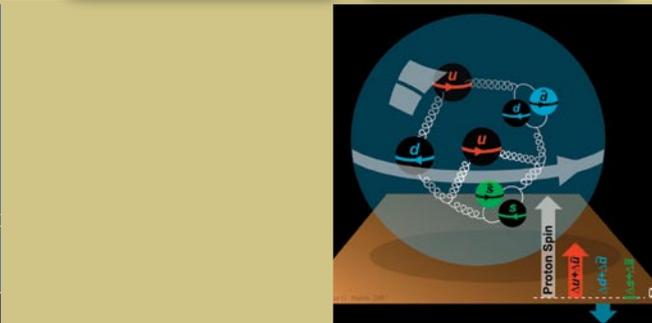
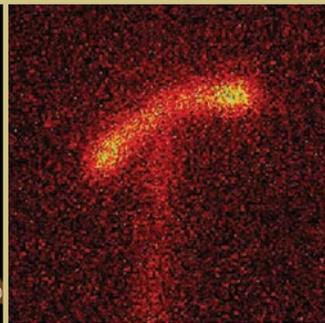
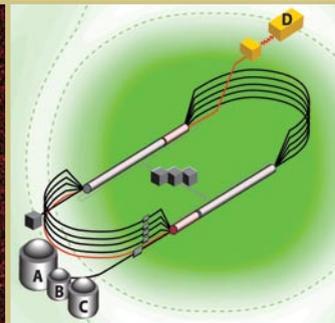
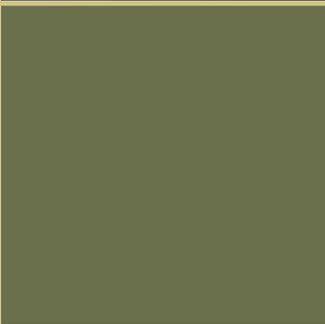
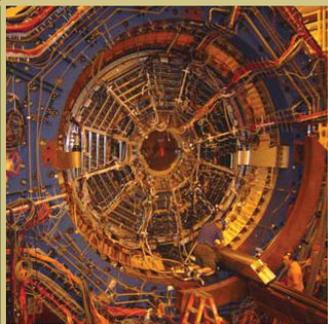
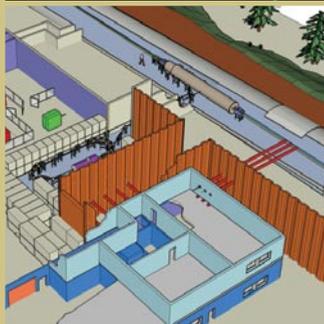
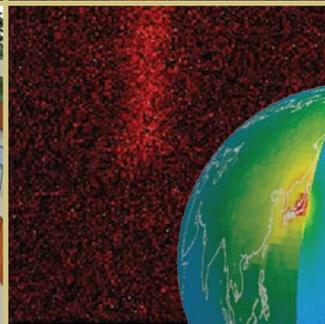
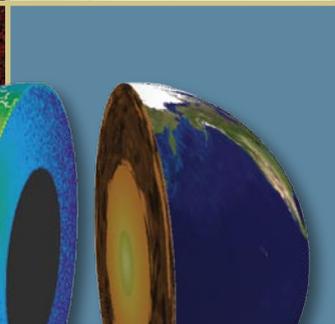
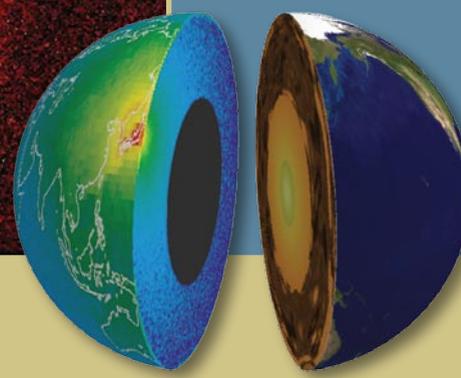

# The Frontiers of Nuclear Science
### A LONG RANGE PLAN

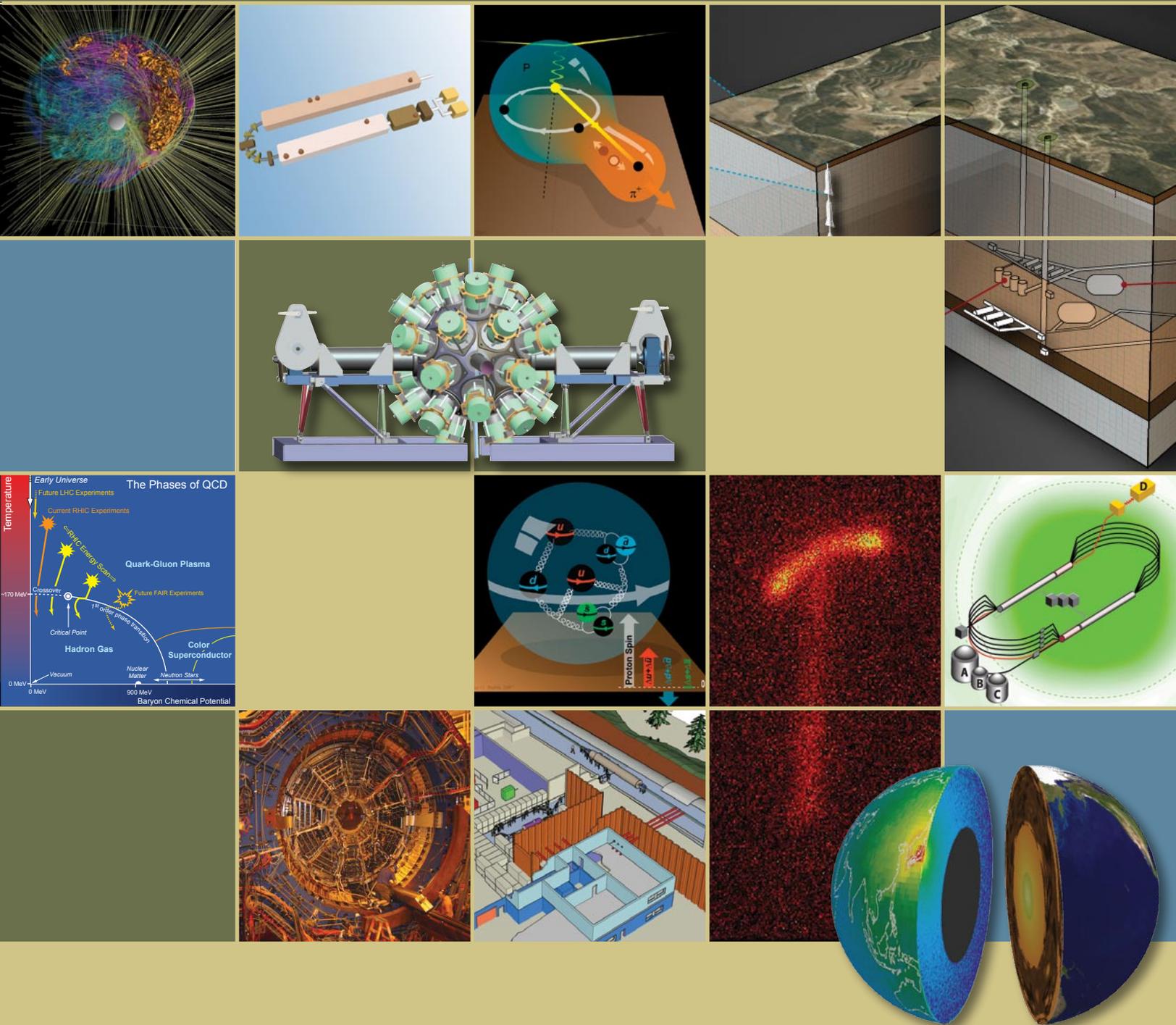

# The Frontiers of Nuclear Science
## A LONG RANGE PLAN

# Contents

# The Nuclear Science Advisory Committee

| | |
|---|---|
| Robert Tribble (Chair) | Texas A&M University |
| Douglas Bryman | TRIUMF, University of British Columbia |
| David Dean | Oak Ridge National Laboratory |
| Charlotte Elster | Ohio University |
| Rolf Ent | Thomas Jefferson National Accelerator Facility |
| Thomas Glasmacher | Michigan State University/National Superconducting Cyclotron Laboratory |
| Ulrich Heinz | The Ohio State University |
| Xiangdong Ji | University of Maryland |
| Roy Lacey | State University of New York – Stony Brook |
| I-Yang Lee | Lawrence Berkeley National Laboratory |
| Naomi Makins | University of Illinois |
| Richard Milner | Massachusetts Institute of Technology |
| Michael Ramsey-Musolf | University of Wisconsin |
| Heino Nitsche | University of California–Berkeley |
| Guy Savard | University of Chicago/Argonne National Laboratory |
| Susan Seestrom | Los Alamos National Laboratory |
| Thomas Ulrich | Brookhaven National Laboratory |
| Ubirajara van Kolck | University of Arizona |
| John Wilkerson | University of Washington |
| William Zajc | Columbia University |

# Preface

In a letter dated July 17, 2006, the Department of Energy's (DOE) Office of Science for Nuclear Physics and the National Science Foundation's (NSF) Mathematical and Physical Sciences Directorate charged the Nuclear Science Advisory Committee (NSAC) to "conduct a study of the opportunities and priorities for U.S. nuclear physics research and recommend a long range plan that will provide a framework for coordinated advancement of the nation's nuclear science research programs over the next decade." This request set in motion a bottom-up review and forward look by the nuclear science community. With input from this community-wide process, a 59 member working group, which included the present NSAC members, gathered at the beginning of May, 2007, to develop guidance on how to optimize the future research directions for the field based on the projected resources outlined in the charge letter from DOE and NSF. A new long range plan—*The Frontiers of Nuclear Science*—grew out of this meeting.

For the last decade, the top priority for nuclear science has been to utilize the flagship facilities that were built with investments by the nation in the 1980s and 1990s. Research with these facilities has led to many significant new discoveries that have changed our understanding of the world in which we live. But new discoveries demand new facilities, and the successes cannot continue indefinitely without new investment.

Based largely on the research of the past decade, the path that will lead to continued progress and future discoveries in nuclear science is now clear. Following the path will require the new and upgraded facilities and detectors outlined in this long range plan. The U.S. nuclear science community is committed to this course. Without new investments, future research in the field will be dominated by scientists in Europe and Asia. Already Japan has spent over $1.5 billion on new and upgraded nuclear science facilities in the last decade. Similarly Europe, led by Germany and France, has spent or committed close to $2 billion on major new projects. China is now building new nuclear science research facilities, and India promises to launch its own new construction projects soon. The world's investments in new and upgraded nuclear science research facilities will soon top $4 billion. During this same period, the United States has had no major new construction projects in the field.

By following through on the commitment to double the funding for physical science research over the next decade, the United States can provide the resources needed to maintain its competitive edge in nuclear science. New funds for the field would be devoted to much-needed facility and detector construction. Operations at existing facilities would continue at approximately the same level as today. The size of the permanent research staff would remain roughly constant. But growth in the number of graduate students would be accommodated to produce the future workforce for the country. Projections indicate that an increase of about 20% is needed over today's Ph.D. production in order to meet the demands to fill new positions in industry, medicine, and national laboratories and to fill replacement positions from retirements (particularly at universities and national labs). Furthermore, the investment into new and upgraded facilities and detectors promises to yield applications in many areas including energy, medicine, homeland security, and materials research.

The new plan presented here, *The Frontiers of Nuclear Science*, gives the guidance that, if followed, will guarantee that the United States continues to be a leader in nuclear science research into the foreseeable future.

# 1

**Overview and Recommendations**

## INTRODUCTION

The visible matter in the universe—the material that makes up stars, planets, and us—has existed for billions of years. But only recently have we had the knowledge and the tools necessary to begin to explain its origin, evolution, and structure. Nuclear scientists around the world have made tremendous strides toward this goal over the past few decades. And through that effort, we have developed applications that make our world safer and healthier. But the task is a great challenge—the strong force that creates matter behaves like no other force in nature. To fully understand it requires unique techniques and tools. The United States has long played a leading role in this quest, and with the new investments outlined in this Long Range Plan, we will continue to be a leader in the field well into the future.

Our evolving understanding of nuclear science has been closely intertwined with the multiple revolutions that transformed physics in the 20th century. Following the initial discovery of radioactive decay in 1896, a theoretical foundation for the field soon began to emerge with the introduction of quantum mechanics, together with Einstein's insight that mass and energy were truly interchangeable—a fact that would prove to have huge implications for electrical power generation and national security. In the middle decades of the century, the independent particle Shell Model provided new insights into nuclear structure. The observation of parity violation in beta decay gave key information that led ultimately to the electroweak Standard Model. The realization that nuclear reactions provide the energy that drives the cosmos revolutionized our understanding of the universe. And the discovery of point-like structure inside the proton and neutron confirmed that these particles are made of more fundamental building blocks—quarks and gluons—which in turn led to the development of a theory of the strong interaction called quantum chromodynamics (QCD).

Today, nuclear science is developing knowledge that could lead to new revolutions in the 21st century. Through strong recommendations by the Nuclear Science Advisory Committee (NSAC) in previous Long Range Plans, and support from the funding agencies, two flagship facilities were completed in the 1990s: the Continuous Electron Beam Accelerator Facility (CEBAF) at the Jefferson Laboratory in Newport News, Virginia, and the Relativistic Heavy Ion Collider at Brookhaven National Laboratory. The experiments performed at these two facilities have fundamentally changed our knowledge of QCD and how it defines nuclei as they exist today, as well as how it determines the properties of matter that existed in the hot-dense plasma soon after the Big Bang and in the dense cores of neutron stars. Elsewhere, nuclear scientists' recent measurements of neutrino properties have played a key role in verifying that neutrino oscillations occur—and hence, that neutrinos have mass. Together with the still unexplained predominance of matter over antimatter in the universe, this startling discovery provides our first look beyond the Standard Model. Nuclear physics is now poised to discover key ingredients of the New Standard Model through exquisitely sensitive neutrino studies and fundamental symmetry tests that will complement new particle searches at the Large Hadron Collider (LHC).

Theoretical advances using next-generation computing facilities, coupled with new experiments on rare isotopes at National Science Foundation's (NSF) flagship facility, the National Superconducting Cyclotron Laboratory at Michigan State University (upgraded in 2001), are advancing our understanding of nuclei to new levels—and opening up the real possibility of developing a truly predictive theory to describe nuclei and their interactions. Meanwhile, the past decade has produced major advances in our knowledge of the evolution of the universe and the origin of the elements. With these advances, we now know that nuclear reactions on unstable nuclei are extremely important in astrophysical environments. In some explosive situations, such as a supernova, they dominate. Nuclear scientists around the world are developing beams of unstable nuclei in part to allow us to better understand their role in stellar evolution. We are just beginning to tap the potential of this new technology in nuclear astrophysics and nuclear structure.

Building on the foundation of the recent past, nuclear science is focused on three broad but highly related research frontiers: (1) QCD and its implications and predictions for the state of matter in the early universe, quark confinement, the role of gluons, and the structure of the proton and neutron; (2) the structure of atomic nuclei and nuclear astrophysics, which addresses the origin of the elements, the structure and limits of nuclei, and the evolution of the cosmos; and (3) developing a New Standard Model of nature's fundamen-tal interactions, and understanding its implications for the origin of matter and the properties of neutrinos and nuclei.

Each nuclear science frontier is guided by overarching questions that serve to define it. These questions are listed here and expanded upon in the science sections of the Long Range Plan.

**Quantum Chromodynamics**
- **What are the phases of strongly interacting matter, and what roles do they play in the cosmos?**
- **What is the internal landscape of the nucleons?**
- **What does QCD predict for the properties of strongly interacting matter?**
- **What governs the transition of quarks and gluons into pions and nucleons?**
- **What is the role of gluons and gluon self-interactions in nucleons and nuclei?**
- **What determines the key features of QCD, and what is their relation to the nature of gravity and spacetime?**

**Nuclei and Nuclear Astrophysics**
- **What is the nature of the nuclear force that binds protons and neutrons into stable nuclei and rare isotopes?**
- **What is the origin of simple patterns in complex nuclei?**
- **What is the nature of neutron stars and dense nuclear matter?**
- **What is the origin of the elements in the cosmos?**
- **What are the nuclear reactions that drive stars and stellar explosions?**

**Fundamental Symmetries and Neutrinos**
- **What is the nature of the neutrinos, what are their masses, and how have they shaped the evolution of the universe?**
- **Why is there now more visible matter than antimatter in the universe?**
- **What are the unseen forces that were present at the dawn of the universe but disappeared from view as the universe evolved?**

## RECENT ACCOMPLISHMENTS

Since the last Long Range Plan was written, many significant new discoveries have been made in nuclear science. Only a small fraction of the exciting work from the past six years is discussed in the science sections that follow. Here we have reduced this to an even smaller set simply to provide a flavor for what nuclear science is about today.

### Near-Perfect Liquid

In one of the most surprising discoveries of the past few years, the hot matter that is formed when gold nuclei accelerated by the Relativistic Heavy Ion Collider (RHIC) collide seems to expand like an almost perfect liquid. A quantity used to characterize the resistance of liquids to flow—the ratio of the viscosity to entropy density—is smaller than for any other known liquid. It is so small, in fact, that it appears to be very close to a lower limit recently derived from the techniques of string theory, indicating that the quark-gluon plasma formed in the collisions at RHIC is a strongly coupled plasma and not a dilute gaseous plasma as originally expected. This finding opens a completely new area of physics—the study of extremely high-energy high-density plasmas whose microscopic and collective properties are dominated by quantum phenomena. One of the most important scientific challenges for the next decade is a quantitative exploration of this new state of matter.

### Spin of the Nucleon

Protons and neutrons—the nucleons at the heart of every atom—are complex, dynamic objects composed of quarks and gluons that are bound tightly together by the strong force. Each feature of the nucleon's rich internal structure that we have discovered has taught us much about QCD. Theory and experiment working in concert can teach us, for example, whether the quarks form clusters such as mesons within the nucleon, and how quarks of different flavor may be differently distributed in space and/or momentum. One property of intense interest is the intrinsic angular momentum—the "spin"—of the nucleon: it is precisely 1/2 (in units of $h/2\pi$), and if we are to claim any understanding of QCD, we must be able to identify how this value arises from the nucleon's internal structure. Recent measurements at Jefferson Lab (JLAB) and at DESY in Germany have shown that the up quarks have their spin aligned with that of their parent proton, while the down quarks' spins prefer the opposite direction, and the strange quarks have little polarization at all. Summing all flavors together, the quark spins account for only about 30% of the proton's spin. Recent, much-anticipated measurements at RHIC have narrowed down the contribution from the spin of the gluons, leaving the sharing of the nucleon spin still as an open puzzle. Its solution could lie within the grasp of future experiments at RHIC and at JLAB after its 12 GeV CEBAF Upgrade. However, with current technology we cannot access all of the regions where quarks and gluons reside in the nucleon. Should Nature have chosen to build the nucleon's spin with large contributions from regions inaccessible to current experiments, a complete understanding of the proton spin would require development of new tools to attack this fundamental problem.

**Jet Quenching**

QCD jets occur in all kinds of very-high-energy collisions, arising from the hard scattering of incoming quarks and gluons and their subsequent breakup into a characteristic spray of particles that are measured in a detector. Jets have been observed at RHIC in proton-proton, deuteron-gold, and gold-gold collisions. But a dramatic change occurs in the jet distributions from gold-gold collisions. When collisions are nearly head-on, the typical pattern of back-to-back jets, which are clearly seen in proton-proton and deuteron-gold collisions, disappears and only one jet is observed. This jet quenching indicates that the scattered quarks and gluons that would produce the second jet undergo large energy loss—corresponding to a density that is 100 times that of normal nuclei—as they traverse the matter formed in the collision. Recent results further suggest that the energy lost by the high-energy jets may appear as a collective "sonic boom." If this suggestion proves correct, it will allow determination of the speed of sound in this new matter. The effect of jet quenching may open a new way for us to probe the gluon density distribution of the matter—the quark-gluon plasma—formed in these violent collisions.

**Charge Distribution of the Neutron**

The neutron, as its name implies, is an electrically neutral particle. But the neutron has magnetic properties that are similar to its electrically charged counterpart, the proton, which suggests that its internal charge distribution is quite complex. A sustained effort worldwide over the last decade (including experiments at JLAB and the Bates Laboratory at the Massachusetts Institute of Technology) that utilized new polarized beams and targets has resulted in a much clearer picture of the neutron's charge distribution. The core of the neutron is positively charged. The neutron becomes electrically neutral due to the significant cloud of negative charge produced by virtual mesons that surround the core. These new results provide very strong constraints for theory—particularly lattice QCD calculations—that aim to reproduce the electric and magnetic properties of the neutron.

**Probing the Limits of Stability**

How many neutrons can we add to a stable nucleus before it cannot hold any more? Answering this question will provide critical information for theoretical predictions where experiments cannot reach, and it will yield experimental constraints for theories of fission that are relevant to future energy technologies. Today we know the answer to this question only for the lightest of the elements. We also have a good understanding of the converse question—how few neutrons can a nucleus hold and remain intact. As we push measurements to more neutron-rich isotopes, we learn new features about the strong interaction, which ultimately dictates where this limit, called the neutron drip line, occurs in nature. Recent measurements at the National Superconducting Cyclotron Laboratory (NSCL) of very neutron-rich isotopes of Al and Mg indicate that the drip line for these elements is likely further from the line of stability than previously thought. Mapping the neutron drip line will provide a wealth of information on how the strong interaction saturates. It is also vital in order for us to fully understand the origin of the elements. Very heavy elements that exist in nature are thought to have formed during stellar explosions in an environment where rapid neutron capture occurs on an element until it approaches the neutron drip line. Through a sequence of neutron captures and nuclear beta decays, the very heaviest elements, such as uranium, can be made. This r-process nucleosynthesis is one of the key astrophysical processes that we believe must occur—but so far, we have little nuclear data to constrain it.

**Stellar Reaction Rates**

In recent measurements, the $^{14}$N(p,γ)$^{15}$O reaction cross section that controls hydrogen burning in stars via the Carbon-Nitrogen-Oxygen (CNO) burning cycle has been measured to be about a factor of 2 lower than previously thought. This increases the derived ages for globular clusters and related limits for the age of the universe, by about a billion years. Meanwhile, pioneering measurements of reaction rates with the present generation of stable-beam and rare-isotope-beam facilities have employed new techniques to constrain a number of key reaction rates in nova explosions. This has reduced uncertainties in modeling the hot CNO and Neon-Sodium burning cycles and thus the production of long-lived radio-isotopes such as $^{18}$F and $^{22}$Na—nuclear decays that space-based gamma-ray telescopes look for in novae remnants. In a breakthrough for nuclear theory, *ab initio* nuclear calculations are achieving sufficient precision to be able to determine nuclear reaction rates and resonance lifetimes.

**Neutrino Oscillations and Neutrino Mass**

Just before the 2002 Long Range Plan was published, the SNO collaboration completed the first measurements of both charged- and neutral-current neutrino scattering for solar neutrinos. These results verified, beyond doubt, that electron neutrinos produced in the Sun were changing into a different type of neutrino, thus explaining the puzzling shortage of events seen in previous solar neutrino detectors and confirming the model predictions for solar energy generation. Similar oscillation effects had been found a few years earlier for atmospheric muon neutrinos. Now recent observations from KamLAND have shown that antineutrinos produced in a nuclear reactor also undergo oscillations. The fact that neutrino oscillations occur, and therefore that neutrinos have mass, has had an enormous impact on both nuclear and particle physics. Together with cosmological determinations of the matter and energy composition of the universe, these observations have provided the first direct evidence for physics beyond the Standard Model. In just a few years, the neutrino mass mixing matrix has been established, and the three mixing angles are now known or constrained. Nuclear scientists are now working to determine the absolute mass scale for neutrinos and to ascertain if neutrinos, unlike any other particle in nature, may be their own antiparticle. If so, neutrinos may hold the key to explaining the predominance of matter over antimatter in our universe.

**Precision Electroweak Studies**

Precise measurements of electroweak interactions of leptons and nuclei can uncover footprints of the fundamental symmetries of the New Standard Model. The E821 collaboration recently completed the world's most precise measurement of the muon anomalous magnetic moment at Brookhaven; the result deviates from the best Standard Model prediction now available. The E158 experiment at SLAC has carried out the first measurement of parity violation in electron-electron scattering, providing the most stringent test to date of the energy dependence of the weak mixing angle, one of the most important parameters of the Standard Model. Together with increasingly refined measurements of neutron and nuclear beta decay, these precision measurements have stimulated considerable theoretical progress. Nuclear theorists have substantially improved the precision of Standard Model predictions and completed extensive new calculations using supersymmetry and other candidates for the New Standard Model.

## PLANNING FOR THE FUTURE

As nuclear scientists have pushed the boundaries of technology with new accelerators, more powerful computers, and new detectors, we have made major discoveries about the nature of the strong and electroweak interactions. This new Long Range Plan focuses on the new science we must confront to further improve our understanding of these basic interactions, as well as on the tools needed to carry out this program and open up the path for even more discoveries.

Long-range planning by the NSAC began nearly 30 years ago. Today the process is community driven. It starts with Town Meetings that are organized by the Division of Nuclear Physics of the American Physical Society. In developing this Plan, four Town Meetings were held to cover the science in our three frontier areas (studies in QCD are so broad that two meetings, with joint and parallel sessions, were held at the same time and location to cover it), and a separate Town Meeting was convened to discuss education issues. In addition, sessions were organized at each Town Meeting to discuss education and applications of nuclear science. White Papers, which were written following the Town Meetings, served as input to the 59-member working group of nuclear scientists, which included NSAC members, who gathered in Galveston, Texas, from April 30 to May 4 to develop the recommendations for the Plan. The outcome of the Galveston meeting is a set of four recommendations that provide a roadmap for the development of new and existing facilities that, when implemented, will maintain U.S. leadership in this critical area of science.

> **RECOMMENDATION I**
>
> **We recommend completion of the 12 GeV CEBAF Upgrade at Jefferson Lab. The Upgrade will enable new insights into the structure of the nucleon, the transition between the hadronic and quark/gluon descriptions of nuclei, and the nature of confinement.**

A fundamental challenge for modern nuclear physics is to understand the structure and interactions of nucleons and nuclei in terms of QCD. Jefferson Lab's unique Continuous Electron Beam Accelerator Facility has given the United States leadership in addressing this challenge. Doubling the energy of the JLAB accelerator will enable three-dimensional imaging of the nucleon, revealing hidden aspects of its internal dynamics. It will complete our understanding of the transition between the hadronic and quark/gluon descriptions of nuclei, and test definitively the existence of exotic hadrons, long-predicted by QCD as arising from quark confinement. Through the use of parity violation, it will provide low-energy probes of physics beyond the Standard Model, complementing anticipated measurements at the highest accessible energy scales.

> **RECOMMENDATION II**
>
> **We recommend construction of the Facility for Rare Isotope Beams (FRIB), a world-leading facility for the study of nuclear structure, reactions, and astrophysics. Experiments with the new isotopes produced at FRIB will lead to a comprehensive description of nuclei, elucidate the origin of the elements in the cosmos, provide an understanding of matter in the crust of neutron stars, and establish the scientific foundation for innovative applications of nuclear science to society.**

We now have a roadmap to achieve the goal of a comprehensive and unified description of nuclei. Essential new data on exotic isotopes, which only FRIB will provide, will allow us to understand the nature of the forces that hold the nucleus together, to assess the validity of the theoretical approximations, and to delineate the path toward integrating nuclear structure with nuclear reactions. Advances in astrophysics and astronomy are driving the need for new and improved information on rare isotopes, including those near the very limits of nuclear stability that will be available for the first time with suitable rates at FRIB, to understand the chemical history of the universe and the synthesis of elements in stellar explosions. Rare isotopes are needed to test the fundamental symmetries of nature, and are essential for the many cross-disciplinary contributions they enable in basic sciences, national security, and societal applications. To launch the field into this new era requires the immediate construction of FRIB with its ability to produce ground-breaking research, and effective utilization of current user facilities, NSCL, HRIBF and ATLAS.

> **RECOMMENDATION III**
>
> **We recommend a targeted program of experiments to investigate neutrino properties and fundamental symmetries. These experiments aim to discover the nature of the neutrino, yet-unseen violations of time-reversal symmetry, and other key ingredients of the New Standard Model of fundamental interactions. Construction of a Deep Underground Science and Engineering Laboratory is vital to U.S. leadership in core aspects of this initiative.**

The discovery of flavor oscillations in solar, reactor, and atmospheric neutrino experiments—together with unexplained cosmological phenomena such as the dominance of matter over antimatter in the universe—calls for a New Standard Model of fundamental interactions. Nuclear physicists are poised to discover the symmetries of the New Standard Model through searches for neutrinoless double beta decay and electric dipole moments, determination of neutrino properties and interactions, and precise measurements of electroweak phenomena.

The Deep Underground Science and Engineering Laboratory (DUSEL) will provide the capability needed for ultra-low background measurements in this discovery-oriented program. Experiments also will exploit new capabilities at existing and planned nuclear physics facilities. Developing the New Standard Model using the breadth of new experimental results will require enhanced theoretical efforts.

> **RECOMMENDATION IV**
>
> **The experiments at the Relativistic Heavy Ion Collider have discovered a new state of matter at extreme temperature and density—a quark-gluon plasma that exhibits unexpected, almost perfect liquid dynamical behavior. We recommend implementation of the RHIC II luminosity upgrade, together with detector improvements, to determine the properties of this new state of matter.**

The major discoveries in the first five years at RHIC must be followed by a broad, quantitative study of the fundamental properties of the quark-gluon plasma. This can be accomplished through a 10-fold increase in collision rate, detector upgrades, and advances in theory. The RHIC II luminosity upgrade, using beam cooling, enables measurements using uniquely sensitive probes of the plasma such as energetic jets and rare bound states of heavy quarks. The detector upgrades make important new types of measurements possible while extending significantly the physics reach of the experiments. Achieving a quantitative understanding of the quark-gluon plasma also requires new investments in modeling of heavy-ion collisions, in analytic approaches, and in large-scale computing.

## FURTHER INTO THE FUTURE

Gluons and their interactions are critical to QCD. But their properties and dynamics in matter remain largely unexplored. Recent theoretical breakthroughs and experimental results suggest that both nucleons and nuclei, when viewed at high energies, appear as dense systems of gluons, creating fields whose intensity may be the strongest allowed in nature. The emerging science of this universal gluonic matter drives the development of a next-generation facility, the high-luminosity Electron-Ion Collider (EIC). The EIC's ability to collide high-energy electron beams with high-energy ion beams will provide access to those regions in the nucleon and nuclei where their structure is dominated by gluons. Moreover, polarized beams in the EIC will give unprecedented access to the spatial and spin structure of gluons in the proton.

An EIC with polarized beams has been embraced by the U.S. nuclear science community as embodying the vision for reaching the next QCD frontier. EIC would provide unique capabilities for the study of QCD well beyond those available at existing facilities worldwide and complementary to those planned for the next generation of accelerators in Europe and Asia. While significant progress has been made in developing concepts for an EIC, many open questions remain. Realization of an EIC will require advancements in accelerator science and technology, and detector research and development. The nuclear science community has recognized the importance of this future facility and makes the following recommendation.

> **We recommend the allocation of resources to develop accelerator and detector technology necessary to lay the foundation for a polarized Electron-Ion Collider. The EIC would explore the new QCD frontier of strong color fields in nuclei and precisely image the gluons in the proton.**

## INITIATIVES

As part of the planning process, a number of initiatives were discussed. Three of those merited special consideration by the working group and are considered to be very important to the future of the field.

### Nuclear Theory

Experimental and theoretical developments in nuclear science go hand in hand, sometimes led by experiments and other times by theory. A very strong theory program is critical to the success of our field. Recent successes in areas such as the internal structure and interactions of hadrons, the properties of hot and dense matter, the explosion mechanism of supernovae, the internal structure of nuclei, and the nuclear equation of state have come from the application of large-scale computing to problems in nuclear theory. We strongly endorse the ongoing programs at Department of Energy (DOE) and National Science Foundation (NSF) to provide resources for advanced scientific computing.

Beyond continued support for a healthy base program, **we recommend the funding of finite-duration, multi-institu-**

tional topical collaborations initiated through a competitive, peer-review process. In addition to focusing efforts on important scientific problems, these initiatives are intended to bring together the best in the field, leverage resources of smaller research groups, and provide expanded opportunities for the next generation of nuclear theorists.

### Accelerator Research and Development

Much of the experimental research in nuclear science is based at particle accelerator facilities. The future program is tied strongly to developments in accelerator technology. Accelerators also are used in many applications outside of nuclear science. We must support the development of new capabilities that will enable future discoveries. Advances in accelerator science and technology and the ability to attract students to this field will be strongly enhanced by **targeted support of proposal-driven accelerator Research and development supported by DOE and NSF nuclear physics.**

### Gamma-Ray Tracking

Gamma-ray detector arrays with high efficiency and high resolution have had a major impact on our understanding of nuclear structure over the past decade. But conventional arrays do not maintain their power when the particles undergoing decay move at high speeds. Now a new technology that allows for tracking of gamma rays promises to make the same high resolution, and even higher efficiency, possible for fast particles. A $1\pi$ detector array, GRETINA, is being built based on this new technology, and its first element has verified that the device will work as planned. The implementation of a $4\pi$ gamma-ray tracking detector, GRETA, will revolutionize gamma-ray spectroscopy and provide sensitivity improvements of several orders of magnitude. Thus **the construction of GRETA should begin upon successful completion of GRETINA. This gamma-ray energy tracking array will enable full exploitation of compelling science opportunities in nuclear structure, nuclear astrophysics, and weak interactions.**

## THE PROGRAM TODAY

Realization of the program that follows from implementing the principal recommendations of this Plan will take nearly a decade. During this time, we must maintain an active program by effectively utilizing our existing U.S. facilities and those available elsewhere in the world.

### Research and Operations

Progress in nuclear science depends critically on a healthy research program, and on the operation of the broad range of facilities needed to carry out that program. Research is carried out by individual faculty members, large university groups, and groups at national laboratories. Each of these components plays an important role in the research enterprise. In particular, the integration of research and teaching at universities provides a natural environment for the education and training of the future scientific workforce.

University-based accelerator facilities support a compelling and diverse portfolio of research in nuclear structure, nuclear reactions, nuclear astrophysics, and fundamental symmetries. The infrastructure at universities—either in support of operations and research at local accelerator facilities or participation in development of detectors, instrumentation, and equipment for collaborative research projects—enables university scientists to contribute significantly to new initiatives in the field. Support for the university facilities and research infrastructure is essential to enable them to remain significant partners in the nation's scientific endeavors.

Operation of the major user facilities is an indispensable component of the U.S. nuclear science program. Strategic investments in new and upgraded facilities have positioned the nation in a world leadership role in many areas of nuclear science. Exploiting these facilities provides extraordinary opportunities for scientific discoveries, preserves scientific output, and educates young scientists vital to meeting national needs. Specifically, it is critical: to sustain effective facility operations at CEBAF and RHIC in support of their world-leading research in QCD and at levels that allow equipment upgrades to proceed in a timely way; to optimally operate NSCL, taking advantage of its world-leading capabilities with fast rare-isotope beams; and to invest in the ATLAS and HRIBF user facilities with their unique low-energy heavy-ion and rare-isotope beams.

### International Collaboration

For many decades, nuclear scientists in the United States have sought the best facilities in the world to carry out their research. Some examples of this include measurements done by U.S. researchers studying QCD at high-energy physics facilities such as DESY in Germany, Fermilab, and SLAC; experiments carried out on nuclei far from stability at facilities including ISOLDE at CERN, GANIL in France, RIKEN in Japan, and GSI in Germany; experiments involving relativistic heavy-ion beams at the SPS at CERN; and the underground laboratories throughout the world that have housed neutrino detectors. Certainly the United States has benefited from the influx of researchers from abroad who have participated in, and often led, experiments at U.S. facilities.

As nuclear science facilities become larger and more complex, the costs of construction and operation escalate. It is not only natural but necessary that we maintain a global view of our field in order to maximize the potential for future discoveries. New facilities that will impact nuclear science are coming online in Europe and Japan. In Europe the LHC at CERN will soon be observing Pb-Pb collisions in a totally new energy regime. U.S. nuclear scientists already are involved in preparing several of the large detectors for these experiments. Construction has now begun at GSI on the Facility for Antiproton and Ion Research (FAIR). FAIR will open new possibilities for research in rare isotopes and in QCD. Also underway in Europe is an upgrade of the facilities at GANIL in France which will greatly extend the rare-isotope research capabilities there. In Japan, the Radioactive Ion Beam Factory (RIBF) at RIKEN has begun initial operations. As the beam power at RIBF is increased, it will become the premier facility for producing nuclei far from stability until FRIB is completed. Also in Japan, the JPARC facility, which features high-power, high-energy proton beams, is nearing completion. These new facilities, along with existing ones such as ISAC at TRIUMF, provide opportunities for nuclear scientists in the United States to carry out research abroad that is not possible with our existing facilities.

**Education**

Education and outreach are fundamental underpinnings that support the mandate of DOE and NSF to advance the broad interests of society and to help ensure U.S. competitiveness in the physical sciences and technology. The role of educating the next-generation workforce in nuclear science is crucial since the United States is facing a potentially serious shortage of qualified workers in pure and applied nuclear science research, nuclear medicine, nuclear energy, and national security. A key to increasing the number of new nuclear science Ph.D.s is to reach potential students while they are still undergraduates. Outreach by nuclear scientists also is vital. The public, and even scientists in other disciplines, are often uninformed or misinformed about nuclear science and its benefits. The nuclear science community should endeavor to increase the number and diversity of students who pursue a graduate degree in nuclear science and to effect a change in the understanding of the field by the public, through:
**(1) the enhancement of existing programs and the inception of new ones that address the goals of increasing the visibility of nuclear science in undergraduate education and the involvement of undergraduates in research; and
(2) the development and dissemination of materials and hands-on activities that demonstrate core nuclear science principles to a broad array of audiences.**

**Applications**

The world we live in today is a reflection of the enormous advances in technology of the past century. Nuclear science has contributed directly to key areas of this technological revolution, including energy production and medicine, and indirectly through training of a scientific workforce. New applications derived from basic research in nuclear science continue to be developed. As a recent example, nuclear scientists are directly involved in developing techniques based on basic nuclear science research for scanning cargo containers for dangerous materials as they enter our ports or pass our border stations. While we cannot say what new applications will be developed in the future, we can say, based on past history, that there will be many.

Through applications, nuclear science provides a return on the federal investment made to support the program of basic research. Recognizing this, we welcome closer ties between basic research and the applications of our trade.

## RESOURCES

NSAC was provided budget guidance by DOE and NSF in the charge letter that requested this new Long Range Plan. Specifically, the letter stated: "The projected funding for DOE is compatible with implementing the 12 GeV CEBAF Upgrade, and starting construction of a rare-isotope-beam facility that is less costly than the proposed Rare Isotope Accelerator (RIA) facility early in the next decade. At NSF the process has been put in place for developing a deep underground laboratory project and bringing this project forward in a funding decision."

RIA was the highest priority for new construction in the 2002 Long Range Plan. It was an ambitious project that, if built, would have given the United States clear leadership in studies of the physics of nuclei and nuclear astrophysics. It was tied for third for new construction in the DOE report *Facilities for the Future of Science: A Twenty-Year Outlook* published in 2003. But by early in 2006 it became clear that the cost of RIA, which was estimated to be about $1.1 billion in 2005, exceeded funding that would be available in future budgets. An NSAC subcommittee considered the options for a new facility for rare-isotope beams—FRIB—that could fit within future budget projections and in a recent report recommended that DOE and NSF proceed with a facility based on a 200 MeV/nucleon, 400 kW superconducting heavy-ion driver linac.

In assembling this Plan, the working group considered carefully the need to balance the requirements of the ongoing program with the development of new research tools. Previous Long Range Plans have recommended substantial increases in research funding in order to develop and maintain the infrastructure needed to build and operate new instrumentation. These increases have not occurred at DOE, but rather research has been funded at close to a constant effort for the past two decades. At NSF, funding for research in nuclear science has been close to a constant dollar level and consequently has eroded with inflation. Thus the need to increase research funding remains. When faced with a choice of improving research funding or developing our facilities, the consensus, as exemplified in the recommendations, was to maintain a near constant level of effort for the research program and facility operations, based on the FY2008 President's budget request, and to invest additional resources in the tools needed to make new discoveries in the future.

Implementing the four principal recommendations of this Plan can be accomplished with a funding profile consistent with doubling the DOE's Office of Nuclear Physics budget, in actual year dollars, over the next decade, together with NSF funding for DUSEL including some of the equipment for experiments to be carried out in DUSEL. Following a staged approach, construction funds for the upgrade of CEBAF at JLAB, Recommendation I, will begin in FY2009 and continue until FY2015. Work toward the construction of FRIB, Recommendation II, will begin in FY2009 with major construction funding occurring between FY2012 and FY2017. An NSF panel recently chose the Homestake mine as the site for DUSEL. Preparation of a full proposal for this new facility, which is part of Recommendation III, is the next step in making it a reality. The experiments in fundamental symmetries and neutrino physics that are part of Recommendation III will need both DOE and NSF support. Detector upgrades at RHIC, which are part of Recommendation IV, will proceed over several years, and the accelerator modifications needed to implement beam cooling, which will significantly increase the RHIC luminosity, will be carried out between 2012 and 2015. Following this course, nuclear science in the United States will continue to have world-leading facilities into the future.

As part of the charge to develop this Plan, NSAC was asked to provide information on "what the impacts are and priorities should be, if funding available provides constant level of effort (FY2007 President's Budget Request) in the out-years (FY2008–2017)." Since starting work on the Plan, the President's Budget Request for FY2008 has been submitted to Congress. For purposes of discussion, the FY2008 request level has been used for projection, at constant effort, into the out-years.

Constant effort funding falls far below the level needed to carry out the four recommendations in the Plan. The staged approach of upgrades and new facility construction that has been put forward already delays projects ready to be carried out sooner if funding were available. The U.S. nuclear science program will erode without significant new capital investments. At present, this need is most acute in research programs that require intense beams of rare isotopes—essential for advancing our understanding of both the physics of atomic nuclei and nuclear astrophysics. Maintaining a U.S. leadership position in this vital subfield requires the generation of significant new capabilities for rare-isotope beams on a timely basis. If budgets were restricted to constant effort, proceeding with any of the new initiatives presented in this Plan would be possible only by reduced funding for operations and research, with clear adverse and potentially dire consequences for core components of the U.S. nuclear physics program. Since nuclear science, like all areas of basic research, evolves in time, it is impossible now to forecast what strategy would minimize damage to the field if future budgets dictated such stark choices.

We have witnessed many major new discoveries in nuclear science over the last decade that were the direct result of the construction and operation of new facilities and detectors during the 1980s and 1990s. We also have seen a growing use of exciting new technologies developed in nuclear science both in well-established areas of application, such as medicine, and in important new areas such as homeland security. Continuing this growth and reaping the benefits it provides will require new investments. With these investments, the United States. will maintain its present world-leading position in nuclear science. And we will continue to contribute to the economic growth, health, and security of our nation.

# 2 The Science

# Quantum Chromodynamics: From the Structure of Hadrons to the Phases of Nuclear Matter

A fundamental quest of modern science is the exploration of matter in all its possible forms. Over the past century, as this quest has taken us further and further inward, scientists have discovered that all matter is composed of *atoms*; that each atom contains a tiny, ultra-dense core called the *nucleus*; that the nucleus is composed of particles called *protons* and *neutrons* (often referred to collectively as *nucleons*); and that these nucleons are members of a broader class of particles called *hadrons*, which are complex bound states of nearly massless *quarks* and massless *gluons*. The quest has even yielded a set of mathematical equations known as *quantum chromodynamics* (QCD), which gives us a theoretical framework for understanding how quarks, gluons, and all these hadrons behave.

Today, however, we stand at a new frontier: understanding precisely how the quarks and gluons are assembled to form the nucleons.

This problem is fundamental and unique in the history of science. First, the forces described by QCD are far stronger than the familiar forces of gravity, electricity, and magnetism. In everyday objects such as cars, tables, and hairbrushes, for example, the mass of the whole can be found very easily: just add up the masses of all the parts. This approach works as well, to very good approximation, for molecules, atoms, and nuclei. But that does not work at all for a proton or neutron, where the masses of the quarks and gluons account for only about 1% of the total mass. The rest arises from the fiercely strong energy of interaction among the particles, via Einstein's famous relation $E=mc^2$.

Second, the forces described by QCD are highly nonlinear. So unlike photons, which carry the electric and magnetic forces and which basically ignore one another, the gluons that carry the strong force can interact among themselves and spawn additional gluons. One bizarre result is that the strong forces behave in a way that is exactly the opposite of what happens with gravity, electricity, and magnetism: the force is weakest when the quarks and gluons are close together. (The 2004 Nobel Prize in Physics was awarded for the demonstration of this property, known as *asymptotic freedom*.) This result does allow the predictions of QCD to be rigorously tested in high-energy scattering experiments, where the particles do get very close together, and where standard calculation techniques appropriate to weak interactions can be applied. But it also means that the forces become stronger and stronger as the particle separations approach the size of a proton—at which point the gluons proliferate and begin to produce quark-antiquark pairs. Indeed, the forces become so strong that the quarks and gluons are effectively confined inside the nucleons and can never get out. A quantitative understanding of this confinement defies traditional mathematical approaches, even though we know the underlying theory. This poses an enormous theoretical challenge—to find new ways of solving the QCD equations in this regime—as well as a matching experimental challenge: to devise innovative methods to infer the properties of particles that can never be isolated from one another.

Another, closely related problem on the frontier of nuclear science is to understand what happens when nucleons "melt." QCD predicts that nuclear matter can change its state in somewhat the same way that ordinary water can change from ice to liquid to steam. This can happen when nucleons are compressed well beyond the density of atomic nuclei, as in the core of a neutron star, or when they are heated to the kind of extreme temperatures found in the early universe. In particular, QCD predicts that the universe spent the first 10 microseconds after the Big Bang as a *quark-gluon plasma* (QGP), with unbound quarks and gluons roaming freely. QCD calculations utilizing a technique known as *lattice gauge theory* suggest that the quark-gluon plasma condensed into ordinary hadrons only after the temperature fell below a critical value of about two trillion degrees Celsius.

The calculations also suggest that this transition was accompanied by a profound change in the nature of the "empty" space in which the quarks and gluons moved. To physicists, space is never truly empty—quantum mechanics dictates that even the vacuum is filled with quarks and gluons that appear for a fleeting moment and then disappear, giving it structure. Below the critical temperature, the "color" force that binds quarks is long ranged and permanently confines them into hadrons; above it, the force is screened, and quarks as well as gluons are unshackled. This change is accompanied by another drastic modification in the structure of the vacuum. Below the critical temperature, the vacuum forces a quark to constantly change the corkscrew orientation of its spin, known as chirality, from left to right and back. Or to put it another way, the chiral symmetry of the vacuum is "broken." But above that temperature the chiral symmetry of the vacuum is restored, and the chirality remains unaffected. General theoretical arguments dictate that only massless particles can have perfect chiral symmetry; thus, the breaking of this chiral symmetry by the "ordinary" vacuum we see below the critical temperature is intimately connected to the generation of mass of the ordinary hadrons around us.

Understanding the strong force and its implications for the behavior of matter is a major component of the enterprise of nuclear science. Six questions guide this study:

- **What are the phases of strongly interacting matter and what roles do they play in the cosmos?** As discussed above, QCD predicts that soon after the birth of the universe, a sea of quarks and gluons—the quark-gluon plasma—coalesced into protons and neutrons. Can we replicate that transition in the laboratory by creating a high-temperature, high-density environment that temporarily frees quarks from their normal confinement within protons and neutrons? Experiments to address this most fundamental question have produced tantalizing indications of just such a transition. Further studies will lead to an understanding of matter in the early universe and provide important clues on matter as it now exists in the interior of compact stars.
- **What is the internal landscape of the nucleons?** For many years, we have known that the nucleons are composite particles made up of quarks and gluons, and we have partial answers concerning the internal structure of protons and neutrons from years of measurements with high-energy probes. New experiments will provide an unprecedented, tomographic view of the quarks and their motion inside the nucleons, and map the distributions of quarks and gluons in space, momentum, type of quark, and spin orientation.
- **What does QCD predict for the properties of strongly interacting matter?** A critical step in the quest to understand strongly interacting matter is to confront the results of experiments with the quantitative implications of QCD. Doing so is exceedingly challenging because the strong force cannot be accurately described at the relevant scales by means of analytical calculations. Future progress will require extensive numerical simulations on a scale that has never before been undertaken.
- **What governs the transition of quarks and gluons into pions and nucleons?** Nucleons, and the pions that bind them together into atomic nuclei, must emerge from nearly massless quarks and gluons. The process that transforms deconfined matter into hadrons and nuclei remains poorly understood. Dedicated measurements are needed to launch a new stage in understanding both how quarks accrete partners from the vacuum or debris of high-energy collisions to form hadrons, and how the interaction among protons and neutrons arises from QCD.
- **What is the role of gluons and gluon self-interactions in nucleons and nuclei?** QCD predicts that gluons behave in ways unlike any other known particles. The self-interactions of gluons are, in fact, so strong that they confine quarks inside the proton and contribute substantially to its mass, and possibly its spin. The excitations of the confining glue may produce a whole new spectrum of "hybrid" particles that have yet to be seen. And a universal ensemble of densely packed gluons may exist in all strongly interacting particles, including nuclei, but is yet to be observed.
- **What determines the key features of QCD, and what is their relation to the nature of gravity and spacetime?** QCD provides the framework for understanding the detailed workings of a theory that combines strong interactions, relativity, and quantum mechanics. It is a rigorously defined theory in which questions have unique, although often hard to obtain, answers that can be compared to experiment. Understanding some of its features has been one of the richest sources of information used in building fundamental theories beyond QCD. Understanding them more deeply could elucidate recently uncovered intriguing deep connections between QCD and quantum theories of gravitation.

In the sections below, we outline recent progress in our efforts to answer these questions, and we provide a roadmap for how we plan to make major progress in this field over the next decade.

We expect to find many new and unpredicted phenomena as we move forward in this quest. From the study of normal, atomic matter we have learned that even a fundamentally simple force, like the electromagnetic interaction, can give rise to an enormous variety of different materials with many different phases, which often have surprising properties. The greater inherent complexity of the strong nuclear force, together with the special role of the vacuum, similarly promises a wealth of matter states built from quarks and gluons, some of them with entirely novel, previously unobserved properties.

# QCD and the Structure of Hadrons

## OVERVIEW

Understanding of the fundamental structure of matter requires an understanding of how quarks and gluons are assembled to form the *hadrons*: the family of strongly interacting particles that includes protons, neutrons, and other entities found in atomic nuclei, in compact stars, and in the early universe. Nuclear physicists can probe the arrangement of quarks and gluons inside the nucleons by accelerating electrons, hadrons, or nuclei to precisely controlled energies, smashing them into a target nucleus, and examining in exquisite detail the final products. The "low momentum transfer" particles—meaning those that are hardly affected in direction or energy by the scattering process—provide a kind of wide-angle, low-resolution image of the structure. This information allows the experimenters to map the static, overall properties of the proton and neutron, such as their shapes, sizes, and responses to externally applied forces—all of which can then be compared to predictions from QCD-inspired models, and to numerical solutions of QCD that rely on advanced computer simulations.

Meanwhile, a higher-resolution probe of the structure comes from particles that have scored a near-direct hit on something hard and granular inside the nucleon—a quark—and been deflected through a relatively large angle. Such hard-scattering events typically arise through electron-quark interactions or quark-antiquark annihilation processes. Either way, the events can be visualized by picturing the nucleon as a large and ever-changing number of constituents, or *partons*, having appropriate distributions of momentum or spin. Generalized parton distributions (GPD) provide a unified framework for connecting the gross properties encountered at low resolution and the partonic description applicable at high-momentum transfer.

The Jefferson Lab (JLAB) facility in Newport News, Virginia, which features the Continuous Electron Beam Accelerator Facility (CEBAF) and a full complement of experimental tools, is a world-leading laboratory for research into the fundamental building blocks of atomic nuclei. Augmenting JLAB are dedicated experiments at the Relativistic Heavy Ion Collider (RHIC) at Brookhaven National Laboratory, which uses polarized proton-proton collisions, as well as experiments at DESY, Mainz and CERN. By now, the use of sophisticated polarized beams, polarized targets, large-acceptance detector systems, and high duty-factor beams has become standard practice. Major conceptual and technical advances have allowed hadron structure studies to flourish in recent years, with understanding spurred by results from CEBAF and RHIC.

In this chapter, we outline the recent and expected progress in revealing the internal structure of hadrons.

## PROBING THE NUCLEONS: RECENT ACHIEVEMENTS

Some of the most important accomplishments since the 2002 Long Range Plan are:

- **Recent measurements further constrained the quark-gluon origin of the nucleon spin.** JLAB and DESY experiments have found that the up quarks have their spin parallel to the nucleon polarization, while the down quarks have their spin antiparallel—and the sea quarks have very little polarization at all. Experiments at RHIC point to a relatively small gluon polarization. These measurements indicate that the solution of the spin puzzle—how the various ingredients of nucleon structure contribute to its spin—still remains incomplete.

- **The charge distribution of the neutron was mapped precisely and with high resolution.** The measurements confirmed that the neutron has a positively charged core and a negatively charged pion cloud.

- **The era of precision predictions from numerical solutions of QCD—lattice QCD—in the quark confinement region was launched.** This has been facilitated by improvements to technical aspects of the calculations and establishment of dedicated multi-tera-flop computer installations, following up on the Large-Scale Computing Initiative outlined in the 2002 Long Range Plan.

- **Precision measurements of mirror symmetry (parity) violation in electron scattering set tight upper constraints on the contributions of strange quarks to the electric and magnetic properties of the proton.** These results provide one of the most precise comparisons of experiment with lattice QCD. The precision achieved in these measurements paves the way to a generation of experiments in search of the New Standard Model, with sensitivities comparable to those expected at the much higher energies of the LHC.

- **Pioneering measurements have produced initial constraints on generalized parton distributions within the nucleon.** The GPDs have been connected theoretically to six-dimensional space-momentum maps of the nucleon's internal structure, linking the existing spatial maps of proton structure at low energies and parton momentum distributions at high energies. Measurements at DESY and JLAB have demonstrated the experimental access to these GPDs. A roadmap was introduced to map the quark orbital angular momentum from the next generation of such measurements.
- **Sizable spin-orbit correlations were found to affect both the quark distributions within the proton and the process by which hadrons form from quarks.** Unexpectedly large left-right asymmetries were measured in meson production from transversely polarized protons in experiments at DESY and RHIC. This indicates large spin-orbit correlations of quarks within the proton and in the hadron-formation process. Vigorous theoretical work has led to a new framework for treating the long-neglected components of parton motion transverse to the proton's momentum.
- **Three-nucleon short-range correlations in nuclei were directly observed.** Clear evidence for two-nucleon and three-nucleon short-range correlations was observed in inelastic electron scattering from $^4$He, $^{12}$C, and $^{56}$Fe at JLAB. A fundamental tenet of nuclear physics is that these correlations dominate the short-distance structure of complex nuclei, but their short-range character has been hard to quantify.

## THE FUTURE: THE JLAB 12 GEV UPGRADE

Remarkable progress has been made in addressing both the theoretical and experimental challenges over the past decade, enabled in large part by the long-term U.S. investment in two premier accelerator facilities—CEBAF and RHIC—dedicated to the study of complementary aspects of QCD matter. In the immediate future, exciting results will continue to emerge from the continuation of the JLAB program with up to 6 GeV beam energies and the RHIC spin physics program. Still greater progress is fueled by planned upgrades of CEBAF at JLAB, RHIC, and by a future new Electron-Ion Collider (EIC) facility.

The highest recommendation of this Long Range Plan is the rapid completion of the energy doubling of JLAB, the 12 GeV CEBAF Upgrade. This will offer 100% duty-factor electron and tagged photon beams with high intensity, polarization, and broad kinematic reach, opening new opportunities for discovery in our field. The combination of these beams with greatly enhanced detector and electronics technology will improve data rates by orders of magnitude, providing critically needed data for the investigation of a broad range of QCD phenomena that simply have not been experimentally accessible. Exploiting these features will make possible the full exploration of the valence structure of nucleons and nuclei and promises the extraction of full "tomographic" images. Crucial questions concerning the existence and properties of hypothesized new states of matter, the exotic mesons, will be answered definitively, and these answers will guide our understanding of the underlying mechanism of quark confinement. It will be possible to probe nuclear structure at its most fundamental level, in terms of the underlying quarks and gluons. Lastly, using the exceptionally high-quality beams for which JLAB has become famous, one will seek the subtle signals of new physics beyond the current Standard Model of nuclear and particle physics.

In the long term and as the next exciting step, an electron-ion collider with a center-of-mass energy in the 30 to 100 GeV range and a luminosity of at least $10^{33}$/A cm$^{-2}$s$^{-1}$ is envisioned. This collider would complement the detailed explorations of the valence regime, which will be completed by the 12 GeV CEBAF Upgrade, by enabling precise and detailed studies of the nucleons and nuclei in the regime where their structure is dominated by gluons and sea quarks.

## THEORETICAL ADVANCES

Experimental progress is paralleled by major advances in QCD theory. Numerical calculations on a discrete four-dimensional spacetime lattice are the only known way to solve QCD rigorously in the strong coupling regime relevant to nuclear physics. However, the current formulation of lattice QCD cannot solve all hadron physics problems, even in principle. For some problems, approaches in which the physics at different distance scales are systematically separated and solved one scale at a time are more appropriate. For a large class of problems, such as hadron formation from quarks and the structure of hadron excited states, phenomenologi-cal methods and modeling are indispensable. Advances in all these directions have been significant in the past years to address the question: **what does QCD predict for the properties of strongly interacting matter**, or in particular, how can the distributions of quarks and gluons be treated theoretically and related to measurable properties of hadrons, and what are the detailed mechanisms by which mass is generated from nearly massless particles?

**Lattice QCD**

The combination of advances in lattice field theory, developments in computer technology, and the investment by DOE in terascale facilities for lattice QCD has given rise to dramatic progress toward first-principles calculations of hadronic physics. Lattice QCD calculations that include the full range of quantum fluctuations have become standard, and their results can be realistically compared with experimental data after theoretical extrapolation to physical quark masses.

With modern theoretical and numerical techniques, fundamental properties of the nucleon were calculated, such as the nucleon axial charge and the electromagnetic form factors, and were found to be in good agreement with experimental data. Increasingly realistic calculations for the excited states of nucleons and mesons give new hope for understanding the physical origin of hadron resonances. Hadron-hadron scattering is now being explored on the lattice through clever theoretical formulations, promising an exciting future to better understand the force that binds nuclei. These and many other quantities pertinent to the hadron structure, spectrum, and interactions have been explored with increasing levels of sophistication.

**Perturbative QCD and Factorization**

The asymptotic freedom of QCD makes it possible to use perturbation theory to treat interactions of quarks and gluons at short distances. In high-energy scattering, short-distance and long-distance physics may be separated to leading power in momentum transfer—an approach known as "factorization." For example, the cross sections for hard electron-proton or proton-proton collisions that transfer large momentum can be expressed as a product of a short-distance partonic (quark or gluon) cross section, calculable in perturbative QCD, and parton distribution functions that encode the long-distance information on the structure of the proton. Measurements of the cross sections, when combined with perturbative QCD calculations of the partonic hard-scattering cross sections, will therefore provide insight into the structure of the proton. Much of what we know today about the proton's substructure is based on this approach, which is a cornerstone of the ongoing and future programs at JLAB and RHIC.

Traditionally, the spatial view of nucleon structure provided by lattice QCD and the momentum-based view provided by the parton picture stood in stark contrast. Recent theoretical developments have clarified the connection between these two views, that of momentum and that of spatial coordinates, such that ultimately it should be possible to provide a complete "space-momentum" map of the proton's internal landscape. These maps, referred to as generalized parton distributions, describe how the spatial shape of a nucleon changes when probing differing ranges of quark momentum. Projected along one dimension, GPDs reproduce the form factors; along another dimension they provide a momentum distribution. With enough information about the correlation between space and momentum distributions, one can construct a full "tomographic" image of the proton. The weighted integrals, or moments, of the GPDs contain information about the forces acting on the quarks bound inside the nucleon. These moments can now be computed using lattice QCD methods. Not only do they have an attractive physical interpretation, but they can also be compared directly to unambiguous predictions from the underlying theory. Higher-order QCD corrections for many observables sensitive to GPDs have now also been calculated, strengthening the theoretical underpinning for extraction of GPDs.

**Effective Field Theory**

Effective field theories provide a powerful framework for solving physical problems that are characterized by a natural separation of distance scales. They are particularly important tools in QCD, where the relevant degrees of freedom are quarks and gluons at short distances and hadrons and nuclei at longer distances. Indeed, at energies below the proton mass, the most notable features of QCD are the confinement of quarks and the spontaneous breaking of QCD's chiral symmetry. Chiral perturbation theory is an effective field theory that incorporates both; when applied to mesons it is a mature theory. Perhaps the most striking advances in chiral effective field theory have come in its application to few-nucleon systems. This has yielded precise results for nucleon-nucleon forces and also produced consistent three-nucleon forces. This opens the way for precision analyses of electromagnetic reactions on light nuclei, e.g., the Compton scattering reactions on systems having two or three nucleons that will be explored at the High-Intensity Gamma-Ray Source (HIγS) facility at Duke University.

**The Roles of Phenomenology and Model Building**

Models have historically played an important role in the development of nuclear and hadronic physics as stepping stones to more fundamental theories. The spin-orbit interaction discovered by Goeppert-Mayer was the underpinning for the nuclear shell model. The Veneziano model for hadron-hadron scattering amplitudes spearheaded development of string theory. The nonrelativistic quark model and the bag model embody salient qualitative features of hadron spectra and structure. Model building is a valuable tool for developing intuition about complex systems and for connecting results of *ab initio* computations and experimental data. Figure 2.1 shows the interplay between models and lattice QCD, and confirms that almost all of the constituent quark mass in a proton arises from the gluons that propagate with the quark.

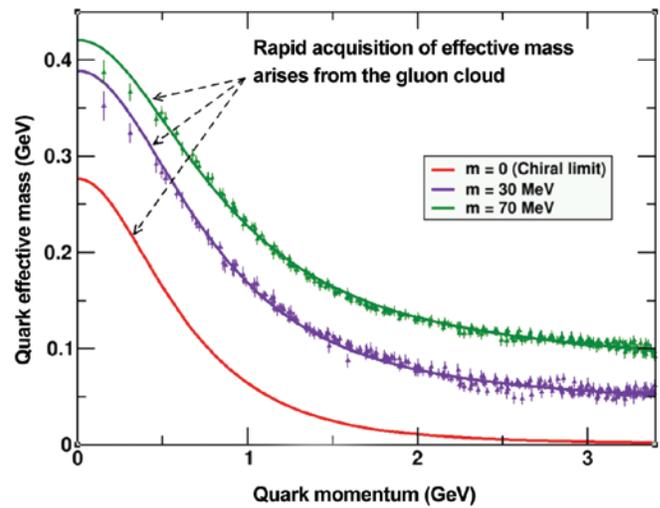

**Figure 2.1:** Mass from nothing. In QCD a quark's effective mass depends on its momentum. The function describing this can be calculated and is depicted here. Numerical simulations of lattice QCD (data, at two different bare masses) have confirmed model predictions (solid curves) that the vast bulk of the constituent mass of a light quark comes from a cloud of gluons that are dragged along by the quark as it propagates. In this way, a quark that appears to be absolutely massless at high energies (m = 0, red curve) acquires a large constituent mass at low energies.

# From Lattice QCD to Hadronic Structure and Hot Nuclear Matter

**Structure from QCD**

How does the observed structure of nucleons and other hadrons arise from QCD? How does the spin of the nucleon arise from its constituent quarks and gluons? What role do the strange quarks play in the nucleon's structure? What is the spectrum of excited states of the nucleon predicted by QCD? What is the nature of the nucleon-nucleon and hyperon-nucleon interaction? What happens to nuclear matter when you compress and heat it? These questions go to the heart of our understanding of the strong interaction and can now be addressed numerically by lattice QCD.

**Quantum Hopscotch**

QCD is the accepted theory of the strong interactions, describing matter in terms of quarks and gluons. Unfortunately, while the QCD equations are simple to write down, they are very hard to solve in any purely mathematical way, largely because the quarks and gluons interact strongly with each other. Lattice QCD is a technique to solve QCD numerically instead. The idea is to replace continuous spacetime with a grid, or lattice, of points, and to model the motion of the quarks and gluons as a series of hops from one lattice point to the next. In lattice QCD one can vary parameters such as the quark masses, or the number of quark colors, and see how hadrons respond—"experiments" not possible in the real world.

**Back to the Real World**

By making the lattice fine enough and by adjusting quark masses to their physical values, physicists can learn about hadrons in the real world of smooth and continuous spacetime. They can then test the accuracy of these calculations by comparing lattice results with experimental data on, say, the charge distribution in the nucleon (see page 20, bottom left), or the contribution of quarks to the nucleon spin. Scientists can also make predictions for hadronic properties not readily accessible to experiment.

## Putting the Heat On

By using asymmetric lattices, theorists can simulate systems of quarks and gluons at nonzero temperature. In its early days, for example, lattice QCD predicted the existence of the quark-gluon plasma, which would announce its existence through a dramatic jump in the energy density of QCD matter at a "critical" temperature of about 170 MeV (~2×10$^{12}$ K; see bottom, right). This prediction motivated the start of an experimental heavy-ion program to discover this form of matter. Advances in lattice QCD techniques and computing power have since enabled increasingly accurate determinations of the transition temperature, as well as the quark-gluon plasma equation of state and a variety of observables that provide a detailed microscopic picture of the matter created in heavy-ion collisions. A recent breakthrough has been the development of algorithms to simulate the properties of matter with nonzero net baryon number density.

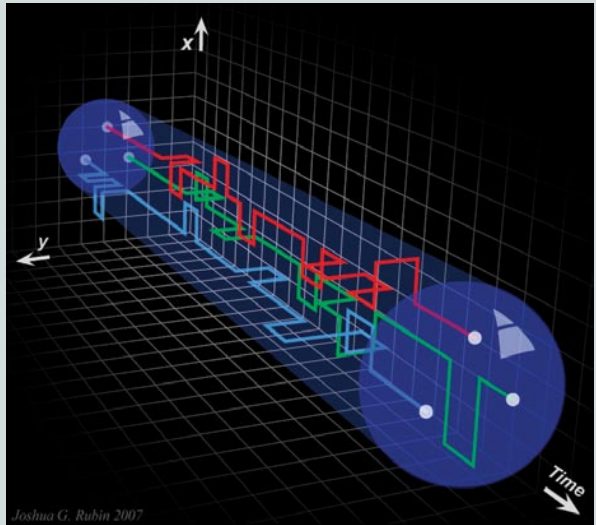

Illustration of the lattice QCD approach to calculating nucleon properties. Using next-generation computing facilities, theorists will be able to calculate the nucleon's internal quark substructure using a grid that is fine enough to accurately simulate our world's spacetime continuum.

## Physicists Innovate Supercomputer Technology

Reliable predictions of nucleon structure and the properties of hot and dense QCD matter require calculations with physical (that is, small) quark masses and with large, fine-grained lattices. This can only be done on supercomputers in the 10–100 Teraflops class, capable of executing tens of trillions of arithmetic operations per second. To achieve this goal, DOE has supported supercomputers dedicated to lattice QCD calculations, including both large computer clusters and an innovative computer, the QCDOC machine, designed and built by physicists for QCD. The unique and innovative technical solutions to key design problems in the QCDOC, developed in close collaboration with scientists at IBM, turned out to be extremely useful also for more general applications. They were adopted in the development of the world's currently most powerful commercially available supercomputer, the Blue Gene/L, by IBM. Advances in lattice QCD have thus helped to re-establish U.S. leadership in the critically important field of high-capability computing and thereby enhance our international competitiveness.

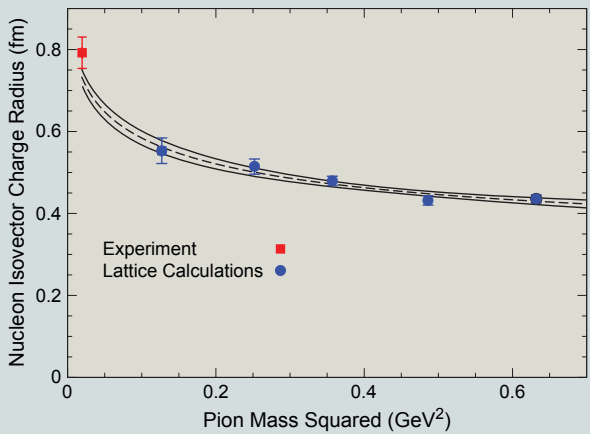

The charge radius of an isovector (proton minus neutron) nucleon, computed in lattice QCD for a variety of pion masses. The curves represent the theoretical extrapolation of the lattice data to the real pion mass and agree well with the measured experimental data.

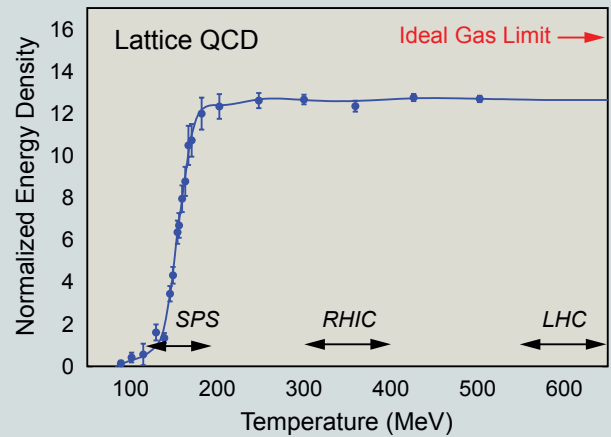

The deconfinement phase transition in QCD matter. The graph shows the energy density from lattice QCD as a function of temperature, normalized to the fourth power of the temperature in order to exhibit the rapid change of the effective number of massless degrees of freedom during the phase transition. Also indicated are the temperature ranges explored by heavy-ion experiments at the SPS, RHIC and LHC.

Recent advances in the phenomenology of hadron dynamics have been stimulated both by lattice QCD computations of hadron properties and quark confinement, and by the efforts to understand the high-quality data emerging from JLAB and RHIC. The interplay between lattice calculations and model studies has resulted in a description of the confining force between color charges, with force lines representing a gluon field contained within a narrow, tubular region.

Excitations of the gluon fields lead to a predicted spectrum of hybrid mesons with exotic quantum numbers—that is, hadrons having properties that cannot be found in ordinary mesons. Their experimental discovery is an important goal of the 12 GeV JLAB physics program.

A new phenomenology has emerged as a result of the correspondence between conformal theories and string theory. The correspondence provides new classes of models that relate QCD-like observables to string observables in weak-coupling gravity. In addition to its intriguing applications to the matter created in ultra-relativistic heavy-ion collisions, this mapping was utilized to reproduce high-energy scaling of parton scattering amplitudes and to make encouraging predictions for meson wave functions and the resulting light meson spectrum.

## THE DISTRIBUTION OF QUARKS AND GLUONS INSIDE NUCLEONS

A key effort in nuclear science is to infer the properties of quarks and gluons that can never be isolated from one another. Over the last decade, major progress has been made to address the question: **what is the internal landscape of the nucleons?** Recent maps of distributions of quarks and gluons inside protons and neutrons in momentum, quark type, spin orientation, and space have been revealing, but puzzles remain. Pioneering measurements have illustrated the potential to link the maps in momentum and space. These efforts will be addressed in turn below.

### The Valence Quark Structure of Protons and Neutrons

Some three decades after the inception of QCD as the accepted theory for strong interactions, major challenges and mysteries remain. At high energies, the property of QCD known as asymptotic freedom, which causes quarks to interact very weakly at short distances, allows for an efficient description of observables in terms of quarks and gluons, or partons. In contrast, at low energies the effects of confinement and spontaneous chiral-symmetry breaking imply a more efficient description in terms of mesons and baryons, or hadrons. Despite this apparent dichotomy, scientists have observed a striking similarity between data measured at high and low energies. This is referred to as "quark-hadron duality." Data accumulated at JLAB have shown that quark-hadron duality occurs at much lower values of momentum transfer, in more observables, and in far less limited regions of energy than hitherto believed. These data provide vital clues to the long-standing challenge of QCD to describe the forces at large distances, comparable to the size of hadrons (~1 fm).

In deep-inelastic scattering (DIS) a high-energy electron or muon is directed at a target (a proton, for example) and scatters from it via the exchange of a highly virtual photon. This photon serves as a surgical laser beam, one of such precision that it strikes a single quark within the target. One of the important findings from JLAB over the last five years is that the quark parton model, developed to describe high-energy (> 20 GeV) deep-inelastic electron scattering, is remarkably successful in describing scattering data at incident energies of order 5 GeV. With the 12 GeV CEBAF Upgrade, a rich field of investigation then becomes possible in the valence quark region. The three valence *up* and *down* quarks of the nucleon comprise the backbone of nucleons on which the gluons and quark-antiquark pairs are formed. The distribution of the high-momentum valence quarks in the proton and neutron remains quite uncertain but is essential to our understanding of nucleon structure. It is especially important to determine the momentum carried by the valence down quarks as well as the distribution of spin carried by both the up and down quarks in the valence region.

Experiments following the 12 GeV CEBAF Upgrade will indeed define the spin and flavor dependence of the valence quark distributions with high precision. For example, measurement of the unpolarized cross sections will determine the ratio of *down* to *up* quarks, $d(x)/u(x)$, as a function of quark momentum fraction $x$, as shown in figure 2.2. Measurements of the inclusive spin asymmetry for DIS from high-momentum valence quarks in the proton and the neutron will provide a precise determination of the polarized valence parton distributions $\Delta u$ and $\Delta d$. It is equally important to measure the ratio of antiquark distributions, $\bar{d}(x)/\bar{u}(x)$, which is extremely sensitive to the origin of the light quark sea. Fermilab experiment E906 is key to accomplishing this goal. Using proton beams, it will provide uniquely clean access to the antiquark distributions and measure $\bar{d}(x)/\bar{u}(x)$ at high $x$ values where we have no information at all about this ratio. Together with advances in lattice QCD, these measurements will allow nuclear physicists to reach a comprehensive understanding of the distribution of quarks in the valence region of the nucleon.

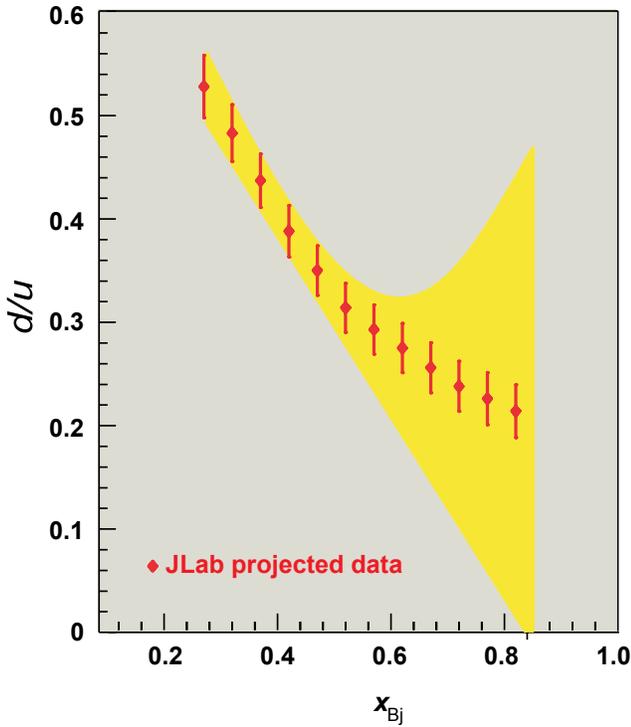

**Figure 2.2:** Projected measurement of the ratio of *d*- and *u*-quark momentum distributions, *d(x)/u(x)*, at large quark momentum fraction *x*, made possible by the 12 GeV CEBAF Upgrade. The shaded band represents the uncertainty in existing measurements arising from the motion and binding of nucleons in a deuterium nucleus.

### The Spin Structure of Protons and Neutrons

One of the great early successes of the quark model was its apparently simple account of the intrinsic magnetism of the proton as arising from the spin alignment of three point-like valence or "constituent" quarks. This naïve view is in stark contrast with the results of two decades of subsequent measurements of the deep-inelastic scattering of spin-polarized electron or muon beams from charged partons inside a spin-polarized proton. These experiments demonstrate clearly that the spins of all the quarks and antiquarks combine to account for only about 30% of the proton's overall spin. The quest for the origin of the missing spin—in this partonic view, it can arise in principle from spin alignment of gluons, or from orbital motion of quarks and/or gluons—is ongoing and involves experiments over a broad range of energies with both lepton and hadron beams. Solving the puzzle of the proton's missing spin is essential to understanding how the constituent quarks of the naïve quark model are related to the actual quarks and gluons probed in high-energy experiments. Several important milestones have been achieved since the 2002 Long Range Plan, but final resolution of the puzzle will require quite a bit more data and even a new accelerator facility. Along the path to a solution, great advances have been made in delineating the distinct information about the nucleon structure and the dynamics of hadron formation provided by probing quark spin preferences both along and transverse to the proton's motion.

**Quark and Antiquark Helicity Preferences.** There have been two significant recent achievements in clarifying how quark and antiquark preferences for spin orientation in the direction of the nucleon's motion compare with that of the parent nucleon itself. At JLAB, DIS from a polarized ³He target has improved measurements of valence quark contributions to the neutron's spin structure by an order of magnitude. The results (see figure 2.3) imply that the up (down) quark's preference to spin opposite to the spin of its parent neutron (proton) persists to higher momentum fractions than originally anticipated. The nucleon structure models that best reproduce these new results attribute an appreciable portion of the nucleon spin to orbital motion of valence quarks, neglected in the naïve quark model. These models will be tested more extensively in similar experiments that require the 12 GeV CEBAF Upgrade.

In the HERMES experiment at DESY the coincident production of pions and kaons in polarized deep-inelastic scattering has suggested that *strange* quarks in the nucleon's "sea" contribute little to the spin of the nucleon. The same data also provide the first information on the *differences* in helicity preferences between up and down antiquarks in the sea. However, these data are not yet sufficiently precise to provide a stringent test of structure models that predict significant differences. More definitive measurements of these flavor-dependent polarization differences in the sea should be provided by future measurements at RHIC of intermediate vector boson production in polarized proton-proton collisions.

**Gluon Helicity Preferences.** Since the last Long Range Plan, the era of direct measurements of gluon spin preferences has been launched. Because the gluons carry no electrical charge, reactions directly sensitive to QCD interactions are required. This is accomplished by making use of high-energy polarized proton collisions at RHIC, and studying the production of jets of hadrons (signaling a scattered quark or gluon), individual pions, or high-energy photons. It has been demonstrated that the measured probabilities for each of these processes to occur is quantitatively understood with perturbative QCD, thus providing a robust theoretical basis for extracting gluon polarizations from measured spin effects. Preliminary results from the most recent measurements in 2006, shown in figure 2.4, clearly rule out previous theoretical speculations that the small net quark polarization may have accompanied a gluon contribution exceeding 100% of the proton spin. However, these data still allow a broad range of possibilities. The CERN experiment COMPASS supports these RHIC spin results using a different process (muon-induced production of hadron pairs) to access the gluon helicity. Considerably improved constraints are anticipated with future RHIC measurements at high collision energies involving coincidences between pairs of jets or photons and jets. It is conceivable that an appreciable fraction of the nucleon spin resides in weakly polarized, but very highly abundant, gluons that carry less than 1% of the nucleon momentum. Such soft gluons are beyond the kinematic reach of RHIC, so that testing this possibility awaits a future Electron-Ion Collider.

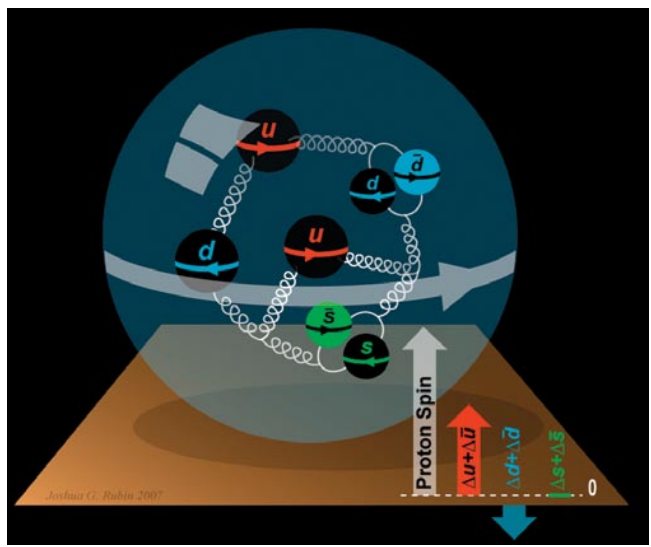

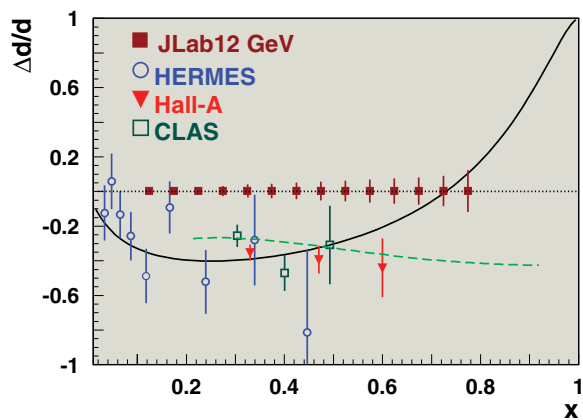

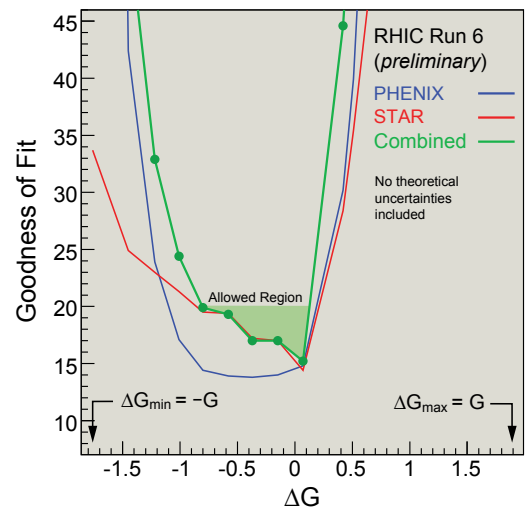

**Figure 2.3:** The top panel illustrates the preferred spin orientations of the up, down, and strange quarks and antiquarks within a polarized proton. The bottom panel shows a sample of the data from which we have gained this insight: the fractional polarization of down quarks in the proton is negative (i.e., pointing against the proton's spin direction) at each value of the quark's momentum fraction $x$ at which measurements have been made. The curves correspond to different model expectations. The solid squares indicate the projected precision of the measurements enabled by the 12 GeV CEBAF Upgrade and JLAB. It is widely expected that the down-quark polarization will change sign as $x$ reaches unity; the 12-GeV data will tell us whether our models are correct or not.

**Figure 2.4:** Constraints on the gluon contribution to the proton's spin from data collected in 200-GeV polarized proton collisions at RHIC in 2006. The curves show the quality of QCD fit as a function of gluon spin contribution to the proton (in units of spin where the proton's spin is 1/2) based on results from neutral π production from PHENIX (blue curve), jet production from STAR (red curve), and combined data sets (green curve). The analyses are done for one model of the gluon polarization as a function of momentum fraction. Uncertainties from the theoretical analysis are not included. Assuming that the model is correct in the regions of momentum fractions outside the range covered by the present data, the net gluon spin contribution seems to lie between small positive and sizable negative (opposite proton spin) values. Future RHIC spin measurements are expected to improve the constraints further.

**Spin Sum Rules.** On general grounds, sum rules connect the static properties of a system to a weighted sum of its dynamical excitation spectrum. The derivation of sum rules is often model independent and serves as a powerful tool to investigate the underlying theory. In the case of the proton or neutron spin, a well-known example is the Bjorken sum rule, which at very short distance scale (or large momentum transfer scale) connects the first moment of the difference between the proton and neutron spin structure to the nucleon axial vector coupling constant as measured in neutron beta decay. The Bjorken sum rule offered one of the cleanest tests of QCD, and its verification was the highlight of the deep-inelastic scattering experimental programs at CERN and SLAC in the previous decade. At very long distance scale (or zero momentum transfer), this sum rule is replaced by the more recently measured Gerasimov-Drell-Hearn sum rule, where the first moment is related to the difference between the anomalous magnetic moments of the proton and the neutron.

More recently, the precision results from JLAB on the proton and the neutron over a wide range of momentum transfers provided, for the first time, a comprehensive landscape of the nucleon (longitudinal) spin structure. The data show a smooth transition from intermediate to short distance scales, with a transition to values consistent with perturbative QCD at a much larger distance scale than expected. At the largest distance scales, chiral perturbation theory makes predictions for the spin sum rules that are being tested with these new data. The first moment of the nucleon's *transverse* spin structure was predicted to be zero for all distance scales, which is consistent with the data within the experimental uncertainties. Higher moments will allow access to the nucleon's "color polarizabilities" at large-distance scales and spin polarizabilities at small-distance scales. These intrinsic nucleon properties describe the ability of the nucleon's constituents to generate a color magnetic field along the direction of its spin, or to resist a change of motion under an external electromagnetic field. They will be measured with two newly planned experiments at JLAB and be computed using lattice QCD techniques.

**Quark Transverse Spin Preferences and Orbital Angular Momentum.** Gluons, like the photons in a laser beam, can have their intrinsic spin pointing along or opposite but not transverse to their motion. Gluon spin alignment thus cannot contribute to the transverse spin of a moving proton.

## The Glue at the Heart of Matter

### Gluons are Majority "Silent Partners" in Ordinary Matter

Ironically enough, much of the mass of the visible universe comes from particles that are not only massless themselves, but can never even be observed in isolation. These "gluons" play an apparently selfless role as the "glue" that binds quarks together inside protons, neutrons, and all atomic nuclei. But they have the unusual property of interacting strongly with themselves—a fact that gives rise to the many unique features of quantum chromodynamics, the fundamental field theory for quark and gluon interactions. In particular, the gluon's self-interactions cause them to become overwhelmingly abundant inside protons (see below), with a cumulative energy that dominates the proton's total mass and energy.

### What QCD Giveth, QCD Taketh Away

This gluon abundance shows up in the experimental results quite consistently, growing rapidly as protons are probed with increasing spatial resolution and with increasing sensitivity to gluons carrying smaller fractions of proton energy. As their spatial density increases, moreover, the probability that two gluons recombine into one becomes comparable to that for one gluon to split in two. This sug-

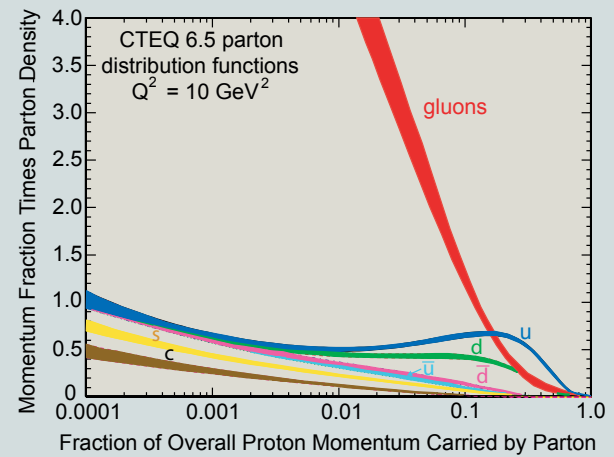

The number density of gluons and of various types of quarks and antiquarks inside the proton, as a function of momentum fraction carried by the quark or gluon (parton). The curves are from fits to high-energy scattering data. The width of each band represents the uncertainty.

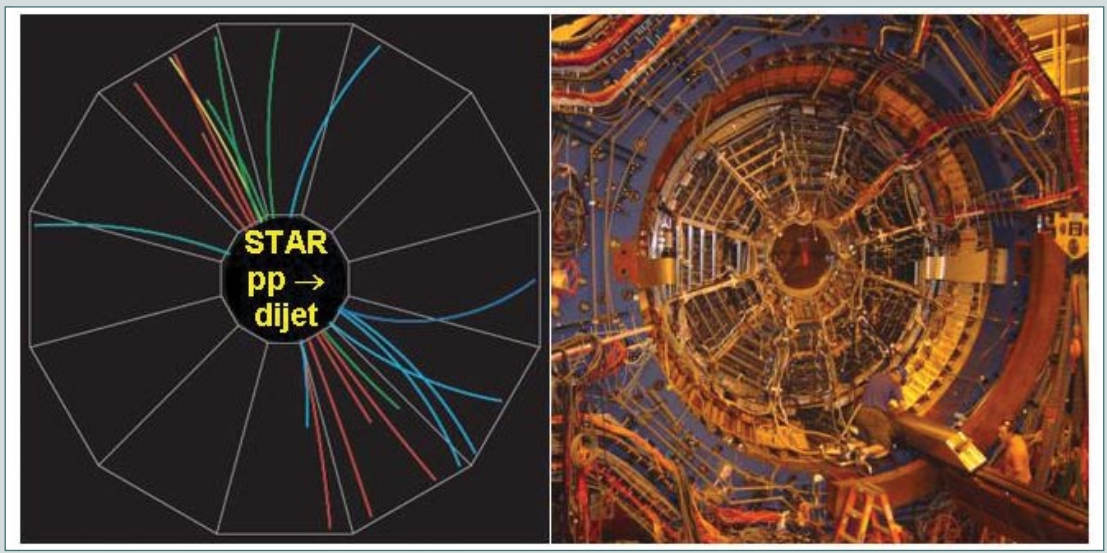

End views of a RHIC proton-proton collision event display (left) containing charged-particle tracks (colored curves) reconstructed from the STAR detector (right). The back-to-back "jets" of particles (with positive and negative charges curving oppositely in the magnetic field of STAR) signal an event in which a quark or gluon from one spin-polarized proton undergoes a simple hard scattering from a quark or gluon provided by the other proton beam. Such events are used to measure the contribution of gluons to the proton's spin.

gests that the competition between these two processes will eventually lead to a saturation of the gluon density—a prediction that suggests in turn that protons, neutrons, and all atomic nuclei should have the same appearance when viewed at sufficiently high energy. In effect, they would each become a glob of glue behaving as a single entity with universal properties. Discovering this universal gluonic matter is a major long-term goal of nuclear physics. Achieving that goal will require a future Electron-Ion Collider. In the meantime, however, our existing QCD laboratories have begun to clarify two other manifestations of gluon physics: their contribution to the proton's spin and their role in excitations of matter.

**The Glue in the Proton's Spin**

Quarks, gluons, and protons all have intrinsic spin, which is a kind of internal angular momentum. For example, the proton's spin is the source of its inherent magnetism, which is exploited in MRI imaging. But where does the proton's spin come from? Presumably it is just the sum of all the contributions from all of its quark and gluon constituents. Yet recent experiments show that no more than 30% of the total can come from spins of the quarks and antiquarks alone. So how much of the rest comes from the gluons? Since the last Long Range Plan, several experiments have been launched to help answer that question. At RHIC, this quest utilizes multi-tasking detectors built to analyze the daunting debris when two ultra-relativistic heavy nuclei collide. The gluon spin studies focus on such simple events as in the figure above, where two proton beams with controlled spin orientations produce high-energy quarks or gluons. In complementary studies at HERA and CERN, electron or muon beams yield photons that may fuse with gluons in a spin-polarized proton target. Results over the next few years may reveal how much of the missing spin is quantitatively accounted for by gluons.

**Gluon Vibrations**

The profound role of gluons remains hidden in ordinary matter because the quantum numbers of the particles found there are not altered by the presence of glue. But the exposure can be improved in certain "exotic" mesons, where QCD predicts that gluon "vibrations" built on quark-antiquark ($q - \bar{q}$) states lead to otherwise impossible quantum number combinations. These exotic particles should be excited most cleanly by beams of photons, which couple to mesons with parallel $q$ and $\bar{q}$ spins. Their discovery is a prime motivation for the new experimental Hall D planned as part of the 12 GeV CEBAF Upgrade.

But quarks can have a transverse spin preference, denoted as *transversity*. Because of effects of relativity, transversity's relation to the nucleon's transverse spin orientation differs from the corresponding relationship for spin components along its motion. Quark transversity measures a distinct property of nucleon structure—associated with the breaking of QCD's fundamental chiral symmetry—from that probed by helicity preferences. The first measurement of quark transversity has recently been made by the HERMES experiment, exploiting a spin sensitivity in the formation of hadrons from scattered quarks discovered in electron-positron collisions by nuclear scientists in the BELLE Collaboration at KEK in Japan.

Fueled by new experiments and dramatic recent advances in theory, the entire subject of transverse spin sensitivities in QCD interactions has undergone a worldwide renaissance. In contrast to decades-old expectations, sizable sensitivity to the transverse spin orientation of a proton has been observed in both deep-inelastic scattering experiments with hadron coincidences at HERMES and in hadron production in polarized proton-proton collisions at RHIC. The latter echoed an earlier result from Fermilab at lower energies, where perturbative QCD was not thought to be applicable. At HERMES, but not yet definitively at RHIC, measurements have disentangled the contributions due to quark transverse spin preferences and transverse motion preferences within a transversely polarized proton. The motional preferences are intriguing because they require spin-orbit correlations within the nucleon's wave function, and may thereby illuminate the original spin puzzle. Attempts are ongoing to achieve a unified understanding of a variety of transverse spin measurements, and further experiments are planned at RHIC and JLAB, with the aim of probing the orbital motion of quarks and gluons separately.

The GPDs obtained from deep exclusive high-energy reactions provide independent access to the contributions of quark orbital angular momentum to the proton spin. As described further below, these reaction studies are a prominent part of the science program of the 12 GeV CEBAF Upgrade, providing the best promise for deducing the orbital contributions of valence quarks.

### The Spatial Structure of Protons and Neutrons

Following the pioneering measurements of the proton charge distribution by Hofstadter at Stanford in the 1950s, experiments have revealed the proton's internal makeup with ever-increasing precision, largely through the use of electron scattering. The spatial structure of the nucleon reflects in QCD the distributions of the elementary quarks and gluons, as well as their motion and spin polarization.

**Charge and Magnetization Distributions of Protons and Neutrons.** The fundamental quantities that provide the simplest spatial map of the interior of neutrons and protons are the electromagnetic form factors, which lead to a picture of the average spatial distributions of charge and magnetism.

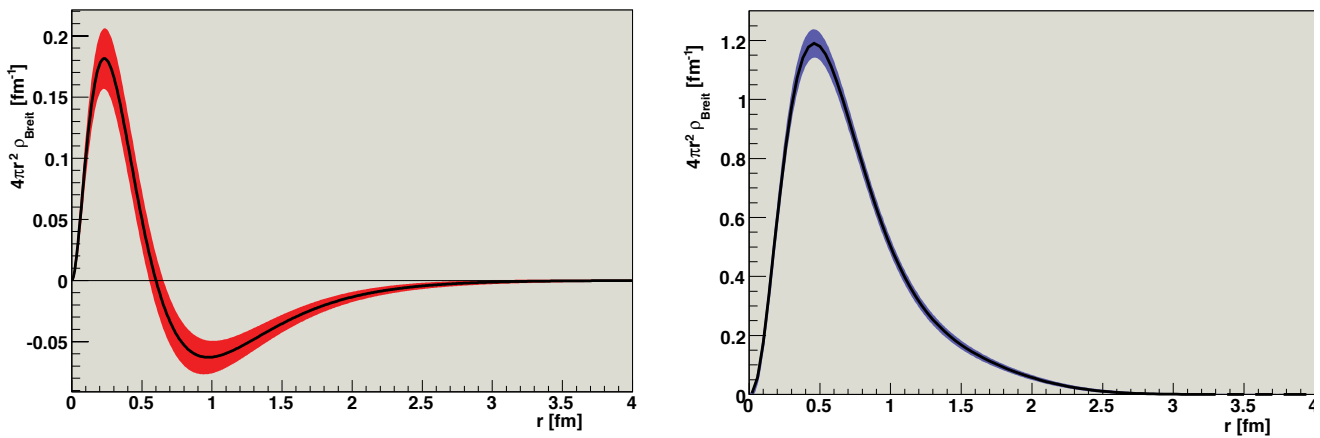

**Figure 2.5:** On the left is the distribution of the charge within the neutron, the combined result of experiments around the globe that use polarization techniques in electron scattering. On the right is that of the (much larger) proton distribution for reference. The widths of the colored bands represent the uncertainties. A decade ago, as described in the 1999 NRC report (*The Core of Matter, the Fuel of Stars*, National Academies Press [1999]), our knowledge of neutron structure was quite limited and unable to constrain calculations, but as promised, advances in polarization techniques led to substantial improvement.

Historically, their determination has come mainly from unpolarized electron scattering experiments. In the past two decades, substantial advances in high-intensity polarized beams, high-polarization targets, and polarimetry have led to a new class of precision measurements of the neutron and proton electric and magnetic form factors. At the time of the 2002 Long Range Plan, polarization-based experiments at JLAB had discovered, in stark contrast to what had previously been recorded in textbooks based on unpolarized scattering, that the proton's charge and magnetism distributions are quite different. The discrepancy in the results from the two methods is now thought to arise from subtle "two-photon" contributions, which obscured the interpretation of the unpolarized scattering results. Future experiments comparing the scattering of electrons and positrons with the aim to directly determine the two-photon contributions are planned at JLAB, at the VEPP-3 facility in Novosibirsk, Russia, and at DESY.

A comparably precise determination of the neutron's spatial structure has been much harder to achieve because of the absence of a free neutron target. But, the combined results of polarization experiments at JLAB, the Mainz Microtron Institute, and the MIT-Bates Laboratory have changed the situation. The neutron's charge distribution (together with that of the proton for comparison) is shown in figure 2.5. These results clearly identify the neutron's positively charged interior and negatively charged halo, consistent with the view that the nucleon's structure is strongly influenced by a cloud of paired-up, light quark-antiquark pairs, the pions. As we look toward the next decade, experiments will probe ever shorter distance scales, going into a regime where the details of, for example, the quark orbital motion will play a more significant role. Such measurements remain the only source of information about quark distributions at small transverse distance scales. The differences between proton and neutron form factors represent an important benchmark for lattice QCD calculations.

**The Role of Strange Quarks.** The role of strange quarks in the structure of neutrons and protons has been a subject of intense theoretical and experimental interest over the last two decades. Quark-antiquark pairs and gluons from the sea account for the vast majority of the mass of the nucleon. However, the virtual character of the quark-antiquark pairs means they cannot contribute to properties such as overall charge. They can, however, contribute to the local charge distribution, and thus to the nucleon's form factors. Strange quarks, normally not thought of as being part of a nucleon, can contribute in this way to the nucleon's properties by modifying the internal distribution of its charge and magnetism.

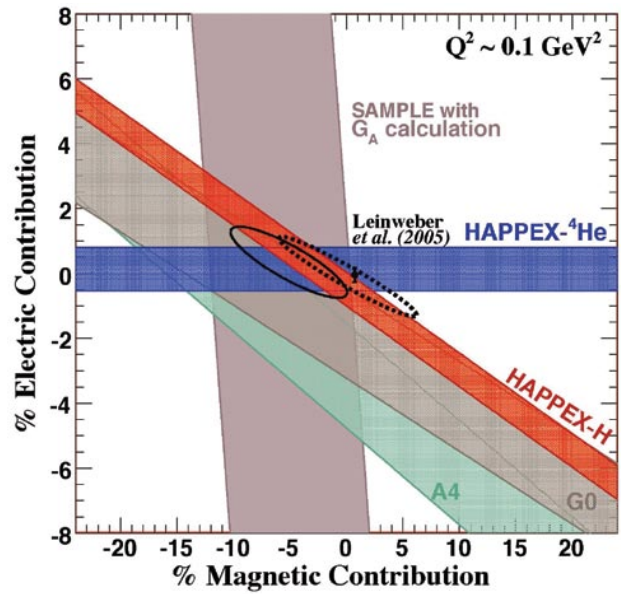

**Figure 2.6:** The worldwide program of parity violating electron-scattering data at $Q^2 \sim 0.1$ GeV$^2$ that constrain the electric (vertical axis) and magnetic (horizontal axis) contributions of strange quarks to the proton's charge (magnetism) at large spatial distances (low $Q^2$). The solid ellipse represents a fit to the data shown, incorporating Standard Model radiative corrections. The dashed ellipse includes additional data at short distance and removes Standard Model constraints on the nucleon axial form factors. Also shown, as a plus symbol, is a calculation that incorporates methods from lattice QCD.

Experiments have been underway to identify strange-quark contributions, using polarized electrons scattered from unpolarized targets. The experiments look for tiny changes in the scattering of electrons when the beam spin is reversed, a violation of the parity, or "mirror," symmetry that is attributable to the weak force. They thus measure the weak force equivalent of the charge and magnetism distributions, which can be combined with the precision electromagnetic data to disentangle the strange-quark contributions. JLAB has carried out the most sensitive of these experiments to date. The combined results allow the possibility of separating the electric and magnetic contributions for the first time, as shown in figure 2.6. Taken together, they place tight constraints on the strange-quark contributions: less than 5% of the spatial

# Femto-photography of the Proton: Catching the Quarks in Action

**Quarks in Orbit**

How do the quarks and gluons account for the proton's spin? We know that the quark spins account for at most a third of it. Yet recent data suggest that the gluons' contribution is also small. If that is the case, then where is the rest of the spin? Only one possibility remains: the *orbital* angular momentum of the quarks and gluons. There is considerable evidence that quarks do indeed reside in organized orbits within the proton. But the challenge is to catch them in action and actually measure this motion. Experimenters are approaching the problem using two different techniques.

**Swing to the Left, Swing to the Right**

In 2003, the HERMES experiment at DESY observed a strange phenomenon: when a high-energy electron beam crashes headlong into a proton whose spin is pointing upward, the positively charged pions produced in the collision prefer to head off to the beam's left. Other final-state particles have different preferences. Vigorous theoretical work over the past five years has established a rigorous connection between these left-right asymmetries and the orbital motion of the quarks within the proton. In effect, one is seeing that motion firsthand: fast $\pi^+$ mesons (whose quark substructure is $u\bar{d}$) obtain their impetus to fly off to beam-left from the orbital motion of up and antidown quarks around the proton's spin direction. New data obtained from RHIC, JLAB, DESY, and CERN provide clearer pictures of just what the quarks are doing within the proton. However, to map out these new spin-orbit distribution functions with any sort of precision will require a great deal more data—data that only the 12 GeV CEBAF Upgrade and the proposed Electron-Ion Collider can provide.

**Photographs at the Femtometer Scale**

One of the most fascinating possibilities of modern electron-scattering experiments is that they can measure the *spatial distribution* of quarks and gluons, providing actual three-dimensional images of the proton at the femtometer scale. The reconstruction of spatial images from scattering experiments by way of Fourier transform of the observed scattering pattern is a technique widely used in physics, e.g., in X-ray scattering from crystals. Recently, it was discovered how to extend this technique to the spatial distribution of quarks and gluons within the proton, using processes that probe the proton at a tiny resolution scale. The spatial distribution of the partons is encoded in the generalized parton distributions, a rich formalism that both unifies existing descriptions of proton structure and takes a giant step into unknown territory. Mapping out the GPDs is an ambitious program that requires the energies and high luminosities of the 12 GeV CEBAF Upgrade and, for the gluons, a future Electron-Ion Collider. Are the up quarks, the down quarks, and the gluons in the proton equally distributed in space? Are the quarks more polarized in the center than at the periphery? Answers to questions like these await and will revolutionize our knowledge of proton structure. Finally, encoded in the GPDs is yet another secret: the orbital angular momentum of the quarks and gluons. With the GPDs, we not only obtain still photographs of the quarks, but we catch them in action as well.

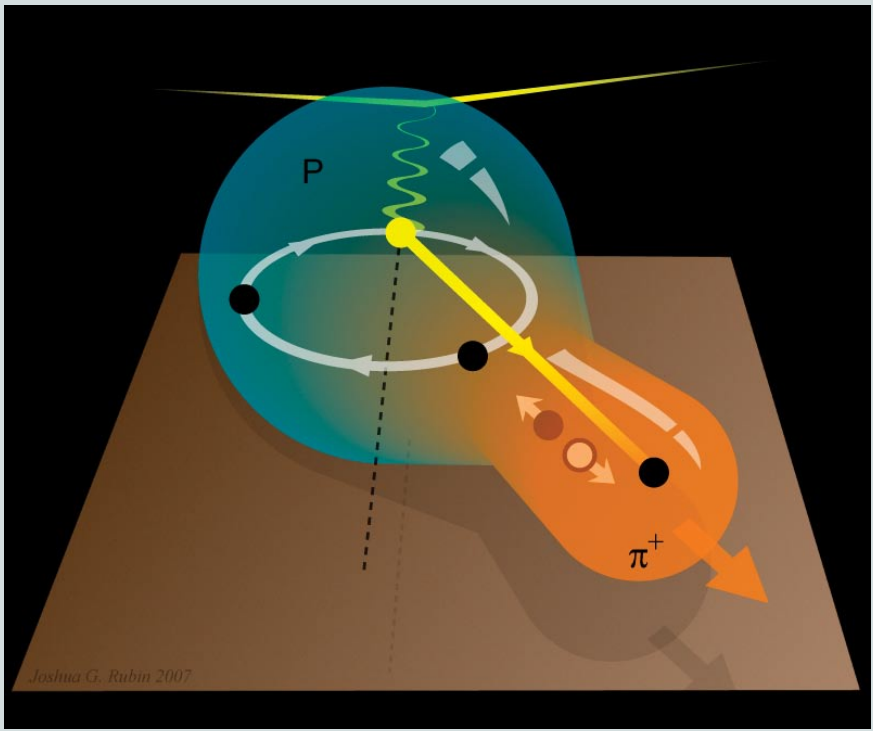

Illustration of a possible mechanism by which the orbital motion of an up quark inside a proton causes positively charged pions ($u\bar{d}$) to fly off predominantly to beam-left.

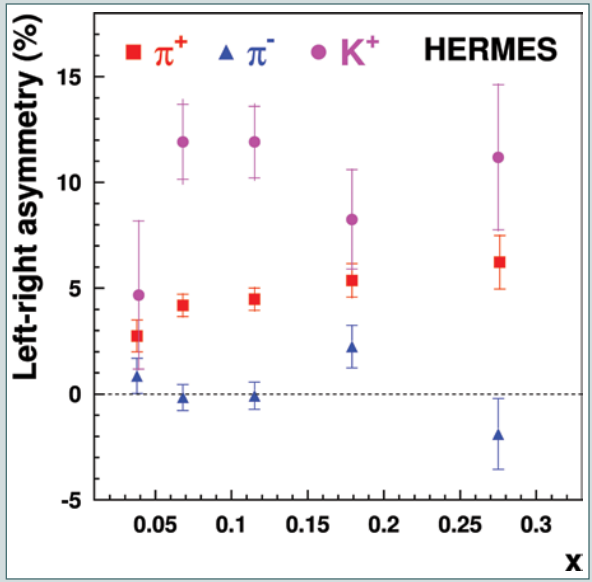

Left-right asymmetries sensitive to quark orbital motion for $\pi^+$ ($u\bar{d}$), $\pi^-$ ($\bar{d}u$), and $K^+$ ($u\bar{s}$) mesons produced from a transversely polarized target. The asymmetries are shown as a function of the fraction $x$ of the proton's momentum carried by the struck quark. The difference between the $\pi^+$, $\pi^-$, and $K^+$ asymmetries reveals that quarks and antiquarks of different flavor are orbiting in different ways within the proton.

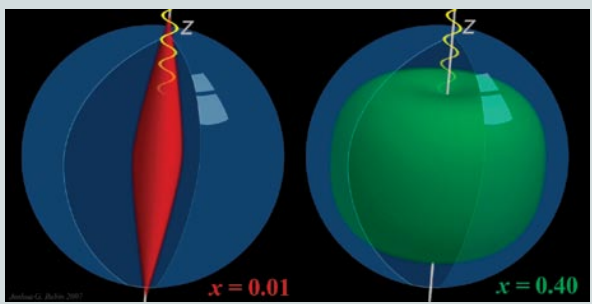

Illustrations of the spatial distribution of up quarks that will become accessible with the 12 GeV CEBAF Upgrade. The red and green forms outline the regions of highest quark density in a proton viewed by a beam traveling along the z axis. The images are for two different values of quark momentum fraction x, showing the expected difference in the spatial distribution of high and low $x$ quarks.

distributions of the proton's magnetism and significantly less of its charge. New data will soon place limits on these contributions at smaller spatial separations. These measurements, and the technological developments that made them possible, also enable the next generation of even more precise experiments that use parity-violating electron scattering to search for evidence of interactions beyond the Standard Model in a way that is complementary to the sensitivities expected from high-energy experiments at the LHC.

**Polarizabilities.** Another important property of the nucleon is its electromagnetic polarizability—the ability of its internal constituents to orient themselves in response to external electric and magnetic fields. The most direct method of determining such polarizabilities is Compton scattering, the direct scattering of a photon from the nucleon. This provides stringent tests of calculations that link the effective low-energy description of nucleons to QCD. As with the nucleon electromagnetic distributions, the formalism to describe the polarizabilities can be extended to probe differing distance scales, using the technique of virtual Compton scattering. Collectively, the results indicate that the nucleon's paramagnetic (or intrinsic) polarizability is of opposite sign to its diamagnetic (or induced) response. The next generation of such experiments will be carried out at the HIγS facility.

**Linking Space and Momentum Maps.** While much of what is known about the spatial structure of the nucleon comes from the above experiments, these measurements do not provide details of how fast the quarks move or how much momentum and energy they carry, or how their spin and angular momentum contribute to the nucleon's spin. Access to this information is one of the main goals of the high-energy hard-scattering experiments described above. Two experimental processes used to reveal correlations between space and momentum are deeply virtual Compton scattering (DVCS) and deeply virtual meson production, in which—similar to the form factor measurements—the recoiling nucleon remains intact after the scattering. In these processes a photon or meson is produced in a short-distance reaction of the electron with a single quark inside the nucleon, which, thanks to the celebrated "asymptotic freedom" of QCD, can be calculated in terms of well-known elementary quark-gluon interactions. The experimental data can thus be used to extract the long-distance information about the distribution of quarks in the nucleon, described by the GPDs, which encode universal, process-independent characteristics of the nucleon.

The feasibility of such DVCS experiments has been demonstrated in a set of pioneering experiments with the HERMES detector at DESY and at JLAB. The results are in qualitative agreement with theoretical predictions based on GPD parameterizations that are constrained by the measured form factors and quark momentum distributions. Full realization of the potential of GPDs will require the 12 GeV CEBAF Upgrade, which provides the unique combination of high beam energy, high luminosity, and advanced detection equipment necessary to study exclusive processes. As with determination of the electromagnetic form factors, this will require an extensive program rather than a single experiment. The crucial new information about the space-momentum correlations will be extracted from measurements of DVCS observables, and meson production will provide information about the spin and flavor structure of such correlations.

Equally interesting is the spatial distribution of gluons in the nucleon. Because the standard electroweak currents "see" only quarks, it is not possible to probe the glue directly in elastic scattering. The gluons can be probed, however, in exclusive deep-inelastic processes at high energies, as well as in the production of hadrons containing charm quarks. A detailed program of "gluon imaging" of the nucleon can be carried out with an Electron-Ion Collider.

## THE HADRON SPECTRUM

Spectroscopy has long provided essential information for scientists trying to understand the nature of our world. The spectroscopy of the hydrogen atom revealed the quantum nature of the world at small distance scales and led to the development of quantum electrodynamics (QED), the theory that describes the electromagnetic interaction. The spectroscopy of the even simpler positronium atom, which consists of an electron and a positron, has provided some of our most sensitive tests of QED.

An analogous development is at the heart of our efforts to understand the strong interaction. The spectroscopy of the hadrons—baryons, which have three valence quarks, and mesons, which have a quark and an antiquark—has provided essential clues that led to the development of QCD, but it has also left deep puzzles. For example, why has no hadron with a valence gluon been found? Nuclear physics is tackling these puzzles via a new generation of spectroscopy experiments and theoretical investigations that take advantage of remarkable developments in technology. The central questions being addressed by this program are: **what are the roles of gluons and gluon self-interactions in nucleons?** Or, more particularly, what is the detailed mechanism by which quarks are confined within hadrons? What are the effective degrees of freedom that define the spectrum of hadrons? What is the role of glue in the spectrum? In order to answer these questions we must experimentally identify and determine the properties of the states of both mesons and baryons. We discuss these efforts, in turn, below.

### Excited States of the Meson—Understanding the Confinement of Quarks

A unique feature of QCD is the confinement of quarks and gluons inside the particles they form. However, an understanding of the origin of the confinement mechanism still eludes us. We believe that gluons play a crucial role, but direct experimental signatures for the role of glue have been elusive. Recent theoretical progress indicates the existence of a new family of exotic particles, "hybrid mesons," in which the role of the glue can be observed more readily. Theory predicts that the gluons not only hold the quarks together, but can also move collectively and contribute more than just mass. The simplest such motion is a rotation of the glue—much like that of a jump rope. The masses of the hybrid mesons are related to the energy in the rotation; thus, information on their masses will provide information on the confining gluon field.

The 12 GeV CEBAF Upgrade provides a new and unique way to produce exotic hybrids using beams of high-energy spin-polarized light. The GlueX experiment has been optimized to explore the existence of exotic hybrid mesons with sensitivity that is hundreds of times higher than in previous efforts and to provide crucial measurements that are expected to help unravel the mechanism of quark confinement.

### Excited States of the Proton and Neutron

Adding just the right amount of energy to a proton can promote it to one of its excited states. The fact that the identified "elementary particles" of the 1950s and 1960s could be described as the excited state spectrum of hadrons built from quarks provided early clues to the underlying quark-gluon structure of the hadrons. Modern experiments not only search for new excited states of the proton and neutron, but also probe the underlying dynamics of the proton's structure by imparting variable amounts of momentum to the proton while producing an excited state. New data on the production of the first excited state of the proton along with QCD-based models indicate that the proton behaves like a small three-quark core surrounded by a cloud of pions.

Simplified quark models of the proton predict the spectrum of excited states. Models in which the three quarks are independent of each other predict a richer spectrum of states than has been observed, while models in which two of the quarks are coupled together explain the existing spectrum but are in disagreement with other observations on the structure of the proton. In order to determine how the three quarks behave, more detailed information is required, including new data with a variety of meson-baryon final states and a variety of polarization observables to disentangle the resonant amplitudes corresponding to excited states with different quantum numbers. The phenomenological analysis tools must also be improved by including the final states in a coupled-channel analysis. A broad effort is underway at JLAB to provide the necessary experimental data and analysis tools. Experiments using polarized photon beams have taken data, and a program is in preparation using highly spin-polarized (frozen spin) targets of both protons and neutrons in conjunction with polarized photon beams.

## THE EMERGENCE OF NUCLEI FROM QCD

Connecting the phenomenology of nuclei to the fundamental theory of QCD is a long-term goal of nuclear physics. The key question to address is: **what governs the transition of quarks and gluons into pions and nucleons,** or in particular, what manifestations of gluon and quark interactions appear in the properties of nuclei and in the interaction among neutrons and protons? Ultimately, answering this will require a series of overlapping effective theories applicable to successively more complex phenomena at increasing length scales. Over the next decade we expect to see a number of exciting developments in this direction: (1) nuclear three-body interactions being largely determined from a combination of lattice QCD results and experimental input to effective theory; (2) the application of these improved three-body interactions in the calculations of light nuclei; and (3) determination of nucleon-hyperon interactions, possibly relevant to the properties of neutron stars.

### Short-Range Correlations in the Nuclear Medium

The nucleon-nucleon interaction is relatively well known and is believed to dominate the structure in complex nuclei. In addition, it is widely believed that a three-nucleon force is necessary to explain the energy spectrum of nuclei. Nevertheless, the direct observation of correlated two-nucleon and three-nucleon effects in the nuclear medium has been evasive. The powerful combination of the multi-GeV electron beam and a large acceptance detector at JLAB has permitted the direct observation of two- and three-nucleon correlations in nuclei. The key to observing these effects is high-energy electron scattering from quarks in nuclei where the struck quark carries more than the momentum of a single nucleon. In figure 2.7, the ratio of the scattering yield of several nuclei to that in $^3$He is shown as a function of Bjorken $x$. As $x$ becomes unity, the quark carries all the momentum of a single nucleon. As $x$ exceeds unity, the ratio rises indicating that more than one nucleon must be involved in the interaction. Similarly, as $x$ exceeds the value of two, the ratio rises again, indicating that more than two nucleons are involved in the process. This pattern appears to be a general feature that is independent of the target nucleus, and it directly confirms our understanding of the underlying dynamics in nuclei. The HIγS facility will contribute to further understanding of the dynamics by providing important new data on photon interactions with light nuclei. In particular, the role of three-nucleon interactions will be explored through unique measurements of the disintegration of polarized $^3$He nuclei by polarized photons.

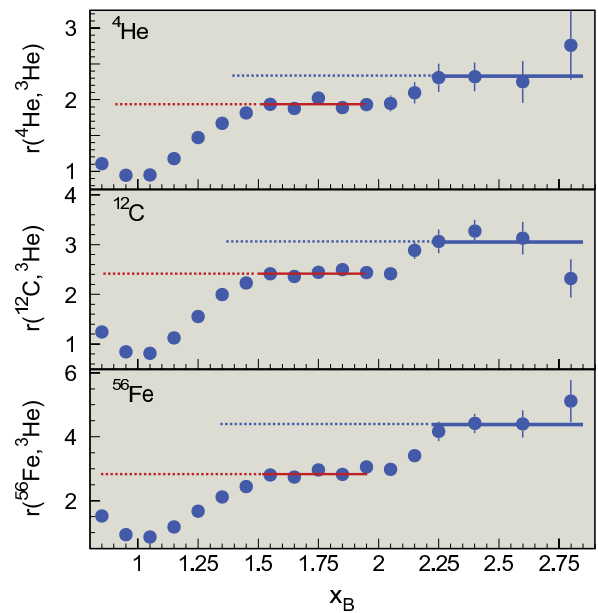

**Figure 2.7:** Ratios of measured inelastic electron scattering yields for $^4$He (top), $^{12}$C (center), and $^{56}$Fe to that of $^3$He are plotted as a function of Bjorken $x$ (or $x_B$). The first plateaus above $x_B=1$ indicate the region where two-nucleon correlations dominate, while the plateaus above $x_B=2$ indicate the region where three-nucleon correlations become important.

**The Role of Quarks in Nuclei**

One of the most intriguing questions in nuclear physics is where the quarks and gluons of QCD become important to understanding the structure and properties of nuclei, which to a good approximation are described as a system of nucleons in a mean field. With the discovery of the nuclear EMC effect some 20 years ago, the momentum distribution of quarks inside the nucleus was found to differ, notably in the valence region, from that of quarks in a free nucleon. Despite a significant worldwide effort in experiment and theory, there is as yet no consensus concerning the origin of this effect. Explanations of it are hampered by the enormous challenge of constructing a theory that can consistently account for nuclear binding as well as the structure of free nucleons. Further difficulties arise from the lack of evidence for the changes to the antiquark distributions predicted to arise from pions being exchanged between nucleons in the nucleus. To make progress in understanding this effect, experiments to measure the spin and the quark-flavor dependence of the EMC effect are essential. There are fascinating suggestions that the spin-dependent EMC effect is even larger than the spin-averaged effect, and its manifestation would have profound consequences for our understanding of nuclear structure in terms of QCD. This will require the 12 GeV CEBAF Upgrade as well as the E906 experiment at Fermilab. The possible modification of the proton elastic form factors in the bound nucleus will also be explored by determining the effective charge of a proton embedded in a nucleus. The deviation of this Coulomb sum from the free proton charge signals modification of the proton elastic scattering process in the nucleus. Perhaps the "smoking gun" is the JLAB result suggesting the ratio of electric to magnetic form factors of a proton in helium is different from that of a free proton.

**Spacetime Characteristics of Fundamental QCD Processes**

The dynamical process by which bound hadrons are actually *formed* is a fascinating process. How can we actually make new hadrons and observe their formation when quarks are eternally bound to each other? The answer is that we can make new quarks, by converting energy into matter. In DIS, a momentum "kick" sends the quark flying away from the proton remnant … but not for long. Both the struck quark and the remnant left behind carry color, and the strong force attracting them together grows rapidly as they separate. A so-called "color string" is formed, and just like a real string, it can break. The string can literally snap. When this happens, a portion of the string's energy is converted into matter—quark-antiquark pairs which emerge from the vacuum. As the struck quark and remnant continue to recede from each other, further string breaks occur and more quarks and antiquarks are made, roughly trailing along behind the leading particles in "jets." Eventually confinement comes into play again when the system has cooled down sufficiently for the quarks and antiquarks to bind together into different hadrons. These final particles are detected by experimenters. All of this happens on time scales of order $10^{-23}$ seconds and distance scales of order $10^{-15}$ to $10^{-14}$ meters (one to 10 fm). Experimenters use an ingenious yardstick of tiny size with which to observe the phases of hadron formation—the nucleus. Stable nuclei, from which practical targets can be made, range in radius from about one to 10 fm. The urgency to understand in-medium hadronization has been impelled by the analysis of heavy-ion data at RHIC.

New DIS data, most notably the recent measurements from HERMES on many nuclear targets and covering a broad kinematic range, have led to a much more refined understanding of hadron formation. The basic picture to emerge from these data is of a three-stage process, each with its own characteristic time scale and featuring different interactions with the nuclear medium. In the first stage, the struck quark propagates essentially freely through the nucleus, losing energy in various ways: deceleration from the color string attaching it to the remnant; gluon radiation; and energy loss due to multiple scattering with the surrounding nucleons. After a certain time, the quark pairs up with an antiquark making a color-neutral "prehadron." This object is expected to be of smaller size than a physical hadron due to the phenomenon of color transparency. This phenomenon predicts that if the momentum transfer of the interaction is high enough, the nuclear medium will become *transparent* and final-state attenuation will disappear. The prehadron size is expected to fluctuate and, along with it, its probability of interaction with the spectator nucleons. Finally, a detectable hadron is formed. The combination of JLAB and DESY data provides evidence for color transparency. This effect can be studied more conclusively with the 12 GeV CEBAF Upgrade. Aside from the very special kinematic regime where color transparency could occur, the net result of these various interactions between the struck quark, the prehadron, and the final hadron with the nuclear medium is that fast-hadron production is suppressed in heavy nuclei. The latest DIS data map out this attenuation as a function of many kinematic variables. Knowledge of these dependences is key to unraveling the different stages and time scales involved, and much has been learned. For example, the latest data from HERMES provide new access to parton energy loss, and the 12 GeV CEBAF Upgrade will permit a study of unprecedented precision.

## Energy Recovering Light Source Technology

The Department of Energy's Jefferson Lab is the site of the world's most powerful tunable laser, the Jefferson Lab free-electron laser (FEL), which has seen wide application in the fields of medicine, materials science, photochemistry, and biophysics. Since becoming operational in 1998, the FEL has produced light tunable from the far-infrared (Terahertz light) through the infrared, visible, and ultraviolet wavelengths, all with several orders of magnitude more power than any other tunable light source. A schematic diagram of the FEL is shown in the figure. The FEL is built on the unique superconducting radiofrequency (SRF) technology developed at Jefferson Lab for DOE's nuclear physics program. In the FEL, electrons are whipped up to high energies by an SRF linear accelerator. A wiggler, a device developed by DOE and DOD during the 1980s for powerful X-ray sources, produces magnetic fields to shake the electrons, forcing them to release energy in the form of photons. As in a conventional laser, the photons bounce between two mirrors and are emitted as laser light. SRF technology allows high average-power electron beams to be produced and operated in a cost-effective manner. In the FEL, it allows the laser to stay on 100% of the time instead of only 1 or 2%. Cost savings are compounded by the design of the FEL's unique energy-recovering linear accelerator. Once electrons have exited the wiggler, they are steered back into the machine's linear accelerator, allowing the machine to recover more than 90% of the energy that is not converted to useful light in a single pass. The strength of an energy-recovering linear accelerator is the economic feasibility of using the beam "only once." This pioneering demonstration of SRF energy-recovering technology is the basis for an entire new class of devices that are currently being designed for basic energy sciences, nuclear and particle physics, and defense applications. The FEL is capable of producing an unprecedented 14.2 kW of infrared laser light.

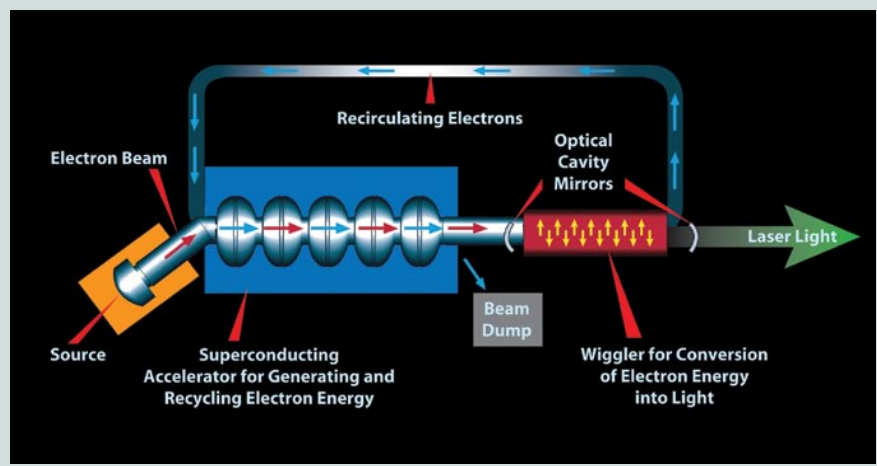

To enable experimenters to probe deep inside the atom's nucleus, Jefferson Lab pioneered superconducting technology for accelerating electrons to high energy. Expanding upon this expertise, the lab developed a one-of-a-kind energy-recovering linear accelerator, which powers its free-electron laser and recycles more than 90% of the electron energy that is not converted to useful light. A schematic diagram of the FEL with the energy recovery accelerator system is shown in the figure.

## OUTLOOK

Technical innovations on both experimental and theoretical fronts have brought us to the threshold of an era of precision studies of the quark-gluon substructure of hadrons. There is now a clear path forward to answer the central scientific questions posed at the start of this chapter. The path involves several accelerator facilities and experimental approaches, as needed to examine, with a wide range of resolution scales, the multifaceted internal structure produced by the strong color force among quarks and gluons.

The 12 GeV CEBAF Upgrade already underway at JLAB will definitively establish the contribution of valence quarks to hadron structure. Their distribution in space, momentum, and spin will be mapped by a combination of deep-inelastic scattering and deeply virtual photon and meson production. The correlations between space and momentum distributions encoded in the generalized parton distributions will provide access to a full tomographic image of the internal landscape of protons and neutrons in the regime where quarks dominate.

A complementary, but narrower, view into the "sea" region, where gluons and quark-antiquark pairs become important, will be provided by polarized-proton collisions with upgraded detectors at RHIC and the E906 fixed-target experiment at Fermilab. Full exploitation of this higher-energy and higher-spatial-resolution capability will determine much of the contribution of gluons and antiquarks to the proton spin. In the longer term, an Electron-Ion Collider would combine the broad capabilities of electron scattering with the energy reach and spin sensitivity to expose hadrons and nuclei in the unexplored regime where their internal structure is completely dominated by self-interacting gluons.

These advanced experimental efforts must be matched by strong and focused theoretical work to provide robust interpretations of the measurements. Major computational facilities to carry out first-principles numerical QCD calculations are essential. Further developments in effective field theory approximations to QCD will provide critical insights into the intricate and subtle ways Nature has chosen to build ordinary matter from unseen particles bound by a force that prohibits their escape. Recent progress makes even the weaker residual force that holds neutrons and protons together inside atomic nuclei seem within the reach of a QCD-based understanding.

# The Phases of Nuclear Matter

## EXPLORING THE QUARK-GLUON PLASMA

The theory of quantum chromodynamics (QCD) predicts that the protons and neutrons found in ordinary atomic nuclei can "melt" under certain conditions. The required conditions are extraordinary, involving energy densities 30 to 100 times those found in a normal nucleus, as well as temperatures above some 2 trillion degrees Celsius. But beyond those values, according to QCD, the nucleons will dissolve and fuse with one another, releasing the quarks and gluons inside to form a *quark-gluon plasma* (QGP).

The Relativistic Heavy Ion Collider (RHIC) at Brookhaven National Laboratory was built to test this prediction and to explore the properties of the QGP should it actually exist. As the collider's name suggests, the idea was to heat nuclear matter beyond the critical temperature by colliding heavy nuclei at very high energy. And indeed, the experiments performed at RHIC since it began operation in 2000 have provided spectacular evidence that the QGP does exist—but that its properties are quite different than expected. QCD calculations were interpreted before 2000 to suggest that any QGP produced by the collider would behave like a dilute gas: a loose assembly of particles that would explode outwards from the collision point like fireworks, in a spherical pattern. In fact, the plasma turned out to behave more like a liquid—a nearly "perfect" liquid of liberated quarks that flow with almost undetectable viscosity. The experiments have also shown that assembly of particles is anything but loose: the plasma is threaded by "color" fields that have a strong effect on how the quarks and gluons move.

The next phase of the RHIC physics program will focus on detailed investigations of this newly discovered state of matter, both to quantify its properties and to understand precisely how they emerge from the fundamental properties of QCD. These experiments will also allow us to explore a wider region of the phase diagram of nuclear matter, where calculations hint at a discontinuous transition between the different phases. In parallel, a similar science program at the Large Hadron Collider (LHC) at the European Center for Nuclear Research (CERN) will explore nuclear matter at even higher temperatures. Before we describe the next steps in the scientific exploration of the phases of nuclear matter, we first discuss the experimental results on which the surprising discoveries made at RHIC are based.

## RECENT ACHIEVEMENTS

The insights into the nature of the QGP summarized above have made the first phase of RHIC a great success. In its first six runs (2000–2006), RHIC has provided four different collisions systems (Au + Au, d + Au, Cu + Cu, and p + p) at a variety of energies, ranging from nucleon-nucleon center-of-mass energies of 19.6 to 200 GeV, with 60% polarization of the proton beams. The largest data samples were collected at the highest energy of 200 GeV, where the accelerator has achieved sustained operation at six times the design luminosity. The ability to study proton-proton, deuteron-nucleus, and nucleus-nucleus collisions at identical center-of-mass energies with the same detectors has been a key contributing factor to systematic control of the measurements. The experiments also measure patterns in particle production that constrain, for each event, the distance of closest approach of the two incoming nuclei (impact parameter) and the orientation of the (reaction) plane formed by their trajectories. This detailed categorization of the collision geometry provides a wealth of differential observables, which have proven to be essential for precise, quantitative study of the plasma. The results obtained by the four RHIC experiments (BRAHMS, PHENIX, PHOBOS and STAR), published in nearly 200 scientific papers, are overall in excellent quantitative agreement with each other.

### The RHIC Discoveries

The initial set of RHIC heavy-ion results has provided evidence for the creation of a new state of thermalized matter at unprecedented energy densities (30–100 times that for normal nuclear matter), which exhibits almost perfect hydrodynamic behavior. Among the results, several discoveries stand out:

- **Near-Perfect Liquid.** The measured hadron spectra and their angular distributions bear witness to the enormous collective motion of the medium. In addition, measurements of electrons from the decays of hadrons containing charm quarks indicate that even heavy quarks flow with the bulk medium. These observations are in agreement with the hydrodynamic expansion of a nearly viscosity-free liquid—often characterized as a "perfect" liquid. The relative abundance of the produced hadrons with transverse momenta below 1.5 GeV/c, the shapes of their transverse momentum spectra, and their anisotropic or elliptic flow

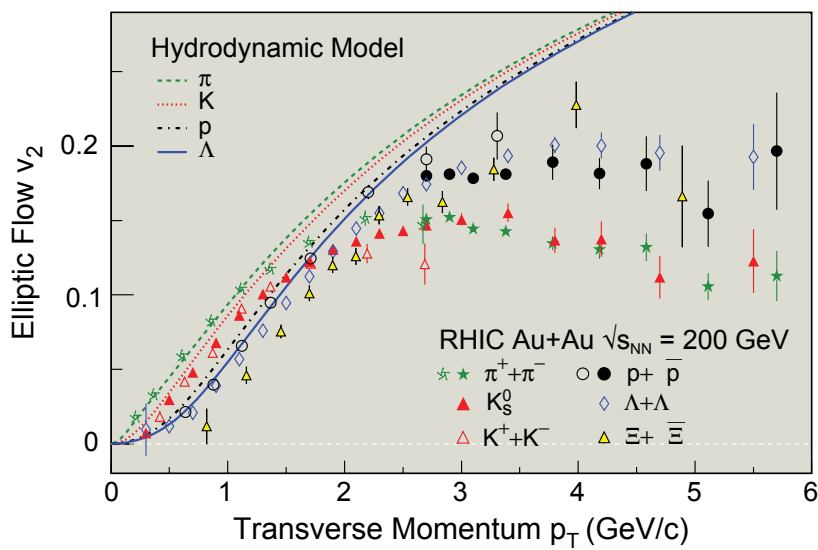

**Figure 2.8:** Elliptic flow strength parameter $v_2$ for different particle species plotted as a function of transverse momentum $p_T$, compared with predictions based on ideal hydrodynamics. The parameter $v_2$ measures the angular anisotropy in particle momentum that can arise for a system that is initially anisotropic in shape, which occurs in off-center nuclear collisions. The connection between the initial spatial anisotropy and the momentum anisotropy measured by $v_2$ is attributable to the pressure of the expanding medium. The different momentum dependence of $v_2$ for particles of different mass is characteristic of particle production from a common source that is an expanding fluid. Note that more than 98% of all produced particles have transverse momenta below 1.5 GeV/c, where the data closely follow the predictions for a perfect liquid.

## Serving the "Perfect" Liquid

Liquids occupy an intermediate place in the world of materials, between gases and solids. Like solids, liquids resist being compressed—as anyone who has made a belly flop into a swimming pool knows very well. It usually takes a very large force to achieve much compression. Like gases, however, liquids can effortlessly change their shape to fit their container—as anyone who has poured water into a glass can verify. This is why both gases and liquids are commonly referred to as *fluids*: substances that can flow.

In general, however, flow is not forever. Friction between different fluid elements tends to slow their relative motion. So without an input of external energy to keep things going, most flow patterns will slow to a halt very quickly; witness the little eddies that trail behind a rower's oars, gradually but inexorably subsiding. Physicists measure this tendency for flow patterns to fade by a quantity known as the *shear viscosity*. The larger the viscosity, the faster the damping. All fluids exhibit at least a tiny amount of viscosity—even helium, which forms a nearly friction-free fluid when cooled to very low temperatures.

To compare different fluids, engineers often divide a fluid's viscosity by its density. The resulting quantity, called the *kinematic viscosity*, turns out to be directly proportional to the distance atoms or molecules can travel in the fluid before interacting with another particle. Somewhat paradoxically, the kinematic viscosity actually *decreases* as the interactions get stronger because the particles in the fluid find it more difficult to communicate over large distances to regions where the flow velocity may be different. Beyond a certain point, however, increasing the interaction strength will cause the kinematic viscosity to rise again; the particles may coalesce into a solid, for example, or they may aggregate into long, polymer-like chains such as those found in honey. This suggests that, in general, the kinematic viscosity of a fluid cannot become arbitrarily small. Indeed, the kinematic viscosity of all known substances exhibits a minimum when measured as a function of temperature.

But how small can this minimum be? Until recently, no one knew an answer to this question. This changed in 2004 due to a surprising insight developed in the framework of superstring theory. Building on earlier work that had generalized the concept of the kinematic viscosity to relativistic quantum systems in which the number of particles constantly fluctuates due to quantum effects—the QGP being a prime example—this new development showed that there does indeed exist an absolute lower bound to the generalized kinematic viscosity. Moreover, that lower bound is proportional to Planck's constant $h$ (the precise value is $h/8\pi^2$)—a fact that can ultimately be traced back to the quantum mechanical uncertainty relation, which limits the rate at which elementary particles can scatter.

By this criterion, then, a "perfect" fluid is a substance whose generalized kinematic viscosity reaches the absolute lower bound. No known substance does patterns (figure 2.8) are well described by relativistic hydrodynamics for a perfect liquid (so-called "ideal" hydrodynamics) with an equation of state similar to the one predicted by lattice QCD. The magnitude of the observed collective flow points to rapid thermalization and equilibration of the matter.

While a "perfect" liquid is defined as a fluid without shear or bulk viscosity (a fluid's internal resistance to flow), one of the exciting theoretical discoveries of the past few years is the insight that there may be a lower bound on the ratio between the shear viscosity and entropy density of any fluid. Thus, in reality, a perfect liquid is a fluid that attains this lower bound. There is mounting evidence from analysis of RHIC data that the matter produced is nearly such a perfect liquid, with a viscosity to entropy density ratio within a factor of four of the bound.

- **Jet Quenching.** QCD jets are ubiquitous in high-energy collisions of all kinds, arising from the hard scattering of incoming quarks and gluons and their subsequent breakup ("fragmentation") into a characteristic spray of particles (pion, kaons, protons, etc.) that can be measured in a detector. Just as the differential absorption of x-rays in ordinary matter can be used to explore the density distribution and material composition inside the human body, so can the absorption of jets in the QGP be used to obtain a direct tomographic image of the gluon density of the plasma. The strong quenching of jets, observed in central Au + Au collisions via the dramatic suppression of particle production and modification of jet correlations at high transverse momentum, is compelling evidence for the large energy loss of scattered quarks and gluons (partons) traversing matter that has a high density of so. To physicists' surprise, however, the fireball produced in a heavy-ion collision comes closer than anything else: the RHIC data show that the generalized kinematic viscosity of a QGP cannot exceed the quantum limit by more than a factor of four (see figure). The reason for this small value is still not fully understood. It may imply that the matter produced at RHIC is a *strongly coupled* QGP, or it may be the result of novel effects associated with strong color fields. A central goal of the ongoing research program at RHIC is to clarify this issue—and in the process, to determine precisely how "perfect" this ultra-hot fluid really is. Achieving this goal will require more detailed measurements and much more sophisticated simulations of the collision dynamics. Of particular interest will be the behavior near the QCD critical point, where direct analogy to ordinary fluids suggests that the RHIC fluid will come closest to "perfection."

In the meantime, nuclear physicists have led the way in applying the insights derived from string theory and RHIC collisions to a very different realm. As surprised as they were to find that the hottest, densest matter ever studied is also the most perfectly fluid, their colleagues in atomic physics have been just as surprised to observe a similar perfection in ultra-cold matter, when a gas of very slowly moving atoms is confined in a magnetic "trap." By varying the magnetic field applied to such a system, these low-temperature researchers are able to tune the interactions between the atoms to the largest possible values allowed by quantum mechanics. At this special value of the magnetic field, they find that the flow of the atomic gas is just about as perfect as a QGP's, with a kinematic viscosity only about a factor of three or four above the lower bound. (By no coincidence, perhaps, the gas is also fairly close to its own critical temperature.)

The viscosity to entropy density ratio plotted versus a reduced temperature $(T-T_0)/T_0$ for various materials as indicated. The conjectured lower bound is $\eta/s = 0.08$. The range of values inferred from the RHIC data (red point) is rather close to this bound.

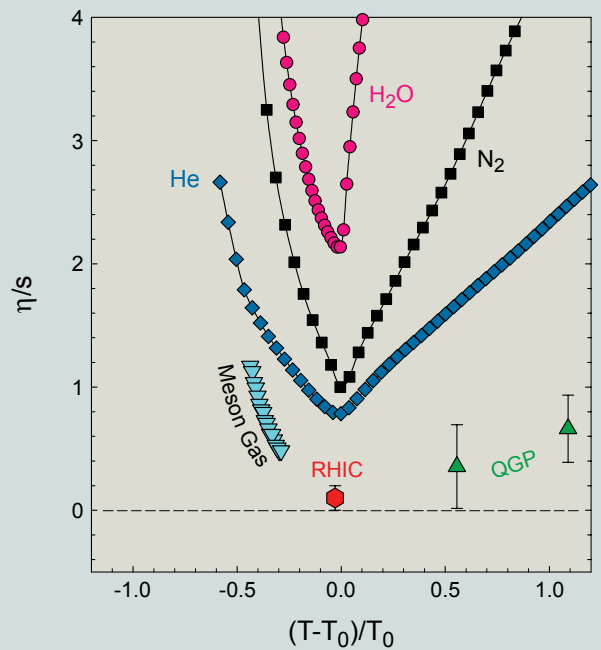

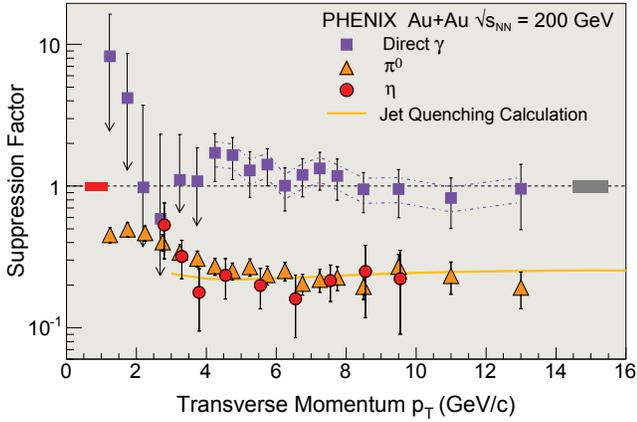

**Figure 2.9:** Suppression factor $R_{AA}$ of the production rates of high transverse-momentum ($p_T$) photons ($\gamma$) and $\pi^0$ and $\eta$ mesons measured in central Au + Au and p + p collisions, scaled to be unity if nuclear collisions are a simple superposition of p + p collisions. Mesons are suppressed in Au + Au collisions ($R_{AA} < 1$), while direct photons are not ($R_{AA} \approx 1$). This distinction is indicative of jet quenching. The solid line shows $R_{AA}$ values for a model calculation that incorporates jet quenching.

collision energy. Such "hard probes" have accurately calculable production rates and well-understood interactions with the medium, and they provided a primary motivation for constructing RHIC with high center-of-mass energy. This strategy has been spectacularly validated by the discovery of jet quenching and its ongoing development as a quantitative tomographic probe of the QGP. Figure 2.9 illustrates this discovery, showing the suppression of both $\pi^0$ and $\eta$ meson production in central Au + Au collisions compared to expectations from measurements in p + p collisions. The strong suppression of jet "fragments" (a factor of five) stands in distinct contrast to similar measurements of direct photon yields, which show no suppression and are consistent with perturbative QCD calculations of their initial production rate in all systems.

These different suppression effects are notable since photons are color neutral while high-$p_T$ hadrons (such as $\pi^0$ and $\eta$) are the remnants of color-charged jets. The figure shows conclusively that strong suppression occurs only for objects carrying color charge; in other words it must arise from interactions with a medium that itself contains a high density of color charges.

Dramatic additional evidence for the color opacity of the medium is seen in studies of the angular correlation of the radiation associated with a high momentum "trigger" particle. In p + p and d + Au collisions, a hard recoiling particle frequently occurs at 180° to the trigger, reflecting the back-to-back nature of jets that are color charges. (Color charge is carried by quarks and gluons and is analogous to the more familiar electric charge of electrons and ions.) Recent results suggest that the response of the medium to such large energy loss may also be hydrodynamic: the energy lost by high-energy jets may appear as a collective "sonic boom."

The rates for jet production and other hard-scattering processes grow rapidly with increasing produced (to leading order) in pairs. In sharp contrast, central Au + Au collisions show a strong suppression of such recoils (figure 2.10, left), accompanied by an enhancement and broadening of low-momentum particle production (figure 2.10, right). Recent analyses indicate that the momentum of the recoiling jet may generate a collective "shockwave" in the medium, analogous to the V-shaped wake of a powerboat traveling on a still lake.

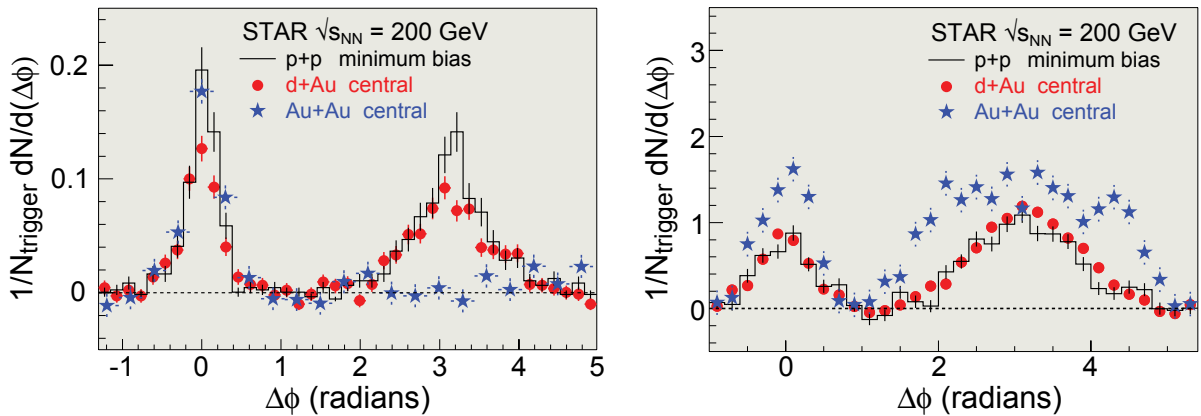

**Figure 2.10:** Correlations in azimuthal angle of pairs of particles in p + p, d + Au, and central Au + Au collisions. In all cases, the "trigger" particle of the pair has high-momentum ($p_T > 4$ GeV/c). *Left:* "Associated particles" recoiling with high momenta ($p_T > 2$ GeV/c) exhibit strong suppression in Au + Au. *Right:* "Associated particles" with low recoil momenta ($p_T > 0.15$ GeV/c) are strongly enhanced in Au + Au.

- **Novel Particle Production Mechanisms.** Evidence that the medium is composed of deconfined, thermalized, and collectively flowing quarks comes from detailed measurements of a wide variety of particle species, which reveal surprising patterns in heavy-ion collisions that are not seen in collisions that are more elementary. Particle species, which contain three "valence" quarks (baryons), show yields that are strongly enhanced relative to those containing a valence quark and antiquark (mesons) at "intermediate" transverse momentum ($p_T$~2-5 GeV/c) in nuclear collisions, compared to elementary systems such as p + p. This observation is well described by models in which baryons and mesons are generated by the coalescence of quarks drawn from a collectively flowing, equilibrated "partonic liquid." The elliptic flow patterns of baryons and mesons also show precise scaling (figure 2.11) when expressed in terms of the number of valence quarks. Ideal hydrodynamics predicts a complex "fine structure" in the behavior of the elliptic flow strength parameter $v_2$ for different mass particles as a function of transverse momentum $p_T$ (figure 2.8). The same calculations predict a simple scaling behavior for $v_2$ when analyzed as a function of "transverse kinetic energy" $KE_T \equiv (m^2 + p_T^2)^{1/2}-m$, which incorporates the effect of mass, resulting in two distinct branches, one for mesons and the other for baryons (figure 2.11, left). When both $v_2$ and $KE_T$ are scaled by the number of valence quarks (two for mesons and three for baryons), the two branches merge into a universal curve for all species (figure 2.11, right), suggesting that the dominant features of the flow pattern are developed at the quark level.

- **Initial Conditions.** Central Au + Au collisions at RHIC generate about 4500 final-state particles. Curiously, this strikingly large multiplicity is *lower* than theoretical expectations for it before the start of RHIC operations. This observation, together with the suppression of high-transverse momentum particles at forward angles in d+Au interactions, may indicate saturation of the density of low-momentum gluons in the wave function of a heavy nucleus relative to that of a proton. The scale $Q_s$, which defines the region in which saturation occurs, is expected to grow with the atomic mass. The gluon field in a nucleus below the saturation momentum may be described quasiclassically and is predicted to exhibit universal behavior, i.e., at very high energies the gluon distribution is the same in all hadrons and nuclei. This component of the wave function is called the *color glass condensate* (CGC) (the name derives from the expected glass-like dynamical behavior of the high-density, randomly oriented color fields at low momentum in nuclei). The CGC can be probed in detail in the forward kinematic region of d + Au and Au + Au collisions at RHIC that is sensitive to the small-momentum part of the gluon wave function in the Au nucleus (momentum fraction $x < 0.01$). The CGC hypothesis is indeed consistent with measurements of high transverse momentum particle yields at forward angles. It also agrees with the observed dependence of particle multiplicity on centrality and beam energy in Au + Au collisions, supporting the existence of saturation effects in the gluonic matter inside the colliding Au nuclei at RHIC energies.

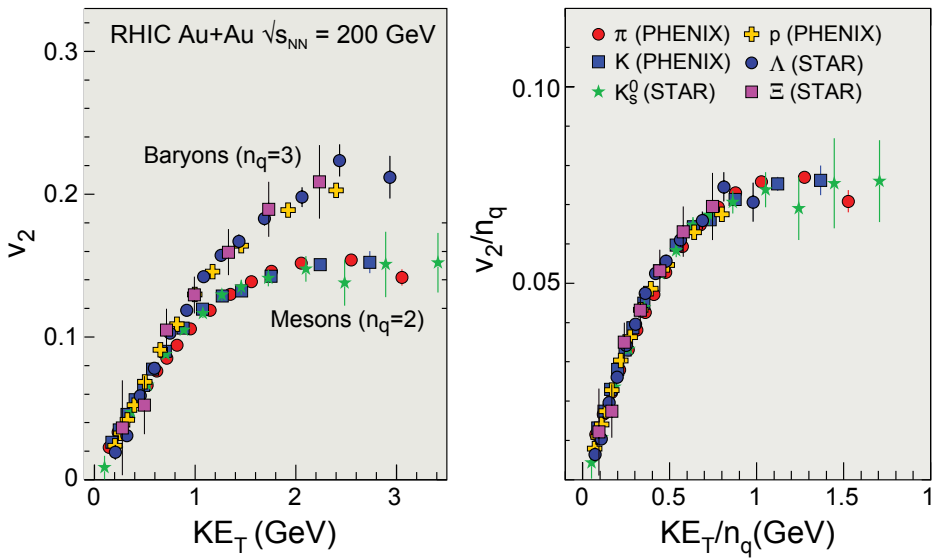

**Figure 2.11:** Scaling features of the elliptic flow parameter $v_2$ (defined in figure 2.8). The left panel shows $v_2$ for a variety of meson and baryon species plotted against kinetic energy of the particle ($KE_T$). The common behavior at low $KE_T$ is characteristic of hydrodynamic flow, while the splitting into a baryon and a meson branch at higher $KE_T$ suggests a particle production mechanism that depends primarily on the number of valence quarks in the particle. The right panel shows the same data with both axes now scaled by the number of valence quarks in each particle species. The surprisingly precise scaling of $v_2$ for a wide range of particles (differing in mass by a factor ~10) suggests that these particle distributions are driven primarily by the flow pattern of deconfined quarks.

These discoveries stand out, but many other results contribute additional important aspects to the overall picture that an equilibrated QCD medium of unprecedented energy density is formed, with novel and unexpected properties. In particular, heavy quarks are seen both to flow and to lose energy in the medium. This is notable since heavy quarks can be thought of as "boulders" in the flowing QGP stream— even they appear to be dragged along by the swift current of QGP expansion. The RHIC experiments have also confirmed important features of ultra-relativistic heavy-ion collisions seen previously at lower energies, such as the accurate characterization of all hadron abundances by a chemically equilibrated distribution with chemical freeze-out temperature of 160–170 MeV, independent of collision system and centrality. First results on charmonium production reveal a striking similarity of suppression in the medium to results at much lower beam energy, contrary to many expectations.

**Advances in Theory**

The wealth of high-quality experimental data from RHIC has motivated major advances in the theory of relativistic heavy-ion collisions and QCD matter. Quantitative modeling of nuclear collisions is based primarily on relativistic ideal fluid dynamics for the thermalized QGP phase, though technically more challenging viscous calculations are coming within reach. Microscopic hadronic Boltzmann cascades are used to model the final, more dilute and viscous hadron gas phase of the collision, which disintegrates into free-streaming hadrons. Theorists have developed sophisticated models describing the energy loss of energetic quarks and gluons and their conversion into hadrons in an expanding fireball medium. Our understanding of how the fireball thermalizes has greatly benefited from recent advances in the CGC description of the ultra-dense gluonic matter contained in the nuclei prior to the collision.

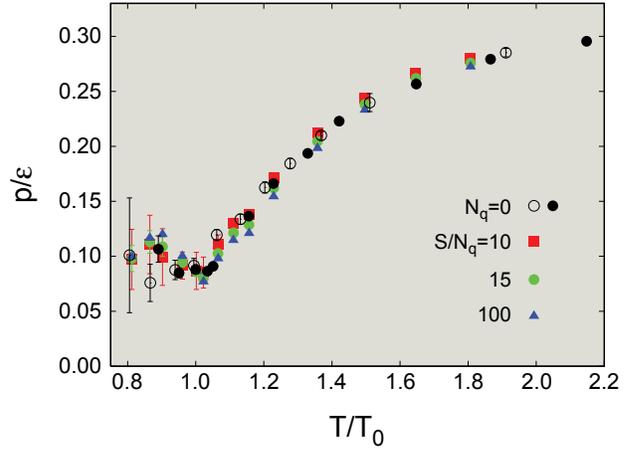

**Figure 2.12:** Lattice QCD calculation of the equation of state (EOS), or ratio of pressure to energy density, as a function of temperature. Results are shown for two different calculations for zero quark (baryon) number $N_q = 0$ and three different ratios of entropy to baryon number $S/N_q$ as indicated. The temperature is in units of the critical temperature for the quark-hadron phase transition, $T_0$. The EOS is an important input for hydrodynamic simulations of relativistic heavy-ion collisions. For a gas of weakly interacting massless quarks and gluons, or for a strongly interacting scale-invariant liquid, this quantity (closely related to the speed of sound) has the value 1/3. The small value near $T_0$ implies that the QGP at RHIC has a soft equation of state.

These phenomenological models rely on input from calculations of fundamental properties of hot QCD matter. The past five years have witnessed major improvements in the accuracy and reliability of such calculations:

- **Lattice QCD** enables accurate numerical calculation of strongly interacting bulk matter properties in thermodynamic equilibrium directly from the fundamental theory of QCD. Until recently, lattice QCD calculations were limited to equilibrium properties of systems with equal densities of quarks and antiquarks, though new techniques have extended the *ab initio* study of the phase diagram to nonzero net quark density. At zero net quark density, relevant for RHIC and the early universe, the equation of state of QCD matter has now been calculated with a controlled extrapolation to realistic quark masses (figure 2.12). Significant progress has also been made in the calculation of spectral functions, which describe the real-time response of the hot and dense medium to external perturbations. A recent example is the first computation of the electrical conductivity of the QGP. Spectral functions of charmonium, recently calculated by several groups, suggest that the ground state J/ψ (a bound state of charm and anticharm quarks) might survive up to temperatures of at least 1.6 $T_0$. These findings differ from earlier estimates based on perturbation theory and, if confirmed, would lead to a major revision of the theoretical scenarios for charmonium suppression in nuclear collisions.

- **AdS/CFT Correspondence.** The experimental discovery that the QGP produced in RHIC collisions is a strongly coupled plasma with low viscosity, not a dilute gaseous plasma, poses a challenge to theorists. While lattice QCD is the proper tool for understanding the static *equilibrium* thermodynamics of such a strongly coupled plasma, it does not allow us to calculate its *dynamical* evolution in a heavy-ion collision. A novel tool called "the AdS/CFT correspondence" (Anti de Sitter space/Conformal Field Theory), originally developed in the framework of superstring theory, has recently been brought to bear on this problem. Theorists have been able to calculate physically relevant quantities, such as the shear viscosity, the jet-quenching parameter, the drag coefficient for a heavy quark, the photon emission rate, and even the velocity dependence of the color screening length for strongly coupled plasmas associated with a variety of gauge theories that, while significantly different from QCD, nonetheless reproduce some if its key features at high temperature. These calculations have already yielded qualitatively new insights. For example, they suggest that there may exist a fundamental lower bound on the ratio of shear viscosity to entropy density. They also indicate that (heavy) quark energy loss at strong coupling may occur via sound wave emission rather than via gluon radiation. This finding may explain some of the features in the RHIC data, which are reminiscent of the Mach cone associated with supersonic motion of a projectile, in this case an energetic quark or gluon.

- **The Color Glass Condensate.** At high collision energy the production of particles with small longitudinal momentum can be thought of as the liberation of partons from the saturated gluonic matter that exists in each of the colliding nuclei. The predicted universal properties of this state, together with the large scale given by its saturation momentum $Q_s$, offer the hope for controlled, first-principle calculations of the initial state properties of the dense matter created in nuclear collisions at RHIC and LHC. This "shattering of two colliding sheets of color glass" creates most of the entropy observed in the final state, converting the initial, coherent nuclear wave functions into a disordered state in which partons move through strong remnant color fields. Like the more common electromagnetic plasmas, such a state is predicted to have severe plasma instabilities whose exponential growth leads to rapid randomization of the parton momenta, possibly explaining the rapid thermalization of QCD matter observed at RHIC.

- **Cold Dense Quark Matter.** QCD has been shown to provide rigorous analytical results, even at a nonperturbative level, for the properties of cold nuclear matter squeezed to arbitrarily high density. It has long been known that cold dense quark matter, which may occur in neutron star cores, must be a color superconductor. Recent theoretical progress has made this subject both richer and more quantitative. An analytic *ab initio* calculation of the pairing gap and critical temperature at very high density has been done, and the properties of quark matter at these densities have been determined: though color superconducting, it admits a massless "photon" and behaves as a transparent electric insulator; it is a superfluid with spontaneously broken chiral symmetry. At densities that are lower but still above that of deconfinement, color superconducting quark matter may be (in a particular sense) crystalline, with a rigidity several orders of magnitude above that of a conventional neutron star crust.

## CHALLENGES AND FUTURE OPPORTUNITIES

**Exploring the Strongly Coupled Quark-Gluon Plasma**

The most important scientific challenge for the field in the next decade is the quantitative exploration of the new state of nuclear matter discovered by the RHIC experiments. We must understand how its remarkable "perfect" liquidity emerges from its microscopic structure, and measure its physical properties with controlled accuracy: its equation of state, conductivity, speed of sound, and radiation spectrum. This section describes the experimental and theoretical path to this goal. Key future measurements in this program at RHIC will utilize rare (i.e., low cross section) probes, which rely on the enhanced detector and luminosity capabilities of the upgraded RHIC facility known as "RHIC II." The unprecedented flexibility of the RHIC accelerator complex also enables lower energy runs to search for the phase transition point where the QGP just begins to form. The phase diagram at yet higher temperature will be explored by utilizing the enormous increase in beam energy at the LHC. Theory opportunities rely on the continued advance of computational capabilities, on conceptual advances such as those described in the previous section, and on the development of a general framework for the quantitative modeling of relativistic heavy-ion collisions.

**Hard Probes: General Considerations.** Nuclear reactions at collider energies generate an important set of processes called *hard probes*, which arise from relatively rare violent scatterings of the quark and gluon constituents of the incoming nuclei. Hard probes are key diagnostic tools of the QGP, for two reasons: their production rates are calculable using the well-established techniques of perturbative QCD, and

### Quark-Gluon Plasma Meets Spintronics

In a striking example of cross-disciplinary collaboration, the confluence of condensed-matter theory and the theory of relativistic heavy-ion collisions has led physicists to propose a new class of nanoscale electronic devices.

On the nuclear physics side, the insight came from attempts to understand the large masses of neutrons and protons. The most famous constituents of these particles, the quarks, turn out to be nearly massless by themselves. So are the individual gluons that hold the quarks together. Instead, most of the nucleon mass seems to arise from the way the quarks and gluons act as a group, collectively coalescing into a "chiral condensate." The name reflects the condensate's origin in the spontaneous breaking of an approximate chiral symmetry in the equations of quantum chromodynamics. But in any case, the chiral condensate melts at the kind of high temperatures and densities achieved by the Relativistic Heavy Ion Collider at Brookhaven National Laboratory (BNL).

On the condensed matter side, meanwhile, the same kind of chiral symmetry turns out to govern the dynamics of certain exotic materials at the nanometer scale. Graphene, for example—a single layer of carbon atoms arranged in a honeycomb lattice—can produce a chiral condensate when an external magnetic field is applied.

This coincidence has now inspired physicists at BNL to investigate an application based on graphene-magnet multilayers (GMMs). The resulting U.S. Patent 60/892,595 (pending) "Graphene (Antiferro) Ferro-Magnet Multilayers" awarded to BNL scientists I. Zaliznyak, A. Tsvelik, and D. Kharzeev describes the concept of rewritable nanoscale spintronic processors and storage devices that could be possible using GMM technology.

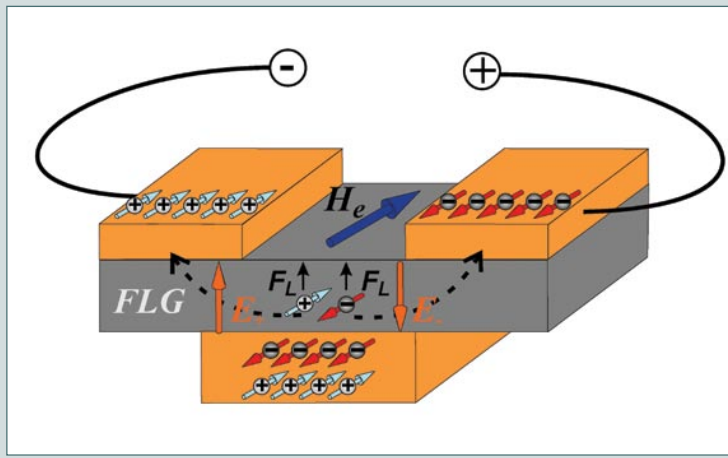

A spintronic transistor based on electric transport by polarized charge carriers in a graphene-magnet multilayer. The magnetic field created by the (anti) ferromagnet layer polarizes electrons and holes in graphene, creating spin-polarized electric current.

their sizable interactions with the hot QCD medium provide unique and sensitive measurements of its crucial properties. The most common hard probes are collimated jets of particles at large angles to the beam. Such jets are usually due to the scattering of gluons or light quarks (*up*, *down*, or *strange*); less frequently, they contain heavy *charm* or *bottom* quarks. On occasion, a high-energy prompt photon or a massive bound state of a heavy quark-antiquark pair (*quarkonium*) will be created. Each of these reaction channels plays a distinct and important role in understanding the physics of the QGP, as we now detail.

**Jet Quenching.** As discussed in the RHIC discoveries section, absorption of jets in the QGP ("jet quenching") provides a direct tomographic measurement of the gluon density of the plasma. Future jet-quenching measurements at both RHIC II and the LHC will utilize their extended ranges of jet energy and much higher statistical precision to explore the interaction of jets with the dense matter in quantitative detail. Measurements will incorporate many-particle correlations and even complete jet reconstruction—as routinely done in p + p collisions—to make multiple, complementary measurements of jet-quenching effects, turning the current qualitative conclusions into precise, quantitative statements about the gluon density, the energy loss coefficient, and other key properties of the QGP.

An important future measurement is the comparison of heavy-quark jet quenching with that of light quarks and gluons. Standard QCD theory predicts that moderate energy-heavy quark jets will experience much reduced energy loss in the QGP. In contrast, present RHIC data show substantial energy loss by heavy quarks, though only indirectly and with limited precision through measurements of inclusive electrons from heavy-quark decay. The planned RHIC detector upgrades and higher beam luminosity will yield direct measurement of high-quality heavy-quark data to elucidate the energy loss mechanism in detail.

The most precise jet-quenching measurement will utilize the QCD Compton scattering process, in which an incoming quark and gluon scatter hard to create an outgoing quark jet recoiling from an energetic photon (or at the LHC, a Z-boson). Since the photon or Z-boson almost never interacts with the dense matter, it gives a precise constraint on the jet energy and enables a highly controlled measurement of the modification of the recoiling jet by the plasma. This process is rare, however, and robust measurements of it at RHIC are only possible following the luminosity upgrade.

Attention has turned recently to the key question of how the hot QCD medium itself responds to jet energy loss. Measurements indicate that some jets lose so much of their energy that they equilibrate with the thermal medium, in the process possibly generating shockwaves or other dynamical effects. The jet shape contains crucial information on the energy loss mechanisms in the dense medium, and its measurement will test key aspects of our theoretical understanding of the interaction between hard probes and dense high-temperature QCD matter. These measurements and their interpretation are difficult, however, and such studies will benefit greatly from the high statistics provided by RHIC II and the LHC.

This many-faceted experimental program, combined with the accurate theoretical modeling needed to interpret the data, will measure essential properties of the QGP in a highly controlled fashion. The *qualitative* conclusion from the present jet-quenching measurements, of large color opaqueness of the QGP, will be turned into precise, *quantitative* statements about its essential properties.

**Quarkonium Suppression.** The hallmark of the QGP is *deconfinement*, meaning that quarks and gluons at high temperature are not confined to the interior of composite particles, such as protons and neutrons, but are able to propagate independently. Quarks and gluons carry color charge, which is analogous to the more familiar electric charge of electrons and ions. Deconfinement of quarks and gluons at high energy density, and thus the formation of a QGP, results from the neutralization (or *screening*) of their color charges, much like the electric charges of electrons and ions are screened in an ionized gas or plasma. Nature has provided a tool to study deconfinement and the screening of color charge experimentally through the measurement of particles called *quarkonia,* which consist of bound pairs of heavy quarks (charm or beauty). Screening of color charges neutralizes the attractive force binding the pair together, preventing quarkonium states from forming. Unless obscured by other effects, deconfinement is therefore signaled by the *suppression* in the measured production rate of quarkonia.

Nature has been even kinder, providing us with several different quarkonium states that probe color screening at different distance scales. Lattice QCD calculations indicate that some quarkonium states are tightly bound and survive to high plasma temperatures (J/Ψ, ϒ(1S)), while others are more loosely bound and dissociate at relatively low temperature (χ$_c$, ψ′, ϒ(3S)). The definitive demonstration of deconfinement in high-energy nuclear collisions will be the measurement of the suppression pattern of several of these states, allowing us to constrain the initial temperature in the collision to a narrow region. A good start on this program is underway at RHIC through measurements of the J/Ψ particle, which is the most copiously produced quarkonium state. Consistent with some theoretical expectations, J/Ψ suppression has indeed been observed. However, this measurement alone does not allow unambiguous interpretation. Similar measurements of other, much rarer quarkonium states are critical but are more demanding, requiring the RHIC II luminosity upgrade and significant RHIC running time.

**Uranium-Uranium collisions.** The RHIC accelerator has already achieved its maximum energy in heavy-ion collisions. Nonetheless, one can explore matter at significantly higher densities than currently generated in central Au + Au collisions by colliding Uranium beams. The acceleration of Uranium ions in RHIC becomes feasible with the new Electron Beam Ion Source (EBIS), which is under construction. Uranium nuclei have a strong elliptic deformation in their ground state, with the long axis being almost 30% longer than the short one. By using the zero-degree calorimeters and multiplicity detectors of the RHIC experiments to select central U + U collisions where the long axes of the two nuclei are aligned ("tip on tip"), a larger number of nucleons collide with each other within a smaller transverse area than in central Au + Au collisions, resulting in a 50–60% larger energy density. This may actually cover a fraction of the energy density gain made possible by colliding spherical Pb nuclei at the 30 times higher LHC energy, thus providing an important bridge across the energy gap between these two accelerators. On the other hand, by selecting central U + U collisions with the long axes of the nuclei parallel to each other ("side on side"), one generates an elliptically deformed fireball region with a transverse area and an energy density that is twice as large as in peripheral Au + Au collisions, which produce similarly deformed fireballs. These features enable the study of elliptic flow to test the perfect liquid behavior of the QGP at significantly higher energy densities than in Au + Au collisions, and the investigation of the path-length dependence of parton energy loss in a deformed QGP fireball of much larger size and density than achievable in peripheral Au + Au collisions.

**Modeling Heavy-Ion Collisions.** A heavy-ion collision at high energy is a highly dynamic process. We seek to study the fundamental new states of matter generated in such a collision, but in practice our view is distorted by the complex evolution of the fireball as it expands and cools. The art of heavy-ion experimentation is to identify those measurements that are especially sensitive to properties of the hottest, densest phase of the collision, but their interpretation in terms of the fundamental properties of matter also requires sophisticated modeling of the collision dynamics. Such modeling must provide a detailed and calibrated description of the evolution of the fireball in space and time, from the initial conditions in terms of strong color fields, through the dynamical processes leading to thermalization, to hydrodynamical expansion, and final breakup of the system into the particles seen in the detector.

The best current approach to such modeling utilizes relativistic ideal hydrodynamics augmented by hadronic Boltzmann transport processes. This has provided the basis for our current qualitative understanding of the matter generated in RHIC collisions as a strongly interacting perfect liquid, but the approximations and limitations inherent in such an approach limit its utility for precise and systematically well-controlled measurements of the medium properties. Progress in this area requires the development of three-dimensional viscous relativistic hydrodynamics, as well as detailed simulations of the propagation of hard probes through the matter and their effect on the medium. The efforts of a broad community of theorists interested in interpreting the data in terms of basic material properties, such as the equation of state, viscosity, stopping power, heavy-quark diffusion constant, and color screening length, will increasingly rely on the availability of sophisticated and validated modeling tools of this kind.

**Lattice QCD.** There are many new opportunities in lattice QCD. These include a fully controlled calculation of the equation of state and *ab initio* calculations of microscopic properties of QCD matter such as fluctuations of conserved charges, density correlations, plasma excitations, and transport coefficients. It will become feasible to map out the phase diagram of QCD at finite temperature and moderate net baryon density, and determine the location of the critical end point of the phase boundary (see below). This informa-tion will be vital for the success of a future low-energy RHIC run, as well as for the experimental program at GSI/FAIR. Detailed lattice studies near the phase transition temperature will be important benchmarks for the comparison of data from the RHIC and LHC experiments.

Lattice calculations of spectral functions, which describe the dynamic response of the equilibrated medium to external probes, are still in their infancy. To date, almost all such calculations have been done in the quenched approximation, i.e., neglecting the effect of dynamical quarks. To have quantitative relevance for RHIC phenomenology, such calculations must be performed with dynamical light quarks. This will soon become feasible due to the expected increase in computer resources such as the 100 teraflop Blue Gene supercomputer at BNL and the one petaflop Blue Gene installation at ANL. Improved calculations of the meson correlators at non zero momentum will enable quantitative estimates of transport coefficients, in particular the heavy-quark diffusion constant, and will clarify the dependence of quarkonium suppression on its velocity with respect to the plasma.

**Analytical Approaches to Strong Coupling.** In view of the paucity of analytical methods for dynamical problems in strongly coupled quantum field theories, the value of AdS/CFT calculations as a tool for gaining qualitative insight into the physics of strong coupling is already well established. To better assess the semiquantitative agreement of analytical predictions with experimental results, we need to understand which properties of strongly coupled gauge theories are universal properties, independent of "microscopic details," and thus relevant to the strongly coupled QGP. This question must be addressed by extending the AdS/CFT calculations to more observables and to more (and more QCD-like) gauge theories. If evidence accumulates that the QGP of QCD and of theories with a dual string theory description are in the same universality class, the motivation to address more challenging calculations via AdS/CFT methods will increase. A nonzero chemical potential can be added. One can envision implementing finite volumes of QGP with more and more realistic geometries, or incorporating longitudinal and radial expansion and elliptic flow. Finally, the equilibration process can be studied at strong coupling.

However, the drive for increased insight into the correspondence between AdS/CFT-motivated calculations and QCD is not only a quest for improved quantitative agreement. Just as importantly, these efforts have revealed a deep connection between gauge theory and quantum theories of gravity. As such, it opens the exciting possibility to exploit the unique features of QCD, in particular our ability to study it in both weakly coupled and strongly coupled regimes, to better understand the quantum nature of gravity and spacetime.

## EXPLORING THE QCD PHASE DIAGRAM

Lattice QCD simulations show that the transition between ordinary hadronic matter and the QGP occurs smoothly at high temperature in a matter-antimatter symmetric environment, with many thermodynamic properties changing dramatically but continuously within a narrow temperature range. In contrast, if nuclear matter is compressed to higher and higher densities without heating it up—a feat accomplished in nature within the cores of neutron stars—several discontinuous first-order transitions between various phases of nuclear matter and color superconducting quark matter are expected.

Studies, which put these insights and expectations together into a map of the QCD phase diagram, predict that the continuous crossover currently being explored in heavy-ion collisions at the highest RHIC energies will become discontinuous if the excess of matter over antimatter is larger than a certain "critical" value. This *critical point*, where the transition changes its character, is a fundamental landmark on the QCD phase diagram. While new lattice QCD methods have enabled significant progress in the past five years toward the goal of locating the QCD critical point, its precise location and even its existence remain an open and difficult theoretical problem. Definitive study of the QCD critical point requires both new theoretical developments and new experimental measurements. Its discovery in future low-energy runs at RHIC is a distinct possibility, and the FAIR facility at GSI will complete these measurements at yet larger net baryon density.

### Finding the Color Glass Condensate

Exploration of gluon saturation and the CGC requires probes that are sensitive to the low-momentum gluons inside a hadron or nucleus. The predicted novel and universal properties of the CGC manifest themselves at momenta below the saturation momentum scale $Q_s^2 = Q_0^2(A/x)^{1/3}$, which grows with increasing mass $A$ of the nucleus and decreasing fraction $x$ of the longitudinal momentum carried by the gluons, and where $Q_0$ is the overall momentum scale to be determined

# Search for the Critical Point: "A Landmark Study"

When ordinary substances are subjected to variations in temperature or pressure, they will often undergo a *phase transition*: a physical change from one state to another. At normal atmospheric pressure, for example, water suddenly changes from liquid to vapor as its temperature is raised past 100° C; in a word, it boils. Water also boils if the temperature is held fixed and the pressure is lowered—at high altitude, say. The boundary between liquid and vapor for any given substance can be plotted as a curve in its *phase diagram*, a graph of temperature versus pressure. Another curve traces the boundary between solid and liquid. And depending on the substance, still other curves may trace more exotic phase transitions. (Such a phase diagram may also require more exotic variables, as in the figure).

One striking fact made apparent by the phase diagram is that the liquid-vapor curve can come to an end. Beyond this "critical point," the sharp distinction between liquid and vapor is lost, and the transition becomes continuous. The location of this critical point and the phase boundaries represent two of the most fundamental characteristics of any substance. The critical point of water, for example, lies at 374° C and 218 times normal atmospheric pressure.

The schematic phase diagram shown in the figure shows the different phases of nuclear matter predicted for various combinations of temperature and baryon chemical potential. The baryon chemical potential determines the energy required to add or remove a baryon at fixed pressure and temperature. It reflects the net baryon density of the matter, in a similar way as the temperature can be thought to determine its energy density from microscopic kinetic motion. At small chemical potential (corresponding to small net baryon density) and high temperatures, one obtains the quark-gluon plasma phase; a phase explored by the early universe during the first few microseconds after the Big Bang. At low temperatures and high baryon density, such as those encountered in the core of neutron stars, the predictions call for *color-superconducting* phases. The phase transition between a quark-gluon plasma and a gas of ordinary hadrons seems to be continuous for small chemical potential (the dashed line in the figure). However, model studies suggest that a critical point appears at higher values of the potential, beyond which the boundary between these phases becomes a sharp line (solid line in the figure). Experimentally verifying the location of these fundamental "landmarks" is central to a quantitative understanding of the nuclear matter phase diagram.

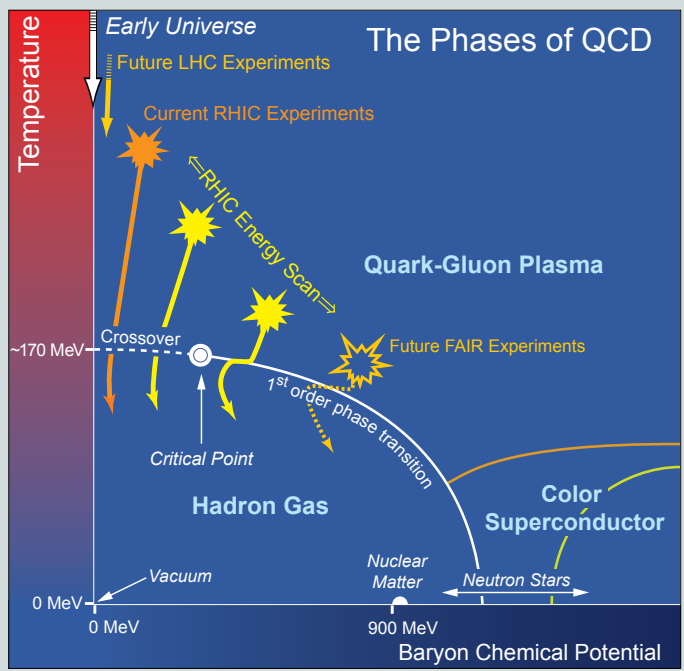

Schematic QCD phase diagram for nuclear matter. The solid lines show the phase boundaries for the indicated phases. The solid circle depicts the critical point. Possible trajectories for systems created in the QGP phase at different accelerator facilities are also shown.

Theoretical predictions of the location of the critical point and the phase boundaries are still uncertain. However, several pioneering lattice QCD calculations have indicated that the critical point is located within the range of temperatures and chemical potentials accessible with the current RHIC facility, with the envisioned RHIC II accelerator upgrade, and at existing and future facilities in Europe (i.e., the CERN SPS and the GSI FAIR). Indeed, the recent discovery of the quark-gluon plasma at RHIC gives evidence for the expected continuous transition (dashed line in the figure) from plasma to hadron gas. Physicists are now eagerly anticipating further experiments in which nuclear matter will be prepared with a broad range of chemical potentials and temperatures, so as to explore the critical point and the phase boundary fully. As the experiments close in, for example, the researchers expect the critical point to announce itself through large-scale fluctuations in several observables. These required inputs will be achieved by heavy-ion collisions spanning a broad range of collision energies at RHIC, RHIC II, the CERN SPS and the FAIR at GSI.

The large range of temperatures and chemical potentials possible at RHIC and RHIC II, along with important technical advantages provided by a collider coupled with advanced detectors, give RHIC scientists excellent opportunity for discovery of the critical point and the associated phase boundaries.

by experiment. Gluon saturation effects are therefore most prominent in probes of heavy nuclei at small $x$, corresponding to particle production at forward angles. Saturation effects also increase with collision energy and are expected to manifest themselves in the central region of heavy-ion collisions at the LHC in a similar way as at forward angles at RHIC.

Tantalizing evidence for CGC effects in current data is discussed in the RHIC discoveries section. These findings have motivated a strong research program in forward angle physics, including detector upgrades to STAR and PHENIX. A central focus of the upgraded RHIC program, and ultimately of a future Electron-Ion Collider (see EIC section), will be the determination of the scale at which saturation effects become important, a quantitative understanding of the gluon density at small $x$ in nucleons and nuclei, and verification of the predicted universal properties of the CGC.

**The Energy Frontier.** Exploration of the high-temperature, high-entropy density domain of the QCD phase diagram requires heavy-ion collisions at higher energies. At 30 times the energy of RHIC, lead-lead collisions at the LHC are expected to create a rapidly thermalized QGP at 15–20 times higher initial energy density and twice the initial temperature (i.e., 4–5 times the critical temperature for color deconfinement). This provides a significant lever arm for probing the QGP's viscosity and other transport properties as they gradually evolve in the phase diagram from the strongly coupled "perfect" liquid studied at RHIC and RHIC II toward the more weakly interacting, gaseous plasma state expected for asymptotically high temperatures. The large collision energies at the LHC will produce copious rates of energetic hard probes of hot QCD matter, greatly improving the kinematic reach for tomographic studies with "gold-plated" signatures, such as correlations between direct photons or Z-bosons and hadronic jets. Many jets produced at the LHC are so energetic that they clearly stand out from the huge background of soft hadrons arising from the decay of the QGP, allowing full jet reconstruction and a detailed exploration of jet shape and structure over a substantial kinematic range. Such measurements will provide critical tests of our theoretical understanding of the interaction between hard probes and the higher-temperature QCD matter produced at the LHC.

## OUTLOOK

To realize the compelling scientific opportunities described above for the RHIC program, upgrades are required to extend the reach of the RHIC detectors (PHENIX and STAR) and to achieve a 10-fold increase in RHIC luminosity. Details about the machine upgrade, including a new Electron Beam Ion Source, can be found in the section on Facilities. Support is also essential for theoretical studies of QCD matter, including finite-temperature and finite-baryon-density lattice QCD calculations, ongoing phenomenological modeling, and new initiatives. Strategic investments in LHC detector instrumentation and computing, which leverage the capabilities of the existing LHC detectors, are required for significant and timely U.S. participation in the LHC heavy-ion program. Collectively, these investments will result in a dramatic advance in our understanding through quantitative comparison of theory and experiment. The synergies of RHIC and the LHC will lead to a much deeper understanding of the properties and dynamics of dense QCD matter.

### RHIC Detector Upgrades

Ongoing and planned detector upgrades will increase the PHENIX acceptance for tracking and calorimetry, increase the rate capability of STAR, and in both detectors provide high-resolution vertex detection for charm and beauty, as well as enhanced tracking, calorimetry, and trigger capability at forward rapidity.

New capabilities for the event-wise identification of particles containing charm and bottom quarks will allow full exploration of the interaction of heavy quarks and their bound states with the QGP. Dramatic improvements in the measurement of low mass $e^\pm$ pairs will provide new capability for the study of thermal radiation from the plasma and medium-induced modification of mesons, a predicted signature of chiral symmetry restoration. Upgraded particle identification capabilities will enable measurements of particle spectra, resonances, and correlations and fluctuations, at a new level of precision. These latter measurements are crucial for a comprehensive search for a critical point in the phase diagram of QCD matter. Forward detector upgrades will provide greatly improved understanding of the initial state of a heavy-ion collision. Significantly increased data acquisition bandwidth will result in a dramatic improvement in effective utilization of integrated luminosity, extending RHIC's scientific reach for the measurement of rare and hard probes.

Additional upgrades to PHENIX and STAR are targeted specifically at studies of nucleon spin.

**RHIC Luminosity Upgrade (RHIC II)**

RHIC will remain the world's most flexible facility for the study of heavy-ion collisions. Its peak luminosity has already exceeded the design target, but many key measurements require sensitivity to processes that occur at rates below one per hundred million Au + Au reactions; robust measurement at this level remains out of reach at the present facility (see representative examples in figure 2.13). The RHIC II luminosity upgrade will enable the study of

- multiple quarkonium states and their suppression patterns, to explore color interactions in the QGP and measurements of quarkonium generation at hadronization;

- gamma-jet tomography, to pin down partonic energy loss and the response of the plasma to this deposited energy;

- suppression and flow studies of particles containing heavy quarks, to determine their relaxation time in the plasma and measure the plasma viscosity;

- spectra of many rare hadron species, to study how the plasma hadronizes and to quantify viscous effects in the late-stage hot hadron gas;

- very high-$p_T$ hadron production and correlations, to quantify the plasma transport parameters;

- high-$p_T$ and multihadron correlations, to quantify how the plasma affects jet structure and shape; and

- statistically precise measurements of collective excitations of the plasma, to allow extraction of medium properties, such as the speed of sound.

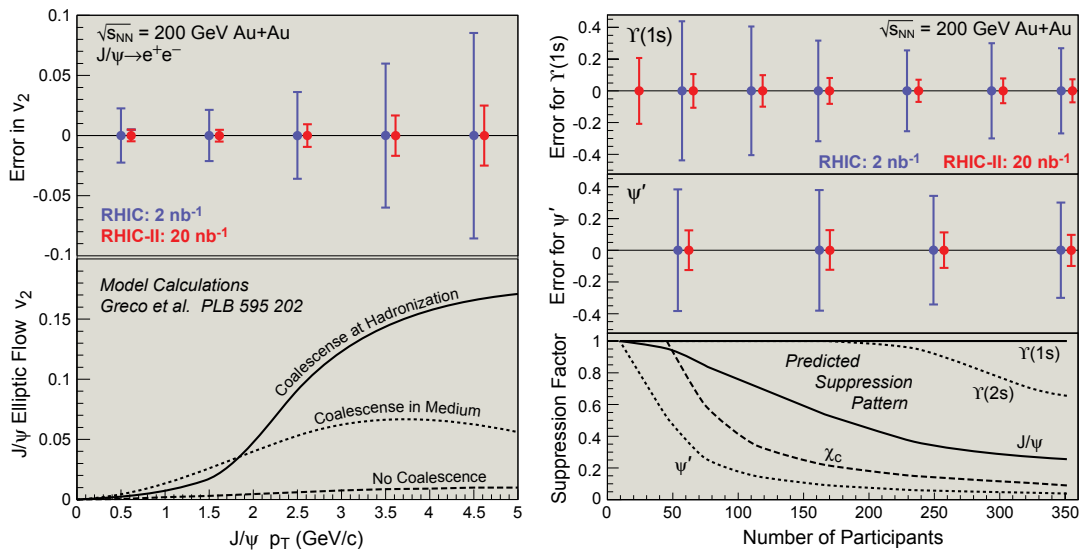

**Figure 2.13:** Quarkonium suppression and flow at RHIC I and RHIC II. The figure shows theoretical predictions for crucial quarkonium measurements at RHIC, together with projected capabilities for the most challenging measurement channels before ("RHIC," in blue) and after the luminosity upgrade ("RHIC II," in red). On the left, the J/Ψ elliptic flow parameter $v_2$ (defined in figure 2.8) is seen to be sensitive to the detailed mechanism of J/Ψ formation from deconfined, uncorrelated charm quarks in the fireball. Very late formation, at the time of breakup of the fireball ("coalescence at hadronization"), generates large J/Ψ $v_2$ due to flow of the charm quarks and can be clearly distinguished from other scenarios only at RHIC II. The right side shows the expected suppression factor $R_{AA}$ for several quarkonium states, arising from their different binding strengths in hot matter. The widely differing behavior of different states enables a precise measurement of the temperature of the hot medium. The experimental projections for the two extreme cases (ψ′ and Υ(1s)) show that sufficient precision is achievable only following the RHIC II luminosity upgrade.

48    The Phases of Nuclear Matter

To fully exploit the upgraded capabilities of the detectors, the RHIC accelerator itself must therefore be upgraded to provide substantially higher luminosity.

Electron cooling in the AGS to increase the RHIC luminosity by a factor of up to 30 at low energies (below ~20 GeV) is also being explored. It would greatly facilitate the search at RHIC for the existence and location of a critical point on the QCD phase diagram, affording maximal flexibility in varying the baryon chemical potential by providing robust luminosity as far down as $\sqrt{s_{NN}}$=5 GeV. This capability would enable a comprehensive energy scan, for instance, to search for critical event-by-event fluctuations near the QCD critical point.

**Heavy Ions at the LHC**

The LHC physics program includes four weeks of heavy-ion physics running per year. The primary collision system for ion studies will be Pb + Pb at 5.5 TeV/nucleon pair, a 30 times greater collision energy than at RHIC. The LHC is currently expected to begin commissioning with proton collisions in mid-2008, with heavy-ion beams commissioned in late 2008 and the first significant heavy-ion running in 2009. The higher collision energy of ion collisions at the LHC is expected to generate deconfined matter characterized by a longer lifetime and much higher initial energy density than at RHIC. It will also result in high rates for a wide variety of hard probes over a very broad kinematic range.

The simultaneous operation of heavy-ion experiments at RHIC and LHC offers an unprecedented opportunity to understand QCD matter in great depth by studying, for example, the same jet-quenching or elliptic-flow measurements on physical systems evolving from vastly different initial states. The comparison of RHIC and LHC measurements promises to give deep insights into the nature of the experimental probes and their interactions with the medium, and consequently the hot QCD medium itself. Many key QCD properties depend logarithmically on energy; a factor 30 in collision energy between RHIC and LHC thus gives important advantage for probing such predicted dependences quantitatively.

Three LHC experiments (ALICE, ATLAS, and CMS) will participate in heavy-ion running, with U.S.-based scientists playing leading roles in the programs of ALICE and CMS. While each detector has its specific strengths, they all have good capabilities for heavy-ion jet quenching and photon and quarkonium production measurements, with di-muon mass resolution sufficient to separate the various quarkonium states. Collectively, they provide the necessary capability for new discoveries and comprehensive comparison of the vastly different initial and final state systems expected at RHIC and the LHC.

**Theory**

The primary goal of the RHIC scientific program in the coming years is to progress from qualitative statements to rigorous quantitative conclusions. Quantitative conclusions require sophisticated modeling of relativistic heavy-ion collisions and rigorous comparison of such models with data of greater precision and extended reach. A successful quantitative interpretation of the heavy-ion data will require close collaboration of the experimental data analysis with the theoretical modeling effort. Without such an effort, the RHIC physics program cannot be successfully completed, and the synergies from the parallel LHC heavy-ion program cannot be adequately brought to bear on the physics program of RHIC.

Thus, an essential requirement for the field as a whole is strong support for the ongoing theoretical studies of QCD matter, including finite temperature and finite baryon density lattice QCD studies and phenomenological modeling, and an increase of funding to support new initiatives enabled by experimental and theoretical breakthroughs. The success of this effort mandates significant additional investment in theoretical resources in terms of focused collaborative initiatives, both programmatic and community oriented.

# The Emerging QCD Frontier: The Electron-Ion Collider

## A NEW EXPERIMENTAL QUEST TO STUDY THE GLUE THAT BINDS US ALL

Without gluons, matter as we know it would not exist. Protons, neutrons, the other strongly interacting particles, atomic nuclei, atoms, and everything made of atoms—all of it would simply disintegrate. Gluons are what keep that from happening, by forcing the quarks to stay tightly bound to one another instead of roaming freely. As described in the theory of QCD, gluons collectively create a binding force that acts on a quark's "color" charge, in much the same way that photons collectively create an electromagnetic force that acts on a particle's electric charge. But unlike photons, which are electrically neutral, gluons have a color charge of their own—which means that they interact among themselves as well as with the quarks.

While these self-interactions make gluons the dominant constituents of matter, they also make the QCD equations extremely difficult to solve. However, recent breakthroughs indicate that systematic solutions may be feasible when the gluons collectively behave like a very strong classical field—that is, when their density is sufficiently high and their color coupling is sufficiently weak. And if that is indeed the case, these developments lead to a striking prediction: all hadrons and all nuclei should appear the same when viewed at high enough energy. In effect, hadrons should begin to look like a universal form of gluonic matter, known as the *color glass condensate* (CGC). Inside this condensate, moreover, the density of gluons would be *saturated*, meaning that the rate for one gluon to split into two would precisely balance the rate of recombination: two gluons merging into one. Indeed, this saturation mechanism is critical to avoid a theoretical catastrophe: high-energy scattering probabilities would otherwise exceed 100%.

Despite the dominant role played by gluons, their properties in matter remain largely unexplored. Physicists have been able to address certain aspects of gluon behavior at RHIC and CEBAF, which are currently the world's premier accelerator facilities focused on QCD, and HERA, which terminated operation this year. In particular, they have found hints of saturated gluon densities in measurements of electron-proton collisions at HERA, and of deuteron-nucleus and nucleus-nucleus collisions at RHIC. Researchers also expect saturation to have a profound influence on heavy-ion collisions at the LHC. But getting to the heart of the matter—unveiling the collective behavior of dense assemblies of gluons under conditions where their self-interactions dominate—will require an Electron-Ion Collider (EIC): a new facility with capabilities well beyond those of any existing accelerator.

The Electron-Ion Collider's usefulness as a gluon microscope is somewhat counter-intuitive since electrons do not directly interact with gluons. However, a high-energy electron passing through a nucleus will interact with individual electrically charged quarks, in a process known as deep-inelastic scattering (DIS). The struck quark can then emit a gluon. Or perhaps the quark could have been created from a gluon in the first place, as part of a quark-antiquark pair. Either way, the presence of the gluon will modify the precisely understood electromagnetic interaction of the electron and quark in ways that allow us to infer the gluon properties. The need to make this inference precise defines the basic specifications of an EIC. It should create collisions at *high* energy since that increases the resolving power of the "microscope." It should allow for *variable* energy so as to study deep-inelastic events over a broad range of energies and scattering angles. And it should utilize *beams of heavy nuclei*, since the density of gluons inside a hefty nucleus is considerably enhanced over the density in a single proton.

In addition, an ability to collide *spin-polarized proton and light-ion beams* with *polarized electrons and positrons* would give the EIC unprecedented access to the spatial and spin structure of protons and neutrons in the gluon-dominated region. When compared to present-generation experiments with fixed polarized targets at CEBAF, HERA, CERN, and SLAC, such a capability would allow the EIC to greatly extend the kinematic reach and precision of polarized deep-inelastic measurements. And that capability would likewise yield tomographic images of the nucleon's internal landscape going far beyond the valence quark region, which will be probed with 11 GeV electron beams at CEBAF. These extensions will be critical to completing our picture of the origin of the nucleon's spin, by providing new sensitivity to the orbital motion of sea quarks and to the spin preferences of very "soft" gluons—those that have low momentum inside the nucleon but are highly abundant.

In sum, the results and measurements emerging from the world's existing QCD laboratories consistently show that the physics accessible at an EIC is the next QCD frontier. EIC measurements will be critical in answering one of the over-arching questions for our field: **what is the role of gluons and gluon self-interactions in nucleons and nuclei?** Self-interacting force carriers exist in interaction theories other than QCD (e.g., for the weak force and various hypothesized forces beyond the Standard Model), but without comparable opportunities to discover unique manifestations of those self-interactions in matter. EIC experiments would also yield missing links needed to fully answer two other overarching questions: **What is the internal landscape of the nucleons? What governs the transition of quarks and gluons into pions and nucleons?**

## THE ELECTRON-ION COLLIDER

The detailed requirements for the machine complex and detectors at an EIC are driven by the need to access the relevant kinematic region that will allow us to explore gluon saturation phenomena and image the gluons in the nucleon and nuclei with great precision. These considerations constrain the basic design parameters to be a 3 to at least 10 GeV energy electron beam colliding with a nucleon beam of energy between 25 and 250 GeV or with nuclear beams ranging from 20 to 100 GeV/nucleon.

While the HERA collider at DESY provided electron-proton collisions at even higher energies, the performance needed at an EIC relies on three major advances over HERA: (1) beams of heavy nuclei, at least up to gold, are essential to access the gluon saturation regime under conditions of sufficiently weak QCD coupling, and to test the universality of the CGC; (2) collision rates exceeding those at HERA by at least two orders of magnitude are required for precise and definitive measurements of the gluon distributions of interest, especially with the complete outgoing particle detection needed for tomographic imaging; and (3) polarized light-ion beams, in addition to the polarized electrons available at HERA, are mandatory to address the central question of the nucleon's spin structure in the gluon-dominated region.

A new EIC facility will require the design and construction of new optimized detectors profiting from the experience gained from those operated at DESY. A central collider detector providing momentum and energy measurements and particle identification for both leptons and hadrons will be essential. Special-purpose detectors that provide extreme forward/backward coverage are also required.

There are currently two complementary concepts in the United States to realize an EIC: eRHIC, which calls for the construction of a new electron beam to collide with the existing RHIC ion beam; and ELIC, which calls for the construction of a new ion beam to collide with the upgraded CEBAF accelerator beam. Both rely on new accelerator and detector technology, and on an allocation of suitable research and development resources for their development. Over the next five years we anticipate sufficient research and development to allow for continued improvements in all key areas. Meanwhile, to ensure that any future facility incorporates the capabilities best suited to answer the most critical scientific questions, there is a need for continuing progress on the theoretical treatment of saturated gluonic matter.

## THE PHYSICS PROGRAM AT AN ELECTRON-ION COLLIDER

### Saturation of Gluon Densities and the Color Glass Condensate

A major discovery of the last decade is the dominant role of gluons in nucleons that are viewed by a high-energy probe. The density of gluons grows extremely rapidly as one probes gluons carrying progressively smaller fractions $x$ of a proton's overall momentum. This growth reflects a QCD cascade in which higher-momentum (harder) parent gluons successively split into two or more lower-momentum (softer) daughter gluons. But this growth cannot continue unabated without eventually violating fundamental rules of physics. Physicists expect the growth to be tamed because at sufficiently high gluon densities softer gluons will again recombine into harder ones. The competition between the splitting and recombination processes should lead to a saturation of gluon densities at small $x$. Recent theoretical breakthroughs suggest that this saturation regime is characterized by new physics that can be best studied at an EIC.

The onset of saturation depends not only on the gluon's momentum fraction $x$, but also on the spatial resolution with which the nucleon is probed. Physicists can select that resolution by restricting their attention to partons that have been given a particular momentum kick $Q$ by the beam electrons. As $Q$ decreases, the resolution of structure details in the plane transverse to the beam direction gets coarser. This increases the sensitivity to the recombination of gluons. This sensitivity also increases when the target nucleons are contained within a heavy nucleus, where many closely spaced nucleons along the beam path can conspire to contribute gluons to the recombination cause, amplifying the gluon density. Combining these two effects yields a predicted *saturation scale*: $Q_s^2 \propto (A/x)^{1/3}$. As the formula shows, this scale grows with increasing nuclear mass number A and with decreasing gluon momentum fraction $x$ (see figure 2.14). When the probe has much finer spatial resolution, $Q^2 \gg Q_s^2$, the physics is described by the well-studied linear regime of QCD, dominated by gluon splitting. But as $Q^2$ decreases below $Q_s^2$ one enters the novel nonlinear regime of saturated gluonic matter.

Gluons are among the class of particles known as *bosons* that are permitted by quantum physics to have more than one particle in the same state. In fact, the occupancy of gluons in the saturation regime is proportional to the inverse of the QCD coupling strength. When $Q_s^2$ is large, this coupling is weak, and therefore the occupancy is large and the behavior of the gluon ensemble is nearly classical. In this limit, theorists predict that one should gain access to a fascinating facet of *all* matter: a dense swarm of gluons, wherein the individual gluons interact weakly but collectively form a coherent classical color field whose intensity may be the *strongest allowed in nature*! This predicted universal facet of matter has much in common with Bose-Einstein condensates (whose study led to the 2001 Nobel Prize in Physics) and with glassy materials, and has thus been labeled the CGC.

To reach CGC conditions one needs a suitable combination of high beam energy (small $x$) and heavy-ion (large A) beams to collide with the probing electrons. A hypothetical observer co-moving with the electron will see a tidal wave of gluons from protons and neutrons along the entire depth of the nucleus. Using heavy nuclei as an *amplifier* of gluon densities allows us to reach this regime with 10 times lower beam energy than would be needed in an electron-proton collider.

The regimes probed in atomic nuclei at high energies are shown in figure 2.14 as a function of the resolving momentum $Q^2$ of the probe, the gluon momentum fraction $x$, and the atomic number A. As suggested by the figure, the CGC regime is expected to be universal across all atomic nuclei and for resolving momenta below the saturation scale. The use of large nuclei and high energies together opens a wide window to explore, with unmatched precision, this novel regime. The figure also shows that the CGC regime of strong gluon fields lies adjacent to the confining regime of QCD, where the interactions among individual gluons grow very strong. The transition between the two regimes may hold the key

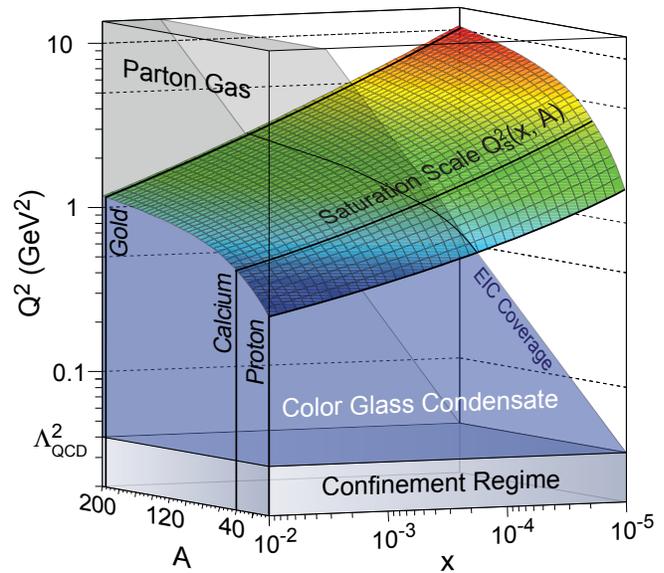

**Figure 2.14:** Regimes of hadronic matter in QCD at high energies, as a function of the resolving momentum transfer of the electron, $Q^2$, the relative momentum fraction of the gluon in the nucleus, $x$, and the atomic number of the nuclei probed, A. The multicolored surface indicates the saturation scale $Q_s^2$ as a function of $x$ and A. The saturation regime ($Q^2 < Q_s^2$) that is kinematically accessible by the EIC is depicted in blue. For very large $Q^2$ matter exists in the form of a parton gas, while the confining regime of QCD is characterized by very low $Q^2$ values. The higher $Q_s^2$ (large A), the larger the accessible $Q^2$ range of the Color Glass Condensate and the more nearly classical its expected behavior. Acceptance at $Q \approx Q_s$ is not essential; coverage of the region at $Q < Q_s$ is. The beam energies of the EIC would allow robust study of the saturation regime for gluons carrying as little as 0.01% of the nucleus' momentum.

to understanding the confining dynamics at the heart of all matter.

Is there, in fact, a universal saturation scale, as suggested by QCD theory? Are the properties of the gluon-dominated matter beyond this scale truly the same in neutrons, protons, and all nuclei? Do such universal properties explain remarkable patterns in particle production via strong interactions that have so far defied quantitative understanding? The discovery of the CGC would represent a major breakthrough in our understanding of QCD and nuclear matter. We turn next to the EIC measurement programs key to such a discovery.

**A Precise Image of Gluons in Nucleons and Nuclei**

The energy dependence of the high energy scattering process of electrons from nucleons and nuclei renders an indirect but precise measure of gluons through the evolution of quark properties with energy scale. Application of this technique established the massive wall of gluons found in HERA data when one probes protons at low x. However, at the moderate resolutions and very small gluon momenta of the saturation region, the situation is much less understood. The conventional linear QCD analysis of energy evolution is expected to break down here, and its application to HERA data yields puzzling results.

In nuclei, the region of gluons for *x* < 0.01 is *terra incognita*, and measurements at an EIC will be the first. Figure 2.15 shows projections of EIC measurements of the ratio of gluon distributions for a lead nucleus relative to that for deuterium. The saturated gluon densities associated with the CGC and those anticipated in linear QCD approaches (labeled as HKM and FGS here) are shown. By combining measurements such as those in figure 2.15 for a variety of nuclear beams with different masses A, one can search directly for a universal scaling behavior that collapses all the data onto a single curve as a function of $Q^2/Q_s^2$. Such scaling would demonstrate the universality of the CGC and directly determine how the saturation scale depends on *x* and A.

The deep-inelastic scattering measurements determine how the gluons are distributed in momentum inside nuclei. It is also important to see how they are distributed in space; e.g., are the conditions for saturation achieved first in localized regions within the nucleus? The spatial distribution can be probed when the scattering electron produces a single high-energy photon, or a meson with the same quantum numbers as the photon, without exciting the target nucleus. The results would provide insight into the role of strong color fields in binding nuclei and in more complex processes of multiple particle production.

Strong gluon fields also enhance the probability for a high-energy probe to interact gently with the nucleus by exchanging a colorless particle comprising multiple gluons—a particle sometimes called the Pomeron. Indeed, experiments at the HERA collider produced the striking discovery that a proton at rest remains intact in roughly one-seventh of its collisions with electrons of energy roughly 25,000 times the proton rest mass. Several models suggest that the amplified gluon fields in heavy nuclei may cause the nucleus to remain intact nearly 30–40% of the time in EIC collisions. This would be truly astounding if confirmed experimentally at an EIC. Measurements for such events place strong demands on EIC detector design.

**The Spin Structure of the Proton**

Deep-inelastic scattering of high-energy electrons and muons from fixed polarized targets have shown that spin alignment of quarks and antiquarks surprisingly accounts for only about 30% of the nucleon's spin. The remainder must be due to the spin of the gluons and/or the orbital motion of the constituents (quarks and gluons). The EIC, with its unique high luminosity, highly polarized electron and nucleon capabilities, and its extensive energy span, will allow access to quark and gluon spin contributions at substantially lower momentum fractions *x* than current and forthcoming experiments at BNL, DESY, CERN, and JLAB.

We currently know much less about the spin preferences than about the number density of low-momentum gluons in a nucleon because there has been no polarized electron-polarized nucleon collider. EIC deep-inelastic scattering measurements would rectify the situation, expanding the kinematic coverage as indicated in figure 2.16. The insert shows quantitative projections for EIC results in comparison with model predictions that make different assumptions regarding the sign and magnitude of the gluon spin contribution to the proton spin. Data from polarized proton collisions at RHIC are now beginning to establish preferences among these particular four models at *x* > 0.03 but will not be able to constrain the shape of the gluon spin distribution at lower *x*, where the density of gluons rapidly increases.

With polarized deuteron or $^3$He beams at an EIC, electrons could also be scattered by polarized neutrons, allowing a precision test of the fundamental Bjorken sum rule, which relates the proton and neutron spin structure via the axial weak coupling strength measured in neutron beta decay. Corrections to this relation are known to higher order in the strong coupling constant $\alpha_s(Q^2)$ than any other quantity in QCD. This measurement could very well provide the best determination of $\alpha_s(Q^2)$ away from the CERN measurements at the $Z$-particle mass and thus constrain the evolution of the strong force with energy to high precision.

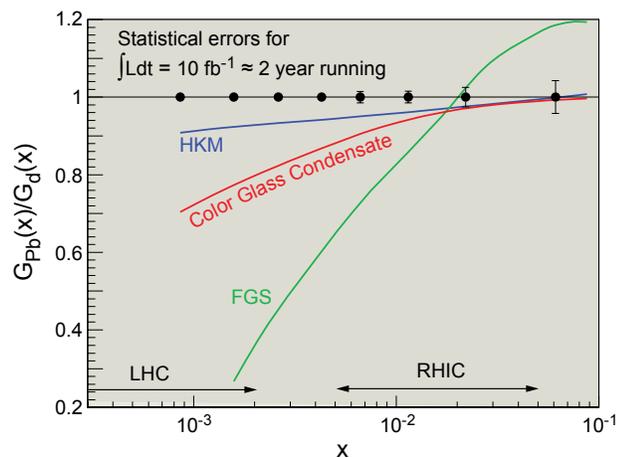

**Figure 2.15:** The ratio of gluon distributions in lead relative to deuterium as determined from projected measurements with an EIC, as a function of gluon momentum fraction *x*. HKM and FGS represent QCD parameterizations of existing data extrapolated linearly to small *x*. The curve labeled "Color Glass Condensate" is a saturation model prediction. Domains relevant to nucleus-nucleus collisions at RHIC and the LHC are shown.

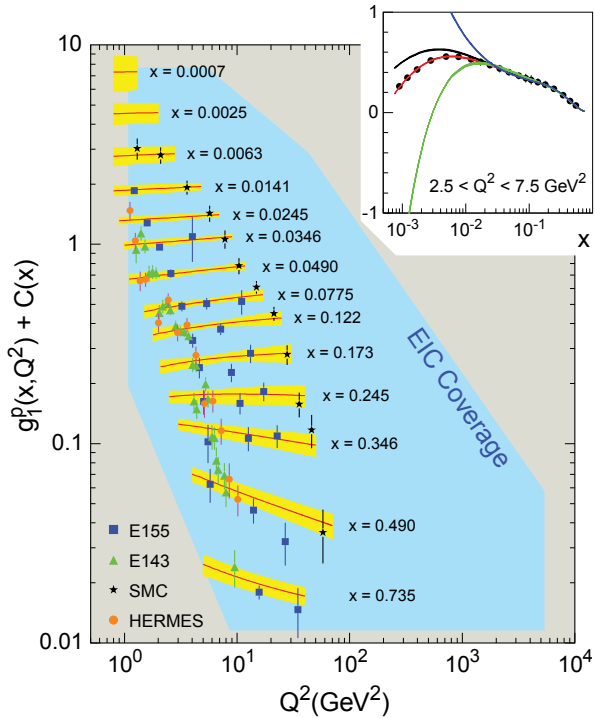

**Figure 2.16:** The world database of polarized deep-inelastic scattering results for the proton, from SLAC experiments E143 and E155, CERN-SMC, and DESY-HERMES. The curves (and error bands) are from a global QCD fit by Boetcher and Bluemlein. The blue-shaded area represents the enlarged ($x$,$Q^2$) area accessible by an EIC. The insert shows one example of the consequent physics reach, comparing projected data as a function of $x$ in one $Q^2$ bin for about one year of EIC running with theoretical predictions based on global QCD fits to the present world data under differing assumptions for gluon spin preferences.

Various avenues have emerged to investigate the contribution of orbital angular momentum to the proton spin. One of them is the study of correlations of the transverse momentum of a quark in the nucleon with the nucleon spin transverse to its momentum. Such correlations have been found to cause surprisingly large left-right asymmetries for the final-state hadrons produced in the scattering process. Measurements at an EIC would allow quantitative investigation of how quarks and antiquarks of different flavor are orbiting in protons and neutrons. An alternative approach will utilize reactions to probe GPDs, to which we turn next.

**The Complete Image: Measurements of Generalized Parton Distributions**

GPDs may be viewed as functions describing distributions of a nucleon's constituents in position and momentum simultaneously. They represent the closest analog to a classical phase space density allowed by the uncertainty principle. The concept of GPDs has revolutionized the way scientists visualize nucleon structure, in the form of either two-dimensional tomographic images (analogous to CT scans in medical imaging) or genuinely six-dimensional space-momentum images.

Measurements of GPDs are possible in select high-energy processes. The experimental study of these rare processes requires high luminosities. While initial maps of GPDs in the valence-quark region will be carried out with the 12 GeV CEBAF Upgrade, the EIC would allow unique access to the gluon, sea quark, and antiquark GPDs, through study of reactions in which an electron scattering also produces a single photon or hadron (e.g., pion, kaon, or $J/\Psi$ particle). The production probabilities encode information about the spatial distribution of the nucleon's constituents. Precise EIC measurements over a wide kinematic range would probe whether gluons are clumped in the nucleon or are uniformly distributed. Such spatial gluon distributions could be measured for varying gluon momenta. In a nucleus, this could be correlated with the hunt for gluon saturation in localized spatial regions. From the trends in measured GPDs, one could invoke the so-called Ji sum rule to infer the orbital contributions of sea quarks and antiquarks to the nucleon spin, thus testing nucleon structure models in which a proton spends part of its time as a fluctuation into a neutron and an orbiting positively charged pion.

**The Formation of Hadronic Final States from Light Quarks and Massless Gluons**

High-energy scattering from nucleons in a collider environment lends itself specifically to study how the creation of matter from energy is realized in QCD when an essentially massless (but colored) quark or gluon evolves into massive (and color-neutral) hadrons. We have known since the work of Einstein that matter can be created out of pure energy, a concept that is at the root of modern physics, but full under-standing of this process remains lacking. The availability of a high-luminosity *polarized* Electron-Ion Collider, using high-efficiency detectors with good particle identification, will facilitate experiments to measure new features of this *fragmentation* process, such as its dependence on quark flavor, spin orientation, and motion, and on passage through nuclear matter. It is already apparent now, from measurements in electron-positron collisions and in fixed-target electron scattering, that correlations exist between the momentum components of hadron fragments transverse to the jet axis and any quark spin preference transverse to its momentum. In addition to systematic exploration of these initial hints at EIC, it may be possible for selected final-state hadrons—e.g., ρ-mesons—reconstructed from their decay daughters to correlate their own spin preferences with the spin orientation of the fragmenting quark. Such measurements are likely to launch a new stage in modeling how quarks accrete colored partners from the vacuum or the debris of the high-energy collision to form colorless hadrons.

**Connection to Other Fields**

EIC measurements will provide critical input to two dramatic aspects of heavy-ion collisions at RHIC and at the LHC. The strong flow observed suggests that the hot fireball formed in a central heavy-ion collisions thermalizes into a nearly perfect liquid in the remarkably short time interval of $\sim 3 \times 10^{-24}$ seconds! Does this rapid equilibration result from "shattering" the CGC in the incident nuclei? EIC measurements of the saturation scale as a function of energy, nuclear size, and position within the nucleus will test this idea further against data at both RHIC and LHC. The observed strong suppression of high-momentum hadrons in the final state of RHIC collisions has been interpreted as evidence of the rapid loss of energy of quarks and gluons traversing the hot, gluon-rich matter. However, recent observations of the strong attenuation of heavy quarks are difficult to explain quantitatively in this scenario. Our understanding of parton energy loss in hot matter can be enhanced significantly by studying similar processes in "cold matter." The EIC will enormously expand the energy range of previous fixed-target measurements, allowing for the first time the study of attenuation of heavy quarks in cold nuclear matter.

A detailed study of collective phenomena in cold QCD matter facilitates comparison with more familiar collective behavior in condensed (QED) matter systems. In addition to the close analogy to Bose-Einstein condensation, saturated gluons also have stochastic properties analogous to that of glassy ("spin glass") systems in condensed matter physics. Further, the behavior of QCD cross sections as they approach saturation are equivalent to that of a wide class of statistical reaction diffusion systems from physics to biology. These connections should blossom as the CGC properties become better delineated.

High-energy studies with polarized electron and proton beams seek to uncover the origin of the proton's spin in terms of the underlying quark and gluon fields. Spin is a fundamental quantum mechanical property of subatomic particles that underlies all of chemistry and explains magnetism. The nuclear physics of spin-½ protons and neutrons explains the origin of the chemical elements and their properties. As a practical consequence, the medical diagnostic technique of magnetic resonance imaging (MRI) is based on the manipulation of nuclear spins in a patient's body. Resolving how spin arises in QCD would settle how an important aspect of nature works at a fundamental level.

## OUTLOOK

The Electron-Ion Collider embodies the vision of our field for reaching the next QCD frontier: the study of the glue that binds all atomic nuclei. A high-luminosity EIC with center-of-mass energy in the range from 30 to 100 GeV with polarized nucleon beams and the full mass range of nuclear beams could be sited either at Brookhaven National Laboratory or at Thomas Jefferson National Accelerator Facility. EIC is the natural evolution of the CEBAF 12 GeV Upgrade, which will be focused on the role of valence quarks, and of the RHIC II upgrade, which will complete the study of the hot and dense matter discovered in RHIC's heavy-ion collisions and will also allow a first look at the gluon contribution to the proton spin. Precision measurements with the EIC, directly interpretable within the framework of QCD, will open a new window to the regime dominated by direct manifestations of the defining feature of QCD: the self-interactions of gluons. These self-interactions lie at the heart of nucleon and nuclear structure and are expected to be essential to the understanding of high-energy heavy-ion collisions.

To develop the most compelling case for EIC in a timely way it is clear that over the next five years significant progress must be made in the conceptual design of the accelerator. It will be important to converge on one accelerator concept so that detailed plans and schedules can start to be developed. This can only happen after essential Research and development is completed in a number of areas including: cooling of high-energy hadron beams, high-intensity polarized electron sources, and high-energy, high-current Energy Recovery Linacs. This research and development effort should be carried out so as to leverage existing expertise and capabilities at laboratories and universities. Research and development for a high-energy EIC will maintain U.S. leadership in an area with important societal applications. The design of collider detectors integrated into the accelerator will also be a key issue requiring detailed physics simulations as well as detector research and development.

Theoretical developments have been essential in shaping the scientific arguments for EIC, have attracted a growing number of young physicists, and have sparked one of the most active areas in all of theoretical high-energy physics. The coming years should see new insights into the universal character of gluon fields when viewed in hadrons at high energies. In addition, first-principles QCD calculations using lattice techniques are expected to provide increasingly realistic predictions to be tested with data. It is essential that theoretical support for EIC-related physics is maintained at a healthy level in the coming years.

When realized, EIC will be a unique facility to study QCD complementary to next-generation accelerators either under construction or planned in Europe and Asia. Thus, EIC has attracted significant interest from physicists around the world, which can be nurtured by timely allocation of resources for continuing Research and development and theoretical progress. It is natural and important that future EIC planning ensure a strong international participation.

# Nuclei: From Structure to Exploding Stars

## OVERVIEW

At the heart of every atom resides the nucleus, a dense kernel of matter. Comprising some 99.9% of the atom's mass—the other 0.1% resides in the outer cloud of electrons—the nucleus is a tightly bound cluster of positively charged protons and electrically neutral neutrons, known generically as *nucleons*. The forces that bind these nucleons are immensely strong, which is why nuclear processes are able to release a prodigious amount of energy; witness the thermonuclear fusion reactions that power our Sun and most other stars in the universe. But the forces between the nucleons are also quite complex, which is why nuclear matter displays a remarkably diverse variety of phenomena.

This chapter describes the experimental and theoretical studies now underway to attain a deeper understanding of these phenomena—a quest that takes nuclear physicists in two directions. Looking inward, they seek a comprehensive description of all nuclei, from the fundamental dynamics of quarks and gluons as described by quantum chromodynamics, to the intricately complex rotations, vibrations, and other such "many-body" excitations found in multinucleon systems, to the marginally stable structure of exotic, short-lived nuclei not naturally found on Earth. Looking outward, meanwhile, nuclear astrophysicists seek to understand how nuclear processes have shaped the cosmos, from the origin of the elements, the evolution of stars, and the detonation of supernovae, to the structure of neutron stars and the nature of matter at extreme densities.

It is still just *one* quest, however: each of these efforts informs the other. Later in this chapter we will look at how this happens in detail, framing the discussion in terms of five overarching questions that are expected to drive the future of research in this field:

- **What is the nature of the nuclear force that binds protons and neutrons into stable nuclei and rare isotopes?**
- **What is the origin of simple patterns in complex nuclei?**
- **What is the nature of neutron stars and dense nuclear matter?**
- **What is the origin of the elements in the cosmos?**
- **What are the nuclear reactions that drive stars and stellar explosions?**

Today these questions are much closer to being answered than they were a decade ago, thanks to substantial theoretical and experimental progress in that time. Nonetheless, we still have a great deal more to learn. For example, we do not yet have a satisfactory microscopic description of the nucleus based on QCD. We do not yet know enough about nuclear binding to determine the limits of the nuclear landscape, where the ratio of protons to neutrons gets so far away from equilibrium that the nucleus can no longer hold together. We cannot even predict the character of nuclei near these limits. Starting from the underlying interactions among neutrons and protons, moreover, we are unable to reliably predict masses or excitation modes of nuclei any heavier than carbon, which in its most common form has only six protons and a like number of neutrons. We cannot provide a satisfactory explanation of how complex nuclei give rise to simple emergent behavior, in the form of simple excitation patterns. We have barely begun to understand the fundamental mechanisms of nuclear reactions and nuclear fission. And we know very little about the nuclear physics underlying many key astrophysical processes—especially stellar explosions and neutron star properties, both of which require an understanding of very rare isotopes.

These research challenges, and the significant investments now being made in interdisciplinary initiatives, have led to growing connections between nuclear physics and other fields. The field of nuclear astrophysics, in particular, has undergone a transformation into a truly interdisciplinary endeavor where astrophysicists and nuclear physicists work together to tackle key problems. Also noteworthy are the significant strides being made in computing the physics of nuclei and stellar evolution by interdisciplinary collaborations with computational scientists.

On the experimental side, however, new data are essential. They will allow us to further explore the nature of internucleon interactions; to assess the validity of the theoretical approximations; to delineate the path toward integrating nuclear structure with nuclear reactions; and to ascertain the validity of extrapolations into new, unexplored regions of the nuclear chart.

Especially critical are experiments with rare-isotope beams. Pioneering efforts in this field have provided impressive first results and have temporarily put the United States in a world leadership position. Yet, the field is still in its infancy and is limited by no access to the rarest isotopes and the beam intensities available today. To address this limitation, physicists have begun planning a next-generation Facility for Rare Isotope Beams (FRIB), which will deliver the highest intensity beams of rare isotopes available anywhere. But FRIB will not be available for a decade. So in the meantime, physicists hope to continue developing a comprehensive picture of atomic nuclei by strengthening operations and carrying out modest upgrades at the National User Facilities (at ANL's ATLAS, ORNL's HRIBF, and MSU's NSCL), as well as by continuing the operations of university-based centers (at Florida State, Notre Dame, Texas A&M, TUNL, Washington, and Yale) and Berkeley's 88-Inch Cyclotron. This effort will go hand in hand with the continued development of state-of-the-art instrumentation and ever-more sensitive experimental techniques and detectors required to best address the science. This approach also guarantees that a suite of first-rate detectors and techniques will be available for initial experiments at FRIB.

Although the U.S. nuclear physics community currently possesses a leadership role in nuclear science with beams of rare and stable isotopes, major projects in this field are moving forward in Europe and Asia. We stand at a critical juncture. FRIB will build on the nation's leadership in this science and provide unprecedented scientific opportunities by opening new vistas in nuclear structure, dynamics, and astrophysics, while delivering in large quantities the isotopes required for testing the fundamental laws of nature and for numerous societal benefits. Without FRIB, we will no longer be the leaders—we will be users at other facilities around the world.

## RECENT ACHIEVEMENTS

Here we list several key achievements made since the last Long Range Plan. They have produced new insights into nuclear structure and nuclear astrophysics, and they highlight the increasingly strong research overlap between the physics of nuclei and nuclear astrophysics.

- **The structure of neutron-rich nuclei at the limits of stability and the origin of the elements.** The charge radii of the neutron-rich, unstable nuclei $^6$He and $^{11}$Li, two "halo" nuclei whose neutron radii are large, were measured with an accuracy of 1% through the determination of isotope shifts. Such precision measurements at the limits of stability provide stringent tests of *ab initio* nuclear model calculations and information on the interactions of neutrons and protons in loosely bound systems. At the same time we continue to discover surprises in very neutron-rich heavier nuclei that challenge our theoretical understanding. Observations of new neutron-rich isotopes of Al and Mg indicate that the neutron drip line, demarking the last bound neutron-rich isotope of a given charge number, may be located further from stability than previously anticipated. In contrast, mass measurements around A ~ 100 and A ~ 140 suggest that in these regions the drip line might be closer to stability than expected. Furthermore, the magic numbers 8, 20, and 28 no longer apply to neutron-rich nuclei. Additional stability occurs for near drip line oxygen nuclei with neutron numbers 14 and 16, as well as for N = 32 neutron-rich Ca, Ti, and Cr isotopes. Critical measurements of neutron-rich nuclei such as the β-decay studies of $^{78}$Ni and $^{130}$Cd, or neutron transfer studies near $^{132}$Sn, established new benchmarks for studying shell structure in extremely neutron-rich nuclei and defined the speed of the astrophysical rapid neutron capture process (r-process) up to the mid-mass region around A = 130. This allows more meaningful model comparisons with the large number of new observational data from very metal-poor stars that now begin to map in detail the enrichment history of our galaxy with heavy elements.

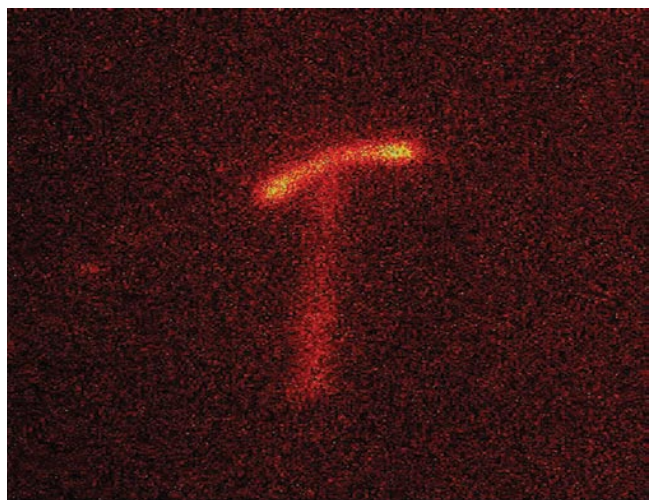

**Figure 2.17:** Digital photograph of the two-proton decay of $^{45}$Fe taken at the NSCL. A $^{45}$Fe nucleus, with the very short half-life of 5 ms, entered the picture from the bottom, was stopped, and then decayed by emission of two protons.

# Nuclear Theory: A Hierarchy of Scales and Models

The most useful way to think about the behavior of strongly interacting matter depends on the energy of the experimental probe and the distance scale. At the highest energies and shortest scales, for example, nuclear matter is best described in terms of its most fundamental building blocks, quarks and gluons, and the fundamental theory of strong interactions, QCD (a). At somewhat larger scales, hadrons (the baryons and mesons) can often be described by the dynamics of the effective (or constituent) quarks, with the gluon degrees of freedom being integrated out (b). At still larger scales, the nucleus is best described as a strongly interacting, quantum mechanical system of protons and neutrons. A common starting point for nuclear physics is an internucleon interaction, represented either by a potential or by a set of meson-exchange forces (c). In this picture, typical single-particle excitations in the nucleus are of the order of the proton (or neutron) separation energy (d). For complex nuclei, however, calculations involving all the protons and neutrons become prohibitively difficult. Therefore, a critical challenge is to develop new approaches that identify the important degrees of freedom of the nuclear system and are practical in use. Such strategy is similar to that used in other fields of science, in particular in condensed matter physics, atomic and molecular physics, and quantum chemistry. Of particular importance is the development of the energy-density functional, which may lead to a comprehensive description of the properties of finite nuclei, as well as of extended asymmetric nucleonic matter. Here, the main building blocks are "effective fields" that correspond to local proton and neutron densities, and to currents of protons and neutrons (e). Finally, for certain classes of nuclear modes, in particular those representing emergent many-body phenomena that happen on a much lower energy scale, the effective degrees of freedom are "collective coordinates" that describe various vibrations and rotations of the nucleus (f), as well as the large-amplitude motions seen in fission.

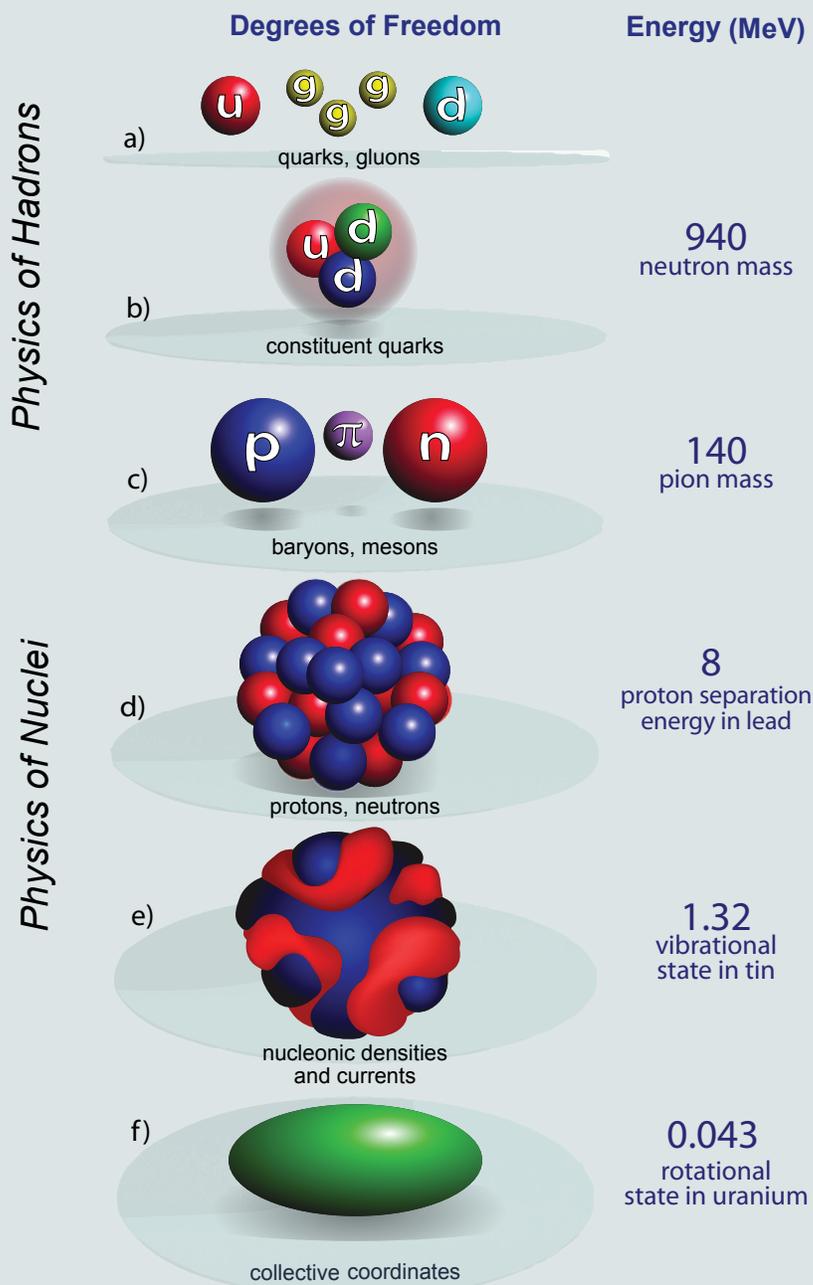

- **New decay modes of extremely neutron-deficient nuclei and the nature of stellar X-ray bursts.** In an experimental tour de force, live digital photography of di-proton emission from the ground state of the extremely neutron-deficient nucleus $^{45}$Fe (figure 2.17) has been achieved, demonstrating this exotic decay mode unambiguously and clarifying its nature. Near the doubly magic neutron-deficient nucleus $^{100}$Sn, where protons and neutrons occupy the same shell model orbits, the fastest known alpha decay was discovered. Together with the discovery of the gamma ray depopulating the first excited state of $^{101}$Sn, these measurements are the first steps to determine the shell structure of extremely neutron-deficient heavy nuclei in a region that also represents the possible endpoint of the astrophysical rapid proton capture process powering X-ray bursts. In connection with Penning Trap precision mass measurements of neutron-deficient unstable nuclei in the critical mass 64–72 region, this has led to an impressive reduction in the nuclear physics uncertainty in advanced X-ray burst models which now include all relevant nuclear reactions.

- **The physics and chemistry of superheavy elements.** We continue to move closer to the predicted superheavy island of stability. Synthesis of superheavy elements up to Z = 118 in fusion reactions was reported, and the chemical properties of elements up to Z = 112 have been studied. In elements with Z = 100–102, energies of isomeric states provided new probes of higher-lying shells, the very shells governing the stability of superheavy nuclei.

- **Collective behavior and symmetries.** Measurements of nuclei in shape transitional regions, near N = 90 and elsewhere, led to an interpretation in terms of quantum phase transitional behavior between different symmetries. A new class of many-body symmetries describing nuclei at the phase transitional point was developed. Pushing the limit of extreme nuclear spin, rotational behavior was discovered at high spins in $^{157,158}$Er, beyond the angular momenta at which many rotational bands terminate, challenging microscopic descriptions of collectivity.

- **Properties of nuclear matter.** Quantitative constraints on the equation of state of symmetric nuclear matter, important to both nuclear physics and astrophysics, were set using heavy-ion collisions. Giant resonance measurements determined the compressibility of nuclear matter to be K = 230 ± 20 MeV.

- **Precision measurements and theoretical description of nuclear reaction rates provide new insight in astrophysics.** Well-understood Big Bang nucleosynthesis models have been used to determine the baryon content of the universe. These results were independently confirmed by the COBE and WMAP measurements of the cosmic microwave background. The $^{14}$N(p,γ)$^{15}$O reaction cross section that controls hydrogen burning in stars via the CNO cycle has been measured to be about a factor of 2 lower than previously thought. This increases the derived ages for globular clusters and related limits for the age of the universe, by about a billion years. Nuclear physics uncertainties in calculations of the neutrino flux from the Sun are being addressed with a number of new measurements of the $^{3}$He(α,γ)$^{7}$Be and $^{7}$Be(p,γ)$^{8}$B reactions. For the first time, *ab initio* nuclear calculations began to address nuclear reactions and resonance lifetimes (see figure 2.18). Pioneering measurements of reaction rates with the present generation of stable-beam and rare-isotope-beam facilities have employed new techniques to constrain a number of key reaction rates in nova explosions. This has reduced uncertainties in modeling the hot CNO and NeNa cycles and the production of long-lived radioisotopes such as $^{18}$F and $^{22}$Na.

- **Understanding nuclear theory of heavy nuclei and supernova explosions.** Nuclear density functional theory provided a global description of empirical proton-neutron interactions. Nuclear theory advances in the description of heavy nuclei have also fundamentally changed our view of the role of electron captures on nuclei in core collapse supernovae and have significantly expanded the range of nuclei thought to participate in such processes. Together with advances in neutrino observations and new simulations, these have been major steps toward the understanding of the physics of core collapse supernovae.

## Magic Numbers: Now You See Them, Now You Don't

Physicists often state that "shell structure" is the cornerstone of any satisfactory description of nuclei. The approach can best be understood by analogy with atomic theory, in which the electrons that surround the nucleus are grouped into "shells" of similar radius and energy. In neutral atoms, the energy it takes to remove the last electron varies with the atomic number, and this quantity decreases markedly each time a major electron shell is filled. The noble gases helium, neon, argon, krypton, xenon, and radon correspond to the filling of major electronic shells requiring a total of 2, 10, 18, 36, 54, and 86 electrons, respectively. The energy it takes to remove the last proton or neutron from the nucleus exhibits similar discontinuities as a function of the number of protons Z and/or neutrons N. From examining the properties of the stable nuclei, it was concluded over half a century ago that nuclei with 2, 8, 20, 28, 50, or 82 protons and/or neutrons have enhanced stability. For neutrons, there is an additional such number, 126. The "magic numbers" then appear since large gaps in energy occur between two of these shells.

During the last decade, it has become clear that the magic numbers are not as immutable as once thought. The suggestion has come mostly from studies of light-mass nuclei with combinations of protons and neutrons that differ greatly from those characterizing the stable ones. The present situation is illustrated in the partial nuclear landscape of the figure. The nucleus $^{12}$Be (Z = 4) was expected to exhibit the properties of a nucleus with magic number N = 8 although none was found. Similarly, the magic character of neutron number N = 20 appears to have vanished in the exotic nucleus $^{32}$Mg (Z = 12). It had been widely

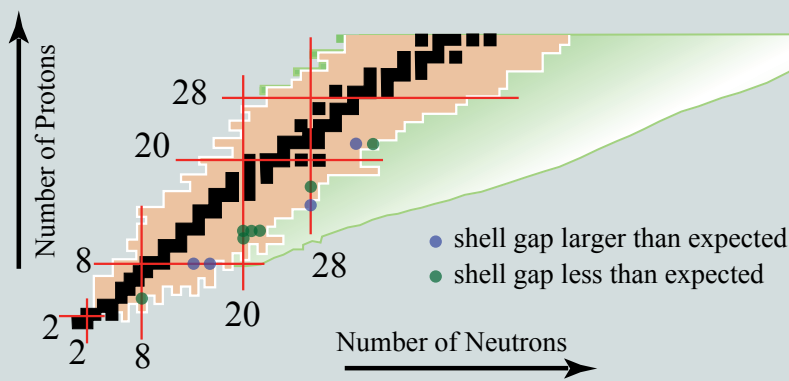

Shell structure of the lighter nuclei. Shown are the original magic numbers (solid red lines) and blue and green dots indicating measurements that point to changes in magic numbers in the neutron-rich nuclei.

speculated that, besides "doubly magic" $^{16}$O (Z = N = 8), the oxygen isotope with 20 neutrons would be particularly stable, but experiments found that this nucleus, $^{28}$O, is not even bound. On the other hand, in other nuclei far from stability strong indications exist that new magic numbers develop. This occurs for N = 14, 16, and 32 as well as at proton number Z = 14. In the neutron-rich isotopes near "doubly magic" $^{132}$Sn, first systematic studies uncovered an unexpected evolution of collectivity with neutron number, and first reaction studies uncovered strong, unexpected enhancement of sub-barrier fusion leading to renewed optimism about the use of n-rich beams for the synthesis of very heavy nuclei.

These experimental observations suggest that some aspects of physics responsible for shell structure in nuclei must not be readily apparent from the properties of stable nuclei but are amplified in exotic systems. With the ultimate goal of providing a satisfactory description of all nuclei, the essential challenge is to understand the mechanism(s) responsible for changes in shell structure as the number of protons and/or neutrons changes. Knowing the magic numbers and the energies of nucleonic orbits is not a mere detail: if we do not understand nucleonic motion, we do not understand nuclei.

Nature does not have the limitation of dealing only with stable nuclei. Stars shine because of nuclear reactions, and nucleosynthesis processes, especially those producing nuclei heavier than carbon and oxygen, often take place in violent stellar explosions which involve exotic nuclei. Hence, an understanding of how the elements were—and continue to be—made in our universe depends on our ability to calculate the reaction rates for their production. These rates in turn depend critically on the shell structure of exotic nuclei—this is one important area where the presence or absence of a magic number has a considerable impact.

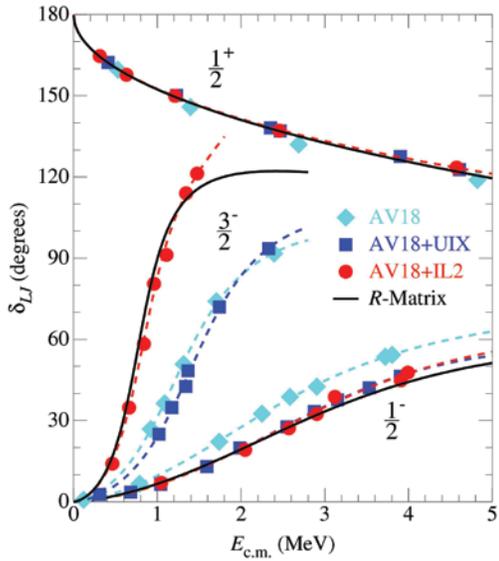
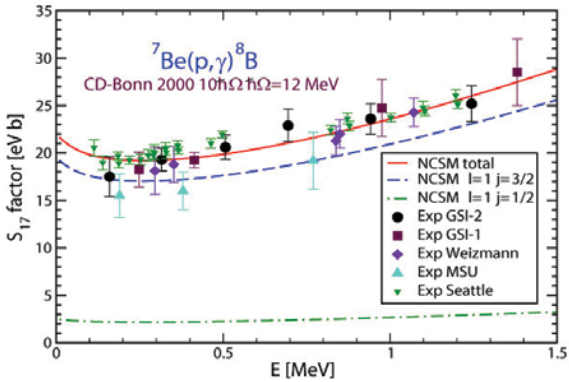
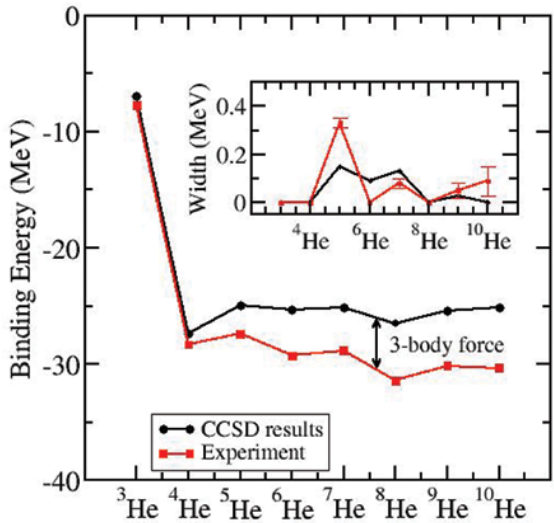

## FUTURE PROGRAM

**What is the Nature of the Nuclear Force that Binds Protons and Neutrons into Stable Nuclei and Rare Isotopes?**

The field is positioned to answer this fundamental question of nuclear physics. We have defined a roadmap to achieve the goal of a comprehensive and unified description of all nuclei and their reactions. This roadmap bridges light, medium, and heavy nuclei. Properties of light nuclei can be calculated with increasing success in terms of protons and neutrons interacting via two- and three-body forces fitted to scattering data and excitation spectra in light nuclei. Medium-mass nuclei are described with nuclear shell model calculations using effective forces either derived from nucleon-nucleon scattering data or extracted from nuclear experiments. Density functional theory (DFT) is the method of choice for heavy nuclei.

The interactions among nucleons in the nucleus emerge from QCD. As such, only specific features of QCD are important for the interactions between nucleons. In recent years, a more-explicit link between QCD and the interactions between nucleons has been made possible thanks to effective field theory (EFT) based on chiral symmetry breaking (in which pions obtain small mass), where a natural power counting provides a hierarchy between nucleon-nucleon (NN) and higher-order multinucleon interactions. In such an approach, as well as for more traditional potential models based on meson exchange, experimental data in the few-body, and in the many-body, sectors are needed in order to constrain the low-energy couplings.

While much is known about NN potentials, the same cannot be said of multinucleon potentials. *Ab initio* studies for light nuclei show that the three nucleon (NNN) force has a significant role in determining what has classically been referred to as the "spin-orbit" properties of the nucleus. NNN forces also appear to be important for the stability of so-called Borromean nuclei like $^{6,8}$He, $^9$Be, and $^{11}$Li, which

**Figure 2.18:** With the aid of terascale and (soon) petascale computing, *ab initio* techniques such as the Greens Function Monte Carlo (GFMC), No Core Shell Model (NCSM), and Coupled Cluster (CC) theory are beginning to probe various aspects of reactions. Pictured are (a) NCSM calculations of the astrophysical S-factor for $^7$Be(p,γ)$^8$B reactions compared to various experiments, (b) recent calculations of phase shifts from GFMC for n + α scattering, and (c) CC calculations for He chain binding energies and resonance lifetimes compared to experiment.

# Linking the Forest and the Trees in Atomic Nuclei

One of the most important challenges in the science of nuclei is understanding how the overall structure of the atomic nucleus (the "forest") changes with the number of protons and neutrons (nucleons) within it and their detailed interactions (the "trees"). The traditional approach to nuclear structure envisions the nucleons orbiting the nucleus subject to an average force. This force gives the benchmark shell structure and magic numbers of nuclei. However, such a picture does not account for important interactions such as pairing, which tends to make nuclei spherical, and the interactions of the outermost orbiting protons and neutrons (p-n interactions), which tend to drive the emergence of collective behavior in nuclei, along with the emergence of elongated or deformed shapes.

It turns out that these p-n interactions can actually be deduced by taking a specific double difference of measured nuclear masses. These double mass differences, called $dV_{pn}$, contain two components, one varying smoothly with the number of particles, reflecting the symmetry energy, and a fluctuating one sensitive to the spatial overlaps of the outermost proton and neutron wave functions. Available $dV_{pn}$ values provide a fresh perspective on structural evolution and a sensitive test of modern microscopic theories.

The left side of the figure shows experimental values of $dV_{pn}$ for the rare earth region of nuclei from $^{132}$Sn to $^{208}$Pb. These empirical p-n interactions show characteristic patterns in heavy nuclei where each major shell has a similar sequencing of orbits, from orbits with high angular momentum lowest in energy to low angular momentum orbits near the end of a shell. Therefore, since the p-n interaction is short range, the largest spatial overlaps of protons and neutrons will tend to occur near the diagonal in the figure where the *fractional* filling of the respective proton and neutron shells is similar. For the same reason the p-n interactions are largest in both the lower left and upper right quadrants, where both protons and neutrons are filling either the first halves of their major shells, or the second halves. In contrast, in the upper left quadrant, the protons are filling the second half and the neutrons the first, and so they occupy orbits that tend to have small spatial overlap, leading to small values of $dV_{pn}$.

This behavior correlates exactly with the well-known development of collectivity in this region, which is also more rapid for nuclei in the lower left quadrant. This correlation provides the first direct empirical link between observed growth rates of collectivity and empirical measures of p-n interaction strengths. Note also that no experimental values of $dV_{pn}$ are known in the lower right quadrant: new mass measurements would give an important test of these ideas.

Advances in theoretical modeling and high-performance computers are beginning to enable quantitative descriptions of nuclei across the chart of the nuclides. For heavy nuclei, the theoretical tool of choice is an approach called density functional theory, long used for studying other complex systems such as molecules and condensed matter. The right side shows results of DFT large-scale calculations of $dV_{pn}$ using an interaction called SkP. The agreement with the data is truly remarkable, in particular, with the preponderance of large values along the diagonal, larger values for protons and neutrons in like quadrants, and much smaller values (blue boxes) elsewhere. This success of theorists in explaining such a key nuclear property augurs well for the future. Indeed, these same DFT calculations provide predictions for nuclei in yet unexplored regions throughout the nuclear chart. The Facility for Rare Isotope Beams will yield many new mass measurements, giving sensitive tests of such calculations, providing input for further development of nuclear models, and helping refine our understanding of the complex nuclear many-body system.

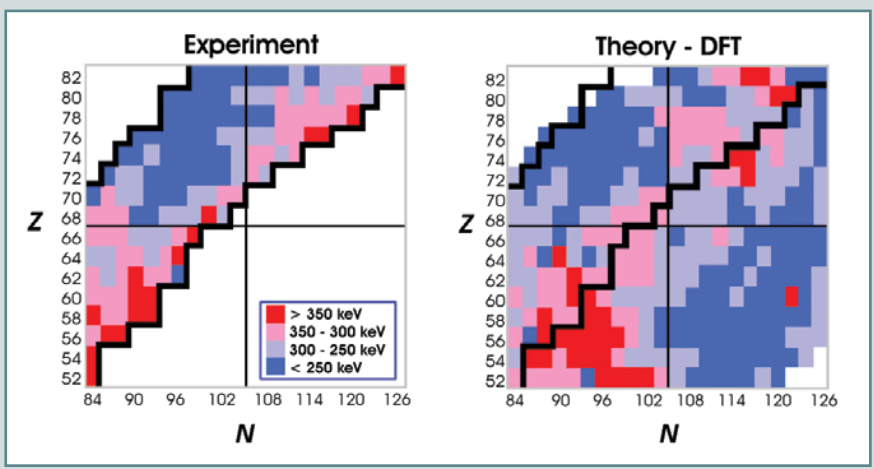

Proton-neutron interactions, density functional theory calculations, and the evolution of nuclear structure. **Left**: Empirical double-mass differences, $dV_{pn}$, values for the rare earth region, color coded by their strengths (largest values are red). **Right**: DFT predictions using an SkP interaction. The zigzag lines facilitate a comparison by demarcating the nuclei whose empirical $dV_{pn}$ values are known.

consist of a normal nuclear core and very weakly bound nucleons orbiting at large radii. Even at the few-body level, there are still significant problems in explaining scattering data, indicating both the need for improved microscopic models of NNN forces and the importance of high-quality few-nucleon data to constrain these models. Theorists use complementary many-body approaches to calculate light- to medium-mass nuclei to understand how these potentials affect nuclear properties.

The interacting shell model has been extended to the regime of nuclei above $^{40}$Ca, and we are witnessing progress toward its implementation in nuclei in the Sr and Sn regions. Extensions using scattering basis states, such as the Gamow shell model, are promising for the description of weakly bound nuclei where nucleons can be scattered into unbound states.

DFT describes heavier nuclei through proton and neutron densities and currents. New ingredients that must now be incorporated include the effects of weak binding, which can lead to new forms of shell structure or even challenge the viability of the single-particle description itself in which the nucleus is viewed in terms of the orbits of individual nucleons and their mutual interactions. These theoretical advances will enable a study of new forms of collectivity, such as modes involving oscillations of weakly bound neutrons (forming a neutron skin) against the core, and coupling to the continuum of scattering states. Key to this progress is a new suite of algorithms that allow for DFT studies free from artificially imposed symmetry constraints.

Much of the information we have on nuclear structure is inferred from reactions with other nuclei. Recent *ab initio* theories (figure 2.18) used to describe the structure of light nuclei will be extended to include dynamic processes, such as light-ion fusion reactions prominent in stars. There has also been great progress in the development of a microscopic theory for multistep direct reactions, which are central to a generalized study of nuclear reactions. They encompass neutron-induced reactions (from capture reactions to all the inelastic scattering channels); transfer reactions, such as deuteron-proton (d,p); as well as breakup reactions. This comes at a critical time when new (d,p) experiments (in inverse kinematics) are being performed on key N=50 and N=82 nuclei, such as $^{132}$Sn.

Experimental determination of the location of the drip line, beyond which neutrons and protons no longer bind, provides a stringent test for nuclear theory. Delineating the location is an extremely difficult goal because the drip lines lie far from stable isotopes. In this context, it will be very important to measure nuclear masses in the neutron-rich Ni and Sn regions and to map the neutron drip line as far as possible in the number of protons Z. The study of these nuclei is certainly one of the key goals of FRIB.

New experimental data are especially needed for nuclei far from stability but yet well within the drip lines where many of the traditional properties of nuclei are expected to change significantly. Recent experimental results provide intriguing indications of these changes. For decades the cornerstone of nuclear structure has been the concept of single-particle motion in a well-defined potential leading to shell structure and magic numbers. We have now learned that the magic numbers are not immutable—they appear to depend on the neutron-to-proton asymmetry and the binding energy. Important areas of future study include shell structure near known or newly discovered magic numbers. Here, the regions near $^{48}$Ni, $^{78}$Ni, $^{100}$Sn, and $^{132}$Sn are critical.

Another frontier lies in defining the limits of the heaviest elements. Indeed there were preliminary indications that the chemistry of element 112 was different from that of its homologue mercury. The most recent, more sensitive experiments, however, indicate that this may not be the case. Further experiments are needed with the extension of the studies to the next possible noble gas, element 118.

The future program to understand the interactions of nuclei, and nuclear decay, is closely connected to the ability to produce and study key nuclei where certain aspects of the nuclear many-body problem can be isolated and enhanced. FRIB is critical to the future since it will provide access to specific key nuclei, such as $^{60}$Ca, which is predicted to be doubly magic and hence will serve as a benchmark. Moreover, detailed studies of a long chain of isotopes of elements like nickel will be possible from mass numbers A=48 to A=83, spanning three doubly magic nuclei and greatly expanding the range of binding energy over which we can test theory.

A related challenge is to develop a first-principle description of nuclear decay. Of all the various nuclear decay processes, nuclear fission, important in r-process nucleosynthesis and the modeling of reactions relevant to the advanced fuel cycle for next-generation reactors, is among the most difficult to tackle. It is a quantum many-body tunneling problem whose typical time scale changes by orders of magnitude when adding just a few nucleons. Despite this difficulty, significant progress is being made by the selection of appro-priate interactions and density functionals combined with methods describing nuclear collective motion. Studies of fission barriers and heavy-element decay will be a key for the further development of theory. FRIB will play a critical role both in providing access to a range of nuclei and the tools to study fission barriers, and also in the production of neutron-rich superheavy elements to allow the study of their lifetimes and decays.

**What is the Origin of Simple Patterns in Nuclei?**

The structure of heavy atomic nuclei is determined by the combined strong and electromagnetic interactions, acting among hundreds of nucleons, which can be thought of as orbiting the nucleus about $10^{21}$ times per second and occupying on the order of half the nuclear volume. Given this environment, one might imagine an extremely chaotic and complex situation, but in fact nuclei often display extremely regular excitation patterns.

These simple patterns range from the regular sequences of single particle levels (resulting in shell structure and magic numbers), to a variety of nuclear shapes and collective modes like vibrations and rotations, phase transitional behavior, and correlated decay modes, such as fission. They can appear either in the excitations of a single nucleus (e.g., the nearly perfect rotational motion exhibited by certain deformed nuclei) or in highly correlated patterns of structural evolution as a function of proton and neutron numbers.

Often, simple patterns reflect underlying symmetries of the many-body system. These symmetries, or geometric descriptions, are certain idealizations that provide an intuitive understanding and yield collective excitations characterized by approximate quantum numbers, transition selection rules, and, often, explicit relations among observables. The models that explore the many-body symmetries, such as algebraic approaches based on group theory and geometrical models (see figure 2.18), correlate large amounts of structural information with an extreme economy of parameters. Sometimes, symmetries offer predictions not evident from detailed microscopic calculations, which are based on nucleonic degrees of freedom, since they describe nuclei in terms of many-body quantum numbers and collective coordinates. Experimental work and theoretical advances to understand the microscopic origin of the simple patterns is one of the great challenges of modern nuclear structure theory.

The discovery and characterization of new collective modes and shape transitional regions in nuclei far from stability are important goals. A key question is whether the characteristics of transitional regions will be similar to that recently found in stable nuclei, e.g., near N = 90. Simple modeling of the interplay of pairing and proton-neutron interactions provides predictions of the locus of possible phase transitional behavior in new regions that can be tested with data from FRIB.

In much of this work, the experimental techniques are well understood and reliable such as Coulomb excitation, inverse kinematics transfer reactions, decay spectroscopy, mass measurements, and laser spectroscopy—all performed in an environment with intensities that are many orders of magnitude weaker than has been traditionally the case. Fortunately, one of the great triumphs of recent years has been the development of sensitive experimental methods that succeed under such constraints. With FRIB and advanced instrumentation such as the next-generation gamma-ray detectors GRETINA and GRETA, the study of simple patterns and excitation modes will continue its recent renaissance.

**What is the Nature of Neutron Stars and Dense Nuclear Matter?**

Neutron stars are among the most fascinating astrophysical objects: their structure and evolution are largely determined by nuclear physics, and they play a central role in many astrophysical events such as supernovae, X-ray bursts, and possibly gamma-ray bursts. The basic properties of neutron stars such as their mass, radius, and cooling behavior are determined by the nuclear equation of state (EOS). The EOS is a fundamental property of nuclear matter. It describes the relationships between the energy, pressure, temperature, density, and neutron-proton (isospin) asymmetry for a nuclear system. It is therefore directly linked to open questions in astrophysics such as the maximum mass of a neutron star and its radius for a given mass. For accreting neutron stars, reaction rates with exotic isotopes on the surface and in the crust determine other important observables such as X-ray bursts and superbursts, a class of rare, particularly energetic long duration X-ray bursts. The nature of these processes needs to be better understood in order to use the observations as probes of neutron star physics.

One of the key questions in this area is the density dependence of the symmetry energy (the change in nuclear energy associated with changing the neutron-to-proton asymmetry) in the nuclear equation of state and the form

# Neutron Star Crusts

Neutron stars are some of the most extreme astrophysical objects ever observed. They pack about 1.5 times the mass of our Sun into a ball of radius only 10 km. The corresponding average density is in excess of the density inside nuclei (~2.5 × $10^{14}$ g/cm³), which generates the repulsive nuclear force that balances the star against gravitational collapse. Anything denser would collapse into a black hole.

Current models of neutron star structure indicate that a significant fraction of its mass and radius is contained in the core at supranuclear density. The nature and composition of this matter remain a mystery and are the Holy Grail of neutron star research. To probe the interior we must rely on observable phenomena that are associated with surface and crust regions of the neutron star.

Near the surface, matter is composed of ordinary iron-like nuclei and electrons—much as in the terrestrial environment. Just a meter below the surface, however, the enormous gravity squeezes matter to densities on the order of $10^6$ g/cm³, at which point the electrons are forced out of the atoms to form a relativistic and degenerate gas. The electron Fermi energy rises rapidly with depth, forcing nuclei to become neutron rich as energetic electrons convert protons into neutrons by the inverse beta-decay reaction: $e^- + p \rightarrow n + \nu_e$. Eventually, these exotic and increasingly neutron-rich nuclei reach a critical point called the neutron-drip density. Beyond this depth, the excess neutrons cannot be bound by nuclei and begin to occupy the space between nuclei. These unbound neutrons become superfluid, resulting in a layer that resembles a superfluid soup containing spherical chunks of "ordinary" nuclear matter. At larger depths, these chunks deform to form noodle-like and lasagna-like structures—which are collectively referred to as the pasta phase. Eventually the pressures are so high that these structures simply merge to form uniform neutron-rich matter. This sequence of increasing neutron-rich phases of matter is depicted in the figure.

The crust and surface of neutron stars provide a host of observable phenomena, which allow us to probe neutron stars directly with astrophysical observations. These include: (1) X-ray bursts and superbursts, which are thermonuclear explosions occurring in accreting neutron stars; (2) glitches, which are sudden rotational spin-ups in otherwise gradually spinning-down isolated neutron stars; and (3) giant flares, which are energetic outbursts seen in highly magnetized neutron stars called magnetars. Thermonuclear phenomena are sensitive to the composition, ambient temperature, and nuclear reactions. Glitches rely on the co-existence of superfluidity and a rigid lattice structure (of nuclei)—a feature naturally realized in the inner crust. And finally, the catastrophic release of magnetic energy needed to power giant flares and the observed ringing in the late time emission pattern requires a solid region like the crust to anchor the wound-up magnet energy. As we try to obtain a quantitative understating of these phenomena, we are challenged to provide an accurate description of the phases of neutron-rich matter in the crust mentioned earlier. This would be within reach through experimental and theoretical studies of exotic neutron-rich nuclei.

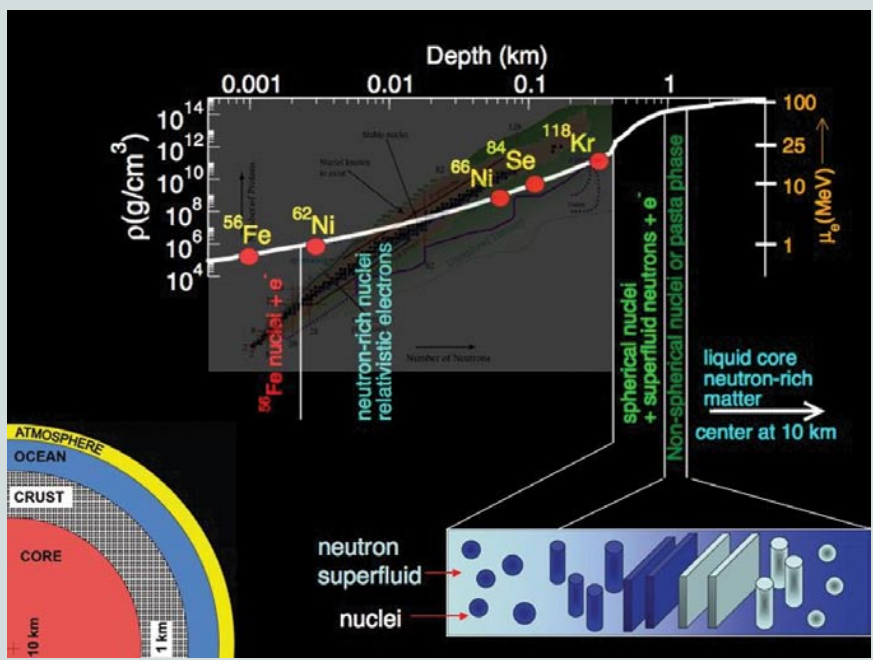

The density and composition profile of the neutron star crust is shown as a function of depth. Beginning with $^{56}$Fe the electron Fermi energy, $\mu_e$, grows rapidly favoring increasingly neutron-rich nuclei from $^{62}$Ni to $^{118}$Kr. At densities greater than about $4.3 \times 10^{11}$ g/cm³ neutrons drip out of nuclei to form a superfluid. Further below, the inlay shows that rod-like and slab-like shapes of nuclei could be favored as a precursor to the transition to uniform neutron-rich matter in the core.

of nuclear matter in the core. The symmetry energy has a major effect on astrophysical processes such as neutron star cooling as it determines the threshold density for the direct Urca process. (An Urca process is a cycle of nuclear reactions where a proton absorbs an electron to become a neutron, which then beta decays to become again a proton. Each step emits a neutrino or antineutrino, respectively. The neutrinos leave the neutron star, and their energy is therefore lost.) The question of whether or not rapid cooling via the direct Urca process occurs in neutron stars is of fundamental importance. Neutron star radii are also particularly sensitive to the symmetry energy at densities just above the normal nuclear densities. The uncertainty of the EOS at asymmetries expected in a neutron star leads to pressure variations that account for nearly 50% of the variations in the predictions of neutron star radii.

Sensitive correlations between the neutron radius of a heavy nucleus and several neutron-star observables have been established from a purely theoretical basis. As the same pressure that supports a neutron star against gravitational collapse creates a neutron skin in a heavy nucleus, an intriguing correlation was established: the thicker the neutron skin of a heavy nucleus, the larger the radius of a neutron star. Additional correlations have emerged between the neutron skin of a heavy nucleus and various neutron-star observables, such as stellar moments of inertia, vibration frequencies of the crust (the 100 m thick outer region of the neutron star), and cooling rates.

The symmetry energy also affects nuclear processes such as the dynamics of intermediate-energy heavy-ion reactions. The determination of the symmetry energy at both low and high densities is a prime motivation for much of the experimental work in this area.

Energetic nuclear collisions provide the only available means under laboratory conditions to compress nuclear matter to several times its equilibrium density. An observable particularly sensitive to the EOS is the flow of particles in a direction perpendicular to the incoming beam (figure 2.19). Energetic collisions of two gold nuclei have been used to constrain the EOS of symmetric nuclear matter squeezed to densities up to and above four times nuclear-matter saturation density. At the highest density, pressures as large as $10^{29}$ atmospheres were extracted. Such analyses place important constraints on theoretical models of the equation of state.

A goal is to achieve dramatic improvements in our knowledge of the EOS of asymmetric nuclear matter and the nature of neutron stars through terrestrial experiments coupled to information from neutron-star observations. Beams from FRIB are crucial to increase the degree of proton-neutron asymmetry in heavy-ion collisions. Complementing the nuclear science measurements will be an assortment of neutron-star observations. X-ray observatories with increased sensitivities will measure neutron-star properties with higher precision.

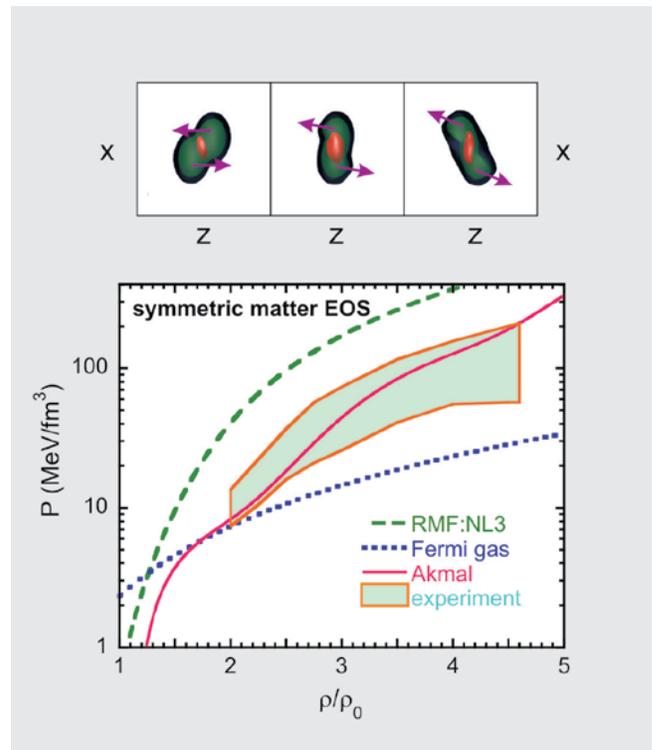

**Figure 2.19:** Upper view: Three stages of a transport calculation of a collision between two Au nuclei are shown. The dark blue, green, and orange contours show the regions where the densities exceed 5%, 20%, and twice the nuclear matter density. The arrows indicate the average velocities of matter is that deflected by the high pressures attained in the central high-density region. Lower view: The shaded region corresponds to the region of pressures in zero-temperature nuclear matter consistent with the experimental flow data. Equations of state for noninteracting nucleons, nucleons interacting with realistic two-body and three-body forces, and nucleons described by relativistic mean field theory are indicated by the dotted, solid, and dashed lines, respectively.

An accurate determination of the thickness of the neutron skin of a heavy nucleus would provide a unique experimental constraint on the symmetry energy. In contrast to all previous hadronic experiments that suffer from uncertainties related to theoretical model dependence, a purely electroweak measurement promises to be both accurate and model independent. Parity-violating electron scattering is particularly sensitive to the neutron density because the neutral weak vector boson couples preferentially to the neutrons. The Parity Radius Experiment (PREX) at the Jefferson Lab aims to measure the neutron radius of $^{208}$Pb via parity-violating electron scattering and will be quite useful in this respect.

Electron captures and pycnonuclear fusion (fusion induced by high pressure) reactions occur in the upper crust of accreting neutron stars where neutron-rich nuclei form a crystalline lattice. These processes have yet to be fully characterized but are clearly important as heat sources that set the thermal structure of the neutron star crust. They are directly linked to observables such as superbursts or the cooling of a neutron star in transients observed once the accretion shuts off. Because the thermal structure of the crust is also affected by its inner boundary and therefore the cooling of the neutron star core, understanding the crust processes might offer an opportunity to constrain neutron star physics through observations of surface phenomena such as superbursts. FRIB will be indispensable for accessing the neutron-rich nuclei that participate in crust processes. Important data that can be determined experimentally are masses, level schemes, and the location of the neutron drip line (maybe up to A~60). Electron capture and pycnonuclear fusion rates will have to be provided by theory, and the major advance in understanding extremely neutron-rich nuclei provided by FRIB will be essential in developing these theoretical predictions.

**What is the Origin of the Elements in the Cosmos?**

While the Big Bang is the source of the "primordial isotopes" such as hydrogen, deuterium, $^3$He, $^4$He, and some $^6$Li and $^7$Li, heavier nuclei are generated in subsequent generations of stellar nucleosynthesis sites from first-generation to present-day stars and in explosive nucleosynthesis sites from novae to supernovae. New observations of abundance distributions in old stars of different metal content and, therefore, age are now beginning to show the history of the buildup of elements in our galaxy and the universe. To address the open questions concerning the origin of the elements in the universe, these observations need to be confronted with nucleosynthesis models that include accurate descriptions of the nuclear processes that create new nuclei.

A particularly important open question is the origin of about half of the heavy elements attributed to the rapid neutron capture process. Sensitive experimental techniques are being developed to clarify the underlying nuclear physics of the r-process and to measure critical nuclear properties such as nuclear masses and decay properties of extremely neutron-rich nuclei. Measurements clarifying the role of neutron capture rates and fission processes are needed, as well as studies of the alpha and neutron capture processes on light nuclei that might provide the seed for the r-process. Neutrino interactions also play an important role in many r-process models, where they create the required neutron-rich conditions. In supernova models, the nuclear EOS, and in particular the nuclear symmetry energy, dictates the properties of the neutrinos emerging from the forming neutron star and therefore affects the r-process. Neutrino-induced nucleosynthesis and the neutrino-induced reprocessing of heavy elements can further modify the final composition.

Experiments have reached the r-process path in a few instances, mostly near mass numbers 80 and 130. With FRIB we will finally take the critical step and perform these experiments along the path of the r-process. Moreover, the theoretical advances enabled by the new data from FRIB will be crucial for reliably predicting the properties of the remaining r-process nuclei. Together, the experimental and theoretical advances will allow us for the first time to predict the abundance patterns produced by theoretical models of the r-process and to compare them with the increasing number of new observations.

The slow neutron capture process (s-process) created roughly the other half of the heavy elements in nature. Reliable s-process predictions are needed to subtract the s-process from the element abundances in the solar system and in many stars in order to define the r-process contribution. This requires precise data on s-process neutron capture rates. Many of these rates on mostly stable nuclei have been studied over the last decades, more recently at facilities like LANSCE at Los Alamos and nTOF at CERN, but in several important cases large uncertainties remain to be addressed. While the s-process is known to take place in red giant stars, the mixing processes leading to the production of neutrons are not well understood.

The production of different nuclei depends sensitively on the temperature and density conditions of the environment at s-process branching points. Future experiments are needed to measure the necessary neutron capture rates on long-lived radioactive targets. This requires high-intensity neutron beams that have become available at new-generation neutron spallation sources such as LANSCE or the Spallation Neutron Source at Oak Ridge. Isotope harvesting at FRIB will provide clean samples for many of the more short-lived

# The Rapid Neutron Capture Process (r-process)

The r-process is responsible for building many of the elements in nature heavier than iron. It is quite mysterious since we do not know with any certainty where it occurs, nor do we understand the actual sequence of nuclear reactions involved. Possible sites of the r-process include matter evaporated from the surface of a hot neutron star forming in a supernova explosion, the merging of two neutron stars into a black hole, and accretion disks in gamma-ray bursts. Whatever the precise details, however, it is clear that extremely neutron-rich rare isotopes play a critical role.

A major step toward a solution of the puzzle has been the discovery of detailed r-process elemental abundance patterns in a few very iron-poor stars located in the halo of our galaxy. These stars formed a long time ago when the galaxy was still iron poor and thus represent a "fossil record" of the chemical evolution of the galaxy. Stars that exhibit r-process patterns can tell us in principle how this process enriched the galaxy with heavy metals over time, step by step and event by event. Large-scale astronomical surveys now underway or planned for the future will attempt to scan millions of stars to find more such "r-process stars," and programs at the largest telescopes will attempt to determine their abundance pattern with high accuracy.

To compare predictions of the various proposed r-process models with the observational record, firm nuclear input is needed. Only with reliable nuclear physics can the observed abundances be interpreted in terms of the astrophysical conditions and the general nature of the r-process site. This requires a concerted effort that includes nuclear experiments with advanced rare-isotope facilities and nuclear theory, and represents a major goal for the future. Experimentally, we are on the verge of a major step in understanding of the r-process. Experiments at existing rare-isotope facilities have already started to reach the r-process in a few cases. With FRIB we will be able to finally produce and study a large fraction of the extremely neutron-rich exotic nuclei along the path of the r-process.

**Astronomical observation**

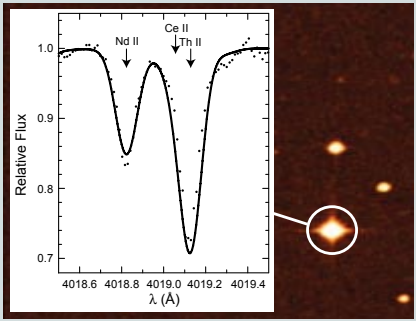

**Radioactive beam experiment**

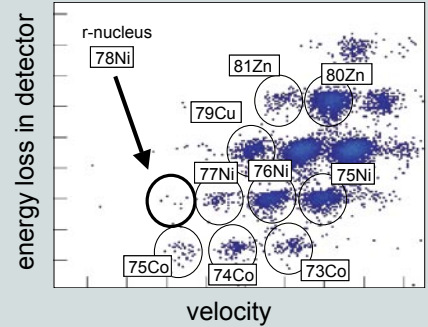

**Left:** A small portion of a high-resolution spectroscopic observation of the very metal-poor halo star CS 31082-001, obtained with the European 8 m VLT telescope, showing one of the 11 detectable absorption lines of the r-process element thorium (Th II) in this star. **Right:** The identification of the r-process nucleus $^{78}$Ni in a radioactive beam experiment at the NSCL. The measured $^{78}$Ni half-life was shorter than anticipated, which accelerates the r-process and constrains nucleosynthesis models. Image of star: copyrighted by the Space Telescope Science Institute (STScI Digitized Sky Survey©, 1993, 1994, AURA, Inc., all rights reserved).

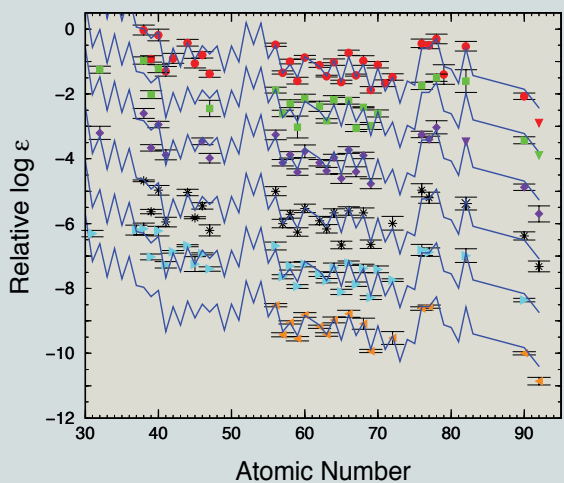

Observed elemental abundances produced by r-process events in the early galaxy are shown from six "r-process stars" displayed with an artificial offset for comparing the relative patterns. The similarity of the pattern from star to star and therefore from r-process event to r-process event is striking and remains to be explained by models. The solid lines are the abundance pattern of the solar system abundances attributed to the r-process. The good agreement for the heavy r-process elements reinforces the remarkable stability of the r-process pattern, while the discrepancies for the lighter r-process elements have been interpreted as an indication for a possible second r-process not observed in these stars but contributing to what is found today in the solar system. The detection of long-lived radioactive elements such as uranium and thorium, which have significantly decayed since their formation, can in principle be used to date the r-process event and to constrain the age of the universe. However, this requires r-process models with reliable nuclear physics to predict the initially produced amount of uranium and thorium.

radioactive materials needed. Of critical importance are also the charged particle reaction rates that generate the neutrons in the s-process. The large uncertainties in the main astrophysical neutron sources $^{13}$C($\alpha$,n) and $^{22}$Ne($\alpha$,n) need to be addressed in order to understand the s-process.

Reliable models of the s-process are important for predicting the seed of the proton process (p-process) in supernovae responsible for the origin of many neutron-deficient stable isotopes found in nature. To model the p-process itself, better nuclear data on ($\gamma$,n), ($\gamma$,p), and especially ($\gamma$,$\alpha$) photodisintegration reactions on mostly unstable nuclei are needed. A long-standing problem has been the origin of the neutron-deficient isotopes of molybdenum and ruthenium, which are found in relatively large abundances in the solar system that cannot be explained with current p-process models. The recently predicted vp-process might offer a possible solution to this problem, as these light p-process species can be produced in proton-rich supernova ejecta via a rapid-proton capture process that is accelerated by ($\nu$,p) reactions, with the free neutrons produced by neutrino captures on free protons. The nuclear physics requirements for modeling this process still need to be fully explored but likely involve the properties of very neutron-deficient nuclei.

In addition to stellar spectroscopy, meteoritic inclusions such as presolar grains have emerged as a major source of information on nucleosynthesis sites. These grains originate directly from red giant stars, supernovae, or novae, and having arrived in the solar system their detailed analysis has provided an unprecedented quantity of new, precise isotopic abundance data that can be directly compared with nucleosynthesis models provided the underlying nuclear physics is understood.

Certain unstable nuclei, whose origin is unknown, such as $^{26}$Al, $^{44}$Ti, or $^{60}$Fe are detected through their decay radiation by space-based gamma-ray observatories or through identification of their decay products in meteoritic inclusions. $^{60}$Fe has also been found recently in deep-sea sediments in layers thought to be polluted with radioactive isotopes from a nearby supernova about 2.8 million years ago. Proton capture rates on unstable nuclei are important to understand the origin of $^{26}$Al and the $^{44}$Ti production in supernovae. They need to be determined experimentally, in some cases through further developments at existing rare-isotope-beam facilities, and in some cases with FRIB. $^{60}$Fe is produced in massive stars by neutron capture, and the corresponding neutron capture rates, including the neutron capture rate on unstable $^{59}$Fe, need to be determined.

**What are the Nuclear Reactions that Drive Stars and Stellar Explosions?**

With FRIB and advances in stable-beam experiments there will be an opportunity to experimentally determine the nuclear physics that comprises the energy sources of stars and cosmic explosions such as X-ray bursts, novae, and Type Ia supernovae. Together with advances in observations, astrophysical modeling, and nuclear theory, major progress toward answering the open questions of the field will occur in the future.

Nuclear reactions in the pp-chains or the CNO cycles determine the evolution of stars during the long hydrogen-burning phase. The fusion of three alpha particles to $^{12}$C followed by the $^{12}$C($\alpha$,$\gamma$) reaction characterizes the red giant phase of stars and sets the stage for subsequent burning phases, such as carbon burning through $^{12}$C + $^{12}$C fusion and oxygen burning characterized by $^{16}$O + $^{16}$O fusion. These key reactions of late stellar burning are the main energy sources of the star and determine the duration of the respective burning phases. The extremely small cross sections of the stellar reaction rates result in the long lifetimes of stars but represent the main challenge to a direct experimental study of these reactions. While great progress has been made in many important cases, stellar models still suffer from large uncertainties in key nuclear reactions such as $^{12}$C($\alpha$,$\gamma$) that then also affect the modeling of core collapse and Type Ia supernovae.

Explosive hydrogen burning occurs in close binary star systems when hydrogen-rich material from a companion star accretes on the surface of a white dwarf or neutron star. Depending on the accretion conditions a thermonuclear explosion is triggered, which is observed as a nova or an X-ray burst. Novae occur through accretion onto a white dwarf, and spectroscopic observations of the ejecta provide detailed information about the nuclei formed in the explosion. As the ejected composition depends sensitively on the nuclear reaction rates, the nuclear physics needs to be understood before one can obtain information about the validity of nova models from comparison with observed abundances. Key questions are the possible contributions of novae to the origin of the elements, and the nature of mixing processes during the explosion that are likely at the heart of the difficulties to describe the observed composition and ejecta masses with current nova models.

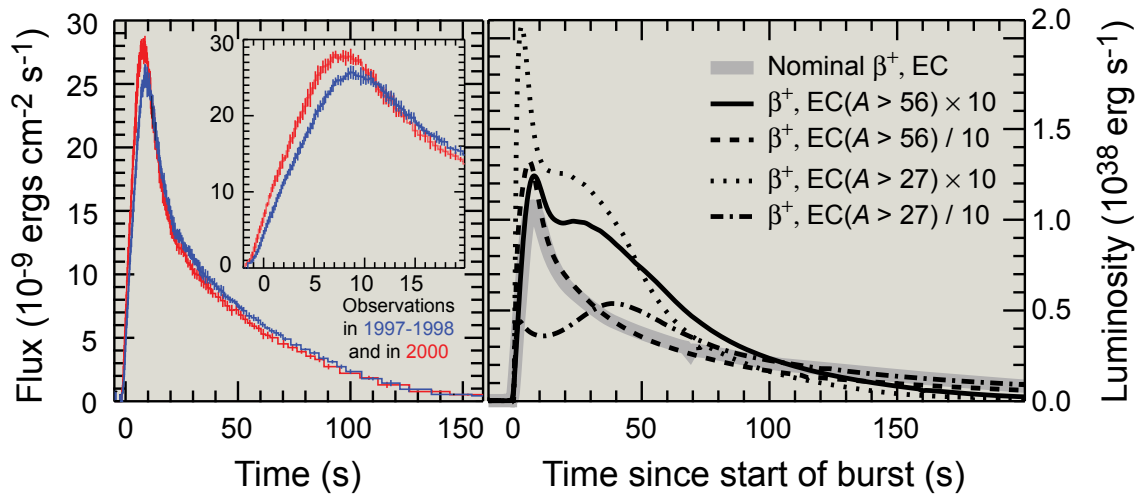

**Figure 2.20:** The left side shows precision observations of X-ray burst light curves from star GS 1826–24 by the RXTE observatory. The observations reveal systematic light curve changes from observations in the years 1997–1998 (blue) to observations in 2000 (red) that are likely related to changes in the accretion rate. Such measurements provide an excellent testing ground for X-ray burst models. The panel on the right shows predictions of X-ray burst light curves from state-of-the-art one-dimensional model calculations. To estimate nuclear physics uncertainties that influence the effective lifetimes of the nuclei in the rp-process, $\beta^+$ and electron capture (EC) rates of two groups of nuclei (with A > 56 and with A > 27) were increased and decreased by a factor of 10.

X-ray bursts, on the other hand, occur through accretion on neutron stars. They represent the most common type of thermonuclear explosions in astrophysics. Recent advances in X-ray astronomy with observatories such as Beppo-SAX, Chandra, XMM-Newton, RXTE, and INTEGRAL have provided a huge amount of new data and raised a number of open questions and puzzles including the increased burst recurrence times at higher accretion rates, the origin of millisecond oscillations and multipeaked bursts, and the mechanism behind the occasionally observed superbursts (figure 2.20). In addition, the interpretation of the multitude of observational signatures in terms of properties of the underlying neutron star is a major goal. In particular, the composition of the ashes of X-ray bursts needs to be understood, as they largely remain on the surface of the neutron stars and set the stage for the processes deeper in the crust. All these direct and indirect observables depend sensitively on the nuclear physics of neutron-deficient nuclei in the αp- and rapid proton capture (rp) processes that govern X-ray bursts.

The experimental determination of stellar reaction rates on unstable nuclei that play a critical role in novae and X-ray bursts has just begun with a few pioneering measurements that in most cases provide only indirect information. A huge step forward will be the development of reaccelerated rare-isotope beams produced by fragmentation and gas stopping at FRIB. This technique will potentially provide a wide range of rare-isotope beams that require only minimal beam development times. At existing facilities, this technique already can provide a much broader assortment of rare-isotope beams, albeit with relatively low intensities. Together with the few higher-intensity beams developed at ISOL facilities, this will in the near future open the door for a wide range of important experiments to further constrain the reaction rates in explosive hydrogen burning. FRIB will provide the beam intensities needed to finally start a full-blown program of measuring stellar reaction rates on unstable nuclei along the rp-process path using direct and indirect methods. Advances in nuclear reaction theory will be needed to fully interpret many of these experiments.

The direct measurement of reaction rates on stable nuclei that require high-intensity beams, and are needed to model stars and novae, also presents enormous challenges. The largest handicap is the small cross section coupled with large natural background, which prohibits the detection of the characteristic reaction signals. The use of underground-based low-energy accelerator facilities, as demonstrated by LUNA at the European Gran Sasso underground laboratory, significantly reduces cosmic-ray-induced background by several orders of magnitude. This approach is complemented by the development of active background-reduction techniques based on event identification or inverse kinematic techniques to reduce the natural radiation and beam-induced background. DUSEL will provide an opportunity for the development of such a facility in the United States. These measurements need to be complemented by extensive theoretical or indirect experimental studies of the resonance and interference structure at stellar energies because of the complexity of the reaction mechanisms near the reaction threshold.

Core collapse supernovae are initiated by gravity, but nuclear physics plays an essential part in the dynamics of the collapse and the nature of the core bounce. More progress is needed in multidimensional modeling of the hydrodynamics and neutrino transport to understand the explosion mechanism. Electron capture rates on stable and unstable nuclei are needed in these models as they affect the dynamics of the collapse and the development of the emerging shock front that drives the explosion. The theories used to predict these rates in the supernova environment need to be further developed, especially for heavier nuclei. To guide this effort with experimental constraints is essential but poses a major challenge for the future. FRIB will be required to address this challenge and will enable the systematic study of weak interaction strength on most of the relevant unstable nuclei using charge exchange reactions with rare-isotope beams.

Type Ia supernovae are the source of most of the elements in the iron region found in the universe and serve as cosmic standard candles. Such supernovae are powered by thermonuclear burning of carbon and oxygen in a white dwarf star that grows in mass beyond a critical limit. The explosion is triggered by $^{12}$C + $^{12}$C and $^{16}$O+$^{12}$C fusion reactions that are not well understood and need to be studied experimentally. Some of the electron capture rates on nuclei in the iron region needed for the study of core collapse supernovae are also required to determine the final nucleosynthesis of Type Ia supernovae, including the amount of $^{56}$Ni, their main power source.

## Atom Trap Trace Analysis

From the monitoring of nuclear-fuel reprocessing plants to the mapping of air and ocean currents, an increasing number of studies depend on scientists' ability to analyze the isotopic composition of extremely small quantities of material. Their most sensitive technique for doing this, accelerator mass spectrometry (AMS), requires that they bring their samples to a central lab. Now, however, researchers at Argonne National Laboratory have developed an ultrasensitive isotope-detection technique that is especially well suited for field deployment. In atom trap trace analysis (ATTA), individual atoms of a chosen isotope are captured and detected with a laser trap. Using current laser trapping methods this technology can be used to perform isotopic abundance analysis of alkali, alkali earth, and noble gas elements. A schematic diagram of an ATTA setup is shown in the figure. A gas sample is injected into the system through a discharge region, where a fraction of the atoms are excited into a metastable atomic level via electron-impact excitation. The thermal atoms are then transversely collimated, decelerated, and captured into a trap with laser beams. A trapped atom scatters photons from the laser beams and appears as a bright dot in the center of the vacuum chamber. A sensitive photon detector is used to measure the fluorescence and count the trapped atoms. This process is isotopically selective because atoms of different isotopes resonate at different frequencies. When the laser frequency is tuned to the resonance of a particular isotope, only atoms of this isotope are trapped and detected.

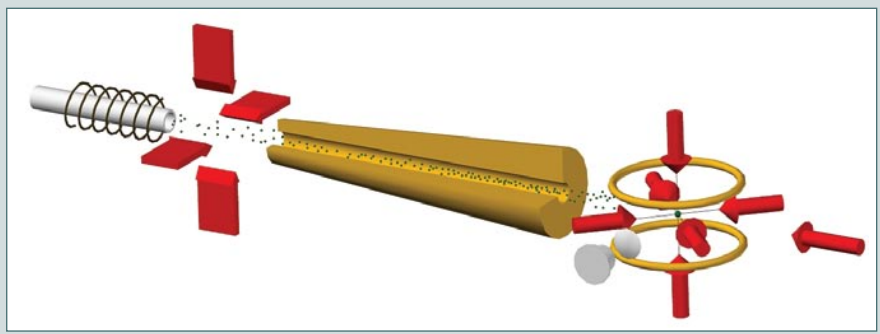

Schematic layout of an ATTA apparatus.

## OUTLOOK

Answering the five main questions for this field will require the capabilities of FRIB. The compelling science case for such a facility received strong endorsement by the National Academies of Sciences in their RISAC report (www7.nationalacademies.org/bpa/RISAC_PREPUB.pdf). The opportunities are numerous, and advances in all aspects of the science depend on critical experiments that can be uniquely performed with specific isotopes that are not available in sufficient quantities at the present time. A detailed list of experiments can be found in the RISAC report. With FRIB we will:

- Develop a comprehensive and coherent picture of the nature of nucleonic matter. FRIB will define and map the limits of nuclear existence, explore exotic quantum systems that inhabit these boundaries, and reveal new phenomena. Key experimental data from FRIB will help guide the development of a unified and predictive theory of nuclei.
- Understand the origin of the elements and the nuclear energy sources in stellar explosions. FRIB will provide key data, such as masses, lifetimes, and reaction rates needed for a quantitative understanding of important nucleosynthesis processes such as the r- and rp-processes.
- Test fundamental symmetries and search for ingredients of the New Standard Model. Key isotopes produced with high intensities at FRIB will provide new opportunities for high-precision tests of time-reversal and other symmetry violations.
- Provide isotopes for applications. FRIB will provide large quantities of new isotopes with unique properties that can be harvested for applications in human health, environmental issues and geosciences, nuclear energy, food and agriculture, material sciences, chemistry, and biology. FRIB will also impact national security, specifically in the areas of stockpile stewardship, homeland security and nonproliferation of nuclear materials, and diagnostics for high-energy density physics facilities.

In the most optimistic scenario, we can expect FRIB experiments to begin in 2017. It is therefore crucial that the existing facilities, at both national laboratories and universities, remain competitive and productive. These facilities focus on the different facets of research in nuclear structure and nuclear astrophysics, and provide a broad and coherent program. They will continue to provide exciting, cutting-edge scientific results, and their research and development efforts have a history of pioneering techniques, equipment, and instrumentation. These facilities and instruments complement future FRIB efforts, and are essential to address the open questions in nuclear structure and reactions, nuclear astrophysics, and the physics of fundamental interactions. The optimal utilization of all the existing facilities is crucial to maintain, and indeed grow, the community of researchers and students that will be ready when FRIB comes online.

The community has identified a path to optimally position it for the start of operations at FRIB. A number of modest upgrades of the national users facilities and of facilities at university laboratories are currently underway, and others are planned. These increments in capabilities will help the field maintain a leadership position in the short term, and a competitive one in the next decade while FRIB is under construction. However, over the last decade, operations at most, if not all, of these facilities have been affected by stagnant or declining budgets. It is important that sufficient funding be provided to operate the facilities and especially the user facilities efficiently and effectively.

Although any new facility should be equipped with state-of-the-art instruments, the situation for FRIB differs significantly from that of other major nuclear physics facilities brought online in the United States recently: the major leap forward of FRIB resides in the large beam power, which will produce more exotic neutron- and proton-rich isotopes at higher intensities, not in the energy or the properties of the exotic beams. Hence, the general requirements for the detectors at FRIB are in many cases the same as those applying to the existing detectors at ANL, HRIBF, and the NSCL. This suite of instruments is second to none and provides the community with many of the tools needed to start up a first-rate research program with exotic beams. In addition, new devices continue to be developed at national laboratories and universities to optimize the use of the available weak, rare-isotope beams. Among these figures the GRETINA project, a $1\pi$ detector that incorporates the new technology of gamma-ray tracking, which will increase the resolving power for certain classes of gamma-ray spectroscopy experiments by at least one order of magnitude.

Adequate funding for the continued development of state-of-the-art detectors at existing facilities for subsequent use at FRIB is an essential step toward ensuring that this new facility achieves its full potential at its start of operations. To this effect, the community of FRIB users has organized itself in a number of working groups focusing on specific instrumentation. Design and research and development activities are ongoing and require further support prior to construction at an existing facility. In this context, the extension of GRETINA to the full, 4π, GRETA array deserves special attention as this spectrometer will revolutionize gamma-ray spectroscopy in the same way that Gammasphere did. The timely and cost-effective completion of this project depends critically on the steady production of Ge detectors and on the continued availability of the highly specialized workforce currently working on GRETINA. Hence, the start of GRETA construction should begin immediately upon the successful completion of GRETINA.

# In Search of the New Standard Model

## OVERVIEW

The quest to explain nature's fundamental interactions, and how they have shaped the evolution of the cosmos, is among the most compelling in modern science. A major triumph in that quest came in the latter part of the 20th century with the development of the Standard Model—a comprehensive and detailed picture of the electroweak and strong interactions. Nuclear physicists have played a key role in that success, starting five decades ago with their observation of parity violation in nuclear beta decay, and continuing to the present day with their increasingly precise experimental tests of the Standard Model's predictions. The model has survived these tests with remarkable resiliency—which is why it is now accepted as the fundamental framework for three of the four known forces of nature.

Despite its successes, however, the Standard Model is at best incomplete. It cannot explain why neutrinos have mass—a fact revealed by the recent discovery of neutrino oscillations. It cannot account for the observed predominance of visible matter over antimatter in the cosmos. It has nothing to say about the nonluminous dark matter that makes up roughly 25% of the cosmic energy density, or the mysterious dark energy that appears to be responsible for cosmic acceleration. And it completely leaves out the force of gravity, which is obviously required for a fully unified theory of fundamental physics.

These and other shortcomings have motivated the search for a larger framework that will address the Standard Model's deficiencies while preserving its successes. Discovering the ingredients of this "New Standard Model" will require progress on the high-energy frontier, where increasingly energetic collisions allow physicists to probe particle structure at smaller and smaller scales. But just as importantly, it will require advances on the low-energy frontier, where nuclear physicists can probe the internal dynamics of particles with exquisite precision. Below we lay out a targeted program of experiments to do just that: a "New Standard Model Initiative" that represents one of the major thrusts in nuclear science for the next decade. This initiative will seek to answer three overarching questions:

- **What is the nature of the neutrinos, what are their masses, and how have they shaped the evolution of the cosmos?**

- **Why is there now more visible matter than antimatter in the universe?**

- **What are the unseen forces that were present at the dawn of the universe but disappeared from view as it evolved?**

Two highlights of this ambitious program—either or both of which could lead to revolutionary discoveries—are the search for neutrinoless double beta decay in atomic nuclei, and the search for permanent electric dipole moments of the neutron, electron, nuclei, and atoms. The neutrinoless double beta decay experiments could determine whether the neutrino is its own antiparticle, and therefore whether nature violates the conservation of total lepton number: a symmetry of the Standard Model whose violation might hold the key to the predominance of matter over antimatter in the universe. Similarly, the discovery of a nonvanishing permanent electric dipole moment would imply the violation of time-reversal symmetry, which could also help explain the matter-antimatter imbalance.

Other experiments in our program will probe neutrino mass and flavor oscillations, low-energy weak interactions of leptons and quarks, and the magnetic moment of the muon—all at unprecedented precision. The results of these experiments, together with the expected direct production of new particles at the soon-to-be-commissioned Large Hadron Collider (LHC), will provide the most important clues into the underlying physics that is the foundation of what we call the New Standard Model. With the guidance from theory, we expect that our experimental program will observe the "footprints" of forces—largely hidden today—that were important at earlier times during the evolution of the cosmos.

## A DECADE OF DISCOVERY

Nuclear science is in a uniquely strong position to pursue these studies at the precision frontier. Recent experimental and theoretical advances achieved by nuclear physicists studying neutrinos and fundamental symmetries have yielded a number of widely recognized results. Among the most noteworthy are:

- **The discovery of flavor oscillations among solar neutrinos.** For more than 30 years the solar neutrino problem perplexed physicists. The initial observation by Ray Davis, Jr., that only about a third as many electron neutrinos came from the Sun as expected from carefully constructed solar models, such as those of John Bahcall, was finally resolved in 2001 when the Sudbury Neutrino Observatory (SNO) collaboration discovered that flavor change was occurring among the solar neutrinos. All of the neutrinos were in fact there, but two-thirds had converted to μ and τ flavors. Indeed, the close agreement of the measured total flux of $^8$B neutrinos with the predictions of the standard solar models implies that those models correctly predict the central temperature of the Sun to 1% accuracy, an astonishing achievement. In 2002, Davis received the Nobel Prize, shared with Masatoshi Koshiba and Riccardo Giacconi, for opening the era of neutrino astronomy. The SNO results show that all three neutrinos participate in flavor mixing; they define the level ordering for the two lightest neutrinos, and in conjunction with other oscillation data and results from tritium beta decay, they have revealed that the average mass of the three neutrinos lies between 0.02 and 2.3 eV. Neutrino mass and oscillations are a clear departure from the SM.

- **The observation of oscillations among reactor antineutrinos.** The KamLAND Collaboration reported the first observation of oscillations from a terrestrial neutrino source in 2003. The KamLAND results isolated the large mixing angle solution among the possible solutions for solar neutrino mixing and showed for the first time a true oscillation pattern, with the electron flavor disappearing and then returning as the ratio of propagation distance to energy increases. The KamLAND experiment yields a very precise measure of the difference between the squares of the masses of two neutrino states.

- **The most precise measurement of the anomalous magnetic moment of the muon.** The muon anomalous magnetic moment, $a_\mu$, provides a uniquely sensitive probe of both the SM and its possible extensions. The SM prediction has an impressively small fractional uncertainty—0.5 parts per million—that has now been matched by the experimental uncertainty obtained by the Brookhaven E821 collaboration. A comparison of theory and experiment indicates a difference $\Delta a_\mu$ of 3.4 standard deviations. The magnitude of the deviation from the SM is striking because it closely follows expectations of supersymmetric models.

- **The most precise determination of the low-energy weak mixing angle in parity-violating electron-electron scattering.** The weak mixing angle is one of the most important parameters in the SM, as it characterizes how the "primordial" force carriers of the electroweak interaction mix to become the photon and neutral weak gauge boson ($Z^0$) that we observe today. The SM predicts how the value of the weak mixing angle depends on the interaction energy. The comparison of the results of the SLAC E158 experiment with the studies of high-energy $e^+e^-$ collisions at CERN and SLAC forms a stringent verification of the SM prediction for the energy dependence of weak mixing angle.

- **New theoretical calculations have sharpened the interpretation of neutrino studies and fundamental symmetry tests, and have delineated their prospective implications for the New Standard Model.** These theoretical advances include both improved SM predictions for low-energy electroweak processes and analyses of their sensitivity to a variety of candidates for the New Standard Model. Theoretical uncertainties associated with strong interaction effects in neutron and nuclear β-decay, the energy dependence of the weak mixing angle, the anomalous magnetic moment of the muon, and matrix elements for neutrinoless double β-decay have been substantially reduced. A comprehensive set of calculations of supersymmetric effects in low-energy precision observables has been completed. Refined computations of the matter-antimatter asymmetry and the related implications of new electric dipole moment searches have been undertaken. The prospective impact of the program of experiments outlined below relies heavily on these and other recent theoretical advances.

## THE NEW STANDARD MODEL INITIATIVE

The recent achievements listed above, together with other experimental and theoretical developments, have begun to sketch the outlines of what the New Standard Model must look like. While we now know that neutrinos have mass and that they oscillate from one type to another, specific questions remain. What are the neutrino masses? Are neutrinos their own antiparticles? Do they violate CP symmetry? Similarly, precise tests of low-energy electroweak processes have yielded impressive agreement with SM expectations, but the next generation of high-precision tests could change this trend. Indeed, the recent experimental result for the muon anomalous magnetic moment differs significantly from SM predictions, and the magnitude of this difference is an important clue for the New Standard Model. Are we on the verge of uncovering a complete supersymmetric particle world? Does the universe extend into extra dimensions? Could another exotic explanation be responsible?

To answer these basic questions will require the pursuit of a select set of sensitive experiments involving electroweak interactions of nuclei, light hadrons, neutrinos, electrons, and muons, experiments whose physics reach complements—and in some cases exceeds—direct searches for new particles at high-energy colliders. Collectively, this set of experiments—the "New Standard Model Initiative"—represents a concerted effort to exploit the unique opportunities at the low-energy precision frontier to discover key ingredients of the New Standard Model. Two experimental programs having outstanding discovery potential anchor the initiative: the search for neutrinoless double beta decay of atomic nuclei and the search for a permanent electric dipole moment of the neutron, neutral atoms, and the electron. The initiative also includes a targeted set of precision studies of neutrino properties and fundamental symmetries: probes of neutrino mass and mixing angles with ordinary β-decay, low-energy solar neutrinos, and reactor neutrinos; and searches for tiny—but significant—departures from SM expectations for the muon anomalous magnetic moment, parity-violating electron-scattering asymmetries, and weak decays of matter. The specific sequence for staging the elements of the precision program will depend, in part, on the availability of experimental facilities. The techniques of high precision and the use of weak probes pioneered by this community of nuclear physicists will also be directed at solving select problems in hadronic physics, for which a few noteworthy examples are included below.

The New Standard Model Initiative of nuclear physics promises to deliver exciting discoveries in the realm of neutrinos and fundamental symmetries. Unlike other components of this Long Range Plan, the initiative is not largely hosted by one major facility. The requirements for each of these sensitive experiments are generally quite unique. Many require the ultralow background only found deep underground. Some neutrino experiments are built near nuclear reactors, while others sit at long distances from particle accelerators. Pion, muon, and polarized electron scattering experiments require the highest beam intensities from our existing nuclear physics accelerators and their planned upgrades, such as CEBAF at Jefferson Lab and the AGS at the Brookhaven Laboratory. The new Fundamental Neutron Physics Beamline (FNPB) at the Oak Ridge Spallation Neutron Source (SNS) will provide cold and ultra-cold neutrons at the highest intensities in the world, creating opportunities that will build on present and future efforts at the National Institute of Standards and Technology (NIST) reactor and LANSCE at Los Alamos. Finally, some experiments can be carried out "in the basement" using ultra-sensitive particle traps. While the breadth of facilities and techniques employed is vast, the physics focus is sharp. Together with the results from the energy-frontier experiments supported by high-energy physics, the outcome of these precision-frontier nuclear physics experiments could revolutionize our understanding of fundamental interactions and consequently define the New Standard Model.

## DEEP UNDERGROUND SCIENCE AND ENGINEERING LABORATORY: DUSEL

A vital component of U.S. leadership in this field will be the construction of a Deep Underground Science and Engineering Laboratory (DESEL) in the United States, together with the suite of low-background experiments—including neutrinoless double beta decay and solar neutrino observations—that require its shielded, clean environment. As these experiments that are central to the New Standard Model Initiative explore unprecedented sensitivity levels, backgrounds of cosmic origin, especially fast neutrons, pose a serious obstacle to further improvement in the physics reach. Mounting the experiments at great depths can lessen these background effects. The need for a dedicated, deep facility was identified in the 2002 Long Range Plan, and that need has only intensified in the intervening years. The nuclear physics community has once again called for construction of DUSEL, as is explained in more detail in the Chapter 6, "Recommendations."

DUSEL will be a world-leading scientific facility with unique capabilities. It will provide around-the-clock access to locations at several depths in which cosmic-ray backgrounds, radioactive contaminants, and seismic disturbances from mining have all been minimized. Experiments will aim to discover key elements of the New Standard Model by searching for tiny signatures of neutrinoless double beta decay,

# The Deep Underground Science and Engineering Laboratory

In 1987, when physicist Michael Moe and his colleagues took their exquisitely sensitive $^{82}$Se double beta decay experiment from a surface lab to a cavern under the Hoover Dam, Moe described it as "like stepping from a busy marketplace into the quiet of a cathedral." The marketplace hubbub, in this case, came from the ceaseless shower of cosmic rays that pours through every square meter of the Earth's surface—20,000 cosmic-ray muons alone every minute, along with neutrons, neutrinos, heavy nuclei, and a host of other species. The rare decay events that Moe and his colleagues were looking for were like the faintest of whispers, lost in all the uproar. Only in the comparative hush under Hoover Dam were they able to pick it out: the first observation of two-neutrino double beta decay of a nucleus.

Today, two decades later, physicists are following much the same strategy. The difference is that the frontier processes they are looking for may now involve no more than a handful of events per year—which means that their experiments will have to be sheltered from cosmic rays not by a dam, but by kilometers of solid rock.

In recognition of this fact, developed nations around the world, including six of the "G8" group, have invested heavily in deep underground laboratories. Those investments in Canada, Europe, Japan, and Russia have been rewarded with extraordinary results, such as the discovery of neutrino oscillations and mass. U.S. scientists Ray Davis and Frederick Reines provided leadership in the early years. Now, U.S. scientists are once again proposing a unique facility, the Deep Underground Science and Engineering Laboratory, which would house an equally unique suite of ultralow-background experiments ranging from neutrinoless double beta decay to solar neutrino observations. Indeed, the need for such a dedicated underground facility is placed in stark relief by the unprecedented sensitivity levels these experiments demand. For instance, the next-generation neutrinoless double beta decay experiments will aim to detect a decay half-life as long as $10^{27}$ years, which will require a detector background rate less than about one count per ton of target material per year. DUSEL can also host new neutrino detectors sensitive to solar neutrinos, supernova neutrinos, and geoneutrinos. Without going deep underground, backgrounds of cosmic origin, especially fast neutrons, are a serious obstacle.

The figure shows the rate per nucleus of interactions of cosmic-ray secondary neutrons with energies above 100 MeV as a function of depth. Those neutrons are the most difficult component of the cosmic rays to shield against. The adjacent panels show the signal rates for WIMP dark-matter particle interactions, for neutrinoless double beta decay, and for solar neutrinos.

The comparison shows how much easier it is to carry out such searches at great depths. Where signals from the new physics will appear is unknown, but the ranges on the scales to the right cover what is expected for three major physics campaigns. As experiments become bigger and more sensitive (moving down on the graph), the need for depth becomes more acute.

In addition to the nuclear physics studies for which it is vital, DUSEL will address other fundamental questions in physics. The eventual development of an intense accelerator neutrino source and a megaton-scale detector would reveal whether CP violation was large in the neutrino sector—in particular, whether it was large enough to explain the baryon asymmetry in the universe. That program awaits a determination of the third mixing angle, $\theta_{13}$. The large detector will also offer a hundredfold advance in sensitivity to proton decay. Finally, the identity of dark matter is still unknown: several DUSEL experiments will be directed toward the resolution of this mystery.

NSF has initiated a process through which DUSEL can be realized. In 2007, the Homestake site was selected for the proposed laboratory. The funding and construction is central to the New Standard Model Initiative and to the U.S. nuclear science program in general, as it will be a premier domestic facility where exciting and pressing research in our and related fields can be conducted.

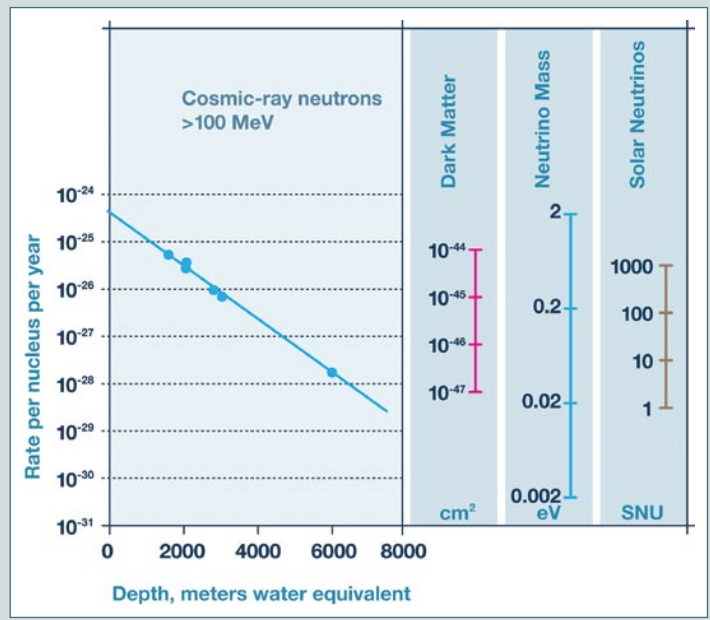

Energetic neutrons from cosmic-ray interactions as a function of depth underground. Signal rates versus depth are shown in the parallel panels for typical dark matter, neutrino mass, and solar neutrino studies.

sterile neutrinos, and dark matter. They will help us explain how the elements were made, measure the energy output of the Sun, and determine the distribution of uranium and thorium in the Earth. Not only physics will benefit, but a wide range of other sciences, including geoscience, geomicrobiology, and engineering, will develop new programs working alongside physicists. Synergies that did not previously exist are already taking shape; geologists and nuclear physicists find themselves tackling the problem of the Earth's internal heat in a collaborative way that was not considered before. The new laboratory will also provide a focus for education and outreach that will draw young people to science at a true frontier.

The National Science Foundation has taken leadership of the process for DUSEL and has recently decided on a site for the proposed laboratory, the former Homestake gold mine in South Dakota. If approved, construction could start in 2011.

## NEUTRINO MASS, LEPTON NUMBER, AND THE ORIGIN OF MATTER

The neutrino is the only known fundamental matter particle that carries no electric charge. Unlike other fermions, it may then be its own antiparticle, a so-called "Majorana" fermion. At present, we do not know whether neutrinos are Majorana fermions or—like the other fermions—have distinct particle and antiparticle states ("Dirac" fermions). If neutrinos are Majorana particles, then their interactions would violate the conservation of total lepton number. The nonconservation of lepton number is a key ingredient in "leptogenesis," a theoretically attractive explanation for the predominance of matter over antimatter involving neutrinos in the early universe. The discovery that neutrinos are Majorana particles would provide strong evidence for the viability of this scenario. It would also support the "seesaw mechanism," a widely held explanation for the unusually tiny scale of neutrino mass. On the other hand, if neutrinos are Dirac particles, then other explanations for the matter-antimatter imbalance, such as electroweak baryogenesis (see below), would be more likely. Explaining the small scale of neutrino mass would require new ideas.

The overall scale of the masses is of great cosmological interest. It is bounded from below by the oscillation data and the fact that no mass can be less than zero, which leads to an average neutrino mass of at least 0.02 eV. As small as that may seem, neutrinos nevertheless outweigh luminous stars in the universe. The upper end of the mass scale is a hundred times larger, bounded by direct mass measurements based on the β-decay of tritium. Cosmology itself yields a somewhat model-dependent upper limit of about 0.2–0.5 eV for the average mass. There is an opportunity for a decisive test of cosmological models if the laboratory and cosmological mass sensitivities can be simultaneously improved.

These are two of the most important experimental questions in physics. Are neutrinos and antineutrinos the same particle? What are the masses of the neutrinos? The most powerful experimental tool for resolving the Majorana versus Dirac nature of neutrinos is the search for the neutrinoless double beta decay (0νββ) of nuclei. In this process, two neutrons (protons) are converted into two protons (neutrons) with the emission of two electrons (positrons) and no antineutrinos (neutrinos). The absence of (anti)neutrinos in the decay implies that total lepton number is violated. Irrespective of the detailed, underlying mechanism, the observation of 0νββ decay would provide the "smoking gun" evidence that neutrinos are Majorana particles and would constitute a major discovery.

Depending on the nature of the underlying lepton-violating mechanism, 0νββ experiments also have the potential to yield important information on the absolute mass scale and, in combination with other experiments such as nuclear β-decay neutrino mass measurements, may be able to provide information on the hierarchy of the masses. For example, if 0νββ decay is facilitated by the exchange of a light, virtual Majorana neutrino, then the decay rate is directly related to neutrino mass as illustrated in figure 2.21. The vertical axis gives $m_{\beta\beta}$, the "effective mass" of a light Majorana neutrino in 0νββ decay, while the horizontal axis is the mass of the lightest neutrino $m_\beta$. The different regions are labeled according to the hierarchy of neutrino masses.

It is possible that the neutrinoless decay could proceed not by the exchange of a light Majorana neutrino but rather by the exchange of some other, much heavier Majorana fermion. Particles of this type are postulated to exist in many candidates for the New Standard Model, such as supersymmetry or models having extended gauge symmetries. The exchange of such a heavy Majorana fermion could yield a 0νββ signal consistent with the one expected for a quasidegenerate or inverted hierarchy, but the masses of the light neutrinos could be much smaller. In such a case one could observe a 0νββ decay signal even when future direct neutrino mass searches gave small masses.

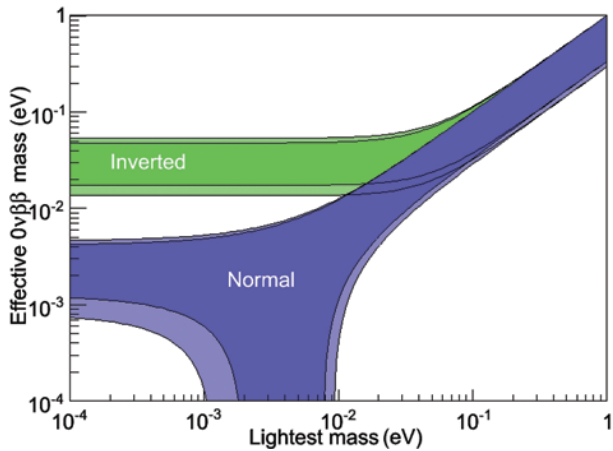

**Figure 2.21:** Relationship between the effective mass for neutrinoless double beta, $m_{\beta\beta}$ (vertical axis), versus the lightest mass neutrino (horizontal axis), for both the inverted and normal hierarchy scenarios. The lighter shaded regions include uncertainties in the neutrino mixing matrix values. The effective mass calculations use neutrino mixing matrix elements and uncertainties from the 2006 Particle Data Group evaluation, with the exception of $\theta_{13}$, where an upper limit of 10° was used. Next-generation double beta decay experiments and the KATRIN nuclear beta decay neutrino mass experiment expect sensitivities in the "quasidegenerate" region above about 0.1 eV.

**A Neutrinoless ββ Program**

During the coming decade, U.S. nuclear physicists will be engaged in a coordinated international program of 0νββ measurements. The program begins with measurements having initial sensitivities to neutrino mass in the "quasidegenerate" region (greater than about 0.1 eV; see figure 2.21), and may be followed by larger-scale, one-ton mass experiments that have a discovery potential near the inverted-hierarchy mass scale, about 0.045 eV. Multiple 0νββ experiments using different isotopes and experimental techniques are important not only to provide the required independent confirmation of any reported discovery but also because different isotopes have different sensitivities to potential underlying lepton-number-violating interactions. The initial experiments should strive to demonstrate backgrounds and the scalability required for future one-ton-scale experiments. In the immediate term, two of the three U.S. 0νββ experiments, CUORE, EXO, and Majorana, have major nuclear physics involvement. EXO is a $^{136}$Xe project supported in part by the DOE Office of High Energy Physics and NSF.

CUORE—the Cryogenic Underground Observatory for Rare Events—is a bolometric detector searching for 0νββ in $^{130}$Te. The Italian–Spanish–U.S. collaboration plans to install and operate $TeO_2$ crystals containing 200 kg of $^{130}$Te at the underground Laboratori Nazionali del Gran Sasso in Italy. Replacing the natural Te with isotopically enriched material in the same apparatus would subsequently lead to a detector approaching the ton scale.

The Majorana collaboration is engaged in an research and development effort to demonstrate the feasibility of using hyperpure germanium (Ge) diode detectors in a potential one-ton-scale 0νββ experiment. The initial Majorana research and development effort, known as the Majorana Demonstrator, utilizes 60 kg of Ge detectors, with at least 30 kg of 86% enriched $^{76}$Ge in ultralow background copper cryostats, a previously demonstrated technology. This Canadian–Japanese–Russian–U.S. collaboration is working in close cooperation with the European GERDA Collaboration, which proposes a novel technique of operating Ge diodes immersed in liquid argon. Once the low backgrounds and the feasibility of scaling up the detectors have been demonstrated, the collaborations would unite to pursue an optimized one-ton-scale experiment.

Several other promising opportunities to carry out sensitive 0νββ experiments exist, and U.S. nuclear physicists have indicated an interest in being involved. The most notable is known as SNO+, a proposed $^{150}$Nd-doped scintillator measurement that would utilize the previous Canadian–U.K.–U.S. investments made in the Sudbury Neutrino Observatory.

The next steps depend on what is observed. If a 0νββ signal is discovered and confirmed in the quasidegenerate region, then we would have proof that neutrinos are Majorana particles and that lepton number is violated. However, the mechanism by which the decay proceeds would remain an open question. Since neutrinos are already known to be massive, there would definitely be a contribution from light Majorana neutrino exchange, but might new, heavy Majorana particles also be present, even dominant? Distinct isotopes vary in their sensitivities to different mechanisms, and a systematic experimental study would identify the most viable options. Theoretical calculation of the decay matrix elements in these isotopes for different mechanisms with significantly reduced uncertainties is a crucial element in defining this phase of the program. Access to advanced computing capabilities is essential for such calculations.

If 0νββ is not observed in the quasidegenerate region, the next phase of the program is to build one-ton-scale experiments having sensitivity to the inverted hierarchy mass region. The U.S. nuclear physics community anticipates playing a leading role in selecting the most promising technologies and isotopes for these next-generation experiments. The time scale for starting one-ton-scale experiments is about 2013, well matched to the schedule for the proposed Deep Underground Science and Engineering Laboratory.

**Nuclear Beta Decay**

Direct information on neutrino mass, complementary to 0νββ decay, comes from the shape of the electron spectrum in ordinary beta decay. The details of the electron spectrum from the nuclei $^3$H and $^{187}$Re near the decay endpoint can reveal the value of $m_v$ independent of whether neutrinos are Majorana or Dirac. The Karlsruhe Tritium Neutrino experiment will be able to discover a 0.35 eV mass at the five standard deviation significance, or rule out a mass above 0.2 eV at the 90% confidence level. The $^{187}$Re experiments using bolometric techniques are not yet at this level of sensitivity but in the future may be capable of attaining improved sensitivity. A nonzero result for $m_v$ at the level accessible to KATRIN would imply that the neutrino spectrum is quasidegenerate. If neutrinos are Majorana fermions, then one would expect a nonzero signal in 0νββ experiments corresponding to several hundred milli-eV. If no such signal is observed, it would suggest that neutrinos have Dirac masses.

## ELECTRIC DIPOLE MOMENTS: CP SYMMETRY AND THE ORIGIN OF MATTER

Although the interactions of Majorana neutrinos in the early universe could have been responsible for the matter-antimatter asymmetry, it is also possible that the asymmetry arose during the era of electroweak symmetry breaking when the universe was about $10^{-11}$ seconds old. According to the SM, the spontaneous breakdown of electroweak symmetry is responsible for generating masses of the elementary particles. Interactions between these particles that violate CP symmetry could lead to the particle-antiparticle asymmetries that subsequently evolved into a net excess of baryonic matter over antimatter; that is, our present matter-dominated universe. This mechanism is called "electroweak baryogenesis." While, we know that CP symmetry is violated in the decays of kaons and B mesons, it is far too feeble to explain the observed matter-antimatter asymmetry. Consequently, if electroweak baryogenesis is responsible, new CP-violating interactions associated with the electroweak scale must exist.

Many candidates for the New Standard Model include CP-violating interactions among particles having masses somewhat greater than 200 times the proton mass, a typical electroweak breaking mass. High-energy collider experiments may discover such particles, but it will be difficult, if not impossible, for collider experiments to determine directly whether the particle interactions exhibit CP violation at a level required to explain the matter-antimatter asymmetry. An alternate and very powerful probe of new electroweak CP violation is to search for a permanent electric dipole moment (EDM) of an elementary particle or quantum bound state. The principles of quantum mechanics tell us that the interaction between an EDM and an applied electric field $\vec{E}$ is proportional to $\vec{S}\cdot\vec{E}$, where $\vec{S}$ is the spin of the particle or quantum system. This interaction is odd under both time-reversal (T) and parity (P) transformations. By the CPT theorem of quantum field theory, a nonzero EDM implies the presence of CP violation.

Searches for the permanent electric dipole moments of the electron, muon, neutron, and neutral mercury atom have been carried out with impressive sensitivities, but so far they have yielded null results. Within the SM, the absence of observed EDMs is not surprising; the SM-based EDMs are very tiny, occurring only through highly forbidden intermediate processes. However, popular SM extensions generally predict the existence of considerably larger EDMs, which opens a broad window for discovery between the present measured limits and the SM "background." Present EDM limits have already ruled out or greatly constrained many theories. Models that survive, such as supersymmetry, must be modified to account for severe constraints imposed by these searches.

The possibility also exists that the strong interaction—described by quantum chromodynamics (QCD)—contains CP-violating effects, parameterized by a quantity called $\bar{\theta}$. On general grounds, one expects $\bar{\theta}$ to have a magnitude of order one, but the null results obtained from the neutron and $^{199}$Hg EDM searches imply that $\bar{\theta}$ is at most of order $10^{-10}$! The most popular explanation for this surprising situation is the existence of a new symmetry of QCD, called Peccei-Quinn symmetry, which is associated with a hypothetical particle called the axion. Apart from its relation to this so-

# Breakdown of Symmetry: Why More Matter than Antimatter?

The Big-Bang Theory of cosmology suggests that the universe should contain equal amounts of matter and antimatter, yet the present universe contains mostly matter. Forty years ago, Andrei Sakharov identified the key ingredients needed to generate the matter imbalance, one of them being a significant violation of charge-parity (CP) symmetry in particle interactions. Although tiny violations of CP symmetry in certain particle decays have been known and studied for decades, the required CP-violation effect remains undiscovered. The Standard Model simply does not allow the observed matter asymmetry to have occurred. Perhaps neutrinos were the culprits (see The Neutrino: Stealthy Messenger). However, another path occurs in some proposed New Standard Models, such as supersymmetry, that postulate large CP-violating interactions in the early universe. A telltale imprint of this or other mechanisms would be a permanent electric dipole moment (EDM) of a basic particle.

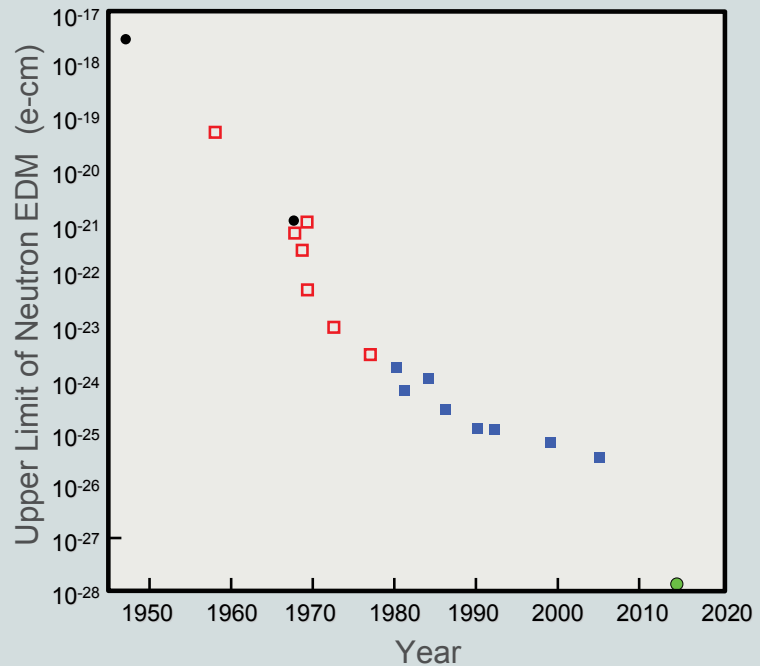

Experimental upper limits on the neutron EDM by year of publication. The green point at the bottom of the figure is the expected sensitivity of a new measurement that is being planned at the Spallation Neutron Source at Oak Ridge.

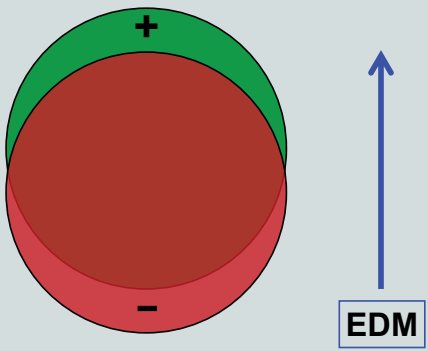

EDM is the electric dipole moment, which for an uncharged particle can be expressed in terms of a small separation of positive and negative charge within the particle.

An EDM is basically a tiny separation between a particle's positive and negative charges, as illustrated in the figure to the left. For the neutron, as well as for the electron and some nuclei, the existence of an EDM can provide the "missing link" for explaining why the universe contains more matter than antimatter.

Nuclear physicists have spent many years reducing the limits for a neutron EDM, as shown in the figure above. The present upper limit indicates that the center of the positive charge must be displaced from the center of the negative charge by less than 0.0000000000003 times the neutron's radius. In other words, if the neutron were the size of the Earth, the charges would be displaced by less than the thickness of a human hair!

A new experiment at the Fundamental Neutron Physics Beamline at Oak Ridge National Laboratory offers the prospect of extending this search by improving the sensitivity more than a hundredfold, as indicated by the green dot off-scale in the plot. The ambitious goal is well within the predicted parameter space of the models that can explain the matter asymmetry. If no EDM is found at this level, radically new explanations will have to be developed!

called "strong CP problem," the axion is an interesting candidate for the relic density of cold dark matter in the universe.

Experimental plans aim to improve the sensitivity of various EDM searches by two to four orders of magnitude. At these sensitivities they could uncover both CP violation associated with the $\bar{\theta}$ term of QCD and/or CP violation associated with new interactions at the electroweak scale. A nonzero EDM would constitute a truly revolutionary discovery. From the consideration of cosmology, one ought to observe nonzero EDMs in the next generation of experiments if electroweak baryogenesis is responsible for the matter-antimatter asymmetry. This point is illustrated in figure 2.22, where the sensitivity of the electron and neutron EDM to CP violation in supersymmetry is shown and compared to expectations based on the observed baryon asymmetry. The sensitivity of the next-generation EDM experiments extends well beyond expectations based on the baryon asymmetry. Moreover, searches for supersymmetric particles at the LHC could probe only a portion of the relevant parameter space in mass, and with limited sensitivity to new CP-violating phases.

As with $0\nu\beta\beta$ experiments, where multiple isotopes are essential, it is important to search for EDMs in a variety of systems to identify the underlying physics. For example, a nonzero result for the neutron EDM would indicate either the QCD $\bar{\theta}$-term or new electroweak CP violation as the culprit. Then, the electron EDM result will point to which one is right—a null electron EDM (eEDM) implies the former, while a nonzero result implies the latter. Additional experiments involving the electron, neutral atoms, and even the deuteron nucleus would be needed to understand the mechanism.

During the coming decade, nuclear physicists will pursue a program using different systems that will exploit the complementary sensitivity to various possible sources of CP violation. Recent technical developments have opened the way for significant improvements in EDM searches of the neutron, neutral atoms, and the electron. We briefly summarize the experimental plan.

**Neutron and Deuteron EDM**

A new experiment is underway at the FNPB to improve the sensitivity to the neutron EDM (nEDM) by a factor of ~100. This involves producing significantly higher densities of ultra-cold neutrons compared to previous experiments. The technique uses superfluid helium, which will also allow the use of higher electric fields that are necessary to further increase the sensitivity. The nEDM experiment is expected to begin construction in 2008. The field of EDM searches is intensely active worldwide, with several new neutron EDM experiments underway in Europe and discussions taking place in Japan. The projected sensitivity of the FNPB experiment is highly competitive with these other projects and relies on considerably different systematic considerations, which are critical to eliminate possible false EDM signals.

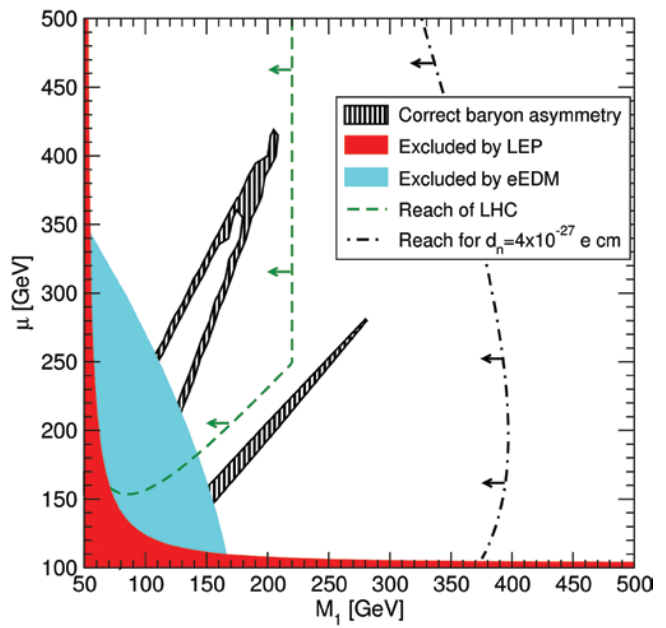

**Figure 2.22:** Experimental sensitivities to new supersymmetric particles and electroweak baryogenesis. The axes give relevant supersymmetric particle masses ($\mu$ and $M_1$). The hatched region indicates allowed values for the observed baryon asymmetry of the universe. The solid red region is excluded by supersymmetric particle searches at the LEP collider, while the solid blue region is ruled out by the current limit on the EDM of the electron. The LHC will explore the region to the left of the green dashed line. A future neutron EDM experiment with a sensitivity of $4 \times 10^{-27}$ e·cm will explore the region up to the black dashed line. The next-generation neutron and electron EDM experiments discussed in this Long Range Plan will reach beyond this sensitivity.

An additional exciting possibility involves a search for the EDM of the deuteron using a storage ring. This concept utilizes the motional electric field created by the confining magnetic field of the storage ring. This new concept appears to have significant sensitivity to the neutron and proton EDM as well as a possible EDM generated by CP-violating forces between the two nucleons. However, additional research and development is needed to determine the ultimate feasibility of this experiment.

**Atomic and Electron EDM**

Other very promising experiments are also under development to search for atomic and electron EDMs. Certain radioactive atoms possessing a large octupole deformation are expected to have greatly enhanced sensitivity to CP-violating forces in the nucleus. Both $^{225}$Ra and $^{223}$Rn show promise as potential high-sensitivity deformed nuclei. Currently experiments using these nuclei are being planned or pursued at laboratories around the world, including Argonne National Laboratory (using a $^{229}$Th source with $^{225}$Ra as a decay product) and TRIUMF in Canada (using a radioactive beam).

Dramatically improved searches for both electron and atomic EDMs may be possible using the ultra-high densities of electrons and nuclei in solid and liquid systems, coupled with superconducting quantum interference devices (SQUIDs). An experiment searching for the electron EDM is being pursued by an Indiana–Yale collaboration using the solid gadolinium gallium garnet. A Princeton experiment is using liquid xenon to search for an atomic EDM.

This program of EDM searches described above will require new advances in nuclear theory. Computations of the neutron EDM using lattice QCD methods are now being refined and could lead to significant reductions on the hadron structure uncertainties. Similarly, nuclear theorists have recently advanced our understanding of atomic EDMs by carefully delineating the effects associated with finite nuclear size and magnetic interactions between the nucleus and atomic electrons. New computations of these effects using state-of-the-art many-body methods will be pursued in the coming years. Substantial advances in computation of the matter-antimatter asymmetry resulting from new electroweak CP violation are underway, which will sharpen the cosmological implications of the new EDM searches.

## NEUTRINO OSCILLATIONS AND INTERACTIONS

The decay width of the $Z^0$ boson establishes that there are three and only three neutrinos that take part in the weak interaction. Neutrino oscillations mean that the three neutrinos have masses and mixed flavor. The flavor basis and the mass basis are believed to have a "unitary" relationship, which follows provided there are no other kinds of neutrinos than the three. In this case, the correspondence between the two descriptions can be *completely defined* with only three trigonometric angles, $\theta_{12}$, $\theta_{13}$, and $\theta_{23}$, and certain CP-violating phases: one if neutrinos are Dirac and three if they are Majorana. The unitarity has been in question for a decade with evidence from the LSND experiment that additional flavors of neutrino may exist. The results from the MiniBooNE experiment in 2007 do not confirm those indications and are consistent with unitarity.

**Precise Determinations of Oscillation Parameters.** The three angles have important implications for astrophysics and fundamental physics. Two angles are large and have now been measured, while the third, $\theta_{13}$, is small and is known only to be less than 10°. The solar neutrino experiments SNO, SAGE, Gallex, GNO, and Cl-Ar yield $\theta_{12} = 33.9 \pm 1.6°$, while the reactor experiment KamLAND fixes the corresponding mass splitting $\Delta m_{12}^2 = (8.0^{+0.4}_{-0.3}) \times 10^{-5}$ eV$^2$. A precise value for $\theta_{12}$ is needed for a test of unitarity based on the Sun, the most precisely calibrated and remote neutrino source, and in the determination of the energy production rate of the Sun, as described below. Still tighter definition of this angle will emerge from SNO, and in fact solar neutrino experiments will be the only source of precision information on $\theta_{12}$ for the immediately foreseeable future. A new determination would be possible by comparison of the fluxes from electron elastic-scattering experiments such as Borexino, KamLAND, and CLEAN, with the flux measured in a pure charged-current experiment such as LENS.

How big is the unknown angle $\theta_{13}$? CP violation in neutrino interactions—with its implications for the origin of matter—is only measurable in neutrinos if $\theta_{13}$ is nonzero. In addition, the value of $\theta_{13}$ is required for a precise measurement of the solar energy production rate, for understanding flavor-change effects in supernovae and subsequent r-process nucleosynthesis, and for planning second-generation long-baseline accelerator experiments. There are two approaches to determining $\theta_{13}$. One exploits the first-generation long-baseline experiments MINOS and NOvA at Fermilab, and T2K in Japan. The other approach relies on reactor antineutrino disappearance experiments, which—unlike accelerator-based efforts—are independent of CP violation and matter effects, and therefore enable a clean determination of $\theta_{13}$. While it is challenging to stage a neutrino disappearance experiment at the required ~1% precision, four such efforts are being actively pursued with sensitivity to $\theta_{13}$ ranging from 6° to 3° (90% CL). They are: Angra in Brazil, Daya Bay in China, Double Chooz in France, and RENO in South Korea.

## Monitoring Nuclear Reactors with Antineutrinos

As nations rush to enhance their global economic competitiveness, the International Atomic Energy Agency (IAEA) is facing an increasingly knotty challenge to global security. The IAEA, which is responsible for monitoring civil nuclear facilities and nuclear inventories in non-nuclear weapons countries, has found that many of those countries are planning to meet at least some of their burgeoning energy needs with nuclear power. The global nuclear power capacity could increase as much as 40% by 2020, according to the agency's studies, with most of the increase occurring in the Middle and Far East and South Asia (for details see http://www.iaea.org/OurWork/ST/NE/Pess/RDS1.shtml). So how can the IAEA monitor all those plants for compliance with non-proliferation treaties?

Meeting this challenge would be a lot easier if the IAEA had an easily field-deployable detector that could peer into an operating reactor and—in real time—give an accurate readout of how much plutonium or uranium was actually present in the fuel rods. And now a potential solution is in hand, thanks to years of basic research on those elusive elementary particles known as neutrinos. Drawing on that experience, nuclear and particle physicists from Lawrence Livermore and Sandia National Laboratories have designed an antineutrino detector that provides two ways of tracking the amount of plutonium-239 and uranium-235 in a working reactor. One method involves examining changes in the total rate of detected antineutrinos over time; the second involves looking at changes in the energy spectrum of the emitted antineutrinos.

A diagram of the one-ton prototype detector is shown in the figure. Its total footprint, including shielding, is about $2 \times 3$ m. This detector has been deployed since 2003 at the San Onofre Nuclear Generating Station in San Clemente, California. The detector is 17 meters below ground in a room about 25 meters from the reactor core. The detector consists of three subsystems: the central detector and two shields. The central detector, in which the antineutrinos are detected, consists of four stainless-steel cells each filled with 0.25 cubic meter of liquid scintillator laced with gadolinium atoms. The signature for the detection of an antineutrino is the flashes of the prompt and delayed light that is initiated by $p + \bar{\nu}_e \rightarrow n + e^+$ reaction. The emitted $e^+$ produces the prompt light from both direct ionization and annihilation with an electron. The delayed light comes about 30 µs later from the gamma rays emitted from neutron capture on gadolinium. The central detector is surrounded on all sides by a passive water shield. This shield attenuates gamma and neutron backgrounds. The active shield placed outside the water shield detects and "vetoes" penetrating cosmic rays.

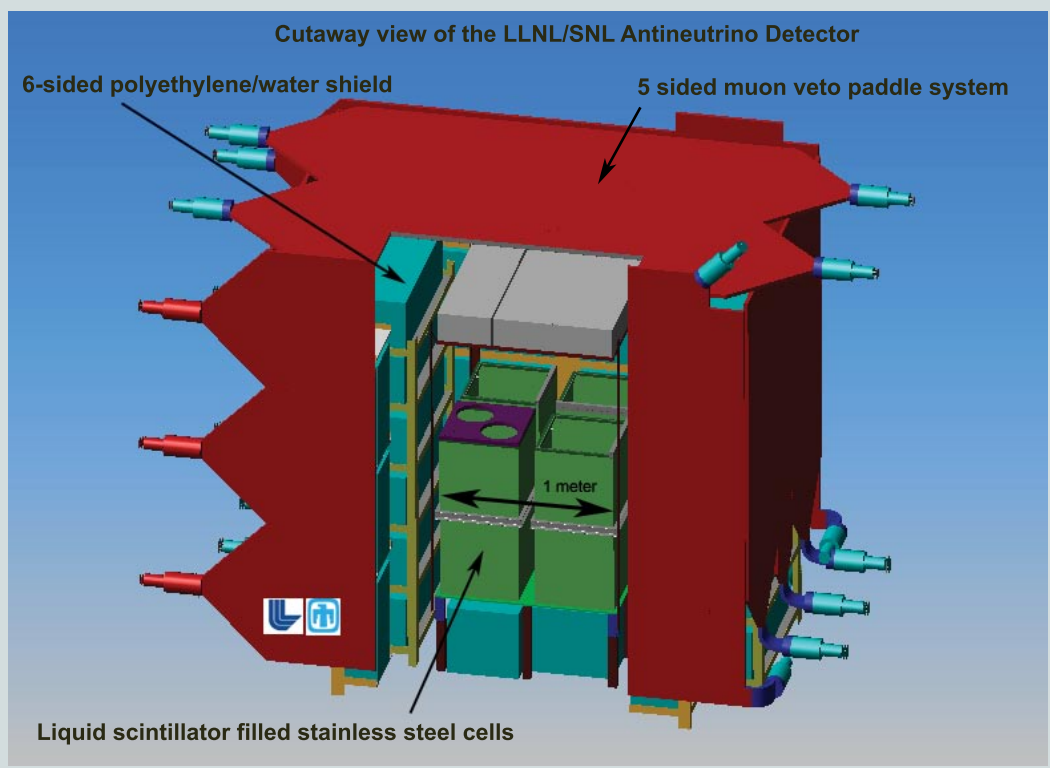

**Cutaway view of the LLNL/SNL Antineutrino Detector**
- 6-sided polyethylene/water shield
- 5 sided muon veto paddle system
- 1 meter
- Liquid scintillator filled stainless steel cells

Drawing of the one-ton prototype antineutrino detector. The prototype detector consists of three subsystems: a central detector and two shields (one passive water shield and an active plastic scintillator). Photomultiplier tubes above the central detector cells detect the light pulses that are produced when an antineutrino interaction occurs in the liquid scintillator in the central detector.

Both Double Chooz and Daya Bay have U.S. participation. The large U.S. component of the Daya Bay collaboration includes several nuclear physics groups having key expertise and experience from KamLAND and SNO.

**Neutrino Cross Sections**

Because neutrino detectors are made of nuclei—typically oxygen, carbon, argon, and iron—knowledge of neutrino-nucleus interactions is needed. Even the relatively simple quasi-elastic channel is not well characterized. Pion-production cross sections, especially by the neutral current, are even less well understood, and they represent a serious background to forthcoming $\theta_{13}$ searches. Interactions involving products from the primary neutrino-nucleon interaction with nuclear matter can also have significant effects on the observed energy of the final states, and thus on the measured values of $\Delta m_{23}^2$ and $\Delta m_{13}^2$. These problems motivate a suite of careful measurements of neutrino cross sections. Two U.S. experiments, MINERvA and SciBooNE, both at Fermilab, are now beginning operation, with nuclear physicists participating. Equally important are neutrino cross sections relevant to supernova physics; these measurements are needed for both detector design and nucleosynthesis calculations. The NuSNS experiment, proposed to be sited near the beam stop of SNS, takes advantage of the intense neutrino fluxes from this megawatt-class accelerator.

# ELECTROWEAK INTERACTIONS OF LEPTONS AND QUARKS

Since the first observation of parity violation five decades ago in the β-decay of $^{60}$Co and in the decay of the muon, nuclear physicists have played a leading role in elucidating the fundamental weak interactions of quarks and leptons. During the coming decade, a targeted program of nuclear physics studies of weak decays, parity-violating electron scattering, and the muon anomalous magnetic moment will search for tiny deviations from SM predictions for electroweak interactions. The pattern of such deviations—or of their absence—provides essential clues to the structure of the New Standard Model.

Historically, tests of fundamental symmetries, such as parity violation in β-decay or muon decay, have advanced our knowledge of the basic forces of nature in numerous ways. The results of the SLAC parity-violating electron-deuteron deep-inelastic scattering in the 1970s led to the correct description of neutral weak interactions, well before the $Z^0$-boson mediator of that interaction was observed directly at CERN. More recently, the comprehensive set of precision electroweak measurements at low and high energies predicted the correct mass range for the top quark prior to its discovery at the Tevatron. During the coming decade, when LHC experiments will search for new particles at the electroweak scale, precise low-energy symmetry tests will continue to help define the symmetries of the New Standard Model.

**Parity-Violating Electron Scattering**

Parity violation in the neutral weak interaction is observed in scattering experiments using polarized electron beams. The rates for scattering electrons whose spin and momentum are parallel and antiparallel are not identical, and this asymmetry is caused by the quantum interference between the parity-violating (PV) component of the neutral weak interaction and the parity-conserving electromagnetic interaction. Recently, the SLAC E158 collaboration has achieved the most precise test to date of the "running" of $\sin^2\theta_W$ by measuring the PV asymmetry in Møller scattering. In the SM, the value of $\sin^2\theta_W$ varies with the energy scale at which it is probed, as indicated in figure 2.23. Given the value of $\sin^2\theta_W$ from experiments carried out at the $Z^0$ pole, one can predict with high precision the value at other energy scales. The present theoretical uncertainty in this prediction is given by the thickness of the line in figure 2.23. The E158 result is in agreement with the SM expectations.

The $Q_{WEAK}$ experiment at Jefferson Lab will measure the PV asymmetry for elastic electron-proton scattering, aiming to achieve an even more precise determination of the weak mixing angle $\sin^2\theta_W$ below the $Z^0$ pole. The PV electron-proton asymmetry can be cleanly interpreted because subtle theoretical uncertainties have been brought under control by a combination of careful nuclear theory and the previous PV experiments that have looked for effects from strange quarks. Since new interactions can lead to distinct effects in PV electron-electron and electron-proton scattering, tests of the running of $\sin^2\theta_W$ in both the purely leptonic and semileptonic processes are sensitive to new physics. For example, a result significantly different from expectation might imply lepton-number-violating interactions in supersymmetry, leptoquarks, or an additional light neutral gauge boson, or Z'. Alternatively, agreement with the SM theory constrains these scenarios. Either outcome will be important.

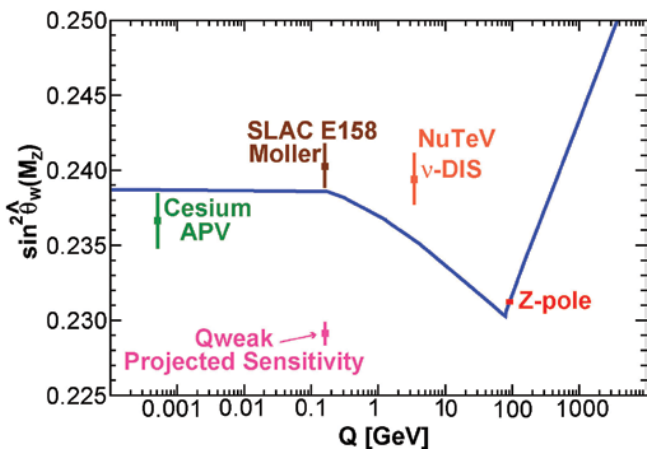

**Figure 2.23:** Predicted scale dependence of the weak mixing angle in the Standard Model. The vertical axis gives $\sin^2\theta_W$ while the horizontal axis gives the energy scale $Q$ at which various experiments have been performed. The error in the predicted "running" is given by the width of the solid blue line and is dominated by the experimental error in $\sin^2\theta_W$ at the Z pole. Results from cesium atomic parity violation (APV), parity-violating Møller scattering, and neutrino-nucleus deep-inelastic scattering (ν-DIS) are shown, along with prospective sensitivities of the JLAB QWEAK experiment.

Two classes of PV experiments that will require investments in new experimental capabilities are envisioned for the 12 GeV CEBAF Upgrade. First, a next-generation Møller asymmetry experiment will exploit the high luminosity of the CEBAF electron beam, the doubling of the energy, and the experience gained in previous efforts, to determine the running of $\sin^2\theta_W$ with a precision comparable to that achieved with the $Z^0$-pole experiments. At present, the values of $\sin^2\theta_W$ obtained from Z-pole measurements at LEP and SLD are in poor agreement, giving significantly different predictions for the mass of the yet unseen Higgs boson. The PV Møller experiment at 12 GeV could help resolve this tension.

A second thrust involves precise measurements of the PV deep-inelastic electron-deuteron and electron-proton asymmetry. These asymmetries are ~100 times larger than the PV Møller asymmetry, making a kinematic survey feasible. The first step in this program will be carried out with the existing 6 GeV beam, followed by additional experiments at 11 GeV. The variation of the asymmetry with both energy and $Q^2$—as well as the use of different targets—will probe a variety of largely unexplored aspects of the nucleon's quark and gluon substructure. Ongoing theoretical activities will provide a comprehensive framework for interpretation of the deep-inelastic asymmetries and delineate their implications for both the SM and its possible extensions.

### Weak Decays: Nuclei, Neutrons, Pions, and Muons

Studying the radioactive decays of nuclei, neutrons, pions, and muons has been a mainstay in nuclear physics for 50 years. The coming decade promises to see a revolution in this field through a combination of increasingly precise measurements of half-lives, correlations between decay products, and rare decay modes. A particular focus will fall on the neutron, where investments in new capabilities using cold neutrons at the FNPB will allow U.S. nuclear physicists to carry out a comprehensive set of high-precision measurements of neutron decay parameters for the first time.

Studies of neutron and nuclear lifetimes and decay correlations can yield important information about the charged-current structure of the SM while probing new physics symmetries in a manner complementary to the PV electron-scattering experiments described above. Within the SM, a comparison of the muon lifetime with the rate for "superallowed" Fermi nuclear β-decay yields the value of $V_{ud}$, the first entry in the Cabibbo-Kobayashi-Maskawa (CKM) quark-mixing matrix. Together with the results obtained from kaon and B-meson decays, the determination of $V_{ud}$ allows us to test the unitarity property of this matrix, a fundamental tenet. Recent theoretical progress has reduced the hadronic physics uncertainty in the extraction of $V_{ud}$ from the β-decay half-lives, thereby sharpening this unitarity test. Current results—in agreement with the SM—place important constraints on candidates for the New Standard Model, including supersymmetry, models with right-handed W-bosons, and extra generations of fermions.

One can also determine the value of $V_{ud}$ by combining measurements of the neutron lifetime and a parity-violating neutron-decay correlation. Doing so provides an important check that unknown nuclear structure effects do not influence the value of $V_{ud}$ obtained from Fermi nuclear decays. A measurement of the beta-asymmetry parameter at LANSCE using ultra-cold neutrons is being pursued with this objective in mind. The future program at the FNPB promises to provide a complementary approach to the same objective. These measurements—together with future correlation studies in neutron decay at NIST and in rare-isotope decays at the proposed Facility for Rare Isotope Beams—will be sensitive enough to hunt for effects outside of the SM.

Pions and muons provide powerful, complementary laboratories for probing the symmetries of the New Standard Model. During the next decade, experiments at TRIUMF and the Paul Scherrer Institut (PSI) in Switzerland involving

# The Neutrino: Stealthy Messenger

Seventy-five years ago, Wolfgang Pauli proposed a hypothetical weakly interacting particle to solve what seemed—based on experimental observations—to be the violation of energy momentum and spin conservation in nuclear beta decay. The particle, called the neutrino by Enrico Fermi, was postulated to be a spin-1/2, electrically neutral particle and to have a very tiny or zero rest mass. The conjectured neutrino was able to reconcile experimental observations with established fundamental symmetry principles and was quickly incorporated into the theoretical framework of nuclear physics. A quarter of a century passed before this elusive particle was finally experimentally observed by Reines and Cowan in the mid-1950s, transforming it from a theoretical construct to a bona fide particle.

Today, unraveling the properties of the neutrino could hold the keys to explaining some of the most basic, unsolved puzzles in physics: Why is there more matter than antimatter in the present universe? How did the fundamental forces of nature evolve from the primeval interactions at the end of the Big Bang? What causes stars to explode in cataclysmic events called supernovae?

In the past decade our knowledge of neutrinos and their role in the universe has undergone a remarkable transformation. We have discovered that neutrinos mutate or "oscillate" from one species to another as they travel from the Sun to the Earth, make their way from the upper reaches of the atmosphere to the Earth's surface, and when they radiate from the cores of nuclear reactors. We have learned that neutrinos do indeed have mass, making them the lightest matter particles known, at least 250,000 times lighter than the electron. And we have learned that, tiny though their masses are, neutrinos play an important role in shaping the largest scales of the cosmos.

Our new knowledge of neutrinos is summarized in the figure below. Neutrino oscillation experiments have measured the differences of the squares of different neutrino masses and isolated two possible orderings of the neutrino levels. These experiments have also told us how the three known neutrinos ($\nu_1$, $\nu_2$, and $\nu_3$) are composed of mixtures of the three elementary "flavors" ($e$, $\mu$, and $\tau$). None of this information appears in the Standard Model, and it has revolutionized our picture of nature's fundamental interactions.

Despite this new knowledge, the most basic properties of the neutrino remain to be discovered. Are they their own antiparticle? Do their interactions violate charge-parity symmetry? If so, neutrino interactions in the early universe could be responsible for generating the predominance of matter in today's universe. How large are their masses? Neutrino oscillation experiments only give us the mass-squared differences, but not the overall scale. And, how do their flavor oscillations affect the shockwaves in exploding stars?

The Standard Model of particle physics answers none of these questions, but new experiments in nuclear physics will provide answers. Searches for the neutrinoless double beta decay of nuclei will tell us whether neutrinos are their own antiparticles. More precise measurements of tritium beta decay will provide the most sensitive direct probe of the mass. And new measurements of flavor oscillations using neutrinos from nuclear reactors and from the Sun will help us better understand how neutrinos affect both the death of stars and the birth of the present universe.

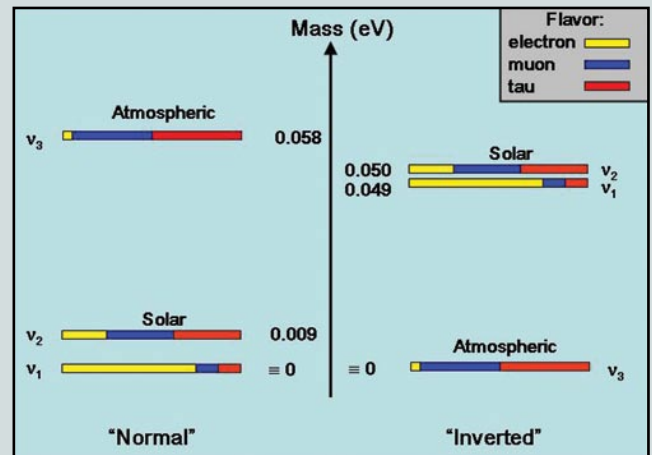

The "normal" and "inverted" hierarchies, allowed orderings for neutrino masses, are illustrated for a case in which the smallest mass is exactly zero (it could be as large as 2.3 eV). As the smallest mass is raised, the levels become more closely spaced and nearly degenerate. The flavor content of each massive neutrino is indicated approximately by the colors. Whether $\nu_3$ contains any electron flavor is not at present known.

A few years after proposing the neutrino as a "desperate way out" to save basic symmetry principles, Wolfgang Pauli exclaimed, "*I have done a terrible thing. I have postulated a particle that cannot be detected.*" Today, we know that neutrinos not only can be detected, but that discovering their basic nature could help resolve some of nature's deepest mysteries.

U.S. nuclear physicists and their international collaborators will measure the ratio of rates for the decays $\pi^+ \to e^+\nu$/$\pi^+ \to \mu^+\nu$ with unprecedented precision reaching ~0.01%. Many candidates for the New Standard Model imply that the SM expectations could break at this level. Pion leptonic decays are particularly sensitive to new pseudoscalar and scalar interactions at extremely high mass scales (up to 1000 TeV). Because this ratio is insensitive to strong interaction uncertainties, it provides the theoretically cleanest test of the universality of lepton interactions.

U.S. nuclear physicists are engaged in new and more precise measurements of muon decay. The TWIST collaboration at TRIUMF has recently completed a set of precise measurements of the positron spectrum in $\mu^+$ decay. A comparison of the observed positron spectrum with SM expectations provides a unique probe of new symmetries, such as right-handed currents or scalar interactions. Measurements of the muon lifetime at PSI by the U.S.-led MuLan Collaboration yield the value of the Fermi constant, which, in the SM, sets the basic strength of the weak interaction.

Most scenarios for the New Standard Model violate lepton flavor symmetry. Two experimental efforts are being pursued that aim to observe charged lepton flavor violation probing high mass scales of order 1000 TeV. The MEG experiment at PSI is searching for the decay $\mu \to e\gamma$ with at least one hundred times greater sensitivity than in previous experiments. A possible experiment at Fermilab to search for charged lepton flavor violation in the neutrinoless muon-to-electron conversion process $\mu + A \to e + A$ is being explored jointly by U.S. nuclear and high-energy physicists. Comparison of results from $\mu \to e\gamma$, $\mu + A \to e + A$, and $0\nu\beta\beta$ could provide important insights into the mechanism for new interactions.

**Muon Anomalous Magnetic Moment: g-2**

Of all the recent achievements in low-energy precision physics, the muon anomalous magnetic moment, $a_\mu$, stands out. Both experiment and theory boast impressive ~0.5 part-per-million (ppm) uncertainties, enabling the sensitive comparison, $\Delta a_\mu \equiv a_\mu(\text{Exp}) - a_\mu(\text{Thy}) = (295 \pm 88) \times 10^{-11}$, a tantalizing 3.4 standard deviation nonzero difference. Apart from the observation of neutrino oscillations and the various components of the cosmic energy density, it is arguably the strongest hint of new physics today. The impact of the work has been significant because the size of the effect supports the predictions of many proposed SM extensions (while being in conflict with, or even ruling out, others). The muon anomaly "test" illustrates the power of the program of precision-frontier experiments proposed in this Long Range Plan. They rely on sensitive measurements and unambiguous theoretical prediction.

The Brookhaven E821 "(g-2)" experiment was statistics limited. A next-generation effort is being developed with the goal of up to a five-fold increase in precision. Nuclear physicists propose to use the AGS high-intensity proton machine, together with a redesigned muon capture and transport system, to greatly increase the data rate and, along with improvements in the systematic errors, thus reduce the overall error on $a_\mu(\text{Exp})$. The existing superconducting muon storage ring will remain the centerpiece of the new experiment, but many support systems will require replacement or upgrades.

Theoretical efforts are underway worldwide to more precisely predict the SM expectation, which involves quantum loop contributions from quantum electrodynamics (QED), weak, and hadronic processes. While the QED and weak contributions are well known, the hadron loops dominate the theory uncertainty. First-order hadronic vacuum polarization (HVP) can be obtained from experimentally measured $e^+e^- \to$ hadron cross sections, coupled with a dispersion relation. New data are coming from the facilities at Novosibirsk in Russia, Frascati Italy, SLAC, and KEK in Japan; therefore, the uncertainty on the HVP terms will be reduced. On the other hand, hadronic light-by-light scattering—dominated by low-energy pion interactions—can only be determined by QCD-inspired models, where nuclear theorists are experts.

What does a muon anomalous magnetic moment differing from the SM prediction by ~300 × 10$^{-11}$ imply? If confirmed, it certainly points to new interactions beyond those of the SM—a major discovery. If those interactions are supersymmetric, then the value of $a_\mu$ will determine the sign of one of the key parameters of a supersymmetric SM—a feat that cannot easily be achieved at the LHC. More generally, the full implications of the muon g-2 experiment will only be understood in the context of other clues such as $0\nu\beta\beta$ decay and EDM searches, weak decays, PV electron scattering, and direct mass and branching ratio measurements from the LHC. These "fingerprints" will collectively shape the form of the New Standard Model. During the time scale of this Long Range Plan, the essential contributions from U.S. nuclear physics in the quest for the New Standard Model will bear much fruit.

**Related Studies**

Nuclear physicists are also making key contributions to research for which other subfields provide the primary capital funding. Reactor neutrino measurements of $\theta_{13}$ and searches for charged lepton flavor violation are two examples discussed above. A third area involves the quest to determine the nature of the cold dark matter (CDM) that makes up roughly one-quarter of the cosmic energy density. Direct CDM-detection experiments involve low-background, cryogenic technology for which nuclear physicists provide key expertise. Similarly, nuclear physicists are participating in indirect CDM searches that entail measurements of high-energy solar neutrinos, as in the IceCube experiment. In a quite different approach to the New Standard Model, searches for extra dimensions and for the connections between the predictions of general relativity and fundamental interactions are being made with exquisitely sensitive torsion balance technology.

## WEAK PROBES OF NUCLEAR PHYSICS, ASTROPHYSICS, AND QCD

Among the motivations for developing a predictive New Standard Model is the power it affords in understanding the structure of hadrons, nuclei, and astrophysical objects. A primary example is the exploitation of neutrinos to unravel first the properties of neutrinos themselves, and then the mechanism by which the Sun produces its energy. A number of important applications of weak-interaction probes being carried out by nuclear physicists are described below.

**Solar Neutrinos**

The success of the standard solar model in predicting the central temperature of the Sun to 1% is a stunning achievement for astrophysical theory and nuclear physics measurements. Yet it may come as a surprise that, while the core presumably burns steadily in quasistatic equilibrium, the convective zone is complex and poorly understood. Solar irradiance variations are suspected to be the origin of the Maunder minimum, a period between 1650 and 1700 when sunspots were absent and the northern hemisphere was exceptionally cold. The Sun's radiant output at the current epoch could differ by a few percent from the energy production rate at the core. A direct measurement of the neutrino fluxes with high precision would yield a value for the true average solar output and whether the present epoch is unusual in any respect. At a time when society is carefully evaluating the energy balances that drive climate change, a precise determination of the Sun's true energy production rate is a matter of importance. Only neutrino physics offers a means to determine this crucially important quantity.

The program requires percent-level absolute measurements of the *pp* or *pep* and $^7$Be fluxes, and a less precise measurement of the CNO cycle (see below). Such measurements can be made by elastic scattering of neutrinos from electrons in detectors such as Borexino (having reported the first direct measurement of the $^7$Be flux, in August 2007), KamLAND (being converted to a solar experiment), or CLEAN (a liquid neon experiment in Research and development). Alternatively, neutrino charged-current reactions on nuclei can be observed in experiments such as LENS (an indium experiment in Research and development). The elastic-scattering experiments enjoy important advantages in this program: the cross section is known, the detected rates are larger than in practical charged-current experiments, and the dependence on the mixing angles is less. Backgrounds are a more serious issue.

The CNO cycle functions as an independent energy production source in the Sun, and is of importance not only in determining the solar energy production rate but also in fixing the presolar metallicity. While laboratory measurements in the LUNA facility at Gran Sasso have resolved uncertainties about the critical $^{14}$N(p,$\gamma$)$^{15}$O reaction rate, the CNO flux is now subject to new theoretical uncertainty due to a controversy over abundances, with recent measurements indicating a 30–40% decrease in the concentrations of light "metals" such as carbon and oxygen.

**Geoneutrinos**

Geoneutrino experiments make it possible to sample the amount and location of heat generation in the Earth, which can provide a strong constraint on mantle convection models and permit a test of a fundamental assumption in Earth formation models, namely that the uranium-to-thorium ratio does not vary. These are important questions and of great significance to geology. They will be answered by nuclear physicists, who have the necessary expertise to perform these neutrino measurements.

**Electroweak Probes of the Strong Interaction**

Precise measurements of PV electron-scattering asymmetries at MIT-Bates, Mainz, and JLAB have been performed on hydrogen, deuterium, and helium targets to derive stringent limits on contributions of the strange quark to the nucleon's electromagnetic structure. This use of the neutral weak interaction gives unique access to poorly understood aspects of hadron and nuclear structure. This approach will continue in the next decade with new experiments using both the 6 and 11 GeV beams at Jefferson Lab. For example, the PV asymmetry for elastic electron scattering from the lead nucleus will provide information about the neutron radius of this nucleus; the "skins" of neutron stars depend on the same quantity. Because the vector coupling of the $Z^0$ to neutrons is roughly 10 times larger than its coupling to protons, PV electron scattering from a neutron-rich nucleus is an ideal means for determining this important input into the nuclear equation of state.

**Hadronic Parity Violation**

While the fundamental weak interaction between quarks is well understood within the SM, its manifestation in nuclei and hadronic interactions remains poorly understood. For example, is there a long-range *weak* nucleon-nucleon interaction—mediated by the exchange of a single pion—analogous to the long-range, strong nuclear force responsible for key aspects of nuclear structure? How does the interaction "dress" the underlying quark-quark weak interaction to generate short-range weak forces between two nucleons? During the next decade, a program of few-body, hadronic PV experiments using polarized neutrons at the FNPB and elsewhere will provide data that can be interpreted with *ab initio*, few-body calculations and effective field theory methods to yield the lowest-order "primordial" PV nuclear interaction. Coupled with progress in lattice QCD calculations described elsewhere in this Long Range Plan, knowledge of the PV nuclear interaction should allow us to explain how the quark-quark weak interaction and low-energy strong interaction conspire to generate weak forces between nucleons.

**Muon Capture**

Muon capture on the proton, $\mu^- + p \rightarrow n + \nu_\mu$, is a fundamental weak interaction process. It is uniquely sensitive to the pseudoscalar form factor $g_P$, which is the least well known of all form factors characterizing the QCD structure of the nucleon in charged-current reactions. Advances in modern effective field theories allow a precise calculation of $g_P$, and its experimental verification represents a fundamental test of QCD at low energies. The U.S.-led MuCap experiment at PSI succeeded in the first precise and unambiguous measurement of $g_P$; 10 times higher final statistics are being collected. Once the question of $g_P$ is settled, the developed precision technique enables the study of the axial current in the two-nucleon system with the process $\mu + d \rightarrow n + n + \nu$. A 1% measurement of $\mu+d$ capture would provide a benchmark result—10 times more precise than all present experiments on weak processes in the two nucleon system—which is closely related to reactions of astrophysical interest, such as solar *pp* fusion and $\nu + d$ reactions as studied by the SNO experiment. Recent EFT calculations have demonstrated that all these reactions are related by one axial two-body current term, parameterized by the low-energy constant $L_{1A}$. Muon capture can determine $L_{1A}$ with precision and, in effect, will help "calibrate the Sun."

**Neutrino Probes of Supernovae**

The observation of neutrinos from the supernova SN1987a in the large Magellanic cloud by both the Kamiokande and IMB detectors confirmed the basic theory of the core-collapse supernova. A detailed comparison, however, remains elusive because only a dozen events were detected from that distant source. It is essential to maintain a continuous capability in neutrino science to detect a supernova in our galaxy should one occur. Neutrinos provide a unique window into the heart of a collapsing star, and play a major role in every phase of supernova evolution from the explosion itself to the ultimate production of the elements. With the termination of SNO, the available worldwide sensitivity to the electron-neutrino component, and also to the neutral-current interactions of all flavors, is now limited. The low rate of galactic supernovae, a few per century, renders problematic the construction of dedicated supernova detectors that do not have additional physics programs. However, opportunities to provide the necessary capability in conjunction with other initiatives should not be missed. The use of lead as a target with good $\nu_e$ sensitivity, and the use of coherent neutral-current scattering, are possible options that could provide the comprehensive readiness for a supernova that is needed. Another important experimental goal is to measure the diffuse supernova neutrino background that presumably fills the universe, a unique cumulative record of all supernova activity.

## OUTLOOK

The targeted program described above addresses pivotal nuclear science questions in fundamental symmetries and neutrino physics: Is lepton number violated? What is the origin of the observed matter-antimatter asymmetry? What presently unseen forces were at work in the early universe? How are stars born and how do they die? What role do neutrinos play in cosmology and astrophysics? Garnering answers to these questions relies on undertaking a new generation of exquisitely designed, precision measurements, supported by strong theoretical guidance and predictions. The proposed program builds on the remarkable discoveries made in the past decade and the growing community of scientists engaged in these exciting research activities. The answers to these questions will have a profound influence on defining the New Standard Model and ultimately on understanding how nuclear physics shapes our world and the universe around us.

# 3. The Tools of Nuclear Science

# Facilities for Nuclear Science

## MAJOR ACCELERATOR FACILITIES

Progress in nuclear science is often driven by new accelerator and other advanced facilities, which allow us to probe ever more deeply into the structure of the nucleus—or even to create entirely new states of nuclear matter. Such facility-driven progress has been particularly notable over the past decade. For example, the Relativistic Heavy Ion Collider (RHIC) facility has led to the creation and study of the quark-gluon plasma: a new state of matter at unprecedented energy densities. The Continuous Electron Beam Accelerator Facility (CEBAF) has provided astonishing glimpses of the inner workings of neutrons and protons. Neutrino-detection facilities have measured neutrino flavor oscillations, resolved the long-standing solar neutrino problem, and reworked our Standard Model picture of particle science. And low-energy facilities have begun to provide a deeper understanding of nuclear structure and the role of nuclei in astrophysical processes.

Now, as we describe elsewhere in this Long Range Plan, the scientific opportunities for the coming decade promise to be even richer—and a new generation of accelerators and detectors will be required to exploit them. Fortunately, the technology to create this new generation is within reach. For example, recent major advances in accelerator technology include the acceleration of electron beams with higher-gradient cavities; beam-cooling techniques to increase the luminosity of heavy-ion colliders; and the production and acceleration of radioactive ion beams. This last advance is particularly important for the construction of a next-generation Facility for Rare Isotope Beams (FRIB), which is expected to have a dramatic impact on many subfields of nuclear science—and an extraordinary effect on nuclear structure and nuclear astrophysics. Meanwhile, the implementation of new detector technologies will also help us reach new frontiers by further extending the science reach of the accelerator facilities, greatly improving gamma-ray tracking, and allowing for precision measurements of fundamental symmetries in the deep underground facility. Finally, none of these advances could be implemented were it not for the tremendous strides made in recent years in advanced computing, with its critical applications to accelerator and instrument control, to data acquisition and analysis, and for computationally intensive theoretical studies. Below we present a brief overview of the present U.S. accelerators starting with those operating at the highest energies.

### High-Energy Nuclear Physics Facilities

There are currently two high-energy nuclear science facilities in the United States—RHIC and CEBAF. Their roles are complementary. CEBAF explores how quarks and gluons are assembled to form the protons and neutrons, while RHIC produces and studies very "hot," high-density forms of nuclear matter.

With its two independent rings of superconducting magnets, RHIC is a highly flexible collider of hadron beams. During its first seven years of operation RHIC has already exceeded the design luminosity for gold-gold collisions by a factor of six, has successfully collided deuterons on gold with both beams at the same energy per nucleon, and has completed an additional run of colliding copper beams with record luminosities. An Electron Beam Ion Source (EBIS) is currently under construction and will be commissioned in 2010. EBIS makes new species available in RHIC, notably uranium. Measurements of U + U collisions are of particular interest since this system can generate initial energy densities about 30% higher than in Au + Au collisions.

As a polarized proton collider, RHIC has had four very successful running periods, reaching a luminosity of $3.5 \times 10^{31}$ cm$^{-2}$ s$^{-1}$ and beam polarization of up to 65%. Most operation with polarized protons has been carried out with beam energies of 100 GeV/nucleon—the gold beam design energy. Additional operation at lower beam energy was also accomplished, again demonstrating the flexibility of RHIC.

CEBAF at Jefferson Lab, which has been operating successfully since 1996, is a superconducting, continuous wave accelerator with a maximum energy of up to 5.7 GeV, 100% duty factor, and excellent beam quality. Three distinct beams, with currents differing by up to six orders of magnitude and a combined current of 200 µA, can be injected simultaneously into the accelerator for delivery to the three experimental halls. This enables simultaneous running of $4\pi$ experiments and high-resolution small-aperture experiments. CEBAF's unique beam capabilities include a broad energy range (360 MeV to 5.7 GeV), small spatial distribution (normalized emittance of less than 1 mm-mrad rms), and very low energy spread (less than $2 \times 10^{-5}$ rms).

An increasing fraction of CEBAF's experimental program requires polarized electron beams. A state-of-the-art high-polarization GaAs photocathode gun has recently demonstrated 85% beam polarization with high reliability and long lifetime. Parity-violation experiments, devoted to measuring asymmetries in elastic electron-nucleon scattering, have achieved unprecedented precision, in some cases better than 100 parts per billion. This precision relies on extremely small correlations between beam helicity and other beam parameters on target, achieved by combining the polarized source capabilities with a detailed understanding of particle optics and careful tuning of the beam transport to ensure adiabatic damping, as designed, over four decades of beam energy.

Near-term capability improvements include further enhancements of the polarized source and a long-term program of refurbishment of the cryomodules with the goal of achieving 50 MV energy gain per cryomodule, compared to the design value of 20 MV.

**Facilities for Nuclear Structure and Nuclear Astrophysics**

The low-energy nuclear facilities currently active in the United States can be broadly divided into two distinct groups: national user facilities and smaller laboratories at universities and national laboratories. They provide unique and complementary beams of both rare and stable isotopes, and form the backbone for nuclear structure and nuclear astrophysics research in the United States.

The Argonne Tandem Linac Accelerator System (ATLAS) is a DOE-funded national user facility for the investigation of the structure and reactions of atomic nuclei in the vicinity of the Coulomb barrier. ATLAS has some capability for rare-isotope acceleration, producing beams mostly by in-flight separation. A major advance in rare-isotope capabilities will be the Californium Rare Ion Breeder Upgrade (CARIBU). Rare isotopes will be obtained from a one-Curie $^{252}$Cf fission source located in a large gas catcher from which they will be extracted, mass separated, and transported to an Electron Cyclotron Resonance (ECR) source for charge breeding prior to acceleration in ATLAS. This will provide accelerated neutron-rich beams with intensities up to $7 \times 10^5$/s, and will offer unique capabilities for a few hundred isotopes, many of which cannot be extracted readily from Isotope Separator On Line (ISOL) type sources. In addition, it will make these

## Polarized Proton Collisions at RHIC

Thanks to the Relativistic Heavy Ion Collider at Brookhaven National Laboratory, which is doing double duty as the first collider of polarized proton beams, physicists have been able to make their best measurements to date of how gluons contribute to the proton's "spin": an intrinsic property that causes the particle to behave like a tiny compass needle with a magnetic north and south pole.

There are two major challenges to producing these polarized proton collisions. The first is to create the polarized beams so that all the protons have their spins pointing in the same direction. This is the task of the Brookhaven Optically Pumped Polarized Ion Source. The process begins with an intense laser beam that polarizes the electrons in a population of rubidium atoms. These polarized electrons are then transferred to unpolarized protons, where they form hydrogen atoms. Then the electrons transfer their polarization to the protons via a coupling between the spins.

Once the polarized protons are in hand, the second challenge is to keep their spins aligned throughout the acceleration process. What makes this difficult is that the electric and magnetic fields used to accelerate the proton beams also interact with each proton's intrinsic magnetic moment, which causes the particle's spin axis to wobble, or "precess," like a spinning top. If the frequency of the spin precession should equal the beam's revolution or orbit oscillation frequencies, this wobble can easily go out of control and destroy the beam's spin alignment. To keep this from happening, the physicists have inserted helical dipole magnets, also called "Siberian snakes," in both the injector and RHIC itself. As a result, RHIC now routinely operates with 100 × 100 GeV proton collisions having up to 65% polarization in each of the two beams. For the first time, moreover, the facility has accelerated polarized protons to 250 GeV.

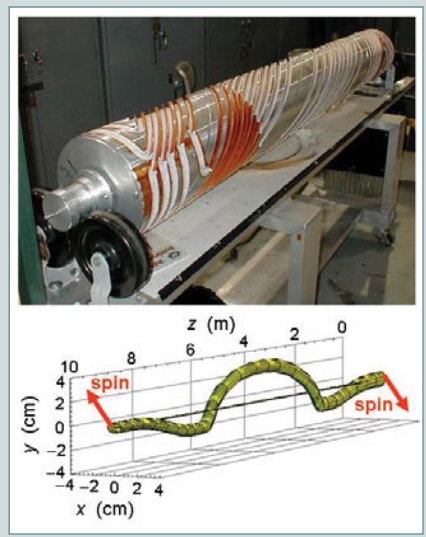

The photo shows the AGS cold-bore snake at an early stage in its helical winding. The lower figure shows the simulated beam path through the snake and indicates that the transverse spin direction reverses during passage through the snake.

accelerated beams available at energies up to 10–12 MeV/nucleon, which are difficult to reach at other facilities.

At the DOE-funded national user facility Holifield Radioactive Ion Beam Facility (HRIBF), radioactive species are produced by the ISOL technique with intense light-ion beams from a cyclotron and post-accelerated by a 25-MV tandem accelerator. More than 175 isotopes, both proton and neutron rich, have been accelerated, with 50 having rates above $10^6$/s. Approximately 30 additional species are available as low-energy (~50 keV) beams. The tandem post accelerator delivers beams at continuously variable energies up to approximately 5 MeV/nucleon for most ions. The ability of HRIBF to deliver reaccelerated beams of neutron-rich fission fragments at energies above the Coulomb barrier is currently unique worldwide. A program is underway to improve HRIBF performance. A High Power Target Laboratory (HPTL) was completed in 2005, providing extensive ISOL production, beam purification, and manipulation research and development capability. A second fully functional ISOL production station (IRIS2) is now being configured, using the shielded space and hardware developed for HPTL.

The National Superconducting Cyclotron Laboratory at Michigan State University is an NSF-funded national user facility with a primary focus on the study of rare isotopes and nuclear astrophysics. The rare-isotope beams are produced by in-flight separation after fragmentation and fission of 50 to 200 MeV/nucleon heavy-ion beams. Rare-isotope beam intensities are currently as high, or higher, than at any facility worldwide. From 2001–2007 the facility delivered over 625 different rare-isotope beams for experiments. Experimental programs with fast beams and reaccelerated beams up to 60 keV, after gas stopping, are in place. An upgrade is underway to add reaccelerated beams to 3 MeV/nucleon following stopping in an advanced gas catcher system and charge breeding.

The Berkeley 88-Inch Cyclotron is a sector-focused, variable-energy machine with axial injection from three ECR ion sources. It is used for nuclear science research and for a variety of applications, including certification of electronics used in avionics. It hosts the Berkeley Accelerator Space Effects (BASE) facility, which provides beams for radiation effects testing. The cyclotron can accelerate all stable ions, producing high-intensity light-ion beams such as protons up to 50 MeV/nucleon, and deuterons and alphas up to 30 MeV/nucleon. Heavier ions are available at lower energy ranging down to 5 MeV/nucleon for uranium.

A new powerful ECR source, VENUS, has been developed at LBNL's 88-Inch Cyclotron for the production of intense beams of highly charged ions. As the first ECR source with optimum fields for operation with a 10 kW 28 GHz gyrotron, it has demonstrated a six-fold increase in beam intensity for $U^{29+}$ over previous sources and has produced 200 eμA of $U^{33+}$ and $U^{34+}$. This increased performance has been incorporated into the design of the proposed FRIB heavy-ion driver, which also reduces the cost of the low-beta linac by utilizing high-charge state ions. The superconducting ECR source was recently added to the 88-Inch cyclotron injection system.

The John D. Fox Superconducting Accelerator Laboratory at Florida State University, which is supported by NSF, is based on a 9 MV tandem electrostatic accelerator with a superconducting linac booster. Unique capabilities include an optically pumped polarized $^{6,7}$Li source and a sputter source dedicated to $^{14}$C beam production. The facility provides in-flight production of radioactive beams with a specially designed beam line (RESOLUT) for selecting the desired species.

The Notre Dame Nuclear Science Laboratory (NSL), which is also supported by NSF, operates three accelerators: the Model FN Tandem and two low-energy high-intensity Van de Graaff accelerators. The radioactive beam program at the NSL is centered on the TwinSol facility, which utilizes two superconducting solenoids.

The DOE supports six university-based Centers of Excellence. Four of these, which are discussed below, have active accelerators. The other two are the Institute for Nuclear Theory at the University of Washington and the Research and Engineering Center at MIT.

The centerpiece of the Texas A&M University (TAMU) Cyclotron Institute is a "K500" superconducting cyclotron, which can produce a wide variety of beams with energies ranging from 70 MeV/nucleon for light ions to 12 MeV/nucleon for heavy ions. Recently, radioactive beams have become more of a focus. At the present, they are produced with the MARS recoil separator and the newly commissioned BigSol solenoid spectrometer. An upgrade to produce high-quality radioactive beams directly from the K500 superconducting cyclotron is underway. The upgrade will utilize a refurbished K150 cyclotron to produce radioactive ions that will be stopped in helium-gas ion guides, passed through a charge-breeding ECR ion source, and then accelerated in the K500 cyclotron. Rare-isotope beams with energies up

# Radiation Effects

The same accelerators developed to advance fundamental nuclear science have also found practical application in the simulation of various radiation environments, from the natural backgrounds found in space, the atmosphere, and at ground level, to the intense fluxes found inside nuclear reactors and future accelerators. These simulations are increasingly important in the aerospace and electronics industries, avionics, large supercomputer and server farms, materials and superconducting magnet development, health physics, and many other fields.

Modern micro-electronics are particularly vulnerable—especially in space, where there is no atmosphere to protect satellites from cosmic rays. A charged particle can cause single-event effects ranging from correctable errors such as "bitflips" in memories, to control failures and destructive burnout. Because these effects are most prevalent for high energy loss along the ion path, accelerator simulations most commonly use heavy ions.

In addition to single-event effects, electronics can also exhibit cumulative effects due to the steady displacement of atoms in the solid-state microchips. This cumulative damage also shows up in photonic components and advanced solar cells. Protons are often used to measure these effects.

While the Earth's magnetic field traps the protons and heavy-ion components of cosmic rays and prevents them from reaching sea level, bombardments with atoms in the atmosphere can generate spallation neutrons of enough energy to make it to sea level, and are a bigger concern at the typical altitudes used by commercial aircraft. Neutron effects on avionics and ground-based electronics result in two types of errors: "soft errors" such as single-event upsets in memory chips, and "hard errors" that require a power cycle or more complicated procedures to correct. The first accelerator simulation experiments were performed by the Boeing Corporation for certification of the Boeing 777 passenger airliner.

Facilities supported by the Department of Energy and the National Science Foundation have provided time for such research on a number of their accelerators, including the heavy-ion accelerators at the 88-Inch Cyclotron facility at the Lawrence Berkeley National Laboratory, the National Superconducting Cyclotron Laboratory (NSCL) at Michigan State University, the K500 cyclotron at Texas A&M University, and the tandem and booster accelerators at the Relativistic Heavy Ion Collider at Brookhaven National Laboratory. Protons are provided by the 88-Inch Cyclotron and at the Indiana University Cyclotron Facility (a former NSF nuclear physics facility). Spallation neutrons are available at the Los Alamos Neutron Science Center (LANSCE). The number of hours provided by these facilities over the course of a year range from less than 100 (NSCL) to 3000 (LANSCE). For heavy ions, Texas A&M University and the 88-Inch Cyclotron combined are providing nearly 5000 hours a year. At LBNL, the U.S. defense agencies provide 40% of the operating budget.

Nonetheless, the need for beam time at accelerator facilities for radiation-effects testing is expected to continue to increase. A broad variety of beam capabilities at U.S. facilities will be needed to meet the evolving requirements for electronic testing. For example, in the last several years the need for higher-energy (≥15 MeV/nucleon) heavy ions has become more important because of the increased use of thick overlayers that must be traversed before reaching the sensitive region of the chip. New technical challenges will be introduced by trends in manufacturing such as reductions in feature size, new materials, and higher clock speeds. Moreover, as testing gets more sophisticated, new physics effects have been found such as geometrical effects and low dose effects in total dose testing.

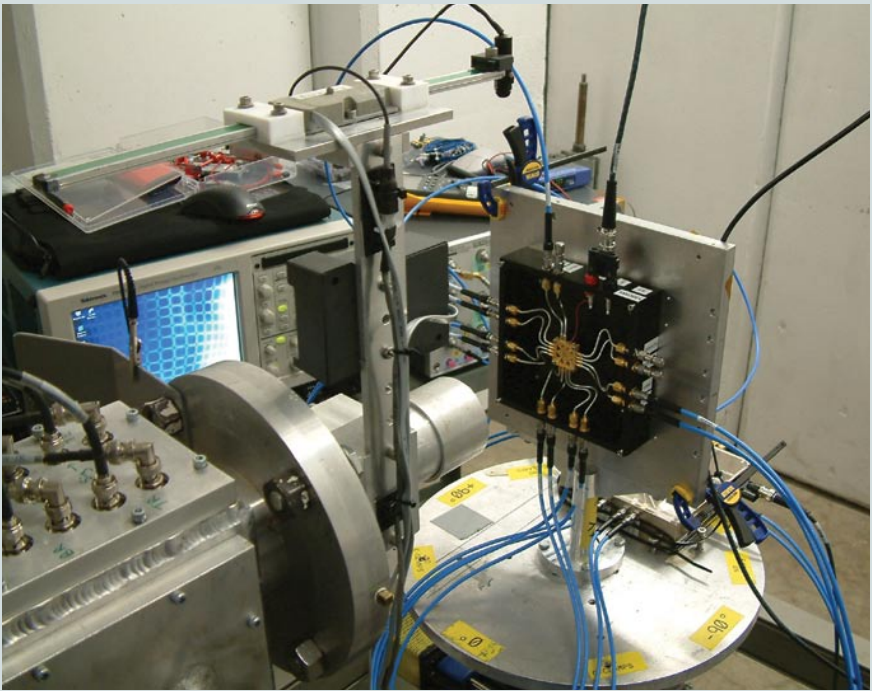

Electronics circuit being tested with a heavy-ion beam at the Texas A&M University K500 superconducting cyclotron.

to 70 MeV/nucleon will be available following the upgrade. The use of the TAMU K500 cyclotron to test the response of electronic components to radiation has grown steadily; it now occupies over 20% of the available beam time.

The Triangle Universities Nuclear Laboratory (TUNL) operates three accelerator facilities: the tandem laboratory, the Laboratory for Nuclear Astrophysics (LENA), and the High Intensity γ-ray Source (HIγS). The first features an FN tandem accelerator equipped with the highest-current DC source of polarized hydrogen ions in the world, along with the high-current, low-energy beam facility (LEBAF). LENA delivers 0.1 to 5 mA beams of protons at energies between 0.2 and 1 MeV for the study of reactions of interest in nuclear astrophysics. The HIγS is operated as a collaborative effort of TUNL and the Duke Free Electron Laser Laboratory. A recently completed upgrade of HIγS allows for increased beam energy up to 250 MeV and increased beam intensities. Upgrades to the linac injector and the free-electron laser could further improve the γ-flux and extend the γ-energy to above the pion-production threshold.

The Center for Experimental Nuclear Physics and Astrophysics at the University of Washington operates an FN tandem Van de Graaff accelerator for research in nuclear astrophysics and fundamental symmetries, with a particular connection to the neutrino physics program. The accelerator is augmented with a terminal ion source and the capability for producing certain radioactive beams.

The A.W. Wright Nuclear Structure Laboratory (WNSL) at Yale University houses a 20 MV tandem accelerator capable of accelerating heavy beams to energies above the Coulomb barrier. The facility has a new suite of instrumentation to carry out a broad program in nuclear structure and astrophysics. The research focuses on structural evolution with neutron number N and nuclear charge Z, the nature of shape transitional regions, and the development of collectivity from both microscopic and geometric, symmetry-based perspectives, as well as the study of reactions of astrophysical importance.

## MAJOR DETECTORS

In addition to the accelerator facilities, detectors and instrumentation are essential for discoveries in the study of nuclei. In the last decade a number of new detectors have been developed. They use innovative ideas and state-of-the-art technology to meet the demands of the experiments. The technical base underlying these is broad, encompassing: solid state, liquid, and gas phase detectors; ionization, proportional, and Geiger-mode detectors; and detection of photon, ionization, and Cerenkov emission radiation. They span scales from submicron resolutions to multikiloton masses. Below, we discuss the major detectors currently in operation.

### Neutrino Detectors

Over the last decade, neutrino detectors have been used to confirm the existence of neutrino flavor oscillations. The SuperKamiokande detector, 1000 m underground in the Kamioka mine, is a Cerenkov detector consisting of 50 kilotons of water enclosed in a tank 40 m in diameter and 40 m high, viewed by 13,000 large photomultipliers. SuperK has been used extensively to study atmospheric and solar neutrino interactions, providing conclusive evidence for neutrino oscillations from atmospheric neutrinos. The Sudbury Neutrino Observatory (SNO), a Canada–U.K.–U.S. collaboration was also a water Cerenkov detector located 2000 m underground in the Creighton nickel mine in Canada. It had a central volume filled with about 1000 tons of heavy water, to distinguish electron neutrinos from other flavor neutrinos via neutrino interactions with deuterium. SNO observed both charged-current and neutral-current events, demonstrating flavor oscillations are the source of the long-standing solar-neutrino "problem." Another solar neutrino detector, Borexino, is sensitive to low-energy neutrinos (<1 MeV) and is located in the Gran Sasso underground laboratory in the Italian Alps.

Oscillations of reactor-produced antineutrinos were observed by the KamLAND experiment, which employs 1000 tons of liquid scintillator viewed by 1879 photomultipliers, all in a large underground tank in Japan. The antineutrinos are produced by several dozen Japanese power reactors located over a range of distances on the order of 100 km from the KamLAND site. KamLAND has also observed neutrinos that are decay products of the radioactive nuclei at the core of the Earth. Further bounding of possible parameters for neu-trino oscillation has come recently from the MiniBOONE detector at FNAL, which did not observe an oscillation pattern for neutrinos consistent with that reported earlier by the LSND experiment at LAMPF for antineutrinos. MiniBOONE is a large tank of 800 tons of mineral-oil liquid scintillator, viewed by 1280 photomultiplier tubes.

**Detectors for Cold Neutron Experiments**

The Fundamental Neutron Physics Beamline (FNPB) at the Spallation Neutron Source (SNS) at Oak Ridge is under construction to provide beams of cold and ultra-cold neutrons. Neutrons are produced by spallation reactions of 1-GeV protons on a mercury target and cooled in side-coupled moderators of liquid hydrogen or other materials. The FNPB guides neutrons some 30 m to experiments via nickel-coated neutron guide tubes and selects cold neutrons in energy and time via a series of rotating-wheel neutron choppers; this also avoids frame overlap between separate SNS pulses at 60 Hz. Ultra-cold neutrons (UCN) are produced using a double monochromator of intercalated graphite to select 0.89 nm neutrons by Bragg scattering, with these neutrons then stopped in liquid $^4$He via downscattering from phonons. The first experiment to use the new facility will be NPDGamma—a search for weak interaction effects in the capture of polarized neutrons on protons by measuring tiny asymmetries in direction of the emitted gamma rays relative to the incident neutrons' spin polarization. A high flux of neutrons bombard a 20 liter liquid para-hydrogen target, with the resulting high flux of gammas detected by a large array of CsI crystals read out by vacuum photodiodes in current mode to search for an effect of 5 parts per billion.

The UCN-A experiment at LANSCE at Los Alamos measures the beta-decay correlation between the neutron's spin and the decay electron's momentum and, thereby, the weak-interaction coupling strength between up and down quarks. The experiment uses polarized and spin-selected ultra-cold neutrons produced by phonon downscattering from a solid deuterium "superthermal" moderator. A 5 m long superconducting solenoid acts as a spectrometer for the decay electrons.

The neutron lifetime has direct implications for Big-Bang nucleosynthesis. A recent experiment using bottled UCNs gave a lifetime well outside the experimental uncertainties of the earlier world average, with potentially profound implications for our understanding of the weak interaction. A new lifetime experiment based on a magnetic trap produces UCNs in situ via downscattering of 0.89 nm cold neutrons from the NIST reactor on phonons of superfluid $^4$He. The resulting neutrons are magnetically trapped in a superconducting solenoid and their decays observed via the scintillation light produced in the superfluid helium by the decay electrons.

**Detectors for Nuclear Structure and Nuclear Astrophysics**

The low-energy user facilities are equipped with state-of-the-art instrumentation designed to explore the unique science opportunities provided by the facilities. At ATLAS the combination of the Fragment Mass Analyzer (FMA) with the Gammasphere detector gives excellent mass resolution and high gamma-ray efficiency, yielding the world-leading facility for the study of exotic nuclei produced by fusion reactions. The use of digital electronics for the focal plane detector of the Recoil Mass Spectrometer at HRIBF makes it a unique instrument for the study of the very short-lived nuclei. The Segmented Germanium Array at NSCL with high efficiency and good position resolution is the best among arrays at fast radioactive beam facilities. Recent additions at the NSCL include the Modular Neutron Array (MoNA) built by a collaboration including many undergraduate institutions for the detection of neutrons and the high-resolution silicon array (HIRA), a large-area silicon strip array backed by a CsI array to detect high-energy scattered light ions. The gas-filled magnetic separator at the 88-Inch Cyclotron has the highest efficiency for the detection of heavy elements.

Unique instruments at the university-based laboratories include devices for the production of and studies with radioactive beams such as the MARS fragment separator and BigSOL at TAMU, the TwinSol system of superconducting solenoids at the NSL, and the RESOLUT spectrometer at the John D. Fox Superconducting Accelerator Laboratory. The instrumentation suite at the WNSL includes the YRAST Ball, a γ-ray detector array, a plunger device for lifetime measurements, the ICEY Ball for conversion electron spectroscopy, SASSYER, a gas-filled spectrometer for spectroscopy of heavy nuclei, and both a superconducting magnet and the Rutgers transient field instrument for magnetic moment measurements.

# Research and Development for the Facility for Rare, Isotope Beams

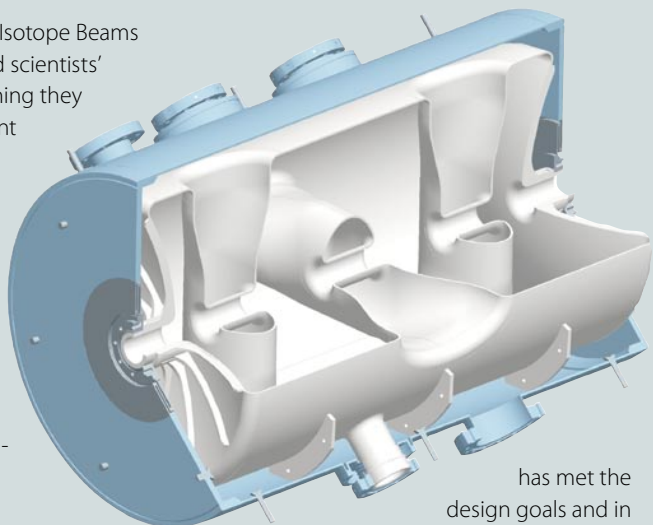

The Facility for Rare Isotope Beams promises to extend scientists' reach far beyond anything they can achieve with current isotope production facilities. But FRIB would never have been viable without advances in key technologies such as new ion sources, new accelerating structures, and new concepts in experimental equipment.

An example of new technology can be found in the figure above, which displays an advanced Superconducting Radio-Frequency (SRF) triple-spoke cavity. This structure was developed for efficient acceleration of heavy ions to half the speed of light. Its design reduces operation and construction costs, yet provides the required operational safety margin.

The ambitious goals of FRIB have necessitated similar advances in many other areas. For example, the goal of achieving beam powers of 400 kW—nearly 10 times higher than any other heavy-ion rare-isotope facility—required a new generation of Electron Cyclotron Resonance (ECR) ion sources based on high-power microwave technology. The VENUS source at LBNL, shown in the figure to the right, has met the design goals and in many instances exceeded them, allowing the cost of FRIB to be significantly reduced.

Efficient use of the new isotopes created by FRIB also requires a new generation of nuclear detection schemes. An example is the GRETINA project, now under construction, and its follow-on full implementation GRETA, which will increase the sensitivity of gamma-ray detection from rare isotopes by up to a factor of 10,000. GRETINA and GRETA, shown in the figure below, are based on a new generation of gamma-ray detectors that include the ability to "track" gamma rays and determine precisely the position in the detector where the photon hits and the direction from which it came.

The advances driven by the goals of FRIB are already finding potential applications in other areas. The SRF developments will have an impact on the research and development work underway in high-energy physics in support of a future International Linear Collider. ECR developments have an immediate benefit at current accelerator facilities and at high-intensity sources of ions for material processing. Detector technology, such as gamma-ray tracking, can provide revolutionary developments in areas such as medicine and homeland security.

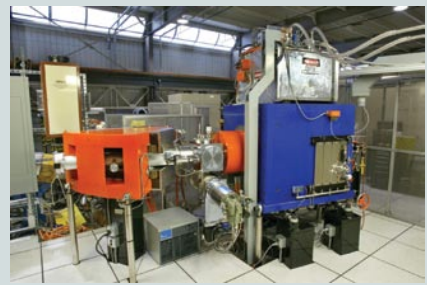

**Figures:**
**Top left:** Cut-away CAD-model view of a niobium triple-spoke superconducting resonator nested in an integral stainless-steel helium vessel developed at ANL. Research for the FRIB accelerator has led to major innovations in superconducting accelerators that will have an impact in many fields.

**Above:** The VENUS ECR ion source developed at LBNL. This source has produced record intensities of highly charged uranium, which will allow FRIB to achieve power levels 10 times what will be available elsewhere.

**Left:** Hand in hand with advances in accelerator physics, new concepts for advanced detection of nuclear decays will also push the reach of new rare-isotope facilities. One of the most significant is the development of very high-efficiency gamma ray detection with tracking capability. The GRETA concept, shown in the figure, will increase the sensitivity of experiments by a factor of 10,000 or more.

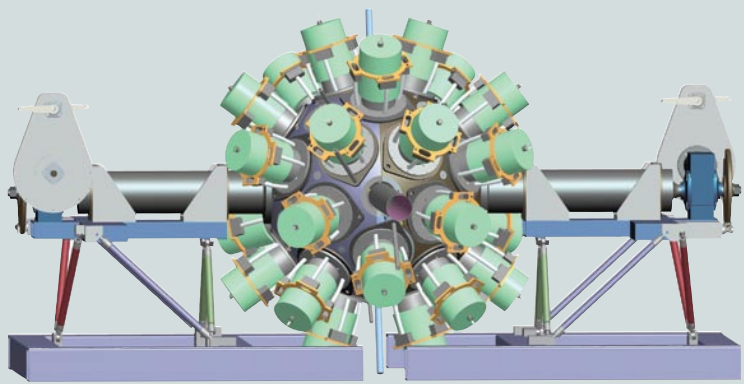

**Detectors at CEBAF: Exploring the Quark Structure of Matter**

Barely Off-Shell Nucleon Structure (BoNuS) is a novel, compact, radial time projection chamber used to detect short-range recoil particles resulting from electron-nucleus collisions. Used with a deuterium target, it provides identification and kinematic constraints for electron-neutron collisions. This yields previously unavailable data on neutron structure functions and resonance cross sections. It is further envisioned to use the recoil-particle identification power of this detector to select interactions in which a helium nucleus is scattered coherently by an electron beam. Such events will provide a low-background means to observe possible gluonic degrees of freedom through the production of mesonic states.

Jefferson Lab and collaborating institutions have developed and are improving a facility for hypernuclear spectroscopy (the HKS and HES) using a specialized beam line and spectrometers. Compared to hypernuclear production with pion or kaon beams, this facility obtains higher resolution (400 KeV full width half maximum), a better determination of binding energies, and higher yields. A new electron spectrometer and beam-line modifications are expected to further improve the resolution and the signal to background ratio in the spectra.

The Qweak experiment at JLAB is a precision test of the Standard Model by the determination of the weak charges of the quarks through parity-violating electron scattering. The apparatus will employ the world's highest power (2400 W) liquid-hydrogen target system, a large (~7 m diameter) precision toroidal magnet, large drift chambers, very high-rate gas electron multiplier detectors, and eight synthetic quartz Cherenkov detectors arranged symmetrically around the eight focusing locations of the toroid.

New Compton beam polarimeters are being constructed in Halls A and C and will be used during the upcoming Qweak and PREx experiments. They will provide non-invasive measurements of the CEBAF beam polarization by scattering circularly polarized light from the electron beam. In Hall C, a 25 W, 499 MHz RF-pulsed fiber "green" laser is being developed by the JLAB injector group. By matching the laser repetition rate and duty cycle to that of the electron beam, the effective luminosity of the laser/electron-beam collisions will be significantly enhanced. In Hall A, an amplified cw green laser is being used. These new Compton polarimeters will provide absolute accuracies of better than 1% and will also be compatible with the 12 GeV CEBAF Upgrade with minimal modifications.

A polarized $^3$He target is used as an effective polarized neutron target in a broad range of spin-structure studies. The targets themselves are glass cells including a long cylindrical target chamber (20–40 cm in length) connected to a spherical "pumping chamber." The contents of the cells include roughly 10 atmospheres of $^3$He gas and a mixture of potassium (K) and rubidium (Rb) metal. The pumping chamber is kept at a temperature of 230° to 240° C to maintain an appropriate density of alkali-metal atoms which are optically pumped with around 100 W of light from high-power diode-laser arrays. Polarization is then transferred through spin exchange collisions to the $^3$He nuclei. The introduction of hybrid mixtures of K and Rb (previously only Rb was used) has made it possible to routinely achieve polarizations of around 50% with an electron beam of 10–15 µA on the target. In a nice example of the interplay between fundamental and applied research, a similar technique is also used to polarize both $^3$He and $^{129}$Xe for MRI studies of human lungs.

**Detectors at RHIC: Probing Hot Dense Matter**

The RHIC detector suite comprises two large detectors, PHENIX and STAR, which together cover the range of predicted and observed signatures for deconfined matter. Two small detectors, PHOBOS and BRAHMS, have been decommissioned after completing their science program. The two active detectors include a pair of zero-degree calorimeters, one on each side of the interaction point, to tag reaction events and provide a common basis for comparison among the two detectors. RHIC has collided Au + Au, d + Au, Cu + Cu, and p + p at energies from 20 to 200 GeV/nucleon pair, operating each year since initial collisions were achieved in 2000.

The PHENIX detector focuses on the detection of leptons, photons, hadrons in selected solid angles, and the rare high-momentum particles emitted from jet fragmentation. It records events at 5 kHz and can examine every RHIC collision with a set of simultaneous rare-event triggers to emphasize electromagnetic and high-momentum probes in its recorded data. The central magnetic spectrometer has an axial-field magnet and two detector arms, each covering one-fourth of azimuth and slightly canted with respect to one another to produce uniform transverse momentum acceptance. Particle measurement is achieved via multi-sampling drift chambers, three layers of pad chambers for 3D

## Nature Revealed through an Imperfect Mirror

The weak force is unique among all the other known forces of nature because it violates a symmetry known as parity. That is, it distinguishes between right-handed and left-handed particles. This handedness was experimentally discovered just over 50 years ago, when a collaboration between nuclear physicists at what was then the National Bureau of Standards, and Prof. C.S. Wu of Columbia University, measured an asymmetry in the radioactive decay of polarized cobalt nuclei. Today, weak interaction experiments have become such a standard part of the nuclear physics arsenal that a program of electron-scattering experiments is using them to probe hadron structure and fundamental symmetries of nature. This program is made possible by two essential ingredients: high-performance polarized electron sources and the control of the beam properties at the nanometer scale.

While the weak force violates parity in a maximum possible way, it is so "weak" relative to electromagnetism that its effect on the scattering of right- versus left-handed electrons is still quite tiny. Indeed, physicists can see the effect only by measuring scattering rates with an accuracy of a few parts per billion. This quest for accuracy drives the requirements for large-acceptance detectors, high-intensity beams with high polarization, and exceptionally good control over small changes in beam properties when the helicity, or handedness, of the beam is reversed. Significant improvements in this last constraint have been achieved through a symbiotic effort by the accelerator scientists and the nuclear physicists performing the experiments. Careful alignment of the optical elements of the laser used to generate the electron beam (as shown in the bottom figure on the adjacent page) minimizes helicity-related shifts in the electrons' position and intensity before they reach the accelerator. Such techniques have been used by the G0 and HAPPEX collaborations at JLAB, and also at SLAC in experiment E158, the best measurement to date of the electron's weak charge.

The tight constraints on the electrons' helicity-correlated parameters also drive accelerator performance, which has several kilometers of beam transport and four decades of momentum gain from injector to target. Helicity-related shifts in the electron beam's position in the injector region are shown in the left part of the figure below. Careful tuning of the beam transport to eliminate coupling between the horizontal and vertical dimensions allows a natural damping as the electrons gain momentum along their path to the experimental area. The result is a factor of 5–30 improvement in these tiny position shifts, shown in the right side of the figure.

The main focus of the present experimental program has been to study low-energy strong interactions, where one can combine the complementary electromagnetic and weak information to disentangle the contributions of up, down, and strange quarks to the proton's charge and magnetism. These experiments now place tight constraints on the strange-quark contributions and provide an important benchmark for calculations of proton structure using lattice QCD or other techniques. The next generation of experiments will rely on these results for precision probes of the New Standard Model.

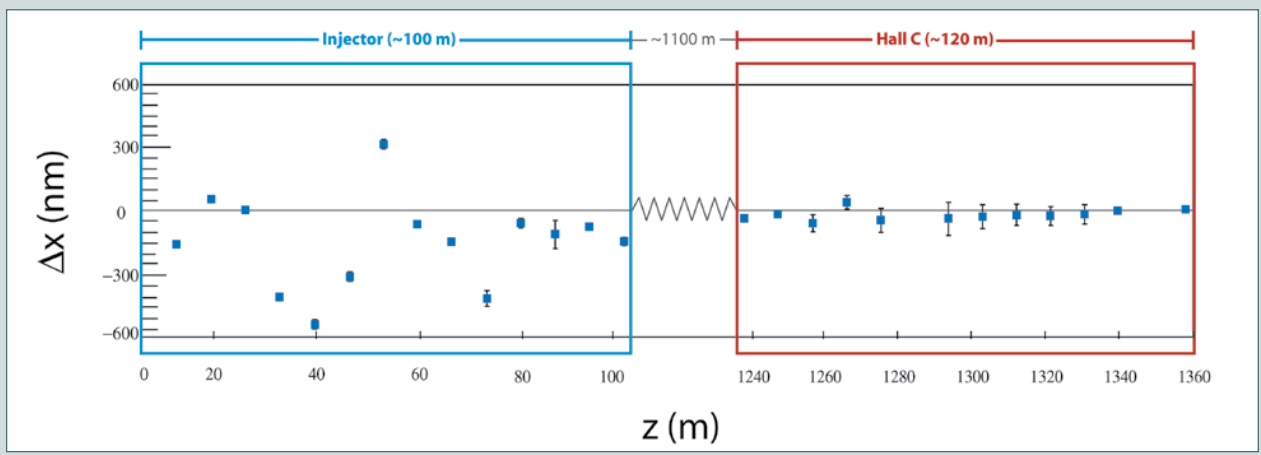

Tiny shifts in the beam position are measured at various points along one pass through the Jefferson Lab accelerators. Precise alignment of the injector laser is required to keep these shifts at the submicron level shown on the left side. Then, careful tuning of beam transport can reduce these shifts substantially, to the nanometer scale, by the time they reach the experimental area after more than a kilometer of travel, through a process known as adiabatic damping.

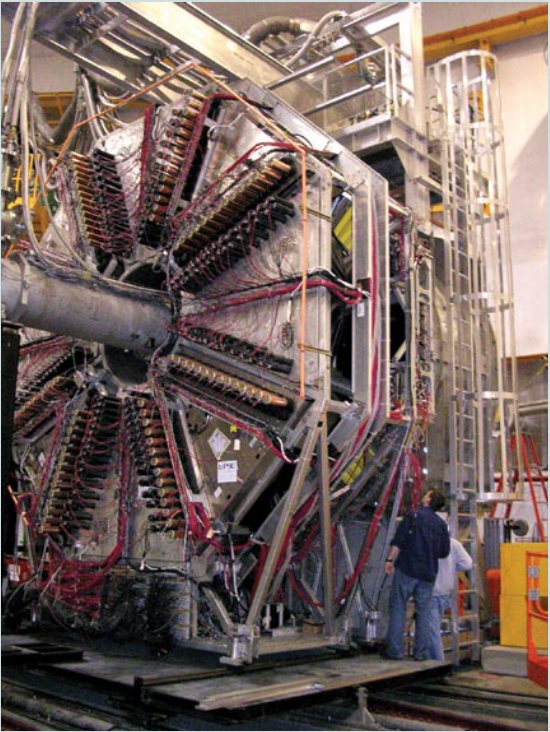

The detector system for the G0 experiment in Hall C at Jefferson Lab. G0 is designed to map the spatial distribution of strange-quark contributions to charge and magnetism in the proton by measuring a small asymmetry in the scattering rate from a liquid hydrogen target when the beam spin is reversed.

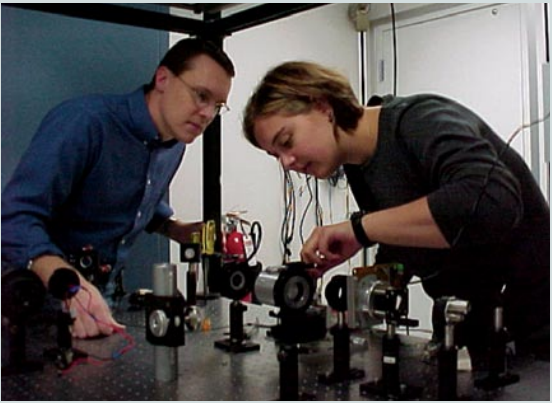

Two members of the HAPPEX collaboration at Jefferson Lab, Kent Paschke and Lisa Kaufman, work on the very precise alignment of the optical elements of the laser beam used to generate polarized electrons. They are able to minimize position and intensity differences in the electron beam before it is injected into the accelerator by careful selection of the laser optics, rigorous alignment procedures, and feedback between the electron beam and the optical elements.

space points, gas ring-imaging Cerenkov detectors tuned for electrons, time-of-flight arrays, aerogel Cerenkov counters to extend $\pi/K$ separation, and electromagnetic calorimeters with fine-grained transverse segmentation. Recent additions include a hadron-blind proximity-focused gas Cerenkov detector for low-momentum electron/positron tagging and fine-grained forward electromagnetic calorimeters. Event tagging and reaction plane are given by two forward segmented beam-beam counters. Two muon arms sandwich the central spectrometers and include lampshade magnets bending in azimuth, cathode-strip tracking chambers, and muon identifier chambers interspersed with 1000 tons of steel plate.

The STAR detector's central element is a large cylindrical time-projection chamber housed in a solenoidal magnet. This provides charged-particle tracking over a very large solid angle and a full reconstruction of charged and certain decaying neutral particles over two units of rapidity and all azimuth. Amplitude sensing on the readout pads adds particle identification via energy loss measurements. This powerful array is combined with a trigger barrel, segmented forward and barrel electromagnetic calorimeters, forward Time Projection Chambers (TPC) for improved kinematic coverage, and a silicon vertex tracker. A time-of-flight sector tuned for high-momentum hadrons extends the particle identification range to high momentum. The very large acceptance of the detector permits observation of global event features, which has proved quite powerful in examining decay patterns of jets.

PHENIX and STAR have embarked on a program of detector upgrades designed to expand kinematic coverage; improve particle identification, particularly for low-energy leptons and short-lived particles; and improve data-taking rates. Research and development on new detector concepts is a major part of the program. Three particular focus areas are large-area silicon-based microvertex trackers based on pixels and interpolating strips, Gas Electron Multiplier (GEM) detectors, and hadron-blind detectors for low-energy electrons and positrons. The silicon detectors being developed are of two types, pixel detectors for inner layers of the trackers and strip detectors for outer layers where hit densities are lower. The pixels are as small as 30 microns square to yield space point resolution as small as 10 microns. This is needed to spot displaced decay vertices of short-lived particles such as D and B mesons and allow their kinematic reconstruction. Very thin silicon substrates, down to 50 microns, are studied to surmount limitations on tracking precision imposed by multiple Coulomb scattering in the detector volume itself; this also helps avoid photon conversions. The strip detectors are both conventional type as well as interpolating type nested strips to use charge-amplitude readout as a means of interpolating hit position to less than a strip width, which is typically 80–100 microns. The GEM detectors are used as components in forward tracking arrays where their position resolution of under 100 microns is needed, in central-barrel tracking arrays where low sensitivity ("blindness") to hadrons is needed to pick out much rarer electrons and positrons, and as elements of trigger arrays where their speed of response and rate capability are matched to RHIC's bunch crossing and collision rates. The blindness is achieved by using gas Cerenkov converters and an evaporated CsI layer to convert the produced Cerenkov photons to electrons that are detected by the GEM.

**Heavy-Ion Detectors at the LHC: ALICE, CMS, and ATLAS**

A new entry into the field of relativistic heavy-ion collisions will be the LHC at CERN, which is planned to have first p-p collisions in summer 2008 and first heavy-ion collisions of Pb + Pb at 2.8 + 2.8 TeV/nucleon in late 2008. One purpose-built heavy-ion experiment, ALICE (A Large Ion Collider Experiment), is under construction together with the ATLAS (A Toroidal Lhc ApparatuS) and CMS (Compact Muon Solenoid) detectors built primarily for p-p collisions. ALICE at LHC has a 7 m diameter TPC as its main tracking detector. This is located inside a solenoid magnet and is supplemented with an extensive inner silicon-based microvertex tracker, a time-of-flight barrel outside the TPC, sectors devoted to high-momentum particle identification devices, a transition radiation barrel for tagging high-momentum electrons and positrons, and a forward muon spectrometer similar to those in PHENIX. A U.S.-led collaboration is adding a large sector of fine-grained electromagnetic calorimeter to aid in jet tagging and identification, with the goal of determining how jets fragment in the environment produced in a heavy-ion collision at the LHC energies, which are 28 times those reached at RHIC. The p-p experiment detectors—CMS and ATLAS—follow the general scheme of hermetic high-energy collider detectors with central microvertex trackers, large spectrometer magnets, calorimetery, and external muon trackers. CMS emphasizes compact design, while ATLAS is based on the world's largest superconducting toroid magnets. Both have unprecedented coverage for charged particles, photons, electrons, and jets as well as high-resolution muon spectrometers, which can identify heavy-vector mesons, such as the upsilon family, via their decay to $\mu^+ \mu^-$. High rate capability is required to handle LHC's expected p-p luminosity of $10^{34}$ cm$^{-2}$ s$^{-1}$. Additions for heavy-ion running include zero-degree calorimeters in both ATLAS and CMS, and high-level triggering in CMS.

## ADVANCED COMPUTING AND ELECTRONICS IN NUCLEAR PHYSICS

The National Supercomputing Centers, such as those at NERSC, ORNL, and ANL funded by the Office of Science and SDSC, and PSC supported by NSF, provide the computational infrastructure to solve a range of problems central to the nuclear-physics program, including aiding in the design of accelerators, *ab initio* calculations of nuclear structure and reactions, supernova simulations, and lattice QCD. These national user centers have been developed in close collaboration with industrial partners, including Cray and IBM. Several groups of nuclear physicists have obtained sizable allocations—many millions of hours of computing time per group—through nationally competed programs such as INCITE to work on problems central to our field. The Office of Science is providing "leadership class" facilities today, with plans to field machines capable of a peak performance of 1 Petaflop/s, by FY2009. The Teragrid network will seamlessly unite these facilities. The development of the appropriate software infrastructure and computational science capabilities, such as that developed under the DOE's SciDAC Initiative, will be essential to maximize the benefit of these facilities to the nuclear science program in areas as diverse as QCD, the physics of nuclei, and theoretical astrophysics.

NERSC also operates PDSF, a farm of Linux computers dedicated to nuclear and high-energy physics problems. In the past five years PDSF has provided about half of the STAR computing capacity and was the main source of computing for a host of other experiments like SNO, KamLAND, and IceCube. There are plans to upgrade the hardware and expand the role of PDSF in the next five years. It will continue to provide STAR with about 50% of its computing, be the major computing center for the ALICE–U.S. collaboration, and remain the computer facility for some existing and future neutrino experiments.

Physics event analysis typically only requires the computing power and storage found in a present-day commodity

PC for a given event, but the very large data sets collected can include billions of events, making a distributed computing architecture a natural choice. BNL and JLAB maintain large data analysis centers, the RHIC Computing Facility (RCF) for the RHIC experiments and the JLAB Computing Center for experiments at CEBAF. Both have thousands of computer nodes and provide support for raw data recording and storage, database storage, raw data reconstruction, and physics analysis. They are joined in this effort by over a dozen centers at universities and labs in the United States and elsewhere. Data transfer among these centers makes use of high-speed Internet connections among the centers and is making increasing use of GRID resources to apply available computing power to analysis projects. Experiments at LHC, specifically ALICE, ATLAS, and CMS which all include a relativistic heavy-ion part of their planned physics running, have based their data reconstruction and analysis plans on a GRID architecture and are making use of such resources worldwide as they prepare for first collisions in 2008/2009.

**Facilities for Lattice QCD**

Theoretical support for the RHIC and JLAB experimental programs relies on numerical calculations. At present the largest demand for computing resources exists for numerical calculations in lattice QCD at high temperature and nonzero baryon number density, closely followed by modeling the expansion of the dense matter created at RHIC using three-dimensional relativistic hydrodynamics.

During the years 2000–2005 the Columbia group in collaboration with IBM and the RIKEN-BNL research center (RBRC) developed the supercomputer QCDOC, out of which the BlueGene/L emerged. Since 2005 BNL has housed two 10 Teraflops installations of the QCDOC computer. One of these installations serves the needs of the RBRC in lattice gauge theory, and the other is part of the U.S. SciDAC effort. At the time of installation it provided most of the computing resources distributed by the U.S. Lattice QCD consortium (USQCD). Today the QCDOC machines still provide about 40% of these resources. BNL and the RBRC intend to continue the close collaboration with IBM that led to the development of the QCDOC machines, with the goal in the next few years of constructing a follow-up machine, which should operate in the 500 TFlops range. The lattice QCD community in the United States has just prepared white papers including an outline of new lattice QCD calculations at finite temperature and density, which would become feasible on a computer in the petaflop/s class, providing the rationale for future developments in computer performance.

The USQCD consortium includes participants from eight university and three national lab physics departments and four university computer science departments. The USQCD effort has obtained large clusters of some 500 nodes each at JLAB and FermiLab as part of the effort to provide platforms for lattice QCD work based on commodity hardware. By selecting the most cost-effective and appropriately balanced combinations of processor and network interconnect—as opposed to the products that individually had the best performance—and by taking advantage of the modest requirements for memory size and disk bandwidth, large-scale clusters have been constructed with better price/performance than any existing general-purpose parallel computing platform.

The U.S. lattice gauge theory community has created a unified program environment that enables its members to achieve high efficiency on terascale computers. Among the design goals were to enable users to quickly adapt codes to new architectures, easily develop new applications and incorporate new algorithms, and preserve their large investment in existing codes. These goals were achieved through the development of the QCD Applications Programming Interface, which has been implemented on the computing hardware at BNL, JLAB, and Fermilab. Community software has been developed by the USQCD collaboration to enable the development of highly efficient code for clusters, the QCDOC, and commercial supercomputers.

**Electronics**

Developments in integrated circuits have played a role in improving experiment capabilities at all nuclear physics facilities. The field takes advantage of advances in computing and networking power in order to collect, record, transmit, and analyze results of experiments. Large increases in capabilities of field-programmable gate arrays, which now routinely include more than one million logic gates, have led to implementations in fast trigger circuits, pattern recognition circuits, real-time tracking and position-determination circuits, and event builders. RHIC experiments in particular developed mixed analog and digital integrated circuits to address challenges imposed by highly segmented detectors, multihundred-thousand channel counts, limits on power dissipation and usage, and need for delay/storage elements while trigger decisions could be formed. Fully pipelined operation was achieved for these circuits, meaning all col-lisions at RHIC could be observed and a real-time decision taken whether to record the event. Upgrades to detectors at all facilities take advantage of these developments coupled with rapid advances in fast digitizers, which can now attain better than 12-bit resolution at greater than 100 M digitizations/second. This has been used in detection of decays of nuclei, which themselves live only microseconds; in reconstruction of detailed waveforms for gamma-ray energy deposit to in turn enable tracking of gamma rays; and in collection of large volumes of tracking and energy-loss data from gas-drift detectors. Data transfer has similarly taken advantage of increases in optical fiber and transceiver performance, with per-link transfer rates in excess of 5–10 Gbits/s now possible. Event building and recording rates have followed, with data-acquisition rates now approaching 1 GByte/s and resulting datasets reaching petabyte scale. The storage and subsequent analysis of these large datasets have posed challenges that are being addressed by developments in distributed computing and GRID technology for distributing raw results to analysis centers and collecting the results. This in turn has motivated development of distributed databases to support this work.

## FUTURE FACILITIES AND DETECTORS

### CEBAF: The 12 GeV Upgrade

The 12 GeV CEBAF Upgrade utilizes the existing tunnel and does not change the basic layout of the accelerator. It assumes that CEBAF has completed its evolution from the original 4 GeV maximum beam energy to the present goal of 6 GeV. A program of refurbishment of the installed cryomodules is well underway and will achieve this goal soon. The basic elements of the 12 GeV Upgrade are: (1) the addition of 10 new modules in space already available in the linac tunnels, (2) stronger magnets in the recirculation arcs, (3) an upgraded cryoplant, and (4) the addition of a 10th recirculation arc permitting an additional "half-pass" through the accelerator to reach the required 12 GeV beam energy followed by transport to the new hall that will be added to support the meson spectroscopy initiative.

Doubling the CEBAF energy requires an additional 0.5 GV/linac energy gain, which will be achieved by adding five new 100 MV cryomodules to each linac. Each cryomodule will have eight seven-cell cavities with a quality factor $Q_0 = 8 \times 10^9$ at the design gradient of 19.2 MV/m. Key elements to meet this cavity performance are a new cavity design, new fabrication method, and new cryomodule assembly procedures. A critical test of a two-cavity cryomodule was recently successfully executed. It demonstrated that the design meets the gradient and $Q_0$ specifications. RF field control of the new cavities is challenging due to the high gradient in combination with the large external Q factor ($>2 \times 10^7$). Successful tests of generator-driven resonator control algorithms have been carried out that met the phase and amplitude control requirements up to a gradient of 20 MV/m. New algorithms with extended capabilities are under development and have shown promising results at high gradient.

As part of the upgrade, research and development is being carried out on solenoidal detectors for the next generation of parity-violation experiments. The parity-violating deep-inelastic scattering research program described in the science section requires a large-acceptance detector capable of handling very high event rates to reach the needed sensitivity to observe the small parity-violating effects. The conceptual design consists of a large superconducting solenoid, equipped with baffles and collimators, GEM, Cherenkov and lead-glass detectors, and a high-power cryogenic target.

### FRIB

The production of rare isotopes, which have previously existed for only fleeting moments in the explosion of stars or in the crusts of neutron stars, is one of the major challenges to progress in understanding atomic nuclei and their reactions. Fortunately, modern, high-power linear accelerator (linac) technology puts many of the key rare isotopes within reach. Indeed, the heart of FRIB is in-flight production of isotopes produced by beams from a superconducting heavy-ion linac capable of producing 400 kW beams of all elements from uranium (with a maximum energy of at least 200 MeV/nucleon) to protons (with a maximum energy of at least 500 MeV). Heavy-ion beams are delivered to a production area with the capability to operate at high power. The resulting rare isotopes will be collected and separated by a high-efficiency fragment separator and delivered to a gas cell for collection, or used directly for fast beam experiments. An isobar separator will purify the ions extracted from the gas cell. These are then used either directly in stopped ion experiments or injected into a charge breeder to be accelerated to energies required for nuclear astrophysics and nuclear structure research. The corresponding production of rare isotopes at rest via spallation or fission of heavy targets by use of the high-power, light-ion capability (acceleration of

# FEL Applications—Carbon Nanotubes

What began in 2001 as an academic investigation—how to make carbon nanotubes with a free electron laser (FEL)—has moved into a new phase. Researchers at NASA's Langley Research Center and the FEL at Jefferson Lab are now producing high-quality, single-wall carbon nanotubes in quantities sufficient for testing in aerospace applications. Many researchers believe that such nanotubes may lead to an entirely new generation of very lightweight materials as strong or stronger than steel.

Carbon nanotubes, cousins of the buckyball, were discovered in 1991. Each one is essentially a sheet of graphite curled up into a cylinder measuring about one nanometer (one billionth of a meter) wide and up to one millimeter long. Laser-synthesized nanotubes tend to have superior properties to those produced by other techniques. The raw material can be very pure, and the tubes themselves are straight, homogeneous, and defect free. The problem is making them in large quantities. Typical tabletop lasers use tens of watts to make nanotubes at around 200 milligrams per hour.

To do better, scientists at NASA Langley Research Center designed a new nanotube synthesis process and apparatus that uses a custom-made, rod-shaped graphite target. A beam of infrared laser light at a wavelength of 1.6 microns from the Jefferson Lab FEL vaporizes layers of the spinning target to create a plume of nanotubes. It has been demonstrated that this process can be used to make about 100 milligrams of nanotubes per minute; that rate can fill a coffee cup with raw material in about 15 minutes. Research is continuing to increase the yield.

The unique features of the laser beam at the Jefferson Lab FEL enable nanotubes to be fabricated differently than with conventional lasers. The Jefferson Lab FEL emits a series of "ultrafast" infrared laser light pulses (each pulse lasts less than a picosecond, or one trillionth of a second) at a high repetition rate of 9.4 MHz, about 9.4 million flashes a second. The ultrafast pulses directly excite the reactants to form the nanotubes, unlike a conventional laser, which heats up a material to produce them. Research continues on the FEL-based synthesis process with emphasis on scaling up to produce sufficient quantities for device fabrication. One interesting finding so far is that the diameter of the nanotubes can be varied by changing the laser's parameters. Since different applications require different sizes, this may turn out to be a big advantage of FEL-synthesized nanotubes. An example of carbon nanotubes fabricated using the Jefferson Lab FEL is shown in the figure.

Initial applications of the nanotubes at NASA Langley will be in fiber-reinforced materials. The research is now focusing on purifying and processing the raw material and will soon shift to incorporating nanotubes into palm-sized test pieces. NASA is interested in high-strength, lightweight materials that are multifunctional (can sense strain, bend and flex themselves, or conduct heat and electric charge). In theory, carbon nanotube-reinforced materials can provide all these functionalities, saving critical weight in a myriad of aero and space applications.

Institutions affiliated with the research include NASA Langley Research Center, Jefferson Lab, National Institute of Aerospace (NIA), The College of William and Mary Department of Applied Science, and Luna Innovations, Inc.

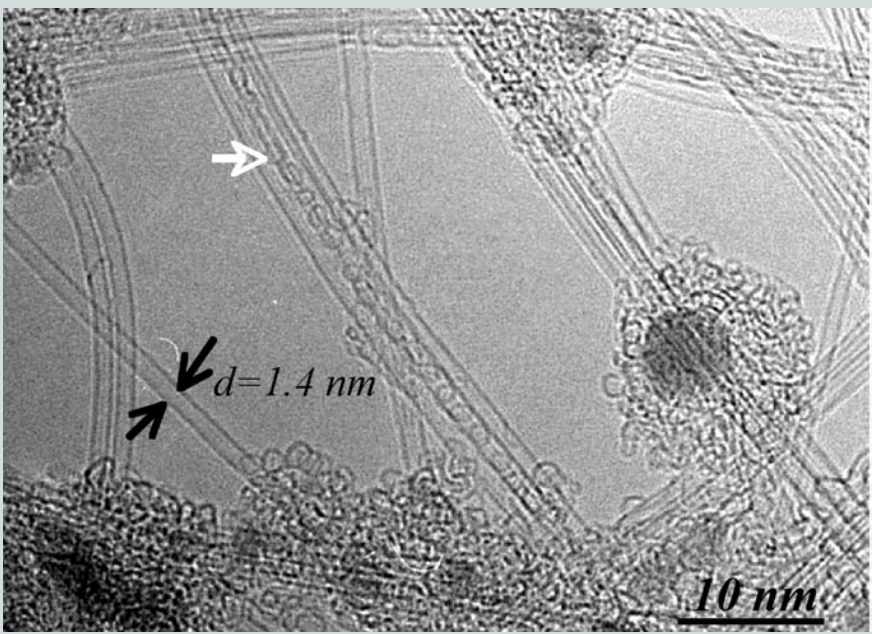

These carbon nanotubes were manufactured with the Jefferson Lab free electron laser. High-quality single-wall nanotubes are clearly visible. The "peapods" are carbon shells trapped within carbon nanotubes (shown by the white arrow); fullerenic carbon shells are seen outside nanotubes. Dark grey specks of target material are also visible.

protons or $^3$He) of the linac also could produce beams via the ISOL approach.

Since the last Long Range Plan a vigorous research and development program on a future RIB accelerator has taken place, involving teams at universities and national laboratories. This effort has dealt with nearly every aspect of research and development, and it has had a major impact on the cost and performance for FRIB. Progress has been substantial on all fronts, and the major technical challenges have been resolved. As a result, the concept currently being considered for FRIB has been demonstrated to be challenging but technically feasible.

The key to the FRIB concept is the high beam power of 400 kW made possible through the simultaneous acceleration of multiple charge states by the driver accelerator. Detailed simulations of the injection scheme from the ion source into the driver accelerator have now been completed and are currently being tested in laboratory experiments. Prototypes of the full set of required cavities for the driver linac have been fabricated, and accelerating fields meeting and generally exceeding expectations have been achieved. In the process, new types of superconducting RF cavities have been developed that have potential for applications in numerous scientific and industrial areas requiring the use of high-intensity accelerators. Also key to the FRIB design is the gas stopper concept in which the radioactive ions of interest, produced through the fragmentation process of 200 MeV/nucleon beams, are slowed down in ultra-pure helium so that they can be brought to a 1$^+$ charge state suitable for post acceleration. Among the main advantages of the approach are the chemical independence and the fast time scale of the technique. Worldwide research and development efforts have demonstrated that a gas cell design for FRIB can achieve high extraction efficiency and will be able to deliver rare-isotope beams of up to 10$^9$ particles per second. Research and development efforts on alternative gas stopping concepts (such as a cyclotron gas stopper) and on various high-efficiency charge breeding mechanisms for post acceleration are also being carried out.

The FRIB concept requires beam powers of up to 400 kW, meaning target cooling, beam dumps, effective beam stripping, and radiation damage are areas of concern. Solutions to these issues based on the applicability of technologies from fusion energy research and the SNS have been demonstrated. Detailed studies have been performed of the required shielding and remote manipulation of production components. A full study of the issues related to the primary beam dump, radiation damage, and component lifetime have also been performed, and no major issues have been identified.

The operation of the facility at 200 MeV/nucleon requires advanced, large-acceptance fragment separators. Satisfactory designs for fragment and isotope separators have been evaluated and will meet the facility requirements. As part of the process, advances have been made in the simulation software used for the design, which is now used worldwide.

**Facilities and Detectors at DUSEL**

The proposed underground laboratory, DUSEL, will explore a wide range of science. Specific basic research topics include solar, atmospheric, long-baseline, supernova, and high-energy astrophysical neutrinos; double beta decay; precision assay of radionuclides, with applications ranging from disarmament treaties to environmental effluent studies; materials science; nuclear astrophysics cross-section measurements; multiple topics in geoscience; and the study of the evolution of biological organisms under harsh conditions. Only a few examples are given here.

At LBNL, research and development is being carried out for CLAIRE (Center for Low energy Astrophysics and Interdisciplinary REsearch). The goal is to provide intense (≥100 mA, two orders of magnitude higher than current state-of-the-art accelerators) low-energy (50 keV to 300 keV) light-ion beams suitable for measuring many key stellar reactions, such as $^3$He($^4$He,$\gamma$) involved in hydrogen and helium burning phases of stars, near or at the Gamow-energy window. Phase II of CLAIRE would increase the acceleration energy to 3 MV and provide a wider variety of ion beams.

Searches for neutrinoless double beta decay capable of reaching the inverted hierarchy mass scale will require detector masses on the order of 1-ton and half-life sensitivities greater than 10$^{27}$ years. The goal of the Majorana Demonstrator project is to establish the feasibility of constructing a scalable 1-ton Ge-based experiment with extremely low intrinsic and external backgrounds. The Demonstrator module will consist of 60 kg of Ge detectors, with at least 30 kg of 86% enriched $^{76}$Ge. The detectors will be enclosed in low-background electroformed copper cryostats, which will be located deep underground within an ultra-low-background shielding arrangement. As part of the research and development evaluation, the collaboration plans to explore using novel p-type point contact Ge detectors as well as segmented n-type detectors. Signal processing from such detectors offers significant benefits in rejecting backgrounds through multidimensional event analysis.

The full spectrum of low-energy neutrinos from the Sun can be measured via the charged-current interaction in a few special nuclei that form an excited isomeric state upon capture, so that a delayed coincidence between the produced electron and a later-decay gamma ray serves as the signature of a solar neutrino-induced reaction. This technique provides a means to overcome the otherwise overwhelming natural backgrounds at energies around 1 MeV. The LENS collaboration is currently constructing a prototype detector, miniLENS, that will contain a fiducial volume of about 60 liters of organic scintillator loaded with 5 kg of indium. Among the goals of miniLENS are to demonstrate the time and position resolution of the segmented lattice design for LENS, to demonstrate the required level of background suppression via isomer tagging using proxy neutrino-induced events, and to establish a clear scale-up route to a full detector.

### RHIC II Luminosity Upgrade

A major luminosity upgrade of RHIC (RHIC II) will be accomplished by beam cooling of the full energy beams, the first such implementation at a high-energy collider. Cooling will increase the heavy-ion luminosity by a factor of 10 at high energy and make RHIC the first collider in which luminosity is limited by the interactions themselves. Cooling at injection energy will increase polarized proton luminosity by a factor of 2–3. Proof of principle of electron cooling at RHIC has been established through detailed simulations benchmarked by a series of experiments on cooler rings, in particular the FNAL Recycler ring cooler, which is closest in characteristics to the RHIC cooler. Intrabeam scattering simulations were benchmarked directly in RHIC. The electron accelerator of the RHIC II electron cooler must produce a high repetition rate of large bunch charges at a low emittance. This is state-of-the-art performance that requires the development of a few new accelerator components. A research and development program aims at the reduction of technical, budgetary, and schedule risks through the demonstration of these components. Major components of the RHIC electron cooler will be tested in a test facility that is currently under construction. Commissioning of the full system could technically be completed by 2012.

### Electron-Ion Collider (EIC)

Two options are currently being considered for an EIC—one uses the existing RHIC rings and adds an electron beam (eRHIC), and the other uses the upgraded JLAB electron beam and adds an ion beam (ELIC). The eRHIC scientific program calls for obtaining an integrated electron-nucleon luminosity of order 50 fb$^{-1}$ over about a decade by colliding 10 GeV polarized electrons on 50–250 GeV polarized protons, up to 100 GeV/nucleon gold-ion beams that are presently available in RHIC and the 167 GeV/nucleon $^3$He ions.

The two design options for eRHIC use either an electron storage ring or a superconducting Energy Recovery Linac (ERL) to provide the 10 GeV polarized electron beam. The ERL design option can achieve a peak luminosity for electron-proton collisions of $2.6 \times 10^{33}$ cm$^{-2}$s$^{-1}$ and has the potential for even higher luminosities. In this design, a high-current, polarized electron beam is generated in a photo-injector and accelerated to 10 GeV in five passes through a 2 GeV superconducting linac. After colliding with the hadron beam in as many as four interaction points, the electron beam is decelerated in the same linac to an energy of a few MeV and then dumped. The energy thus recovered is used for accelerating subsequent bunches. The ERL could provide electrons in the energy range from 3–10 GeV, leading to a center-of-mass energy range from 25–100 GeV. Even higher electron-beam energy of up to 20 GeV is possible. Additional features of this design are a high electron-beam polarization of about 80%, full polarization transparency at all electron energies, and very long straight sections for detectors that are free of beam elements.

ELIC is a high-luminosity, polarized Electron-Ion Collider, which uses CEBAF as the electron accelerator and requires the construction of a 30–225 GeV ion beam complex. ELIC is designed with center-of-mass energy up to 90 GeV and peak luminosity of $8 \times 10^{34}$ cm$^{-2}$s$^{-1}$, using high collision frequency and crab crossing of colliding beams. The green-field design of the ion complex is directly aimed at optimizing the EIC scientific program. The high luminosity is crucial to probe unknown features of the proton landscape, such as the impact of spin-orbit correlations. A "figure-8" design of the collider rings is chosen because it ensures spin preservation and ease of spin manipulation, and is therefore directly aimed at spin physics opportunities. The range of nuclei considered includes polarized light ions ($^1$H, $^2$H, and $^3$He) and medium to heavy nuclei up to A = 208. With this system the distributions of quarks inside the nucleus can be probed over a large range of scales and access modification of gluon distributions in nuclei. ELIC operation is expected to be compatible with simultaneous operation of the 12 GeV CEBAF Upgrade fixed target program.

The future implementation of a high-luminosity Electron-Ion Collider (EIC) will require sustained and focused research and development in accelerator science and technology in several areas. These include simulations of beam-beam effects, including the kink instability and electron-beam disruption, polarized $^3$He production and acceleration, and the development of high-precision, high-energy ion polarimetry. In addition, the EIC implementation at BNL (eRHIC) requires research and development of a high-current polarized electron source, energy recovery technology for high-current beams, and the development of small gap magnets and their vacuum chambers for the return loops. The EIC implementation at JLAB (ELIC) will need research and development of multicell crab cavities, electron cooling simulations with a circulator ring, and simulations of beam stacking with very high space charge forces.

A large fraction of the physics of interest at an EIC depends on auxiliary small-angle detection systems. Following the lessons learned at the HERA collider at DESY, robust techniques need to be developed to allow low-angle electron tagging of quasireal photoproduction events, tagging of nuclei to select forward diffractive events, and tagging of low-angle protons to guarantee the exclusivity of deep exclusive events. Special emphasis needs to be placed on the high-rate environment of the EIC.

The EIC includes in its science portfolio detection of a large range of processes, from abundant inclusive scattering processes to rare exclusive processes. To efficiently select the latter events amongst the background beam gas events associated with the high-luminosity environment provided by the EIC, a multilevel trigger system is foreseen. This trigger system includes the development of trigger algorithms for efficient rare-process and DIS e-trigger selection, most likely requiring the use of tracking at the trigger level for the former.

The performance of the inner tracking detector of an EIC main detector is intrinsically intertwined with the access to the science of interest. The exact specifications of this vertex tracker need to be balanced with detailed physics simulations. Development of a cost-effective large acceptance and compact high-rate tracking solution can then proceed. Cost-effective methods need to be pursued to perform hadron identification over a wide momentum range and to tag recoiling neutrons and heavier nuclei.

The high-luminosity ELIC design relies on very small time between bunch crossings, which in turn require a high-speed data acquisition and trigger system. In particular, trigger, readout electronics, and data acquisition systems need to be developed to pipeline data at the anticipated collision frequency of up to 1.5 GHz. In addition, a trigger with a high rejection rate of hadronic background and an ultrafast digitization system with the required timing properties need to be developed. Building on the recent developments for the CLAS detector, multiprocessing data acquisition systems will be required, and data rates in detectors and electronics will need to be simulated

**GRETA**

New detector technologies are needed to meet the challenges of the next generation of experiments with RIBs. Leading these is the technology of "gamma-ray tracking" which will revolutionize gamma-ray spectroscopy in a way similar to that in which large arrays of gamma detectors did a decade ago. A detector array using this technology can be assembled with closely packed, highly segmented Ge detectors to form a shell covering the target. Signals from the segments are processed digitally to give position and energy for each of the gamma ray interactions. The scattering sequence of the gamma-ray is then determined by tracking algorithms. During the last few years this technology has been shown to be feasible, and GRETINA, a $1\pi$ detector based on the concept, is under construction. However, a $4\pi$ GRETA system will be needed to fully exploit the science opportunities at future RIB facilities and to increase the reach of stable-beam facilities. GRETA will improve the power of GRETINA by a factor of 10–100 for most experiments. In addition, next-generation auxiliary detectors will further enhance the performance of GRETA, which is crucial for achieving the ultimate sensitivity. The gamma-ray tracking technology has important and broad applications for science, medicine, and homeland security.

**Neutron Detectors at the FNPB: Correlations, nEDM, Lifetime, and Spin Rotation**

Neutron-decay correlation experiments examine directional and spin correlations of the final-state proton, antineutrino, and electron. The resulting information bears directly on fundamental interactions such as the ratio of axial-vector to vector weak coupling constants. The antineutrino cannot be detected in any conceivable detector, and the proton's recoil energy is less than 1 keV, making reconstruction of the final state quite challenging. Correlation experiments are based on a central magnet, often a solenoid with an adiabatic field-expansion region to guide the decay products to detectors without destroying their correlations. Detectors can be based on large-area silicon pixels supplemented by electrostatic fields to accelerate the protons into the silicon-depleted region. Readout electronics include fast digitizers to distinguish multiple hits, provide time-of-flight information, and handle the high rates. The magnets include multiple superconducting coils to tailor the field configuration for the particular measurement.

A new search for the neutron electric dipole moment (nEDM) proposes to reach a limit of $10^{-28}$ e.cm, more than two orders of magnitude below the current limit. Polarized cold neutrons would be trapped by downscattering in superfluid $^4$He and contained in a bottle with a controlled magnetic field and a reversible large electric field. The effect to be observed is small—a dipole moment of $10^{-25}$ e.cm in a 10 kV/cm field corresponds to a frequency shift upon electric field reversal of 0.5 µHz. A co-magnetometer based on a small amount ($10^{-10}$) of polarized $^3$He also located in the volume would provide precise monitoring of the magnetic field value and expected spin precession.

Transversely polarized neutrons passing through liquid helium should acquire a parity-violating spin rotation due to the weak interaction. A first experiment to measure this effect used a crossed polarizer and analyzer system, with a pair of target chambers for the He target located between the polarizer/analyzer pair and a central solenoid placed between the two target chambers with field direction aligned to the initial neutron polarization (along x) and strength sufficient to rotate the spin vector 180° (about x) upon passage. A $^3$He detector analyzes the emergent neutrons. By comparison of count rates associated with its two analyzer states, which are configured to modulate between +y and –y, an asymmetry is observed. An upgraded apparatus would use faster detectors and make use of the cold neutron beams from NIST or the FNPB.

# International Collaborations and Facilities

## NUCLEAR SCIENCE: A WORLDWIDE ENDEAVOR

It has often been said that science has no borders. But nowhere is that more true than in nuclear science. No single nation dominates: U.S. nuclear scientists represent approximately 25% of the world's workforce in the field, but many other countries have significant investments as well. Indeed, some 30 to 40 percent of the scientists hosted by the flagship U.S. facilities at RHIC, JLAB, and NSCL are from other countries; their ideas, experiments, and discoveries have been critical to the success of the science programs there. Other U.S. user facilities also have active international participation as do many of the university-based facilities. And U.S. scientists are likewise pursuing opportunities at the many facilities located elsewhere. Key examples of such facilities are the neutrino experiments at SNO in Canada, and at SuperKamiokande and KamLAND in Japan, where the use of unique underground detectors has led to major new discoveries. Other outstanding examples include the nucleon spin physics studies at DESY in Germany and CERN in Switzerland, and the new rare-isotope research program at RIKEN in Japan. Looking into the future, our international colleagues are currently constructing two major multidisciplinary nuclear physics facilities—GSI/FAIR in Germany, and J-PARC in Japan—that will be fully open for worldwide participation.

International collaboration is also more prevalent than ever in nuclear theory. This is particularly important for large-scale numerical computations, such as lattice QCD and supernova simulations, which demand close international collaborations among theorists. The two leading nuclear theory centers, the Institute of Nuclear Theory (INT) in Seattle and the European Center for Theoretical Studies in Nuclear Theory (ECT*) in Trento, Italy, attract scientists from all over the world. Meanwhile, topical centers, such as the RIKEN-BNL Research Center (RBRC) at BNL and the Japan-U.S. Theory Institute for Physics with Exotic Nuclei (JUSTIPEN) at RIKEN represent more targeted international collaborations.

Precisely because nuclear science represents such a widely distributed, worldwide community, maintaining student and postdoctoral fellow exchanges is especially vital. Many nuclear scientists around the world received their Ph.D.s in the United States—but equally important, a large fraction of the U.S. nuclear science workforce has been trained in foreign countries.

More generally, the free exchange of scientific ideas facilitates scientific progress, fosters collaboration among scientists, and advances knowledge. The international free exchange in science has historically benefited the United States, has brought the best and brightest to our nation, and has resulted in cultural understanding among the international community of scientists. For free exchange to be a reality, however, the ability for all scientists to travel to participate in scientific activities is of paramount importance.

## INTERNATIONAL PARTICIPATION AT U.S. FACILITIES

Here we illustrate examples of significant participation from abroad in select U.S.-hosted programs.

*In relativistic heavy-ion physics,* participation from groups outside the United States is vital to the success of the RHIC experimental program. PHENIX and STAR are international collaborations of typically ~500 scientists, which include physicists from more than a dozen countries. Of the 18 different detector subsystems in PHENIX, four were built in Japan, and their operations continue to be supported by Japanese physicists and funds. Six other detector systems were constructed partially in European and Asian countries. The international partners on the STAR program have collectively contributed considerable technical expertise and approximately 15–20 percent of the investment needed to instrument detectors. In addition to their important role in constructing and operating PHENIX and STAR, international collaborators are central to the intellectual activity of the collaborations. Data analyses are performed around the world, with a petabyte of raw data exported for reconstruction and analysis at computing centers in other countries. International collaborators contribute to the scientific leadership, assuming roles such as detector system leaders, physics working group conveners, and deputy spokespersons.

*In hadronic physics,* international collaborators are using the unique beams of electrons and tagged photons at Jefferson Laboratory to carry out a wide range of experimental studies. They contribute through scientific ideas, equipment design, and data analysis. For example, physicists from France are major players in the deeply virtual Compton scattering program at Halls A and B, and they strongly support the G0 program in Hall C. The Italian INFN supports strangeness and baryon spectroscopy physics programs. Japanese teams initiated, and continue to drive, a very effective hypernuclear physics program. Canadian scientists are part of the G0 experiment, and they will participate in the future Hall D GlueX program. As a major contribution to a U.S. facility, RIKEN provided hardware to the RHIC facility that was critical to make it a unique high-energy polarized proton collider and was instrumental in launching the new physics program of measuring the nucleon spin structure functions at RHIC.

*In low-energy nuclear physics,* approximately one-third of the users at the NSCL are from abroad. User groups make significant equipment contributions, such as cryogenic targets and small time-projection chambers, for use in their experiments that exploit the high-rate, rare-isotope beams unique to NSCL. In a typical year, over 40% of the users at the Argonne ATLAS facility are from abroad, with similar numbers for the Oak Ridge HRIBF facility.

*In nuclear theory,* international participation in the INT has always been a key ingredient to the Institute's success. The INT's National Advisory Committee has traditionally had one or more members from countries outside the United States, and the INT programs enjoy heavy participation from international visitors. During the four programs running in a recent year, INT hosted 244 international-visitor weeks, for an average of nearly eight international visitors per program week. INT has hosted numerous international visitors on sabbaticals or on fellowships, and it frequently supports postdoctoral fellows from other countries.

The unique RIKEN-BNL Research Center (RBRC) exists to support young scientists in the study of strong interactions. While the center has no permanent positions, it offers two-year postdoctoral and five-year fellow positions in both theory and experiment. In addition, tenure-track joint fellow-assistant professor positions are offered in collaboration with U.S., Canadian, and Japanese universities. RBRC is based at Brookhaven National Laboratory, and its directorate and administrative staff are drawn from Brookhaven and RIKEN. Areas of current interest include studying the spin structure of the proton, QCD on the lattice—including the construction of two world-class specialized computers for the computations—and studies of nuclear matter at high density. RBRC also holds frequent workshops on these issues, which are organized by RBRC fellows, collaborators, and staff, drawing scientists from around the world. Beyond the center's scientific successes, many RBRC-trained scientists now hold tenured or permanent positions in nuclear and particle physics at major institutions, and six fellows have received a DOE Outstanding Junior Investigator Award. A new international collaboration with Japan in the area of physics of exotic nuclei, JUSTIPEN, began operations in 2006. U.S. scientists visiting JUSTIPEN are supported by DOE. Reciprocal visits to the United States by Japanese colleagues are funded through the JSPS in Japan.

*In education,* U.S. universities continue to train outstanding future Ph.D.s from throughout the world. In 2005 and 2006, 169 Ph.D.s, funded in part by DOE and NSF, were awarded in nuclear science. The breakdown of the DOE-supported group includes 117 experimentalists of whom 51% were U.S. citizens, and 52 theoreticians of whom 37% were U.S. citizens. Traditionally three-quarters of the non-U.S. citizens remain in the United States. Adding to this impact is the number of students at universities outside the United States who carry out their research in the United States. For example, of the more than 100 Ph.D. theses produced since 1996 using the Gammasphere detector, 55 were granted from 22 non-U.S. universities. Within major collaborations such as those at RHIC, the breakdown for awarded Ph.D.s is approximately 56% from U.S. universities, 28% from European universities, and 16% from Asian universities. The international student participation is even greater if master's theses are included. Finally, approximately one-third of the permanent hires in nuclear science at U.S. universities and government labs have obtained their Ph.D.s outside the United States.

## U.S. PARTICIPATION AT EXISTING INTERNATIONAL FACILITIES AND EMERGING OPPORTUNITIES

International investment in nuclear science has grown rapidly, particularly in Europe and Japan where multibillion-dollar facilities for nuclear studies have been built or are under construction. These facilities open up new frontiers in nuclear science research and represent great opportunities for the worldwide nuclear physics community. Other countries such as China and Russia are rapidly ramping up their efforts and are carefully seeking targeted areas where they can have significant impact.

International facilities greatly enhance the opportunities for the science described in this Long Range Plan. For example, *in relativistic heavy-ion physics*, the CERN LHC will include heavy-ion running to study the quark-gluon plasma at the highest possible energy—extending the discoveries and studies made at RHIC. At GSI/FAIR, a heavy-ion program will target the quark-gluon system at high baryon density, where it is expected that a certain tri-critical phase-transition point might exist.

*In hadronic physics,* the German electron (or positron)-proton machine, HERA, has been an important facility for the HERMES experiment, which includes significant U.S. participation. And data obtained from the facility continue to produce very important results in the area of nucleon spin physics. The CERN SPS accelerator is used to create high-energy particle beams, which are delivered to the COMPASS spectrometer that is designed to enable a broad range of measurements in hadron spectroscopy and hadron structure. Smaller facilities—e.g., MAMI at Mainz, Germany, and ELSA at Bonn, Germany—have produced excellent data that test chiral dynamics and yield nucleon strangeness form factors.

*The study of atomic nuclei* with rare-isotope beams began at CERN ISOLDE several decades ago. The ISOLDE facility remains world class through recent and planned future upgrades. Today ISAC at TRIUMF (near Vancouver, Canada) is the world-leading facility to provide reaccelerated beams of rare isotopes produced via the isotope separator online (ISOL) production method. The SPIRAL facility at GANIL (France) provides additional ISOL capability, and the upgrade to SPIRAL-II will provide intense reaccelerated beams of rare isotopes by the middle of the next decade. U.S. scientists are performing experiments at ISAC and SPIRAL, and U.S. scientists have expressed interest in using the SPIRAL-II facility. The Rare Isotope Beam Facility (RIBF) at RIKEN is now being commissioned to provide fast and stopped rare-isotope beams. The anticipated user program will have U.S. participants, and U.S. scientists have collaborated in the first science experiment. The GSI/FAIR facility will have a world-leading rare-isotope program, beginning in the middle of the next decade.

*LHC at CERN.* The LHC is expected to turn on in 2008. It will be the highest-energy collider in the world, expanding the energy frontier by almost a factor of nine compared to the Fermilab Tevatron. The main goals of the two major detector collaborations, CMS and ATLAS, are to discover the Higgs and to search for new physics beyond the Standard Model. Complementing the main LHC program is the accelerator's capability to collide heavy ions at energies approximately 30 times higher than RHIC. A special-purpose detector, ALICE, is being constructed for relativistic heavy-ion physics, and both CMS and ATLAS are sufficiently general in their detection capabilities to contribute to the heavy-ion program with only minor additions to their core designs. Participation in all three heavy-ion experiments includes approximately 100 U.S. researchers at this early stage. The worldwide CMS heavy-ion effort is led by a U.S. institution, and U.S. scientists play leading roles in all heavy-ion programs.

The LHC plans to run eight months of proton-proton collisions and one month of lead-lead collisions per year. The LHC heavy-ion program will be complementary to that at RHIC. At RHIC, the bulk of the hot, thermalized matter produced in the collision's wake is created at midrapidity, well separated from the forward rapidity region where gluon saturation effects at low momentum fraction in the cold initial nucleus should manifest themselves. At the LHC, this kinematic separation should disappear, so that the combined influences of the predicted quark-gluon plasma (hot) and color-glass-condensate (cold) states will have to be unraveled to interpret measured particle spectra. Since the mechanisms for the unexpectedly rapid thermalization and nearly perfect liquid flow of the matter produced in RHIC collisions are only partially understood, it remains to be seen if similar conditions apply to the even hotter matter that should be produced at the LHC. On the other hand, the much higher energies of the LHC promise substantially higher yields of such rare probes of the hot matter as very hard jets or bound states of heavy quark-antiquark pairs.

*J-PARC, Tokai, Japan.* The construction of the Japan Proton Accelerator Research Complex (J-PARC) is well underway. The facility (figure 3.1) is anchored by a new, world-class 50 GeV proton synchrotron and a 3 GeV intense proton booster ring. Pion, kaon, and antiproton beams—derived from the debris emitted in proton-nuclear target collisions—are planned, and a special energetic neutrino beam will be aimed toward the Super-Kamiokande underground neutrino detector, about 300 km away, for the study of neutrino oscillations. The themes of the research program in high-energy and nuclear physics will include the study of hypernuclear states (nuclei having an implanted strange quark), dense nuclear matter, searches for new exotic particles, and high-energy proton scattering. The particle beams to be used in this research have been largely unavailable since the termination of the Brookhaven National Laboratory Alternating Gradient Synchrotron (AGS) fixed-target program. J-PARC is a major three-community facility, providing particle beams for nuclear physics QCD research and neutrino, muon, and kaon beams for tests of fundamental symmetries and neutrino oscillation studies. Japanese nuclear physicists are world experts in hypernuclear physics. They carried out a limited but very effective program at the AGS before the AGS program was terminated. In recent years they have worked at JLAB, bringing their sophisticated detectors to bear on a wide variety of physics problems involving strange quarks.

*GSI/FAIR: Facility for Antiproton and Ion Research, Darmstadt, Germany.* The GSI Laboratory in Darmstadt, Germany, has been the site of a world-class heavy-ion physics program devoted to the study of compressed nuclear matter and nuclear structure. The laboratory is known for the discovery of some of the heaviest elements in the periodic chart. GSI is currently investing in a significant upgrade of their accelerator facilities in a complex of cooler storage rings known as FAIR (figure 3.2). It will have substantial programs in the study of QCD using antiprotons; the study of dense, heated nuclear matter using relativistic heavy ions; tests of quantum electrodynamics using highly stripped atoms; study of the characteristics of dense, hot plasmas driven by heavy-ion beams; and research using beams of rare isotopes. The FAIR facility will produce stored beams of cooled antiprotons, giving new life to a physics program pioneered at the CERN Low-Energy Anti-proton Ring (LEAR), which was closed in the mid- 1990s, and the FNAL antiproton accumulator, which hosted a high-precision study of charm mesons. The focus of the FAIR antiproton-based physics program is the production of ordinary and exotic hadrons containing charm quarks. It is complementary to the GlueX program designed for the 12 GeV CEBAF Upgrade, which will allow the study of exotic hadrons with the light up, down, and strange quarks. The new facility also will produce fast beams of rare isotopes by a technique known as in-flight separation, where nuclear fragments are collected and focused to a beam using a large Fragment Recoil Separator (Super-FRS). The secondary beams at GSI will be produced by fragmentation of heavy projectiles, up to uranium, and collected in the Super-FRS fragment separator. The separated, fast, rare isotopes can be studied directly following the Super-FRS or captured in storage rings to react with internal targets or collide with electron beams.

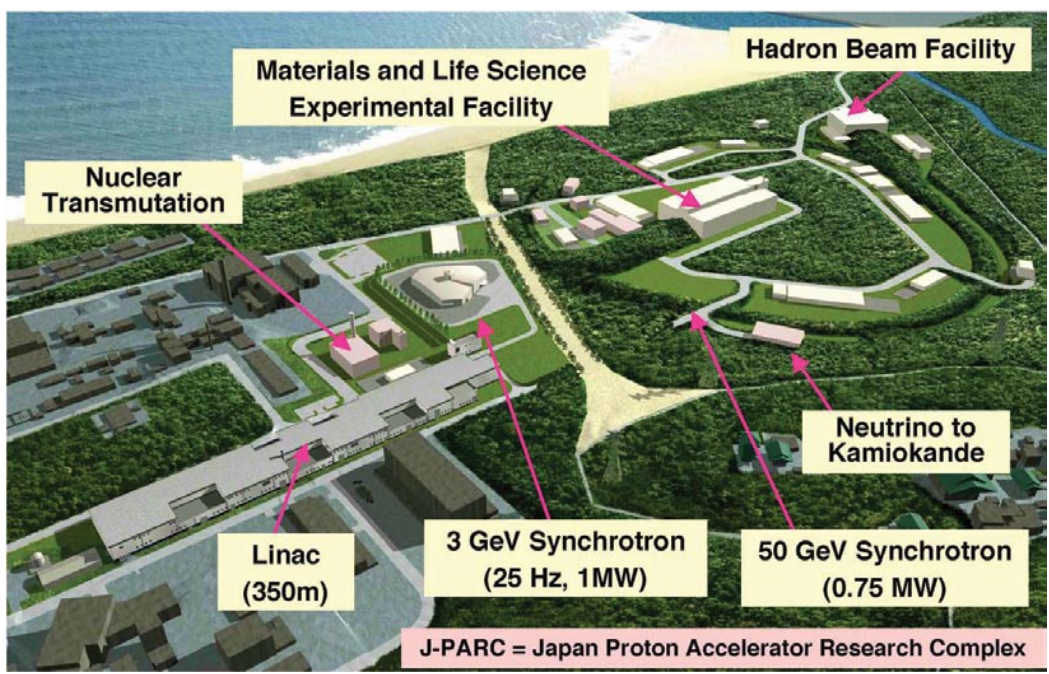

**Figure 3.1:** An artist's rendition of the J-PARC facility being built in Japan at a cost of about $1.5 billion. The facility will provide proton, pion, kaon, antiproton, and neutrino beams for experiments in nuclear science.

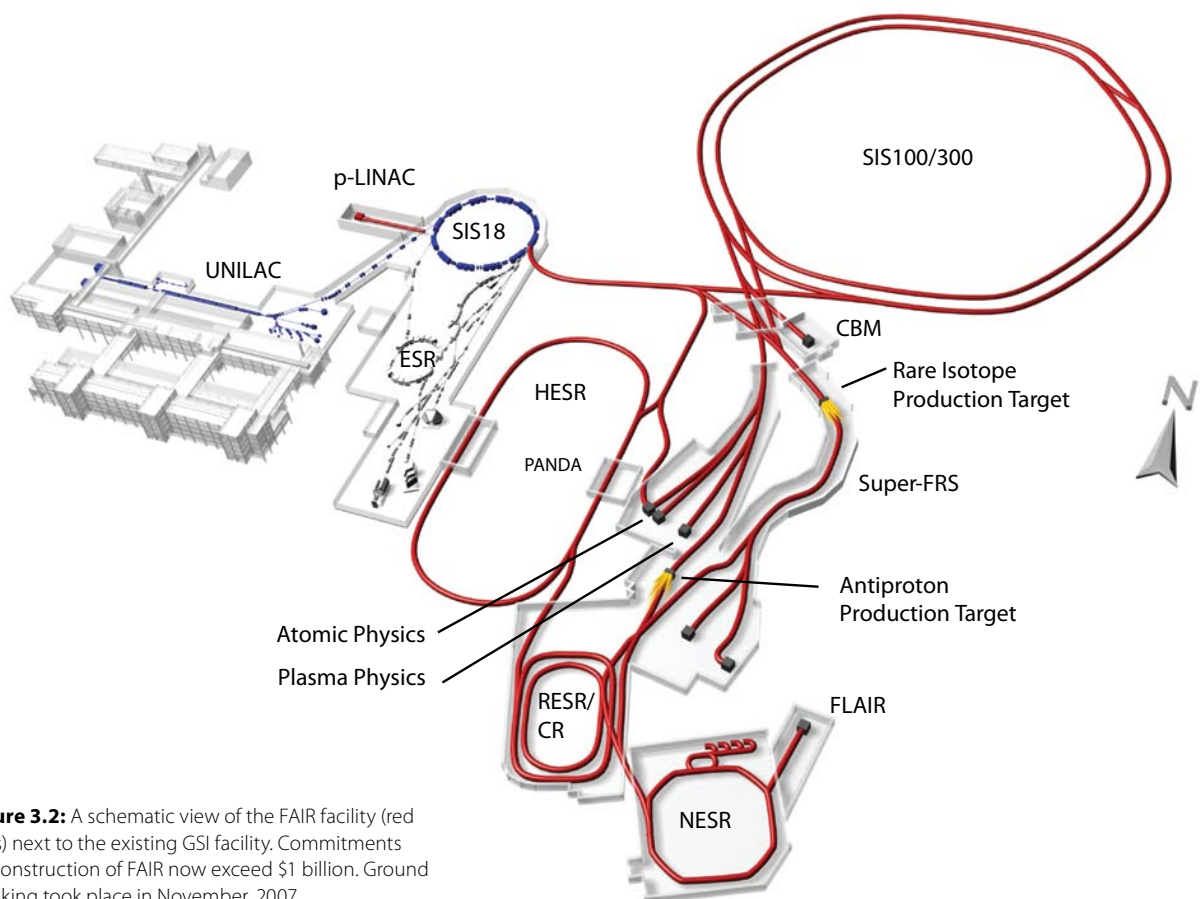

**Figure 3.2:** A schematic view of the FAIR facility (red lines) next to the existing GSI facility. Commitments for construction of FAIR now exceed $1 billion. Ground breaking took place in November, 2007.

*In the area of neutrino physics,* extraordinary discoveries have been made in recent years. SNO, a heavy-water-based solar neutrino experiment, operated 6800 feet underground in a working nickel mine in Sudbury, Ontario. The collaboration is comprised of nuclear physicists from Canada, the United Kingdom, and the United States. The experiment has had a transformative impact on our understanding of neutrinos. Continued involvement by some U.S. groups is expected in a redesigned scintillator-based experiment known as SNO+ that is proposed to be constructed in the existing SNO cavity, using a substantial amount of the original SNO hardware. The KamLAND detector, located underground at Kamoika, Japan, observed the energy-dependent disappearance of reactor antineutrinos over a baseline of 185 km. The Japanese–U.S. collaboration is currently in the process of transforming the detector into a low-energy solar neutrino experiment, capable of observing low-energy $^7$Be neutrinos.

Borexino, a low-background liquid-scintillator experiment, has just started operating at the Gran Sasso Laboratory in Italy. The Italian–French–German–Polish–Russian–U.S. collaboration has made the first real-time observation of $^7$Be solar neutrinos. Based on the initial low backgrounds observed in the detector, the experiment may be able to observe the pep solar neutrinos in addition to searching for geoneutrinos. A reactor-based neutrino oscillation experiment at Daya Bay, China, is an ideally designed neutrino oscillation experiment aimed at measuring the $\theta_{13}$ mixing angle. The Chinese–U.S. collaboration plans to build a number of independent gadolinium-loaded scintillator detectors that will be placed in several different sites near the reactors. The collaboration expects to reach a sensitivity in the measurement to $\theta_{13}$ of ±3°. The large U.S. component of the collaboration includes several nuclear physics groups having key expertise and experience from the KamLAND and SNO experiments. The Karlsruhe Tritium Neutrino (KATRIN) experiment in Karlsruhe, Germany, will search directly for neutrino mass by making precision measurements of electrons from the nuclear beta decay of tritium. The experiment is being constructed at the Forschungszentrum Karlsruhe, which uniquely can handle large amounts of tritium. U.S. nuclear physicists are providing the detector and data acquisition systems. The experiment should have sensitivity to discover a 0.35 eV mass at the five standard deviation significance, or rule out a mass above 0.2 eV at the 90% confidence level. The CUORE neutrinoless double beta decay experiment is located in the underground Gran Sasso Laboratory in Italy. CUORE is an Italian–Spanish–U.S. based collaboration that plans to build a large cryogenic bolometer operating at 10 mK in a dilution refrigerator to search for neutrinoless double beta decays of $^{130}$Te in crystals of $TeO_2$. The United States has committed to provide significant construction funding. The Majorana collaboration is performing research and development aimed at demonstrating the technical feasibility of developing a 1-ton-scale $^{76}$Ge neutrinoless double-beta decay experiment. This Canadian–Japanese–Russian–U.S. collaboration expects to construct a 60 kg demonstration module at the DUSEL in the United States.

*In the area of fundamental symmetry studies,* the Paul Scherrer Institut (PSI) cyclotron at Villigen, Switzerland, is a proton accelerator having a high beam current. At more than 1 MW, it is the highest beam-power accelerator in the world. The proton beam will be used to make an ultra-cold neutron source for use in a new neutron electric dipole moment (EDM) search. The nuclear and particle physics experiments currently hosted by the facility include high-precision efforts in pion decay, muon capture on the proton and deuteron, the positive muon lifetime, and a sensitive search for lepton flavor violation in $\mu \rightarrow e\gamma$ decay. The first three of these experiments are led and largely dominated by U.S. university groups. A 58 MW reactor at the Institut Max von Laue-Paul Langevin (ILL) in Grenoble, France, is used for studies of neutron beta decay and for a series of experiments that have set the present world's most sensitive limit on the EDM of the neutron. The program of fundamental physics measurements at ILL will continue to be competitive with experiments in neutron beta decay and the neutron EDM being mounted in the United States.

# 4 Education: Training the Next Generation

## INTRODUCTION

Education is central to any vision for the future of nuclear science, going hand in hand with the research enterprise. The next generation of facilities and experiments is critical to the health of our discipline, as outlined in the preceding chapters. But so is the next-generation workforce: the talented, dedicated scientists who will run the facilities, perform the experiments, orchestrate the discoveries, and provide a continuing flow of new ideas and challenges. As we look ahead, we must therefore put a high priority on the training of these young scientists.

And nuclear science eduction should not be just for the professionals. In the modern era, the physics of the nucleus impacts everyone. It is more important than ever that the public be informed about science and technology in general, and nuclear technology in particular—both its risks and its opportunities. As more lives are saved through nuclear medicine, for example, more people are affected by the biological effects of ionizing radiation; it is only with a basic understanding that they can make an informed risk-benefit analysis. Meanwhile, as we miniaturize electronics, we are encountering problems due to the effects of ionizing radiation on materials; planners in both the public and private sectors now have to consider the impacts of cosmic-ray damage to communications, weather, or defense satellites. As global climate change becomes a more and more compelling reality, everyone has to consider whether we should use more nuclear power to replace power-generating plants that use fossil fuels. One cannot expect the public to make informed decisions about such pivitol issues—or to accept the decisions made by public officials—without a basic knowledge of the science of the nucleus. In order to improve this very critical aspect of science literacy, we need nuclear scientists in the universities teaching the undergraduates, the graduate students, and the K–12 teachers of tomorrow. As importantly, the full nuclear science workforce needs to participate in reaching out to the general public.

In fact, nuclear scientists have long recognized this need. Even as they have served as the advisors and mentors of the next generation of professionals, they have been highly engaged in the effort to educate teachers and the general public. Moreover, their many efforts to further nuclear science literacy—and their unquestioned acceptance of its importance—are having a profound impact. These activities establish clear evidence that an individual's efforts can make a difference. They also serve as models for new strategies to enhance education in nuclear science, as well as to effectively address outstanding problems for the future.

## RECENT ACHIEVEMENTS

The foundation for the future rests heavily on continuing the outstanding educational opportunities that inspire the next generation of nuclear scientists and on the research at state-of-the-art facilities that drives that inspiration. The many educational activities conducted by university and national laboratory groups within the nuclear science community have been compiled and published on the American Physical Society's Division of Nuclear Physics (DNP) website at http://www.dnp.aps.org/stuinfo/index.html. The programs on that list span the full spectrum from graduate and undergraduate education, to K–12 schools, to reaching out to the general public.

### Graduate education

Since the last Long Range Plan was published in 2002, investment in research and graduate education has led to over 500 new Ph.D.s. The education of these young scientists is not only a measure of success for the field, but an important contributor toward addressing national needs. Based on a survey of Ph.D.s granted in nuclear science "5–10 years out" in 2004, over 40% of scientists who receive Ph.D. degrees eventually work in professions outside the field of nuclear science, as indicated in figure 4.1, contributing broadly to the nation's scientific and technical needs (J. Cerny, *Education in Nuclear Science, A Status Report and Recommendations for the Beginning of the 21st Century*, A Report of NSAC is available at http://www.sc.doe.gov/np/nsac/nsac.html). From a more recent survey of those who received their Ph.D. degree in 2006, almost 30% have already begun a career outside nuclear science basic research, contributing in areas such as medical physics, education, business, government, and industry.

Continued success in the future requires both outstanding education opportunities and a continuing supply of "raw material" for this scientific and technical wellspring. The latter requires reaching out to a diverse population of talented young students—and their teachers—who may not be aware of the rewarding career opportunities for nuclear scientists.

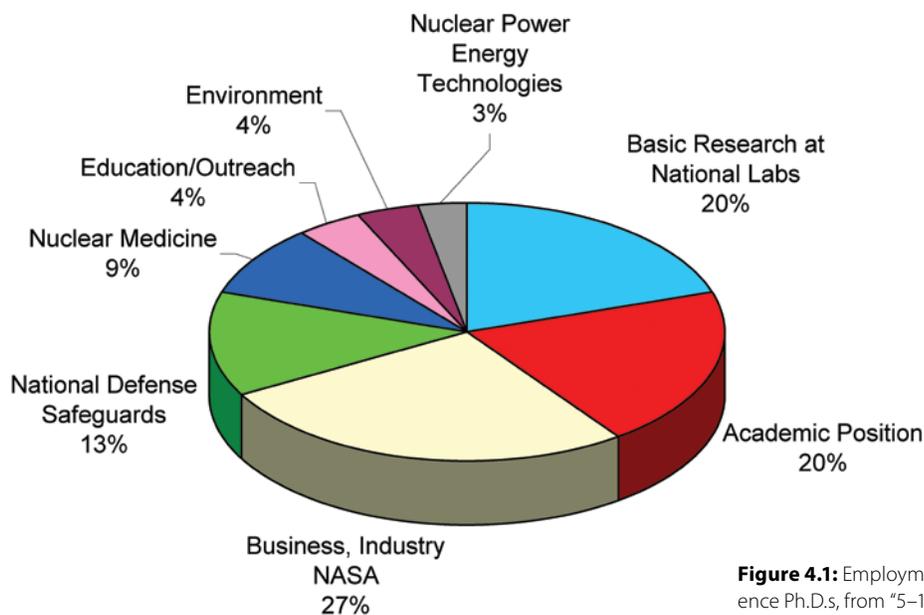

**Figure 4.1:** Employment of nuclear science Ph.D.s, from "5–10 year out" survey in the NSAC Education Cerny Report.

As the following examples show, significant contributions are being made in this direction as well.

### Involvement of Undergraduates in Research

In the past decade nearly a thousand undergraduates have been exposed to nuclear science through active involvement in nuclear science research projects. Currently there are NSF-funded REU sites at 22 universities each of which includes a nuclear science training option. These programs typically engage students from institutions that do not have nuclear science programs, allowing them to do forefront research projects during the summer months. At institutions that do have nuclear science programs, even more students participate in research year round. As an example, the MoNA collaboration, which was created since the last Long Range Plan, includes nuclear scientists at 11 predominately undergraduate institutions and Michigan State University. Together the collaboration has built a state-of-the-art neutron detector and is using it to carry out experiments at the NSCL. The construction of the detector involved a significant number of undergraduate students, and undergraduates continue to be involved in the research being carried out with it. Additionally there are programs at the national laboratories that engage undergraduates in research. Many of these students present their research at the annual Conference Experience for Undergraduates sponsored by the Division of Nuclear Physics. In 2007, over 91 students participated in the conference, bringing the number of undergraduate student participants to over 750 during its first 10 years.

### Undergraduate Curriculum Innovations

Novel courses have been designed that incorporate nuclear science concepts into courses for non-STEM (science, technology, engineering, and math) majors. For example, Clark University designed a course, CHEM007, "Science of Weapons of Mass Destruction (WMD)," for just this purpose. Here, the science behind WMD is discussed in a small seminar environment, and students perform laboratory experiments illustrating the concepts each week. Approximately one-third of the course deals with nuclear science. In the laboratory, students learn to use absorbers to identify different types of radiation, and engage in a "hunt for contraband nuclear material" in which they discover and characterize hidden sources in a simulated homeland security exercise.

### Teacher Training Programs

In many locations, Research Experience for Teachers programs have been developed along with the Research

Experience for Undergraduates programs. Additionally there are teacher programs at some of the national research laboratories, such as the Teacher Academy in Physical Science (TAPS) at Jefferson Lab, which holds a four-week summer classroom and research program for upper elementary and middle school teachers that is designed to build teachers' skills in the physical sciences.

These types of activities have a profound impact. More than a decade ago, a grass-roots local effort—nurtured into a national effort with coordination and funding from the APS Division of Nuclear Physics, DOE, NSF, and the Contemporary Physics Education Project—successfully culminated in publication of the Nuclear Science Wallchart and Teachers' Guide. The Nuclear Science Wallchart is now familiar to educators and students throughout the world, and the Teachers' Guide is in use by teachers and faculty from middle schools to universities to introduce nuclear concepts. This example serves as a success story for the field and compelling evidence of what a targeted national effort can accomplish.

### Direct Involvement of Young Students and the General Public

Many members of the nuclear science community help engage K–12 students in science programs through a broad range of activities, including hosting students as visitors to laboratories and open houses, organizing and promoting science fairs and physics competitions, holding Saturday morning physics sessions, and engaging students in research (e.g., by placing cosmic-ray detectors on high school roofs). National laboratories and major research centers host annual open houses that typically encompass a broad range of science and engineering programs with nuclear science as a component. These programs, which are aimed at the general public, have been an outstanding success. They have become essential elements of the educational effort in the field of physical sciences. For groups funded by NSF, such efforts are encouraged through the broader impact criteria associated with programmatic grants. Within DOE, the national laboratories' education centers have significant support from the Office of Workforce Development within the department's Office of Science. Beyond dedicated resources, however, the success of these efforts relies heavily on leveraging the dedication and commitment of a growing number of nuclear scientists who donate time and effort outside of their own research activities in order to articulate the excitement and importance of nuclear science to a diverse community of stakeholders.

### DOE, NNSA, and DHS Programs

Nuclear scientists have begun working with agencies that often require U.S. citizenship for employment, such as the Department of Homeland Security (DHS) and DOE's National Nuclear Security Administration (NNSA), on new methods of "growing" the next generation of trained workers. Both DHS and NNSA are funding students and university groups to work on specific topics of importance to their mission with the underlying goal of training U.S. students who will ultimately enter the future nuclear science workforce. In addition the Stewardship Science Academic Alliance (SSAA) program, an initiative that provides grants to universities in order to address specific basic science research relevant for the stockpile stewardship program, is currently in its fourth year. The Center of Excellence for Radioactive Ion Beam Studies for Stewardship Science, a Rutgers/ORNL/ORAU/Tennessee/Tennessee Tech/MSU collaboration, is an example of how universities can work with an agency such as NNSA. The SSAA program has been enthusiastically embraced by the young scientists involved and is considered a success by the NNSA defense program. The Department of Homeland Security may follow this model for a new program that would be directed toward nuclear science and engineering, primarily in the area of instrumentation.

### Education Self-Study

At the request of DOE and NSF, NSAC appointed an Education Subcommittee in 2003, chaired by Professor J. Cerny from the University of California–Berkeley, to study, in depth, the issues associated with education and the workforce in nuclear science. The final report of the subcommittee was submitted in 2004. As part of this study, surveys were conducted of current undergraduate and graduate students, postdoctoral fellows, and scientists 5–10 years after receiving their Ph.D.s in nuclear science. The report documented the status and effectiveness of the present educational activities, and projected workforce needs for the future. Based on its findings, the subcommittee made a number of recommendations in the areas of outreach, production and diversity of nuclear science Ph.D.s, and undergraduate and graduate education. The discussion below leans heavily on that report.

## CHALLENGES AND OPPORTUNITIES FOR THE FUTURE

Continued success in the future for our field depends critically on maintaining the commitment to education and striving to further improve our educational activities. While much of the education enterprise is healthy, there are issues that merit special attention, including a decline in the number of new Ph.D.s—particularly U.S. citizens—a lack of diversity in the nuclear science workforce, and a mixed public perception of nuclear science. The first two of these issues are common throughout the physical sciences. Indeed, the challenge of sustaining a trained scientific and technological workforce adequate to the nation's research and technological needs reflects, in part, the downward trend in Ph.D. production in all STEM disciplines in the United States. This trend, which is well documented, threatens U.S. economic competitiveness and scientific and technological leadership.

### The Future Workforce

A significant challenge exists in providing the nuclear science workforce that the nation will need. An accurate assessment of the status of this workforce requires demographic data at all levels—faculty, career research scientist, postdoctoral fellow, and graduate student. It is also important to know the sources and magnitude of the flow of people into and out of the nuclear science "pipeline." Data from the survey of earned doctorates indicate that our field has had a decrease of about 20% in nuclear science Ph.D.s in the past 10 years compared to the previous 10 year period. Tracking Ph.D. production in the near future will be very important to determine if this downturn continues. Certainly a persistent decline at this rate would jeopardize our ability to fulfill the needs projected for the future nuclear science workforce.

The average age of the permanent nuclear science workforce is a genuine concern. Scientists who began careers as faculty or national laboratory researchers during the growth years of the 1950s and 1960s have retired or are nearing that

---

## Training the Next Generation to Meet the Nation's Challenges: Stockpile Stewardship Academic Alliance and Low-Energy Nuclear Physics

The National Nuclear Security Administration within DOE is committed to recruiting a highly talented workforce ready to solve the most challenging scientific problems associated with ensuring the safety, reliability, and performance of the nuclear stockpile in the absence of nuclear weapons testing. To realize this goal, the NNSA initiated the Stockpile Stewardship Academic Alliance program to support the U.S. scientific community in the critical areas of low-energy nuclear science, materials science, and high-energy density physics.

Founded in 2002, the SSAA has supported numerous projects in low-energy nuclear science, from the theory of fission to measurements of neutron-induced

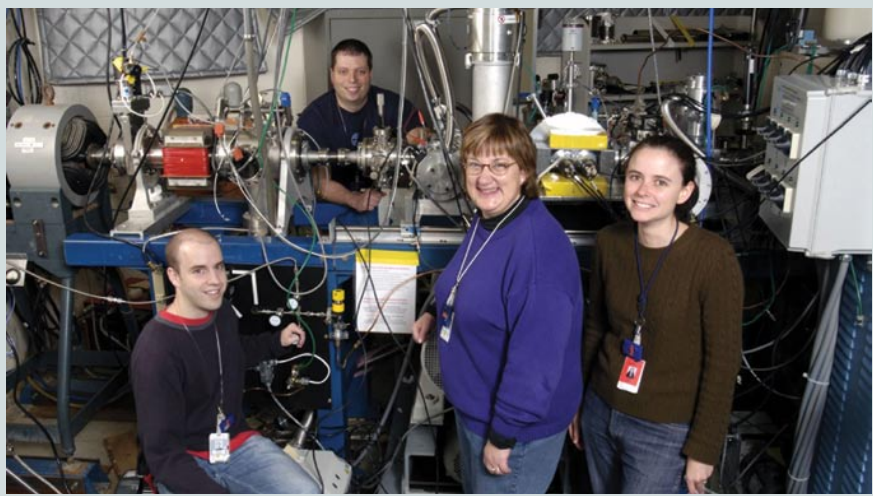

Some of the original participants in the Center of Excellence for Radioactive Ion Beam Studies for Stewardship Science. From left: Jeff Thomas (Rutgers, graduate student), Micah Johnson (ORAU, postdoctoral fellow), Jolie Cizewski (Rutgers, PI), and Kate Jones (Rutgers, postdoctoral fellow). Today Kate Jones is an assistant professor at the University of Tennessee, continuing to do research with the Stewardship Center, and Micah Johnson is a staff member in N-Division at LLNL.

milestone. Those who were students during that period are now relatively mature in their careers. Increased future needs projected by the energy and national security sectors make this problem even more urgent. If the downward trend in Ph.D. production were to continue, the workforce in nuclear science would likely be insufficient in the future to meet the nation's needs in these areas. This is particularly true with respect to the number of U.S. citizens who receive nuclear science Ph.D.s. In 2005, 40% of U.S. graduate students in all physical sciences were foreign nationals. The percentage for nuclear science was comparable. The United States has traditionally attracted the best and brightest scientists from around the world. To sustain U.S. scientific and technological leadership, this tradition must continue. However, the production of U.S.-born nuclear science Ph.D.s needs to remain strong as well, particularly in view of the increased demand in sectors where only U.S. citizens are eligible for employment.

Several recent reports that have examined the future demand for nuclear scientists and engineers suggest increases in Ph.D. production are needed. Precise data are difficult to extract due to the broad spectrum of occupations where nuclear scientists are needed and the uneven level of reporting by universities, commercial concerns, and federal agencies. Even so, all indications are that there will be an increased need for trained nuclear scientists in the future, particularly in fields related to national security. Taking into account retirement of present faculty members, hiring trends at national laboratories, and increased need in areas such as homeland security and nuclear energy, the Cerny report concluded that a 20% increase in annual Ph.D. production was needed. This number may be even higher now that nuclear energy is being formally revisited as a central part of the nation's strategy for a secure energy future.

An opportunity exists in that the number of students receiving bachelor's degrees in physics has begun to increase in the last two years. This, coupled with the exciting opportunities in nuclear science, suggests that the number of Ph.D.s in our field will begin to rise. Support for the research reactions on rare actinides and fission fragments. The Center of Excellence for Radioactive Ion Beam Studies for Stewardship Science is an example of one of these projects. This SSAA—a consortium of nuclear scientists led by Professor Jolie Cizewski of Rutgers University—is developing and using radioactive beams of fission fragments for reaction and structure studies. They are also constructing the Oak Ridge Rutgers University Barrel Array (ORRUBA) of position-sensitive silicon strip detectors. The center provides direct support for undergraduate and graduate students and postdoctoral scholars, and facilitates the participation of other students and postdoctoral scholars. As of fall 2007, 22 undergraduate students, 25 graduate students, and 16 postdoctoral fellows from 19 universities have participated in the activities of this center. All participants are encouraged to attend the annual symposia of the SSAA, as well as play visible roles in the annual review of the center's activities led by scientists from NNSA.

A key component of the center is for all participants to travel annually to either Lawrence Livermore or Los Alamos National Laboratories for a workshop. At these one-day workshops, the graduate students and postdoctoral fellows associated with the center make presentations on their research and have the opportunity to hear firsthand about the challenging opportunities in basic and applied nuclear science. The most successful activity is the poster session where the scientists from these applied science laboratories can personally meet the students and postdoctoral fellows, a first step to recruiting them to future positions at these labs. To date two former postdoctoral fellows associated with the center are doing research at Livermore National Laboratory, with others considering such opportunities as they complete their studies and research.

program as well as the construction of new facilities outlined in other chapters of this Long Range Plan will continue to make nuclear science an attractive choice for these students. Continued improvements of the undergraduate education in nuclear science with regard to both research and course work will also be important to capitalizing on the increasing number of students receiving bachelor's degrees in physics.

**Improving the Undergraduate Experience.** To attract significant numbers of talented new students into nuclear science programs, especially U.S. citizens, it is necessary to recruit from a broad cross section of the student population. Undergraduates are the wellspring of the pipeline. Continued emphasis on increasing the visibility of nuclear science in undergraduate education and the involvement of undergraduates in research will leverage the resources of the community by building on existing programs and the work of college and university departments, national laboratories, and individuals. Introducing the basic principles of nuclear science, and providing advanced courses and laboratories to physics and chemistry students as part of a basic curriculum, has the collateral benefit of exposing more future teachers, engineers, and other professionals to nuclear science, its broad applicability to careers at all levels of schooling, and its value to society.

A three-pronged approach is best suited to address the most important elements:

- Engage undergraduates in **research**. It has been demonstrated that undergraduates who have been involved in research go on to pursue graduate degrees at a higher rate, and are more likely to remain active in research in their professional careers. The early exposure to research also has been shown to decrease the time a student spends in graduate school.
- Ensure that the **education** process exposes undergraduate physics majors to nuclear science concepts as early and as often as possible.
- Ensure the **visibility** of nuclear science for as many undergraduates as possible, in both STEM and non-STEM courses of study.

Undergraduate research is a proven, effective strategy to engage students early in their education. It is also an area in which nuclear science has existing programs and strengths. The graduate student survey in the Cerny report docu-

## Undergraduate Research Comes Full Circle

In the early 1990s Saskia Mioduszewski was introduced to experimental nuclear physics when she participated in undergraduate research programs, first at the University of Tennessee (through the Science Alliance) and subsequently at TUNL (REU program at NCSU). Saskia went on to earn a Ph.D. in nuclear physics at the University of Tennessee and continued her explorations of relativistic heavy ion collisions at BNL. There she established herself as a leader in the RHIC physics community.

Recently Texas A&M University was able to capitalize on a two-body opportunity and attract Dr. Mioduszewski into the physics department as an assistant professor. Texas A&M had hired Ralf Rapp—an outstanding nuclear theorist and her husband—two years earlier. In her position at Texas A&M she is not only able to do cutting-edge science that she enjoys, but also is educating the next generation of nuclear scientists both as a mentor to graduate students and REU students.

Although Dr. Mioduszewski aimed to study science and mathematics, she states that "without those undergraduate research experiences I would not have imagined myself in a career in experimental nuclear physics."

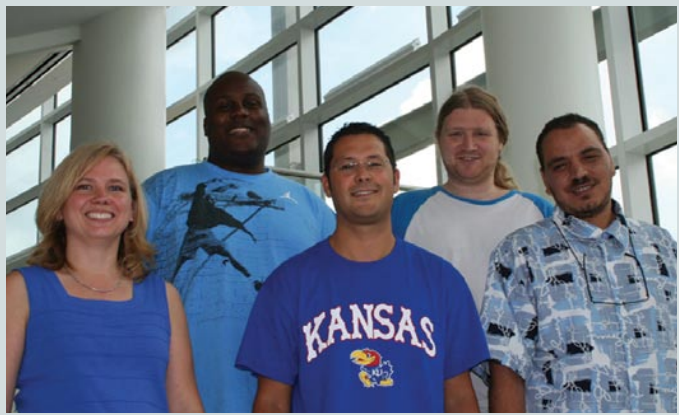

Dr. Saskia Mioduszewski with her current research group members: (from left to right) Saskia Mioduszewski, Martin Codrington (graduate student), Matthew Cervantes (graduate student), Rory Clarke (postdoctoral fellow), and Ahmed Hamed (postdoctoral fellow).

mented that 92% of the female and 88% of the male graduate students surveyed had engaged in undergraduate research experiences. An integrated approach is called for to capitalize on the many exciting opportunities available for undergraduates through the nuclear science research program. Such an approach can identify promising students early in their education, provide them with opportunities throughout their undergraduate experience, and help them with the transition to graduate school. Expanding summer research experiences for undergraduates that leverage existing programs at college and university departments and national laboratories would have a clear positive impact. Integrating such programs at home institutions with conference experiences and a coordinated approach to mentoring within the community would facilitate the transition to graduate school. Pilot efforts already underway in this area are having a very positive impact.

The number of undergraduate courses offered in nuclear physics and chemistry in the United States is decreasing, leaving students who do not have access to such courses largely ignorant of the field. To reach the undergraduates who may become future nuclear scientists, physics teachers, and other science and technology professionals, nuclear concepts must be retained in both modern physics and general physics survey courses. Providing an exposure to nuclear science at schools that have no nuclear faculty is an even bigger challenge. A distinguished lecturer program—a proven success in the field of plasma physics—is one possible approach to increase visibility within physics or chemistry departments at schools with no nuclear science. Another proven strategy for introducing nuclear science concepts to nonscience majors is to develop imaginative courses that satisfy general science requirements as well as provide an introduction to nuclear science concepts.

**The Issue of Diversity.** Diversity of the workforce is another concern for all of the physical sciences. Improving diversity in nuclear science is a complex, challenging problem. Statistics on diversity were gathered as part of the Cerny report as well as for the 92 new Ph.D.s in the 2006 DOE Workforce Survey. These two sets of data are compared in table 4.1, which shows that progress has indeed been made in improving the percentage of women earning Ph.D.s in nuclear science. However, for women in the science and engineering workforce at all levels, work-life balance is a particularly challenging problem. An important issue for the future will be to develop strategies that mitigate the differences in the hurdles faced by men and women who pursue a career in nuclear science. Clearly changes must be made at an institutional level to address this.

The lack of representation of ethnic minorities in the physical sciences is often viewed as a pipeline issue, with a key factor thought to be the lack of exposure and opportunity in the K–12 education process. This is changing, however. Over the past decade, there has been an increase in the participation of women and minorities taking physics in high school, with the percentage of women now at 50% and the number of ethnic minorities increased substantially. The challenge remains to convince this increasingly diverse pool of students, who have been introduced to physics, to consider nuclear science as an exciting and rewarding career choice.

To improve diversity in nuclear science the problem must be attacked at all levels of the pipeline and by all stakehold-

*Table 4.1:* The gender and ethnicity background of U.S. citizens who received nuclear science Ph.D.s. The left column of data is taken from the 2003 NSAC Education Survey of nuclear scientists who had received their Ph.D.s in the period 1993–1998. The right column is data from the 2006 DOE Workforce Survey, for those receiving Ph.D.s in nuclear science in 2005–2006, supported by DOE.

| Source | NSAC education report surveying Ph.D.s 5–10 years prior to 2003 | 2006 DOE Workforce Survey of new Ph.D.s |
|---|---|---|
| **Gender** | U.S. Citizens ||
| Male | 88% | 81% |
| Female | 12% | 19% |
| **Race or Ethnicity** | | |
| Caucasian | 90% | 90% |
| Asian or Asian American | 1.2% | 5% |
| Black or African American | 1.2% | 1% |
| American Indian or Alaskan native | 1.2% | 0% |
| Hispanic | 0.6% | 4% |
| Mixed race or ethnicity | 6.2% | – |

# The MoNA Collaboration: A Multi-institutional Research Collaboration with Undergraduate Participation at its Core

The Modular Neutron Array, a major experimental device located at the National Superconducting Cyclotron Laboratory, was assembled almost exclusively by undergraduate students at 10 different, primarily undergraduate colleges and universities. Following the completion of the array, the MoNA collaboration has continued as an ongoing, multi-institutional research collaboration that includes undergraduate participation as a central feature. Undergraduate students from the MoNA collaboration institutions have participated at the NSCL in every MoNA experiment to date, and MoNA undergraduates are carrying out essential components of the data analysis from these experiments.

Members of the MoNA collaboration have worked hard and creatively to ensure the continuing participation of faculty and students from the undergraduate institutions. Weekly videoconferences, recently enhanced by new hardware provided by a grant from the state of Michigan, provide an opportunity for discussion of data analysis issues, ideas for future experiments, and other topics. Since one of the videoconference sites in the MoNA data collection area is at the NSCL, students can even carry out online data analysis during experiments without traveling to the NSCL. Email distribution lists, including one restricted to undergraduate students, provide avenues for asynchronous communication. Since 2004, members of the collaboration—faculty, graduate students, and undergraduates—have gathered for an annual collaboration meeting, where papers and experiment proposals are written, data analysis issues are addressed, responsibilities are delegated, and the collaboration itself is renewed and reinvigorated.

This multi-institutional model of undergraduate research participation offers a number of important advantages for nuclear science in particular since most experiments are now performed at national user facilities. Regular contact with students and faculty from other institutions helps MoNA undergraduates to see themselves as part of the nuclear science community, and encourages them to continue on this career path.

From its inception in 2001 through January 2007, 62 undergraduates experienced the thrill of participation in cutting-edge research in nuclear physics through participation in research as part of the MoNA collaboration. Forty-two of these had graduated, and 19 (45%) were in graduate school in physics, with five (12%) of those students having already chosen to pursue nuclear physics. Six more students had gone on to graduate school in chemistry or engineering, and three were preparing for high school teaching careers.

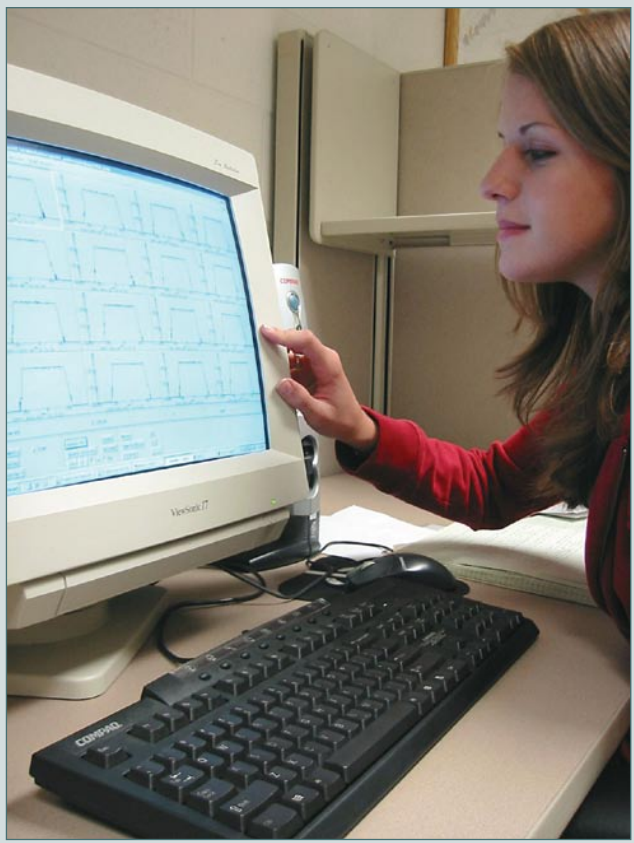

Tina Pike worked on a calibration scheme as a member of the MoNA collaboration while an undergraduate at Hope College. She is now studying medical physics in the graduate program at the University of Wisconsin. Nineteen undergraduates have entered graduate school in physics—five in nuclear physics—after working on MoNA as an undergraduate.

ers. The Cerny report recommended a concerted commitment that would include interventions such as the following:

- enhancing connections with faculty and students of institutions and consortia that serve traditionally underrepresented groups;
- establishing programs to facilitate the transition of early-career scientists into forefront research activities and educational opportunities, e.g., bridge programs between undergraduate and graduate school;
- adopting policies that recognize work-life balance;
- emphasizing the value of recruiting and mentoring members of underrepresented groups;
- enhancing the visibility of underrepresented minorities in the nuclear science community; and
- developing effective models for enhancing the participation of individuals from underrepresented backgrounds and disseminating them via best-practice sessions.

An opportunity exists to expand the talent pool that the field of nuclear physics taps into by building on the activities and talents of some in the nuclear science community that are leading efforts to improve diversity in physics. Nuclear scientists have been active members of the Committee on the Status of Women in Physics (CSWP) and the Committee on Minorities (COM) of the APS. The CSWP sponsors a number of activities including site visits to universities, symposia on women in science, a website for female-friendly physics departments, a compilation of best practices gleaned from the site visits, and most recently a major conference on gender equity in physics. The Committee on Minorities organizes a variety of activities including symposia at APS meetings and scholarships and mentoring for minority undergraduate physics majors. Since the last Long Range Plan, our field has seen the formation of Women Encouraging Competitive Advancement in Nuclear science (WECAN), which has organized events at DNP meetings. Looking toward the future, it is crucial to take advantage of the resources available through the programs developed by these committees. Increased efforts to build and sustain the CSWP, COM, and WECAN will help us to address the issues faced by women and minorities in nuclear science and capitalize on the talents of a broader group of people.

**The Assessment Tools.** One challenge for effective planning for the future of nuclear science is acquiring reliable and complete data on the makeup of the current workforce. Without these data, it is not possible to confidently assess the health of the field or to plan for the future of the field. An ongoing, regular professional assessment of the workforce in nuclear science at all levels needs to be established. Among other statistics, data on citizenship, diversity, and the career paths of Ph.D. scientists need to be tracked. The American Institute of Physics has the experience and tools to perform this task and thus represents a possible resource that could be utilized to meet this challenge. The opportunity exists to design and implement such a process that would track the career paths of nuclear scientists for the future. Having this information available for future planning could prove to be extremely valuable.

**Mentoring.** Enhanced career counseling and mentoring are important to ensure that those who enter the field enjoy a rewarding, fulfilling career. This will not only serve to encourage younger generations considering nuclear science as an option, but it will introduce the next generation of nuclear science Ph.D.s to the full spectrum of attractive careers inside and outside basic nuclear physics research. Regular opportunities for professional development for graduate students and postdoctoral fellows at professional meetings are a key element of this strategy.

### Engaging the Public

To successfully meet the challenges faced by the nation in the future, it is necessary to ensure a broad-based societal knowledge of nuclear science concepts. This will enable informed decisions by individuals and decision-making bodies on a wide range of important topics, including nuclear medicine, energy policy, homeland security, national defense, and the importance and value of nuclear science research. At present, the public, and even scientists in other fields, are often uninformed or misinformed about nuclear science and its benefits. This public perception of nuclear science represents a challenge, the solution to which requires the involvement of nuclear scientists.

To achieve the goal of an informed public familiar with the basic principles and excitement of the subatomic world, creative approaches to outreach are needed that engage the nonscientists intellectually on real-world topics. Training of

# RHIC Communications

**R**HIC has made international headlines since the facility's commissioning in 1999. The very idea of probing the matter that existed during the first microseconds after the Big Bang has sparked the imagination of many. The RHIC story is an excellent example of how new and exciting science can capture public interest if conveyed in an open, comprehensible way.

Conveying this excitement to the public, and to teachers and students at all levels, is critical for the health of nuclear science. Brookhaven National Laboratory's Community, Education, Government, and Public Affairs directorate has worked collaboratively with members of the RHIC community to effectively communicate the excitement of RHIC science, turning controversy over science fiction (the perceived possibility of creating black holes) into positive public anticipation of new scientific discoveries.

The specific audiences targeted for major communications activity include the science community, funding agencies, elected officials, educators and students, the science-attentive public, the general public, and both science and mainstream media. Enhancing the laboratory's image in the local community and promoting science literacy share equal importance with communicating the scientific accomplishments of RHIC to the national and worldwide media.

Over the past decade, numerous studies have pointed to an increasingly urgent need to prepare more U.S. citizens for leadership roles in basic and applied physical sciences. Whether through media reports, summer tours of the collider complex, or numerous other ways of publicizing the machine, the scientists, and the science, keeping RHIC in the spotlight will encourage more students to consider nuclear science as a career choice.

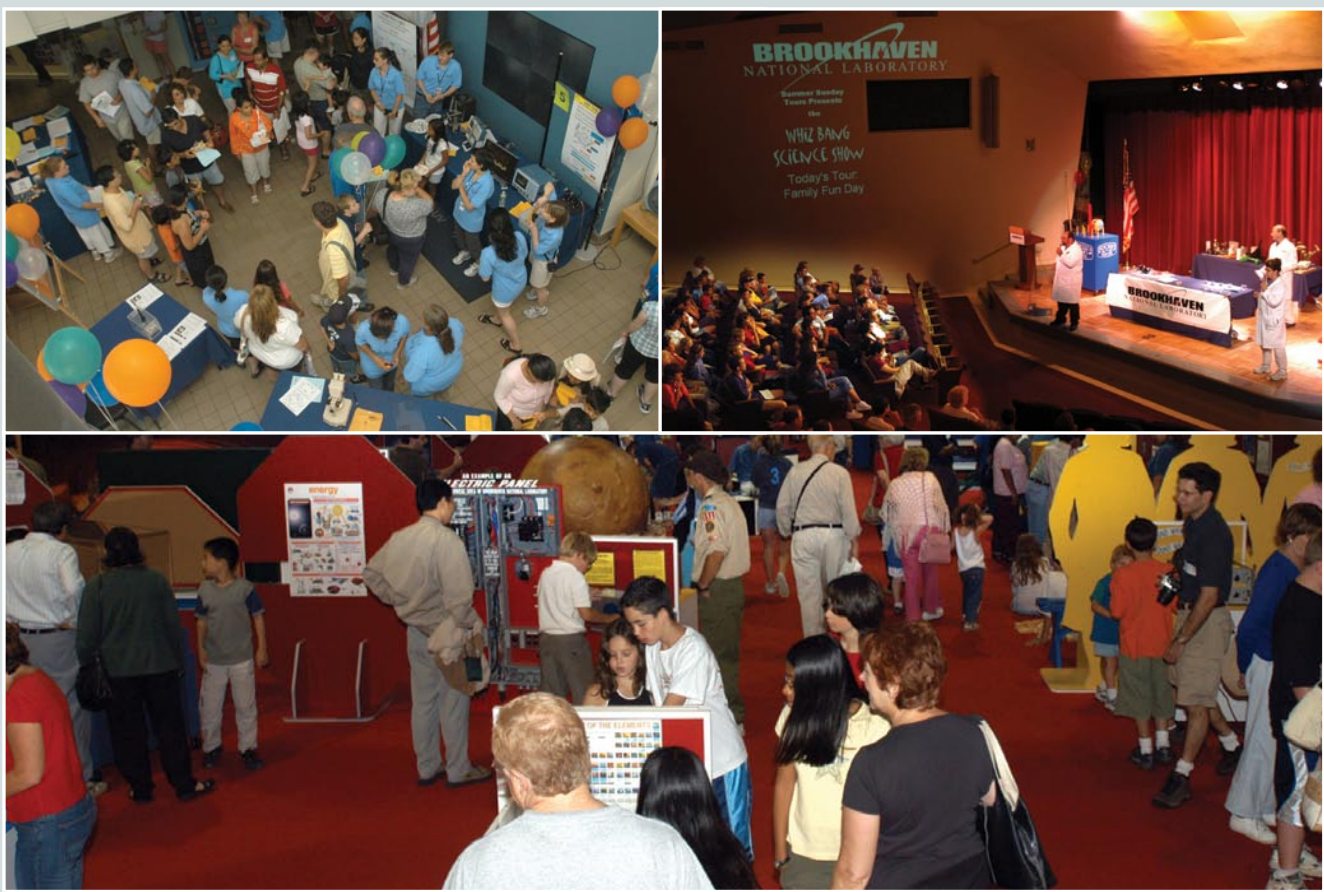

RHIC and BNL Summer Sundays typically attract over 1000 members of the public each August.

high school teachers, and aspiring future teachers, who touch the next generation at a formative stage can make an essential contribution given a basic knowledge of nuclear science concepts, access to nuclear scientists as a resource, and tools for introducing nuclear science in an inquiry-based setting in the modern classroom. Promising high school students can be attracted to participate in research-related activities in order to point them toward the physical sciences as a potential career option, and the curiosity of middle school students can be sparked toward an interest in science during this formative stage of their development.

An opportunity exists to bring together all the various individual efforts to develop a nationally coordinated program of outreach including materials and hands-on activities that illustrate and demonstrate core nuclear science principles to a broad array of audiences. A community-wide effort is needed to provide the public with a clear perception of the value of nuclear science research and its benefits to society. This effort should leverage existing efforts and partner with other communities (such as nuclear energy) to develop and disseminate materials and hands-on activities. The Internet, which has become a significant source of information as well as an inexpensive way of distributing it, can be utilized by developing a nationally coordinated nuclear science website. A nationally coordinated website would serve as a resource center for students, educators, and community members. It would target all levels, from outreach to the general public (e.g., discussions of societal issues) to undergraduate curricula (e.g., sharing of course material). Some aspects of the nationally coordinated outreach program may include the following:

- **Teacher programs.** An additional way to touch the heart and mind of public opinion will be to engage current and future teachers, along with scientists in related research activities, on the development of curricula embodying modern physics concepts. The DOE has recently instituted an Academies Creating Teacher Scientists (ACTS) program at the national laboratories. One example of this is a program being carried out at Jefferson Lab. A partnership between the nuclear science community and education professionals to introduce a component of nuclear science to these programs across the national laboratory system would benefit all stakeholders.
- **Programs for high school students.** Promising high school students should be engaged in discussion as early as possible in order to capture and sustain interest in physics as a possible career choice. This can be accomplished through science camps and research internships. At present, formal programs for high school student research experiences do not exist. However, several laboratories and universities have independently established very successful programs. Individual scientists can and have contributed by mentoring high school students for the Intel competition and other national science fairs. These efforts should be expanded.
- **Public information.** The activities of public affairs offices at national laboratories also provide excellent best-practice models that the nuclear science community can utilize in targeting outreach to specific audiences, including, e.g., the science-attentive public, general public, and both science and mainstream media. An excellent example is the public outreach program at RHIC.
- **Distinguished lecturers.** In addition to its use in sustaining Ph.D. production and improving diversity, a distinguished lecturer series would also provide an effective means of public outreach. Distinguished speakers would be provided with training and access to a library of proven outreach materials, in order to assist them in preparing lectures to inspire the general public.

Successful implementation of outreach programs such as those described here will go a long way toward an enhanced understanding and appreciation of the excitement of nuclear science and its applications in all sectors of society. It will increase the number of teachers incorporating nuclear science into their courses and increase the number of students who are aware of the opportunities for a rewarding career in nuclear science and its applications. From middle school and high school classrooms to the living rooms of everyday households, such activities will have the benefit of enhancing public understanding of nuclear science and its applications and value to society.

## FUTURE ACTIVITIES

While it is necessary to continue to support the diversity of educational programs currently ongoing in the nuclear science community, we believe that there are two particular areas where a concentration of new effort is warranted. In order to address the overall decline in Ph.D. production, the definite drop among U.S. citizens, the lack of diversity in the nuclear science workforce, and the uninformed or misinformed public perception of nuclear science, nuclear scientists are urged to work on the following education and outreach activities:

- **the enhancement of existing programs and the inception of new ones that address the goals of increasing the visibility of nuclear science in undergraduate education and the involvement of undergraduates in research and**

- **the development and dissemination of materials and hands-on activities that demonstrate core nuclear science principles to a broad array of audiences.**

Many possible strategies have been suggested to accomplish these priorities. Ensuring the continued strength of the field and its contributions to society in the future demands that all stakeholders in the future of nuclear science work together to identify strategies that best leverage the strengths of existing and future research programs, as well as the resources to successfully carry them out.

## Academies Creating Teacher Scientists

The Department of Energy Academies Creating Teacher Scientists program is a teacher professional-development program funded by the DOE Office of Science. Each participating teacher makes a commitment to attend a summer program at a DOE national laboratory for three consecutive years. At Jefferson Lab, the ACTS program, serving fifth to eighth grade teachers, is four weeks in duration. It is designed to (1) build teachers' content knowledge and skill base in the physical sciences, (2) equip teachers with more engaging and advanced teaching methods, (3) increase teachers' ability to positively influence students' interest and understanding of the physical sciences, and (4) acknowledge the important role that teachers play in maintaining the educational "pipeline" that develops students with the critical thinking skills needed to solve the nation's future challenges.

Teachers are provided with a rigorous science course (physics in 2007, chemistry in 2008, and geophysics in 2009) taught by an expert teacher from a local high school, lectures and demonstrations from lab scientists, hands-on education workshops and discussions, and a hands-on sampling of the lab's research environment. They spend half of their time working in small groups. All program components address the National Science Education Standards and the Virginia Standards of Learning (SOL).

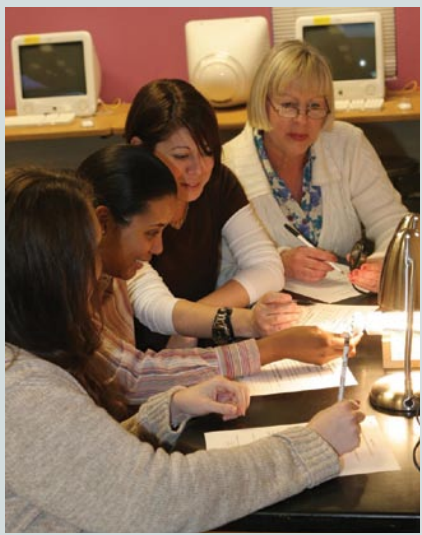

By exposing teachers to the basic research process, Jefferson Lab helps to refine and enhance teacher understanding of basic science inquiry. "I can present this new understanding to middle school students in a manner that gets students invigorated about science as an application to their everyday lives. If science is perceived as relevant to students, then this will help to not only keep their interest, but also serve as a 'seed' for future study and possible career choice," says Carmen Fragapane of Grafton Middle School in Yorktown, Virginia.

The DOE Office of Science provides grants for travel and materials to teachers throughout the three-year program. To address the ACTS leadership goals, all teachers are required to either make a presentation at a regional or national conference or accept a leadership role within their schools during the school year after they complete the three-year program. Through monthly correspondence, Jefferson Lab tracks the teachers' presentations and newly developed leadership roles and helps provide support.

One participant, Christine Ward Diaz, was selected as a recipient of the 2007 Presidential Award for Excellence in Mathematics and Science Teaching. She received her award from President Bush in May 2007 and credited the DOE/Jefferson Lab ACTS program with providing her with critical teaching skills and credentials.

# 5
**The Broader Impacts of Nuclear Science**

# Connections to Other Fields

## INTRODUCTION

The physics of the nucleus is at the heart of our ability to understand the universe, both at the very small scale and the very large. The nucleus emerges from relatively simple interactions between quarks and gluons on a scale of $10^{-15}$ meters. It forms a complex, quantum, many-body system of incredible richness and diversity in its own right. And its properties determine the behavior of red giant stars $10^{11}$ meters in radius. Indeed, nuclear physics encompasses phenomena over an uncommonly wide range—which is why it shares so many themes and challenges with the rest of modern science. Researchers in our discipline are connected through shared concepts, tools, and techniques to high-energy physicists exploring the fundamental building blocks of nature; to the scientists studying complex systems and emergent phenomena in biology, nanoscience, and many other fields; and to the cosmologists and astronomers seeking to understand the origins and workings of the universe itself.

This section explores this plethora of connections between nuclear physics and other sciences, showing how the application of common techniques and technologies can stimulate interdisciplinary contacts and the wider exchange of ideas—and how that cross fertilization, in turn, can lead to more rapid progress and greater vitality on all sides, as well as an improved understanding and appreciation of nuclear science outside the field.

## CONNECTIONS TO THE HIGH-ENERGY PHYSICS AND COSMOLOGY COMMUNITIES

There are many common problems addressed by high-energy and nuclear physicists, who often employ similar theoretical and experimental techniques to do so. These problems range from exploring the nature of the electroweak interaction and physics beyond the Standard Model, to studying how the strong interaction arises from quantum chromodynamics (QCD)—and in particular, investigating how quarks are confined into observable particles.

Nothing better illustrates the synergistic relationship between high-energy and nuclear physics than the recent discovery of neutrino mass and oscillations, the first clear departure from the minimal Standard Model (figure 5.1). In several of the decisive experiments—Super-Kamiokande, Cl-Ar, SAGE, Gallex, SNO, KamLAND, K2K, MINOS, and MiniBooNE—nuclear and high-energy physicists worked together to accomplish a major step forward in physics. Other tests of the Standard Model carried out by nuclear physicists include precise measurements of the neutron beta-decay correlation parameters and lifetime and nuclear superallowed beta decays.

Follow-up studies of even higher precision would be possible at the proposed Deep Underground Science and Engineering Laboratory (DUSEL), which is designed to be a multidisciplinary facility for attacking a wide range of problems in nuclear physics, high-energy physics, astrophysics, cosmology, geosciences, geomicrobiology, and engineering. In the ultra-low-background environment of DUSEL, nuclear and high-energy physicists could collaborate to determine whether neutrinos are their own antiparticles, what their masses are, what dark matter is, and whether the Sun's energy production exactly matches its present-day radiant output. DUSEL would also allow nuclear physicists to engage in many other unusual collaborations. One example is the neutrino tomography of the Earth using geoneutrinos and solar neutrinos. While this concept is still far from realization, it has engaged the interest of two disciplines that had little contact heretofore.

**Figure 5.1:** First step in construction of the New Standard Model. The neutrinos are now identified as states of definite mass, rather than definite flavor. The lower limits for their masses are from oscillation experiments; the upper limits are model-dependent cosmological bounds. Laboratory measurements give an upper limit of 2.3 eV for each neutrino. (Courtesy of Contemporary Physics Education Project. For a discussion of the neutrino mass hierarchy, see "In Search of the New Standard Model.")

| FERMIONS matter constituents spin = 1/2, 3/2, 5/2, ... | | | | | | |
|---|---|---|---|---|---|---|
| **Leptons** spin =1/2 | | | **Quarks** spin =1/2 | | | |
| Flavor | Mass GeV/$c^2$ | Electric charge | Flavor | Approx. Mass GeV/$c^2$ | Electric charge | |
| $\nu_L$ lightest neutrino | $(0-0.13) \times 10^{-9}$ | 0 | u up | 0.002 | 2/3 | |
| e electron | 0.000511 | −1 | d down | 0.005 | −1/3 | |
| $\nu_M$ middle neutrino | $(0.009-0.13) \times 10^{-9}$ | 0 | c charm | 1.3 | 2/3 | |
| $\mu$ muon | 0.106 | −1 | s strange | 0.1 | −1/3 | |
| $\nu_H$ heaviest neutrino | $(0.04-0.14) \times 10^{-9}$ | 0 | t top | 173 | 2/3 | |
| $\tau$ tau | 1.777 | −1 | b bottom | 4.2 | −1/3 | |

Nuclear and particle scientists are poised to make the key contributions leading to a New Standard Model. The search for the long-anticipated Higgs particle at the LHC will be complemented by low-energy studies of flavor physics. Experiments involving lepton number violation in neutrinoless double beta decay may determine the nature of the neutrino and new aspects of CP-symmetry violation. The origin of the matter-antimatter asymmetry may ultimately be revealed through studies of neutrino mass and mixing, the electric dipole moment of the neutron and of atoms, and ultra-rare K- and B-meson decays. Other fundamental symmetries of the New Standard Model may be revealed through precise measurements of the muon's anomalous magnetic moment, parity-violating asymmetries in electron scattering, neutron and nuclear beta decay, and rare muon- and pion-decay modes.

QCD sprang from the Standard Model of particle physics and is believed to be the correct theory of the strong interaction. A non-Abelian gauge theory, it is too difficult for direct calculations as yet because of the difficulty in accounting for interactions among the force carriers. Nuclear theorists

## Taking a Picture of the Earth's Interior with Geoneutrinos

The total heat flow from the Earth is an estimated 40 tera-watts, with a distribution shown in the figure below. Geologists believe that the most significant sources of this heat—and therefore, the likely driving force for plate tectonics, earthquakes, and the geomagnetic field—are the natural decays of uranium and thorium distributed throughout the Earth.

As a step toward confirming this picture, the KamLAND experiment recently published a result claiming the first observation of neutrinos from uranium and thorium decays in the Earth—the so-called geoneutrinos (see figure above). This claim has gotten geologists very excited: because neutrinos pass through the Earth almost completely unhindered, they offer a truly new way to directly probe processes occurring in the depths of our planet. Indeed, geoneutrinos offer the only known method to directly measure the chemical composition at depths greater than a few miles.

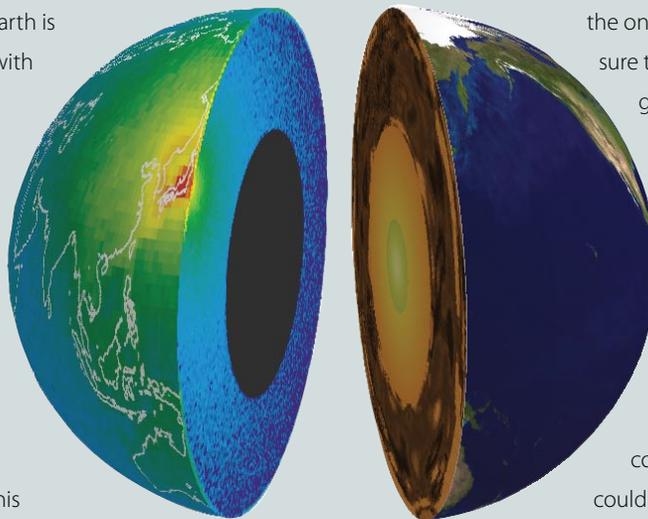

Following up on the results from KamLAND would require a new neutrino experiment located in the Deep Underground Science and Engineering Laboratory, where backgrounds would be extremely low. By precisely measuring the uranium and thorium concentration, such an experiment could revolutionize geologist's understanding of the Earth.

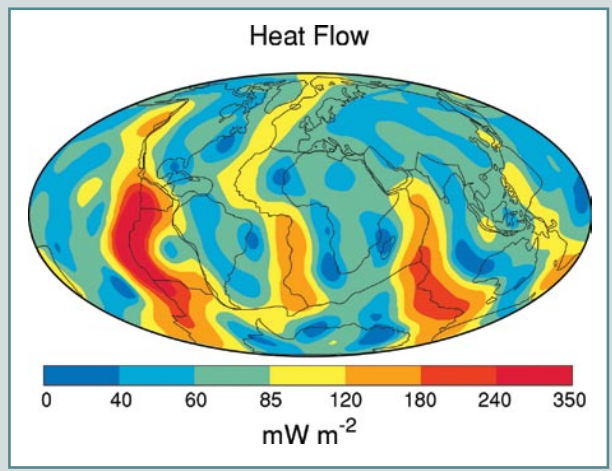

Figures:
Left: Distribution of the Earth's heat flow. (Courtesy of H.N. Pollack et. al, *Reviews of Geophysics* 31, 267 (1993))

Above: Geoneutrinos, anti-electron neutrinos emanating from the Earth, are expected to serve as a unique window into the interior of our planet, revealing information that is hidden from other probes. The left half of this image shows the production distribution for the geoneutrinos detected at KamLAND, and the right half shows the geologic structure. (Courtesy of KamLAND Collaboration: *Nature* 436, 499 (2005)).

have led the way toward practical calculations with QCD by means of lattice gauge theory. Perturbative QCD is another approach that has been successful in the regime of high energies and heavy quarks. A fascinating new connection between nuclear theory and string theory has developed in recent years. In 1998 evidence from string theory suggested that the properties of a strongly coupled quantum field theory similar to QCD could be determined by solving the much simpler classical equations of motion for supergravity in a particular five-dimensional spacetime. This astonishing duality—called the "AdS/CFT correspondence"—has been verified in a number of calculations and makes it possible to indirectly compute observables in the quantum field theory that had previously been incalculable. One of these quantities is the ratio of the shear viscosity to the entropy density for a hot quantum plasma; this calculation has led to the conjecture that the ratio has a minimum allowed value, simply expressed in terms of the Planck and Boltzmann constants. Coupled with experimental evidence suggesting that the quark-gluon plasma produced at RHIC has very low viscosity, there is much interest in the question of how closely the properties of the experimental plasma approximate the ideal system studied through the AdS/CFT correspondence. Currently there is a significant community of string theorists and nuclear theorists collaborating to better understand what other insights might be gleaned from this peculiar and exciting connection between heavy-ion experiments and supergravity in five dimensions.

## CONNECTIONS TO THE ASTROPHYSICS COMMUNITY

The connections between the nuclear physics and astrophysics communities are multifaceted and deep. Indeed, those connections have given rise to the common field of nuclear astrophysics, which investigates stellar evolution, the origin of the elements, big-bang cosmology, solar physics, neutron star and pulsar physics, explosive phenomena such as supernovae, gamma-ray bursts, novae, X-ray bursts, and cosmic-ray physics, among many other phenomena. In order to make progress in addressing these issues, nuclear physicists, astrophysicists, and astronomers must work together.

Beyond the common questions at the interface of nuclear physics and astrophysics, nuclear astrophysics itself has a major impact on a broad range of astrophysics questions.

Large-scale astronomical surveys, together with extensive follow-up studies at the largest Earth- and space-based telescopes, are hugely increasing the sample of metal-poor stars with well-known composition, mapping the chemical enrichment history of the galaxy in unprecedented detail. Almost all of the observed chemical elements were created by nuclear reactions inside stars and during astrophysical explosions; nuclear physics is therefore essential to interpret these data. Once the nuclear physics and astrophysics of nucleosynthesis are understood, they can be used as probes to address broader astrophysical questions. An example is big-bang nucleosynthesis of the light elements, which is used to constrain the baryon content of the universe. The foundation of this understanding is the knowledge of the nuclear reaction rates of light species through beryllium. Another example is the nuclear reactions in the proton-proton chains that power the Sun and produce the neutrinos whose measurement in underground terrestrial detectors revealed that neutrinos oscillate. With the latest measurements of nuclear reaction rates and robust solar models, tight constraints on neutrino properties are now obtained from solar neutrino observations. In the future, nucleosynthesis might emerge as an important tool to disentangle, together with observations, various components of galaxies, addressing fundamental astrophysical questions of structure and galaxy formation.

Radio pulsars, magnetars, X-ray bursters, and the primaries of X-ray binaries are neutron stars and are ubiquitous in the galaxy. Their structure is largely determined by the nuclear force. Satellite and ground-based measurements of their radii, masses, and moments of inertia are of high importance in astrophysics and are also directly relevant to the physics of nuclear matter at high densities. Astronomical observations of radio pulsars and magnetars can also constrain nuclear models of superfluidity in nuclear matter. X-ray bursters are thermonuclear explosions on the surface of neutron stars, and a number of X-ray observatories are providing a vast amount of new observational data. The path of nucleosynthesis actually followed during the burst determines the energy of the burst and, therefore, its observable light curve. The corresponding nuclear reaction rates can only be obtained by experiments using beams of short-lived, neutron-deficient nuclei. Accurate satellite measurements of X-ray bursts and their timing and periodicities can also be used to constrain the equation of state of the underlying neutron star and its crust.

Understanding the equation of state of baryonic matter, namely, the dependence of the energy of the system on density, temperature, and isospin asymmetry, is a complex quantum many-body problem with a close relationship to astrophysics. Gravitational waves, when detected from coalescing neutron stars during their final inspiraling phase, will provide stringent constraints on the equation of state. Conversely, improved models of the nuclear matter equation of state will guide neutron-star observations and will enable accurate predictions of gravitational-wave signals.

Core-collapse supernovae announce the deaths of massive stars and give birth to neutron stars and black holes. Their theoretical study is a paradigm for interdisciplinary physics. It has recently been shown that most gamma-ray bursters occur in an exotic sub-branch of core-collapse supernovae. Such bursters have been seen on the other side of the universe and from the earliest epochs and are of primary interest in astrophysics. The actual mechanism that drives core collapse supernovae has not yet been identified with certainty. This is a key problem in astrophysics. Weak interaction rates are one important piece of the physics that is needed to address this problem. The understanding of the explosion mechanism is also important for modeling the various nucleosynthesis processes in supernovae that are of critical importance for nuclear astrophysics. The situation for models of Type Ia supernovae, which are thermonuclear explosions of entire white dwarf stars, is similar. Here, advances in the understanding of the nuclear processes during the explosion can also have implications for cosmology, where Type Ia supernovae serve as distance indicators probing the expansion rate of the universe. Studies of possible systematic errors in using Type Ia supernovae as standard candles at cosmological distances require accurate supernova models, incorporating critical nuclear physics, and which need to be constrained by nucleosynthesis and abundance observations.

All in all, the scientific interaction between nuclear physicists, astrophysicists, and astronomers has been profound and symbiotic, and nuclear physics continues to be central to progress in all domains of astrophysics and cosmology. Building upon this legacy of past success will be required to further foster and broaden the interdisciplinary connections, such as the Joint Institute for Nuclear Astrophysics (discussed below), and to significantly advance our knowledge of nuclear structure and reactions. The Facility for Rare Isotope Beams (FRIB) will play a critical role in this endeavor and finally will enable the experimental study of a large fraction of the unstable nuclei participating in astrophysical processes.

## CONNECTIONS TO COMPLEX SYSTEMS AND THE MANY-BODY PHYSICS COMMUNITY

The scientific community is witnessing the birth of a new area of physics that supplements traditional macrophysics of large systems and the physics of the microworld, namely mesoscopic physics. This name can be applied to systems that are sufficiently large to display generic statistical behavior but at the same time sufficiently small to allow researchers to study in detail individual quantum states. The general challenge for this interdisciplinary field is to understand the principles of building up complexity out of "elementary" blocks, which, in fact, have complicated structures of their own.

Nuclei are prototypical mesoscopic systems and splendid laboratories of many-body science. Despite the fact that the number of protons and neutrons in heavy nuclei is very finite compared to the number of electrons in a solid or atoms in a mole of gas, nuclei exhibit behaviors that are emergent in nature and present in other complex systems studied by condensed-matter physicists, quantum chemists, and materials scientists. For instance, shell structure, symmetry-breaking phenomena, collective excitations, and superconductivity are found in nuclei, atomic clusters, quantum dots, small metallic grains, and trapped atom gases. Nuclear scientists have made contributions to the field of mesoscopic physics when applying their femto-scale expertise to nano-scale problems. Likewise, nuclear physicists are taking advantage of techniques born in other fields. Some of those interdisciplinary research opportunities, both in terms of tools and phenomena, are outlined in the following sections.

**Understanding the Transition from Microscopic to Mesoscopic to Macroscopic.** Although the interactions of nuclear physics differ from the electromagnetic interactions that dominate chemistry, materials, and biological molecules, the theoretical methods and many of the computational techniques necessary to solve the quantum many-body problems are shared. Examples include basis expansion methods (configuration interaction in chemistry and the interacting shell model in nuclear physics), coupled cluster techniques, quantum Monte Carlo, and the density functional theory. In this context, data taken at FRIB will challenge theory to explain how increasingly complex nuclei can be put together from different combinations of protons and neutrons. Effective field theory and the renormalization group are tools to connect the various descriptions of phenomena at diverse length scales. They were first developed in particle physics and in condensed matter physics, and they have been adopted in many different aspects of nuclear physics as well.

**Quantum Chaos and the Random Matrix Theory.** The origins of the interdisciplinary field of quantum chaos are in nuclear physics theory, specifically in the random matrix theory that was developed in the 1950s to explain the statistical properties of the compound nucleus in the regime of neutron resonances. The nuclear theory has helped to understand the statistical properties of the conductance of quantum dots, yielding new insight into mesoscopic fluctuations. It has also been used to describe spectral properties in both nuclear and condensed-matter systems. In particle physics, random matrix theory has been successfully applied to lattice QCD, with implications for properties of the QCD vacuum and for computer simulations of QCD at finite chemical potential.

**Superconductivity.** Any attractive interaction between fermions at low temperatures generally leads to fermionic pairing analogous to the Cooper pairing of electrons in superconducting metals. It is not surprising, therefore, that pairing lies at the heart of nuclear physics. It is present in finite nuclei and in the nuclear matter of neutron stars, and it is believed to exist in the quark-gluon plasma. Since the number of nucleons can be precisely controlled, various regimes of pairing strength can be studied. It is only recently that such a static-to-dynamic crossover has been seen experimentally in ultra-small, nanosize aluminum grains. Based on the experience learned on a femtometer scale, nuclear theorists could successfully attack the problem of finite-size effects in grains. Another area of cross fertilization concerns various superconducting phases. For example, while neutron and proton pairing fields are essentially decoupled in nuclei near stability, in proton-rich nuclei, they may be mixed, leading to deuteron-like pairs. The idea of such correlations between nonidentical fermions may be special to nuclear physics, but spatially anisotropic pairing fields are also discussed in the context of high-temperature superconductivity in cuprate materials. The ground state of cold baryonic matter at "ultra-high" densities is believed to be a color superconductor whose ground state consists of an equal mixture of up, down, and strange quarks. And while this phase is favored in the limit of infinite density, the nature of the pairing pattern at the densities present in the core of neutron stars is a vigorously debated question. Specifically, one would like to understand the emerging pairing pattern for fermionic species with unbalanced populations. Perhaps here, ultra-cold atoms can come to the rescue.

**Loosely Bound and Open Systems.** Today, much interest in several fields of physics is devoted to the study of small open quantum systems, whose properties are profoundly affected by environment, i.e., the continuum of decay channels. Although every finite fermion system has its own characteristic features, resonance phenomena are generic; they are great interdisciplinary unifiers. Many aspects of open quantum systems that are independent of the system dimensionality have been originally studied in nuclear reactions and are now explored in molecules in strong external fields, quantum dots and wires and other solid-state microdevices, crystals in laser fields, and microwave cavities. In the field of nuclear physics, the growing interest in open quantum systems is associated with exotic weakly bound/unbound nuclei close to the particle driplines. The novel nuclear approaches developed in this context, such as the continuum shell model, are now being applied to studies of other open quantum systems such as coupled quantum dots or complex molecules.

**Dynamical Symmetries and Quantum Phase Transitions.** Nuclear physicists recognized early that the behavior of a complex nucleus is often governed by dynamic symmetries attributed to the features of the interaction between the constituents. The concept of symmetry links diverse physical systems: although seemingly unrelated, they reveal deeply rooted similarities that are described by the same mathematical structure. For instance, algebraic methods introduced in nuclear physics have also been used in the study of molecules and polymers, and, recently, phase transitions and the associated critical symmetries.

The list of topics above is by no means inclusive, and every day brings new developments. An excellent, very recent example is a novel solution to the famous fermion sign problem endemic to the Monte Carlo method, jointly proposed by nuclear theorists and quantum chemists while this Long Range Plan was being developed. Modern nuclear science contributes to, and takes advantage of, exciting and fundamental developments in the interdisciplinary field of many-body physics.

# Nucleonic Matter and Cold Fermions

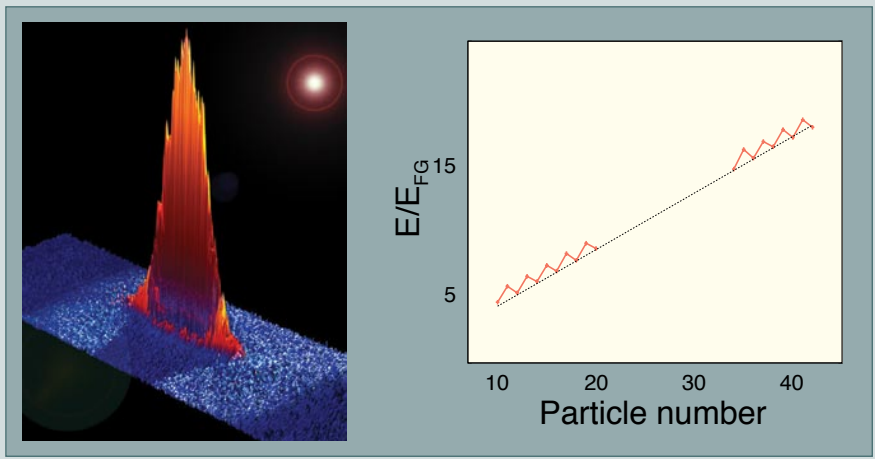

Cold fermions at unitarity. A fermion condensate appears (left) (courtesy of G.B. Partridge and R.G. Hulet, *Science* 311, 503, 2006) as the temperature is lowered. The size of the cloud at low temperatures gives the "universal" equation of state for fermions interacting at short range at infinite scattering length. These measurements can be compared to zero-temperature nuclear quantum Monte Carlo calculations of the energy (in units of Fermi gas energy) versus particle number (right), where the slope of the line gives the ground-state energy and the even-odd energy difference yields the pairing gap in infinite nuclear matter.

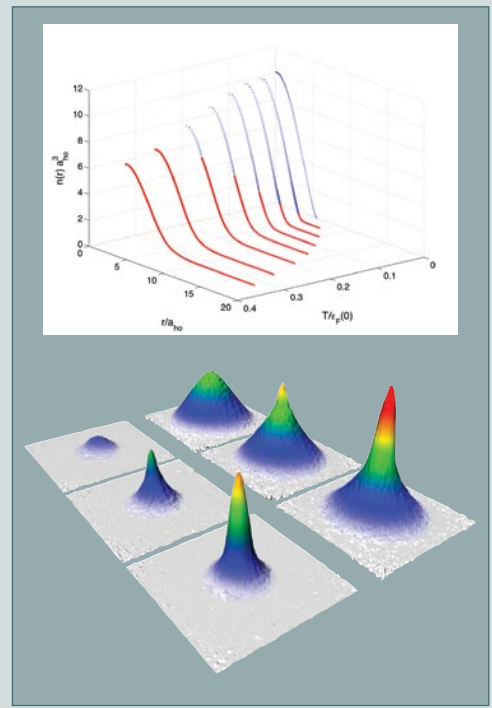

Superfluid to normal transitions. The upper view shows quantum Monte Carlo calculations of the density of a trapped Fermi gas at finite temperatures (front to back; in units of the Fermi energy of a noninteracting Fermi gas) and distance from the harmonic trap center (left to right; in units of the trap's oscillator length), indicating normal (red) and superfluid (blue) regimes. Experiments (bottom view) also study the transition as a function of the imbalance between two fermion species (courtesy of M.W. Zwierlein et al., *Science* 311, 492 (2006)). The images show the density of the majority (right) and minority (left) spin populations as the population difference decreases; the superfluid condensate emerges when the difference is sufficiently small.

Nuclear physicists have made important contributions to the study of strongly coupled superfluid systems like ultra-cold Fermi atoms, which show many similarities to the cold nuclear matter found in the crust of neutron stars. Neutrons, like cold Fermi atoms tuned to resonance, have an interaction that is nearly strong enough to couple them in pairs. Such systems can have an extremely large pairing gap, of the order of the Fermi energy—that is, about 1000 times greater than found in electronic superconductors.

A gas of interacting fermions is in the unitary regime if the average separation between particles is large compared to their size, but small compared to their scattering length. From the theoretical point of view, the properties of a fermion system in the unitary regime are remarkable, often being referred to as universal. Such a system is at the crossroad between fermion and boson superfluids. Nuclear theorists, using sophisticated quantum Monte Carlo techniques, provided the first reliable predictions of the ground-state energy, the superfluid transition temperature, and the pairing gap in the unitary regime. The predictions of the ground state energy (figure above) and the superfluid transition temperature (figure on the right) have been confirmed by atomic experiments.

The transition from superfluid to normal (i.e., nonsuperfluid) state also occurs as the populations of the two fermion species are altered. For large population imbalances, the Fermi surfaces of the two species are too well separated to allow pairing, and a normal state results. As the populations become more equal, a fully paired system occurs in the center of the trap. Future precise experiments measuring the total and difference populations, the density distributions of the two species, and their temperature dependence will provide fundamental tests of theories of strongly paired fermions clearly relevant to the study of superfluidity in nuclei, neutron matter, and quark matter. Future experimental possibilities include the study of pairing in finite systems, such as nuclei and bulk condensed matter, and quantitative investigation of different types of hadrons arising from paired quarks in the quark-gluon plasma.

# Stellar Explosions on a Computer

Explosions of massive stars, also known as core-collapse supernovae, are an important link between the origin of the universe in the Big Bang and the formation and evolution of life on Earth. These explosions are the dominant source of most elements between oxygen and iron, and there is growing evidence that they are responsible for producing half of the elements heavier than iron. These explosions also have the potential to serve as cosmic laboratories for physics at extremes that are inaccessible in terrestrial experiments—a potential that can be realized only by combining realistic, three-dimensional computer models with observations of neutrinos, gravitational waves, and the entire spectrum of photons from supernovae.

As their name suggests, core-collapse supernovae result from the collapse of a star's core, followed by the formation of the outgoing shockwave that is ultimately responsible for the explosion. These shock waves are driven outward through some combination of radiation pressure, acoustic energy, magnetic fields, and neutrinos. They are turbulent. And they involve matter at extremes of density and neutron richness, as well as poorly understood interactions between this extreme matter and neutrinos. Future advances in supernova theory will require the development of sophisticated macroscopic three-dimensional models, sophisticated descriptions of the microscopic physics involved in these stellar explosions, and large-scale computing.

The development of core-collapse supernova models must be based on sophisticated multidimensional simulations (see figure below). In fact, the set of coupled partial differential equations that defines this problem is seven dimensional: three in space, three in momentum space for the radiation transport (for each of six neutrino species), and one in time. Carrying out such simulations will require new algorithms that scale to computer architectures involving tens to hundreds of thousands of processors—architectures that are only now being planned. Hence, the ongoing investments by both DOE and NSF in next-generation computational infrastructures are crucial to lasting progress on this central problem in nuclear astrophysics.

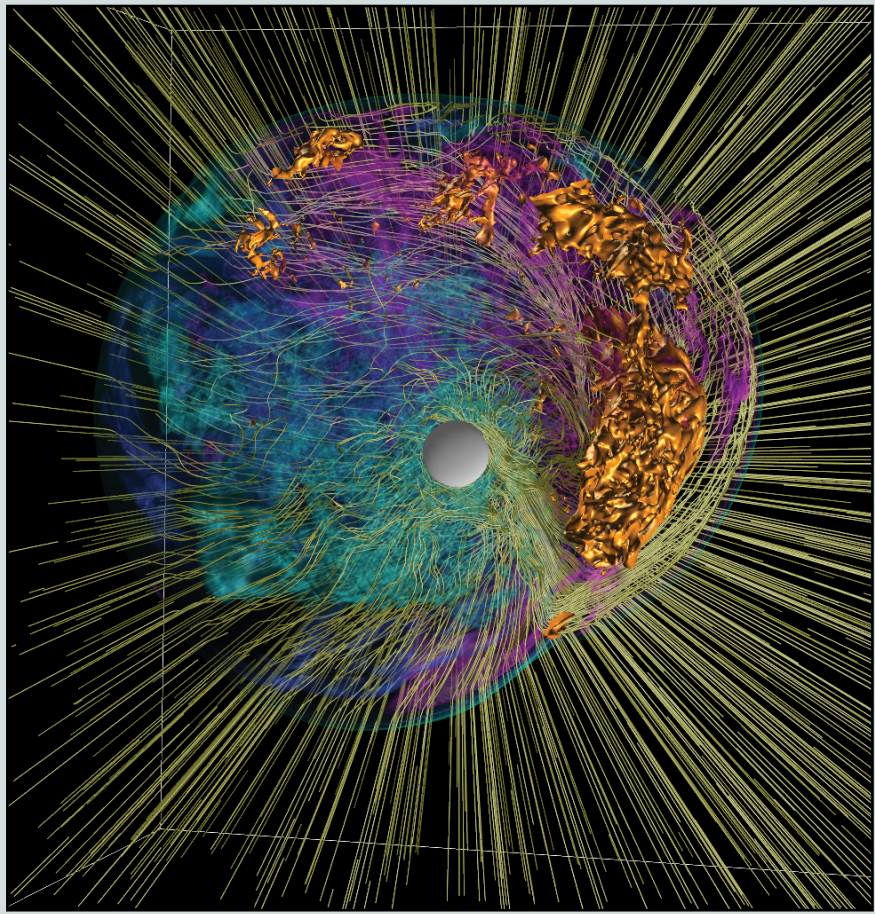

The development of the computationally discovered instability of the core collapse supernova shockwave is shown in a snapshot from a three-dimensional purely hydrodynamic simulation. The instability leads to growing deformations (away from spherical) of the shockwave, which is represented by the surface in this image. The deformations in turn lead to circulating flow below it. Two counter-rotating flows are formed. Streamlines in this image highlight one flow moving clockwise just beneath the shock surface and a second, deeper flow moving counterclockwise just above the protoneutron star surface.

138                                                                                                      Connections to Other Fields

## CONNECTIONS TO THE COMPUTATIONAL SCIENCE COMMUNITY

One of the most striking trends in nuclear physics today is the increasingly important role played by computational science. The definition of "large-scale computing" continuously changes due to power-law increases in raw computational capability, and has led to a common use of tens of teraflop-hours (tera=$10^{12}$) of computing time per month in order to solve some of our current problems. The move to petascale (peta=$10^{15}$) computations will occur during the next two to three years, and exascale (exa=$10^{18}$) computers will provide an unprecedented opportunity for nuclear science within 10 years. Furthermore, nuclear physics experiments generate data sets of unprecedented size, approaching one petabyte per year. These data are analyzed around the world, utilizing state-of-the-art grid technology for data transfers and access.

Computational advances have reached the point where sophisticated numerical studies in nuclear physics are able to directly address outstanding scientific problems and societal applications. In order to encourage advances in computational nuclear physics and other sciences, which share similar challenges and methods, initiatives by both DOE (Office of Science, NNSA, and ASCR/INCITE) and NSF have fostered collaborations between computational nuclear physicists, computer scientists, and other members of the computational physics community. The major efforts in computational nuclear physics are being conducted in three broad areas funded under the DOE's SciDAC and INCITE programs:

- **Nuclear Astrophysics.** The current SciDAC effort being conducted by the Computational Astrophysics Consortium and the recently completed Terascale Supernova Initiative have made great strides in developing extensive multidimensional hydrodynamic and radiation-hydrodynamic simulations of Type Ia thermonuclear supernova explosions, core-collapse supernova explosions, and gamma-ray bursts. The collaborations partner nuclear physicists and astrophysicists with astronomers measuring and modeling optical spectra of supernovae to provide validation of the simulation models. The collaborations develop multi-dimensional simulation tools in collaboration with computer scientists at various national laboratories and universities.

- **Nuclear Structure.** The quantum many-body problem represents one of the great intellectual and numerical challenges of our day for nuclear physics, quantum chemistry, and materials sciences. In the next five years, petascale applications to solve the many-body problem will become commonplace. The complicated nature of the algorithms to solve these problems requires close interactions with computer scientists and mathematicians. In particular the SciDAC Universal Nuclear Energy Density Functional promises to provide a new generation of functionals that will describe heavy and exotic nuclei as well as providing connections with *ab initio* approaches.

- **Lattice QCD.** Developing a solid foundation for nuclear physics in QCD is of paramount importance to the field, and to this end DOE has created the LQCD program, which brings nuclear and particle physicists together with computer scientists in order to develop the techniques necessary on current and future computer architectures to determine the basic structure of nucleons and their interactions.

Many of the future directions discussed in this Long Range Plan will be tied to continuing advances in high-performance computing. Indeed, the efforts listed above are just a few examples of the rich variety of nuclear problems that utilize the large-scale technology available for scientific computing. As in other areas of computer science, the principal challenge is to provide numerical algorithms that scale across tens of thousands of processors.

In addition to answering outstanding scientific questions, connections between computational science and nuclear physics are crucial for addressing numerous critical challenges in areas such as energy, ecological sustainability, and national security. Excellent recent examples are Advanced Fuel Cycle research and the research related to the Science-Based Stockpile Stewardship. These two examples point to the relevance of the computational nuclear theory to other programs of national interest. Here, a key goal is a quantifiable theoretical error estimate.

# CONNECTIONS TO OTHER FIELDS

**Physics and Chemistry of the Heaviest Elements.** The heavy elements provide a laboratory for studying nuclear dynamics and structure under the influence of large Coulomb forces. Research in this area involves ideas that are fundamental to both chemistry and physics. Production of atoms of these elements by nuclear reactions enables chemical studies, and the selectivity of chemistry allows identification of the atomic number of the species. Among the questions being explored in heavy-element chemistry are the form and structure of the Periodic Table and the impact of relativity on electronic motion.

**High Energy-Density Physics.** The behavior of matter at energy densities exceeding $10^{11}$ Joules per cubic meter is a rapidly growing field that spans plasma physics, laser and particle-beam physics, nuclear physics, astrophysics, atomic and molecular physics, materials science and condensed-matter physics, intense radiation-matter interaction physics, fluid dynamics, and magnetohydrodynamics. The compelling research opportunities in this area have recently been discussed in a report *Frontiers for Discovery in High Energy Density Physics*, prepared by the National Task Force on High Energy Density Physics (see www.er.doe.gov/np/program/docs/HEDP_Report.pdf).

Figure 5.2 shows the realm of high energy-density physics as mapped in the density-temperature plane. The quark-gluon plasma entails the highest temperatures and pressures of any laboratory study. But does the quark-gluon plasma exhibit any of the properties of a classical plasma? And what are connections between classical plasmas governed by the electromagnetic interaction and the plasma made of strongly interacting matter? The first indications are that the matter produced at RHIC more nearly resembles a warm and dense plasma, which is strongly coupled, than an ideal weakly coupled plasma. The job ahead is to understand the precise nature of the quark-gluon plasma. This goal is almost identical to the experimental and theoretical work underway to study the physics of warm, dense electromagnetic plasmas. Theoretical and experimental techniques from one area are proving relevant to both areas. Examples include hydrodynamical modeling, molecular dynamics simulations, and multiparticle correlation analyses to measure collective motions.

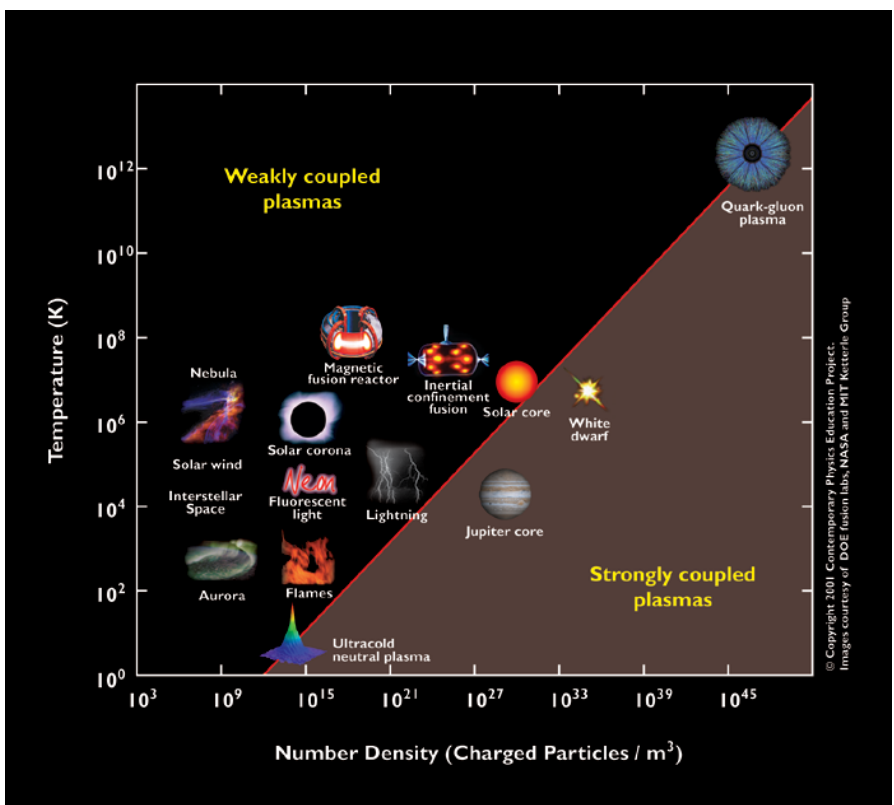

**Figure 5.2:** Different types of plasmas in the temperature-density plane. The red line indicates schematically the boundary between weakly and strongly coupled plasmas. Trapped dense systems of ultra-cold atoms and quark-gluon plasmas are two examples of strongly coupled plasmas in the laboratory, at opposite temperature extremes. The high energy-density physics regime is where pressures exceed 1 Mbar. This corresponds approximately to those parts of the map where temperatures exceed $3 \times 10^6$ degrees Kelvin or densities of more than $10^{29}$ particles/m$^3$ are reached. The quark-gluon plasma holds the record in both directions.

140    Connections to Other Fields

## THE ROLE OF THE INT AND JINA AS INTERDISCIPLINARY FACILITATORS

Fostering connections between nuclear physics and other branches of physics is one of the stated missions of the Institute for Nuclear Theory (INT) at the University of Washington. For the period 2004–2009, for instance, programs and workshops convened or scheduled at the INT include connections to astronomy, cosmology and astrophysics, high-energy physics, atomic physics, and quantum chemistry. In prior years there were strong programs in mesoscopic physics and quantum chaos, which had significant involvement from condensed-matter physicists. The ability to attract scientists to the INT from a variety of disciplines has consistently been a priority of the INT's National Advisory Committee, both to promote cross fertilization of ideas and to maximize the opportunities for nuclear physics to contribute to broader efforts in physics as a whole.

In recent years the field of nuclear astrophysics has undergone a transformation into a truly interdisciplinary scientific endeavor. The Joint Institute for Nuclear Astrophysics (JINA), a collaboration between the University of Notre Dame, Michigan State University, the University of Chicago, and Argonne National Laboratory, has been one of the main drivers of this change. Nuclear physicists, astrophysicists, and astronomers who are members of JINA identify common key scientific questions and address them together in a coherent approach. JINA also plays the leading role in a new effort to create a community-based refereed database for both nuclear and astronomical data of relevance to the field.

## THE FUTURE

There can be no question about the importance of interdisciplinary research that explores connections between traditional fields leading to new knowledge. Today's nuclear science is extremely broad and diverse. It spans the gamut from hot and cold QCD to nuclei, astrophysics, tests of fundamental laws of nature, and myriad applications. Nevertheless, it is characterized by several encompassing themes that reflect the major challenges facing modern science today; hence it has deep links to many other fields. The nuclear science of tomorrow promises to open many new areas that will lead to new connections with other disciplines. Indeed, it is the drive to solve complex questions and problems that leads researchers to meet at the interfaces of disciplines and even to cross frontiers to form new disciplines.

# Applications

## INTRODUCTION

The close connection between fundamental research and economic strength has been well documented in every discipline—nuclear science most definitely included. Basic research into the physics of the nucleus, coupled with research into the chemical properties of radioactive elements, has laid a solid foundation for practical technologies such as nuclear energy, nuclear medicine, particle accelerators, particle detectors, and nuclear weapons. These applications and many others are listed in table 5.1. In the course of this basic research, moreover, nuclear scientists have created a host of tools and instruments that have themselves proved valuable in the marketplace; witness the radioisotope generators used in nuclear medicine. They have compiled a wealth of essential data about the nucleus and the many reactions nuclei engage in. And they have helped train the highly specialized workforce needed to sustain and advance these applied nuclear technologies.

There is every reason to expect this process to continue. Current state-of-the-art research technologies are already offering a number of potential applications, some of which are summarized in table 5.2. And in the coming decade, the technological solutions required to realize the scientific goals of this Long Range Plan promise to advance our national competitiveness even further. For example, accelerator technologies developed for FRIB could improve our capabilities for studying materials through isotope implantation; for producing radio isotopes for medical research; and for making the nuclear data measurements so critical to national and homeland security. Other developments could help address the nation's acute need for radiation detector technologies. At the same time, however, the nuclear physics community itself has a responsibility to be aware of national needs, and

*Table 5.1:* Summary of current applications of nuclear science

**Medical Diagnostics and Therapy**
    Radiography
    Computerized tomography
    Positron emission tomography
    MRI (regular)
    MRI (with polarized noble gases)
    Photon therapy
    Particle-beam therapies
    Instrument sterilization with $^{60}$Co gamma rays
    Linac irradiation treatments
    Radioisotope tagging

**Safety and National Security**
    Airport safety and security
    Large-scale X-ray scanners
    Nuclear materials detection
    Arms control and nonproliferation
    Stockpile stewardship
    Tritium production
    Space-radiation health effects
    Semiconductor performance in radiation environments
    Food sterilization
    Electronic single-event upset testing

**Energy Production and Exploration**
    Nuclear reactors
    Oil-well logging
    Research and development for next-generation nuclear reactors

**Art and Archaeology**
    Authentication
    Nuclear dating

**Material Analysis**
    Activation analysis
    Accelerator mass spectrometry
    Atom-trap trace analysis
    Forensic dosimetry
    Proton-induced X-ray emission
    Rutherfold backscattering
    Ion-induced secondary-ion emission
    Muon spin rotation

**Environmental Applications**
    Climate-change monitoring
    Pollution control
    Groundwater monitoring
    Ocean-current monitoring
    Radioactive-waste burning
    Radon detection
    Smoke-stack monitoring

**Materials Testing and Modification**
    Trace-isotope analysis
    Ion implantation
    Surface modifications
    Flux pinning in high-$T_c$ superconductors
    Free-electron lasers
    Cold and ultra-cold neutrons
    Single-event efforts
    Microphone filters

to maximize the likelihood that the long-term investments made for basic research will also enhance our ability to address those needs. Of particular importance is the continued health of the National Nuclear Data Program, which is supported by the Department of Energy.

Indeed, nuclear science is central to the missions and goals of many organizations within the U.S. government—the Departments of Energy, Defense, and Homeland Security in particular. This section provides some of the many examples of how the knowledge obtained and the technologies developed through basic nuclear science research are being applied to address national needs.

## NUCLEAR DATA: ACCURATE AND ACCESSIBLE

Nuclear data are produced from activities that are motivated either by basic research or by the development of nuclear-based technologies. Globally, programs that generate nuclear data are supported primarily by government agencies. There are three international networks that coordinate nuclear data projects worldwide. Two of these networks, the Nuclear Reaction Data Centres Network (NRDC) and the Nuclear Structure and Decay Data evaluators (NSDD), are coordinated by the International Atomic Energy Agency in Vienna. The third network, the Working Party on International Nuclear Data Evaluation Co-operation (WPEC), is coordinated by the Nuclear Energy Agency of the Organization for Economic Co-operation and Development in Paris.

The dissemination of nuclear data and associated documentation to the consumers of nuclear data is the main goal of these international networks. More information about them can be found at http://www-nds.iaea.org/nrdc.html, http://www-nds.iaea.org/nsdd, and http://www.nea.fr/html/science/wpec.

The hub of the U.S. network, the National Nuclear Data Center (NNDC) at Brookhaven National Laboratory, is the core facility for the DOE-funded U.S. Nuclear Data Program (USNDP). The mission of the USNDP is to collect, evaluate, and disseminate nuclear physics data for basic nuclear physics and applied nuclear technology research. The USNDP includes nuclear data groups and nuclear data experts from national laboratories and academia across the United States. The services provided by this national network of nuclear data groups are essential to organizations with missions that require access to nuclear data. The nuclear data infrastructure provided by the USNDP impacts governmental, educational, commercial, and medical organizations in United States, and is part of the U.S. commitment to the international nuclear data networks.

It is evident that the need for convenient access to nuclear data is rapidly increasing. As shown in figure 5.3, the number of data retrievals from USNDP databases has increased by almost a factor of 10 over the last decade. In 2006 the NNDC web service reached the milestone of one million retrievals from the USNDP databases. The NNDC data retrievals are mainly from users in the United States (42.9%) and Europe (25.7%). The U.S. users are almost equally divided among government, education, and all other types of organizations.

*Table 5.2:* Potential near-term applications of nuclear science.

**Safety and National Security**
- Large-scale neutron-beam scanner
- Large-scale gamma-ray beam scanner
- Large-scale imaging with muons
- Nuclear-reactor monitoring with antineutrino detector

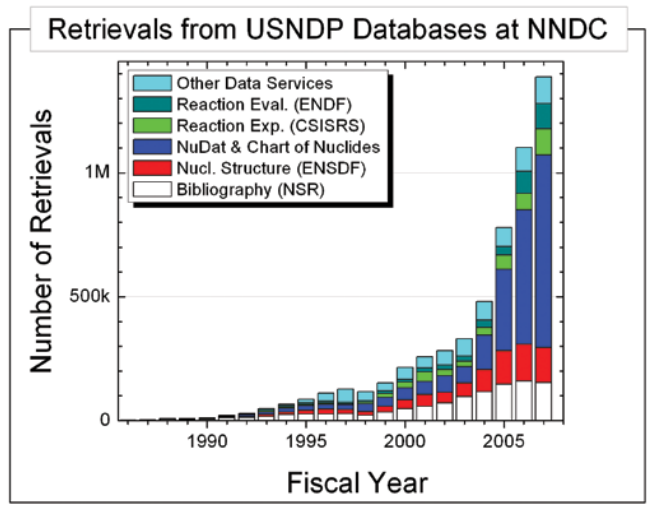

**Figure 5.3:** Number of retrievals from the USNDP databases at the NNDC from 1985 through 2007. This graph is from the USNDP Annual Report for FY2007.

# Muon Tomography

The flow of cargo traffic into the United States is so immense—and is carried in so many different ways, including aircraft, trucks, and ocean vessels—that it is impossible for authorities to search all of them manually for contraband. This fact creates a serious national security challenge: imagine the havoc that terrorists would wreak with, say, one container of shielded nuclear material hidden inside a shipment of normal cargo. What's needed is an automatic way to scan the cross-border shipments as they go by, without having to stop the vehicles, or open them up, or expose their human occupants to any extra radiation.

Making use of particle-detection methods originally developed for basic nuclear science, researchers at Los Alamos National Laboratory (LANL) have invented a new way to do just that. Their method relies upon the natural flux of energetic muons produced by cosmic rays striking the upper atmosphere. These muons propagate downward, bombarding the surface of the Earth at the rate of some 10,000 particles per square meter per minute. And then they keep on going, penetrating tens of meters into the rock beneath our feet.

Years ago, the pioneering nuclear physicist Luis Alvarez realized that this muon flux presented an opportunity: by placing large, muon-sensitive proportional counters underneath Mayan pyramids, he was able to take an unconventional "X-ray" of the pyramids' interiors.

The muon imaging system developed at LANL uses a different detection technique—multiple-scattering radiography—but is in much the same spirit. With muon detectors deployed at major border checkpoints, every vehicle or shipping container entering the United States could be scanned while moving through at a low speed. Objects with large, unexplained masses, especially masses of high-Z material, would create a characteristic perturbation in the muon signal and could thus be identified for manual inspections. An artist's rendition of the muon radiography is shown in the figure above.

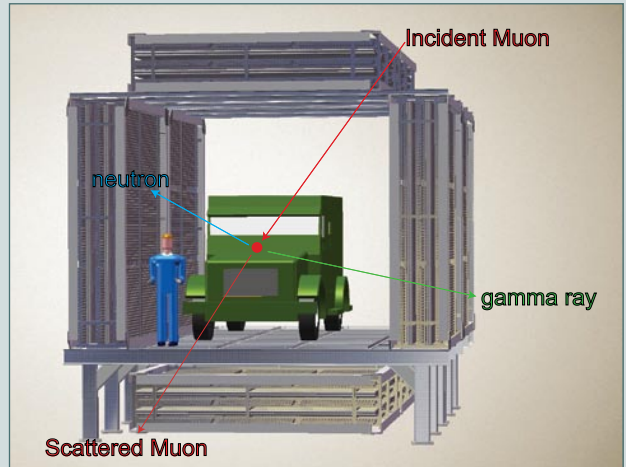

An illustration of a potential muon tomography setup showing incident cosmic-ray muons entering the apparatus from above (red lines) and being deflected by a heavy object in the cargo. Also shown are lines indicating neutrons (blue line) and gamma rays (green line) that might be passively detected from unshielded nuclear material.

To test this idea, Los Alamos physicists used a prototype muon scanner to make radiographs of both an engine and of an engine with a cube of lead hidden beside it. Lead was chosen because its density and atomic number are similar to that of potential nuclear contraband. The data are shown in the figure on the left.

Recently, a cooperative agreement between LANL and Decision Sciences Corp. has been put into place to deploy a muon radiography system for the inspection of incoming cargo.

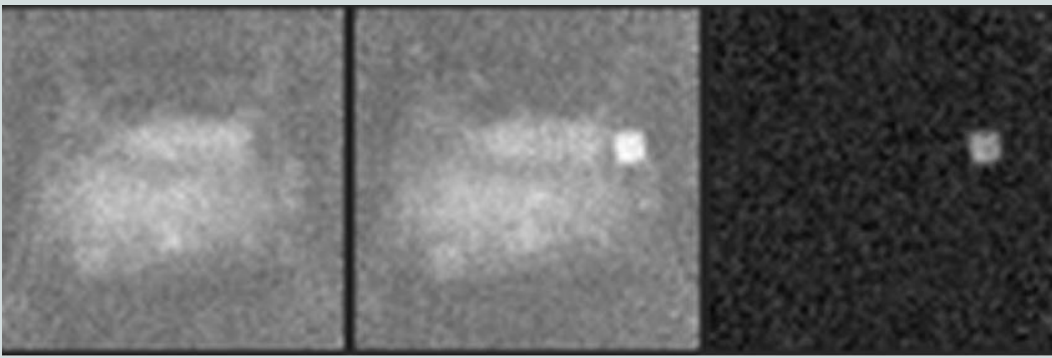

Images showing the mean scattering angle for a slice through the scene. The left panel shows the engine, the middle panel shows the engine plus the 10x10x10 cm³ lead sample, and the right panel shows the difference. The lead is clearly seen.

A priority of the USNDP for the next decade is to produce databases in support of advanced simulation codes. This will be very important for applications such as design studies of Advance Fuel Cycle reactors, as well as nuclear-materials-detection systems for homeland security applications. This effort will also maintain a high level of expertise in the area of nuclear data evaluation to assure the continuation of the nuclear databases with sufficient breadth and quality to meet the requirements of advanced computational applications.

## APPLICATIONS IN NATIONAL SECURITY

In the area of national security, nuclear scientists contribute to a broad range of capabilities and applications. They support the work of assessing and certifying the safety and reliability of the U.S. stockpile that is carried out at labs supported by the National Nuclear Security Administration (NNSA). More recent efforts in building an "attribution" program also require extensive capabilities in nuclear theory, modeling, and experimentation. Nuclear forensics plays an important role in deterring rogue nations from using nuclear weapons and radiological dispersal devices. By extending our forensics capabilities, it becomes easier to accurately identify such nations and address the situation with an informed political process. A main goal of this program is to develop the technologies for using forensics on the postexplosion radiochemical isotope debris from nuclear devices that might be used against the United States to determine the design, fuel types, and possibly the origin of the device.

Short-lived isotopes play an important role in shaping the nuclear products in high neutron fluence environments such as in nuclear-device explosions and supernova. A common requirement of many national security and astrophysics applications is the capability of determining reaction cross sections on short-lived nuclei. The methods being developed in the nuclear physics community are central to these efforts. For example, nuclear theorists developing microscopic methods for calculating cross sections of nuclear reactions involving both light nuclei (e.g., *ab initio* predictions) and heavier nuclei (e.g., in determining nuclear-level densities, fission barriers, etc.) are substantially enhancing our predictive capability. In experimentation, techniques such as high-resolution measurements of gamma-ray energy spectra from ($n,nn\gamma$) reactions induced with pulsed neutron beams are providing ways to determine with significantly improved accuracy neutron-induced cross sections that are important for nuclear security and energy applications. Also, new detectors such as the Detector for Advanced Neutron Capture Experiments and the Lead Slowing Down Spectrometer at Los Alamos, combined with the ability to fabricate radioactive targets for certain longer-lived radioactive isotopes, is enabling important measurements that provide new information on nuclear structure and nuclear reaction dynamics. Another high priority in many applications is a more precise and realistic set of covariance data, i.e., uncertainties and correlations on cross sections. These are needed in stewardship applications, as well as in advanced reactor programs such as the Global Nuclear Energy Partnership (GNEP). Many of the statistical and theoretical methods needed to develop these databases and their maintenance have been supported through the USNDP by the Nuclear Physics Program in the DOE Office of Science. Of particular importance is the USNDP oversight of the ENDF cross-section database, which is the basis of nuclear application simulations in a wide range of areas both nationally and internationally.

Nuclear scientists have developed instrumentation—for example high-energy proton accelerators, magnetic imaging, and detector technologies—that have enabled the most precise penetrating radiographic imaging to date for stockpile stewardship. The technique of proton radiography allows unprecedented studies of macroscopic material dynamics, and in addition to stockpile stewardship applications it has been applied to areas as diverse as performance of turbine engines and performance of the liquid-mercury spallation target at the Spallation Neutron Source at Oak Ridge.

## APPLICATIONS IN HOMELAND SECURITY

A recent Defense Science Board report noted that the risks of nuclear proliferation are severe and difficult to solve politically. The formation of the Department of Homeland Security (DHS) has helped to centralize the nation's efforts to reduce these risks. The country has evolved a multipronged research and development approach to addressing the proliferation threat that is spearheaded by the DHS and involves other agencies such as DOE. The main programmatic goals are to develop new technologies in contraband detection, emergency response, incident assessment, and nuclear forensics. The technical challenges needed to address these missions include:

- passive and active nuclear-detection systems with greater sensitivity and resolution,
- detector materials with greater sensitivity and resolution,
- "pocket" and handheld detector systems with improved ID and enhanced operational effectiveness,
- detection of nuclear materials at longer distance,
- remote emplaced sensors,
- modeling and measurement of operating environments, and
- human-portable and relocatable systems for active interrogation of cargo and improved passive detection performance.

The basic nuclear science programs at DOE and NSF are essential for maintaining and extending the technical base that is needed to advance nuclear-detection technologies. Among the important roles of the science community is the use of basic nuclear-physics knowledge to identify new detection *signatures* for materials of interest—either nuclear materials or chemical explosives. Many of the applied research and

## MRIScreen: Brain Imaging Technology Applied to Airport Security

The U.S. Department of Homeland Security's restrictions on carrying liquids aboard aircraft, as awkward as the rules may be for passengers struggling to fit everything into 3-ounce containers and 1-quart Ziploc® bags, are rooted in a fundamental conundrum: liquid explosives must be stopped before they get on board. Yet current X-ray machines cannot distinguish between benign and hazardous liquids. Now, however, nuclear scientists at Los Alamos National Laboratory have sketched out a proof-of-principle concept, called MRIScreen, that promises to provide a better solution.

The MRIScreen technology is a variation of an ultra-low field (ULF) nuclear magnetic resonance imaging approach that LANL has been applying to brain studies. The original idea was to carry out simultaneous MRI measurements of brain function and anatomy by exploiting a new regime of magnetic fields even weaker than that of the Earth. However, it turns out that the different MRI contrast provided by ULF may also prove useful for distinguishing between liquid materials. With funding from Department of Homeland Security, the LANL group has begun to test a new system based on the brain-imaging design, but optimized for differentiating between liquids at a security checkpoint. They are also working on a second system designed to explore how much further the ULF-MRI technology can be taken for security applications. And they are continuing their brain-function and imaging work under funding from the NIH. Indeed, the researchers recently acquired the first-ever ULF-MRI image of a living human brain.

This application flows out of combining SQUID technology with nuclear expertise in NMR. The scientists involved are also actively using SQUID detectors as part of the nuclear-physics-funded experiment to search for a nonzero neutron electric dipole moment.

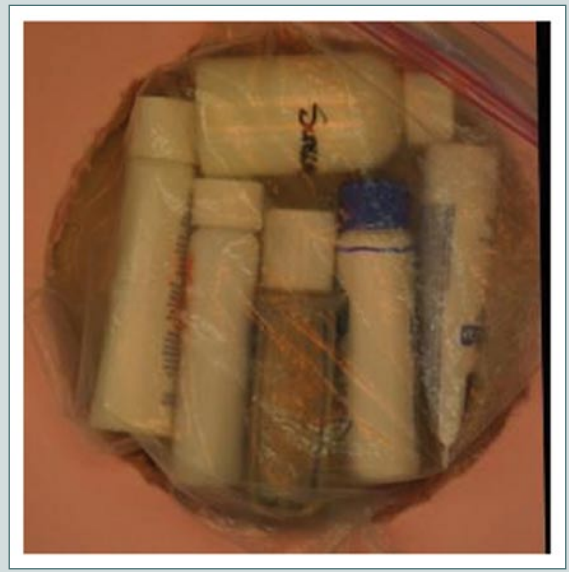

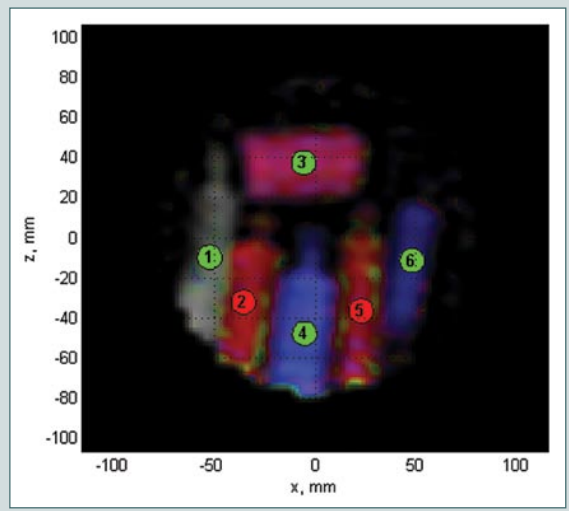

The photo is of a "311" bag with two bottles of threat material hidden among the items. The MRI image with the "dots" is from MRIScreen. The system produces an image that flags safe (green dots) and threat (red dots) objects.

development projects currently underway have parentage in the basic research community. An example can be seen in the sidebar on muon tomography, which has very recently resulted in an industry/national lab partnership to move this technology toward practical application.

Optimizing design parameters for variant detection schemes and event scenarios requires extensive system studies. In some cases, computer modeling is an effective alternative to field testing, which is often time consuming and can be very expensive. Sophisticated computer simulation codes that were born in the basic physics research community have become industry standards for modeling particle interactions in bulk materials. These simulation codes are used to lay the groundwork for proposing and planning new projects and for optimizing the design and analysis of different system configurations. Such computer simulations are essential for interpreting the results from fielded interrogation systems. The accuracy of computer simulations largely depends upon the correct treatment of particle interactions with matter and on the quality of the nuclear database. Evaluated databases, such as the ENDF nuclear data library developed by the USNDP, incorporate detailed information from experiments and nuclear models, thereby enabling transport simulations to accurately model the underlying physical phenomena in complex assemblies.

## APPLICATIONS IN MEDICINE

Nuclear medicine is an essential part of modern healthcare strategies for diagnosing and treating certain illnesses, particularly in the area of oncology. Most modern medical centers have a nuclear medicine department that provides both diagnostics and treatment services. The scientific foundation and many of the technologists working in this branch of medicine are products of basic nuclear science research. Many of the imaging techniques employed in medicine—notably CT, PET, MRI—have their roots in discoveries made by nuclear physicists. Nuclear science research also contributes to the electronics that make these imaging modalities effective. For example, engineers who worked on PHENIX at RHIC have also worked on the electronics that make modern PET scanners work so well—and the detectors used there draw heavily on fast scintillators and the phoswich concept that are widely used in nuclear science experiments. Future improvements to PET scanning will use gamma-ray tracking to improve position resolution. Meanwhile, in addition to diagnostic applications, radioisotope tagging of molecules has greatly helped disentangle how the chemistry underlying biology works. Some examples of recently developed nuclear-based medical diagnostic and treatment technologies are described below.

### Diagnostics

Radioactive atomic nuclei are used extensively in biology and medicine as radiotracers. Because their chemical properties are identical to the stable isotopes of the same element, they can be incorporated into substances that are metabolized by the body, thereby enabling both material tracking and functional diagnoses of organs. Alternatively, the emitted radiation from radiopharmaceuticals can be used for therapeutic purposes. Commonly used radiotracers ($^{99m}$Tc, $^{123}$I, and $^{111}$In) have short half-lives and are typically produced and handled by personnel trained in nuclear science.

**Positron Emission Tomography Imaging.** The use of PET imaging has become a major diagnostic modality. PET imaging provides metabolic information through the use of radioisotope tagged substances that have different rates of uptake, depending on the type and function of the tissue involved. Tomographic images are reconstructed from recording the positron decay of radioactive isotopes (e.g., $^{11}$C, $^{13}$N, $^{15}$O, and $^{18}$F) that are incorporated into radiotracer compounds such as glucose, water, or ammonia, which are metabolized. The $^{18}$F-labeled fluorodeoxyglucose (FDG) has become a primary tracer because the half-life of $^{18}$F is long enough to allow for transportation of a few hours. Other shorter-lived isotopes are produced by on-site cyclotrons. Increasingly often, PET scanners are combined with computed tomography scanners to provide multimodal images that give both metabolic and anatomic information. More than 650 hybrid PET/CT scanners were installed worldwide in 2005. An example shown in figure 5.4 illustrates the enhanced sensitivity in the diagnosis of cancer that is gained by fusing information obtained from CT and PET scans. Hybrid PET/MRI scanners are under development and provide a promising diagnostic modality due to better soft-tissue resolution and lower radiation exposure compared to PET/CT. A commercial head-only PET/MRI fusion system, through Siemens, will be available by the end of 2007.

In the past 20 years the reconstructed position resolution of commercial PET scanners has improved from 8 mm to 4 mm, while the axial extent has increased from 5 cm

# Proton Radiography

The development of nuclear weapons brought with it a need to observe and understand the dynamic behavior of materials when driven by high explosives—a need that has become especially acute since the cessation of U.S. nuclear testing in 1989, which put a premium on accurate predictive models of such behavior. During much of that time, unfortunately, the experimental tools to observe what happens in a high-power explosion remained pretty much unchanged from the Manhattan Project days.

Over the last decade, however, researchers at Los Alamos National Laboratory have developed a new imaging technique that radiographs materials during dynamic experiments using high-energy protons, rather than x-rays. Proton radiography allows researchers to make short movies and obtain much more detailed information on the motions and densities of materials when driven by shock compression than was ever possible before. The penetrating power, or long mean free path, of protons and the ability to focus them is opening up new opportunities for quantitative experiments, accurate model development, and designer training.

The protons used for this work are generated by the 800 MeV linear accelerator that was originally built at LANL by the nuclear physics program. (It was called LAMPF when operated by nuclear physics.) The energy of the protons determines the thickness of the object that can be studied, as well as the spatial resolution that can be obtained. One key technology that enabled this application, the system of magnetic lenses used to focus the protons into an image on the scintillation detectors, had previously been developed and refined by nuclear physicists over a number of years. Another key technology, the high-speed digital cameras that capture the images, was developed specifically for this purpose. This camera technology has received both a research and development 100 award and a Wall Street Journal Technology Innovation award, and is now being developed commercially by Teledyne Imaging Sensors.

The figure above shows sequential frames of a proton-radiography movie of the detonation of a high explosive beneath a tin disk about 5 cm in diameter and 0.6 cm thick. First, the detonation's spherical shockwave bulges the disk. Then, when the compressive shockwave reflects from the disk's upper surface, the shock becomes tensile, dislodging and levitating the spall layer (the "flying saucer"). The shockwave's high pressure and temperature also melt the tin (light gray region connecting the flying saucer and bulge). These radiographs were made with pulses 50 billionths of a second long and are spaced by 1 millionth of a second.

Over 300 dynamic proton radiography experiments have been performed to date. Many focused on studying the detonation of high explosives, the motion of shockwaves through various materials, shock-induced material damage, and shock-driven instabilities for the Stockpile Stewardship program. The proton radiography facility at LANSCE has been also used to study the formation and flow of helium bubbles in mercury for target development for the Oak Ridge Spallation Neutron Facility, coolant flow in engines for the Ford Motor Company, and the deflagration of high explosives.

In addition to its application for studying the dynamical responses of materials

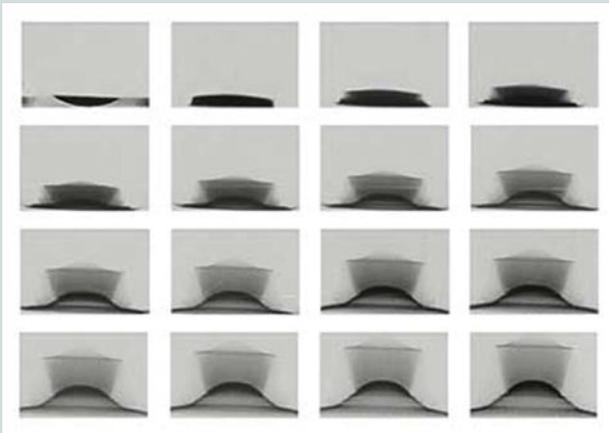

Sequential frames (from top left to lower right) of a proton-radiography movie showing how a tin disk responds to the shockwave produced by detonating a high explosive beneath the disk.

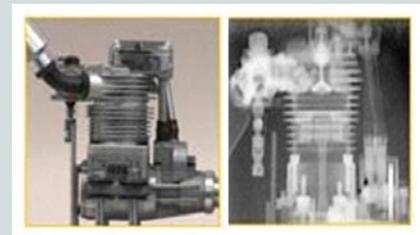

A 150-cc model airplane engine (left), and a proton radiograph of it produced using 800 MeV protons (right). The details of the thicker parts of the engine are clear in the proton radiograph.

in explosions, proton radiography provides the capability of imaging components inside of massive machinery. A proton radiograph of a small engine is shown in the figure above.

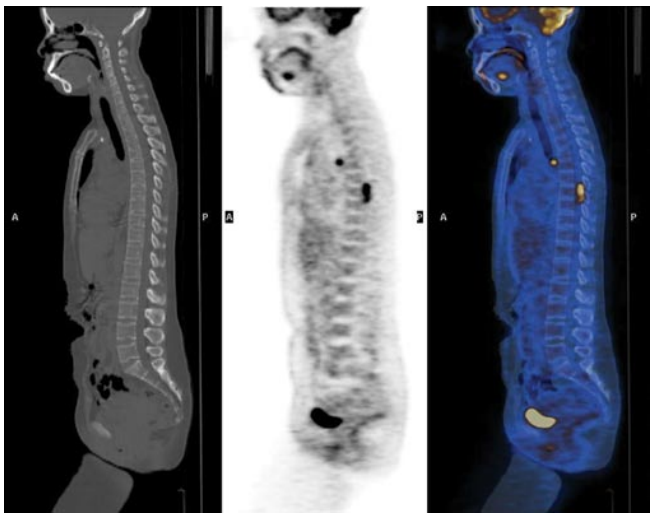

**Figure 5.4:** Images of the same patient taken simultaneously in different modalities: CT (left), PET (middle), and PET/CT fusion (right). While the CT scan displays the patient's anatomy, the PET scan shows (in black) areas of increased $^{18}$F-FDG uptake in the lung and spine of the patient. The fusion image shows the cancerous areas overlaid with the patient's anatomy. (Image courtesy of Kevin Berger, M.D., Michigan State University.)

to more than 15 cm. In addition, new detector materials, many of which were developed for basic research, have led to significantly improved energy and time resolution, which in turn provide more stringent coincidences and reduce random background. Scientists are exploring measuring the time of flight of the photons along the line of flight to obtain improvements to the reconstructed position. Also, new PET isotopes are being explored in research settings. For example, metal complexes of $^{67,68}$Ga, $^{111}$In, and $^{60,61,62,63,64,67}$Cu are being investigated as imaging agents to identify hypoxic tissue in patients following a heart attack or stroke.

**Novel Gamma-Ray Imaging Concepts.** Organ-specific PET imagers developed by nuclear scientists at Jefferson Lab (JLAB) could improve detection sensitivity to certain types of tumors over devices that are currently commercially available. The increased sensitivity should lead to earlier detection capabilities and consequently to more effective treatments. Although the cost per unit is still high relative to current devices, enhancements in the effectiveness of the treatment should stimulate increased use of the new device and eventually lead to a reduced cost to the patient. An example of a mammogram taken with the device is shown in figure 5.5.

This same group at JLAB is making substantial improvements in the resolution of imagers that are based on the detection of single γ-rays by correcting for motion during imaging. These advances in medical imaging technologies were spurred by the experience of this group with radiation detection systems and real-time data acquisition and analysis in support of the basic nuclear physics research program at JLAB.

**Hyperpolarized Gas Magnetic Resonance Imaging.** Research on hyperpolarized magnetic imaging for medicine grew directly out of the developments of polarized targets for nuclear science research at both universities and national laboratories. Magnetic resonance imaging with polarized noble gases allows one to image the internal gas spaces of lungs or sinuses. Due to the limited water content in air spaces, these areas are usually poorly seen with hydrogen-based MRI. High-resolution imaging of the steady-state distribution of laser-polarized $^3$He or $^{129}$Xe enables direct measurement of gas densities in airways. Time-resolved images of gas densities provide functional assessments of gas flow, diffusion, and absorption. Since the last Long Range Plan, phase-2 clinical trials have begun with hyperpolarized xenon imaging in humans, and hyperpolarized xenon production systems have become commercially available. MRI images of hyperpolarized $^{129}$Xe gas in lung tissue of a human during a normal respiration cycle are shown in figure 5.6.

### Hadron-Beam Therapy

Over the past five years, hadron-beam therapy for cancer treatment has evolved from pilot projects at accelerator facilities to well-staffed operations at medical centers, and at dual-purpose facilities for physics research and oncology. The majority of treatment uses protons, but there has been some exploration of neutron therapy, as well. Fast neutrons have a biological advantage over X-rays and are used at neutron facilities at the University of Washington and at Harper Hospital in Detroit. The latter facility employs a small superconducting cyclotron (designed and constructed at the NSCL) that is rotated around the patient.

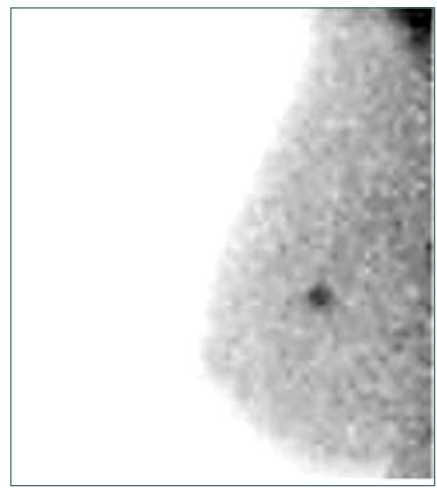

**Figure 5.5:** Image taken of a patient with the gamma-ray imager developed at JLAB. This image clearly shows a suspicious lesion, which was later confirmed by biopsy to be cancerous. A conventional mammogram taken of the same patient did not show evidence of a problem.

As of July 2005 over 48,000 patients had been treated with hadron beams for various forms of cancer worldwide.

**Proton-Beam Therapy.** Because charged particles deposit most of their dose in a narrow range, called the Bragg peak, radiotherapy with energetic protons (>230 MeV/nucleon) allows oncologists to design fine-tuned three-dimensional treatment plans and to achieve better localization of the radiation dose to the target area, compared to photon-beam treatments. This means they can use a larger dose in the tumor while limiting the patient's total dose—a feature that makes proton radiotherapy a superior technology to photon treatments for many applications, especially for pediatric cases. Proton radiotherapy was pioneered at the Harvard Cyclotron Laboratory, which was originally operated for nuclear physics research. Five proton therapy facilities in clinical settings are currently in operation in the United States. A 250 MeV proton therapy superconducting cyclotron was designed at the National Superconducting Cyclotron Laboratory using technology developed for building cyclotrons for basic nuclear physics research. Several of these cyclotrons have been built and installed by the ACCEL Corporation in Europe and are now being offered in the United States through Varian Medical Systems.

Future technological advances in proton radiotherapy are closely tied to innovations in accelerator concepts. Now under development, for example, is a prototype for a compact dielectric-wall accelerator with acceleration gradients of up to 100 MV/m. This new technology would significantly reduce the cost of accelerators, making them more affordable for a broader range of hospitals. Also, concepts of laser-based acceleration are being explored, as is the concept of a scanning proton beam.

**Heavy-Ion Beam Therapy.** Heavy-ion beams provide even better dose localization than do protons and have a similar biological advantage (as measured in the relative biological effectiveness) compared to neutrons. Because the ionization of heavy ions is larger than that of protons, the damage to tumor cells is more severe. This feature makes heavy-ion therapy the modality of choice for slow-growing tumors, which are resistant to proton and photon therapy.

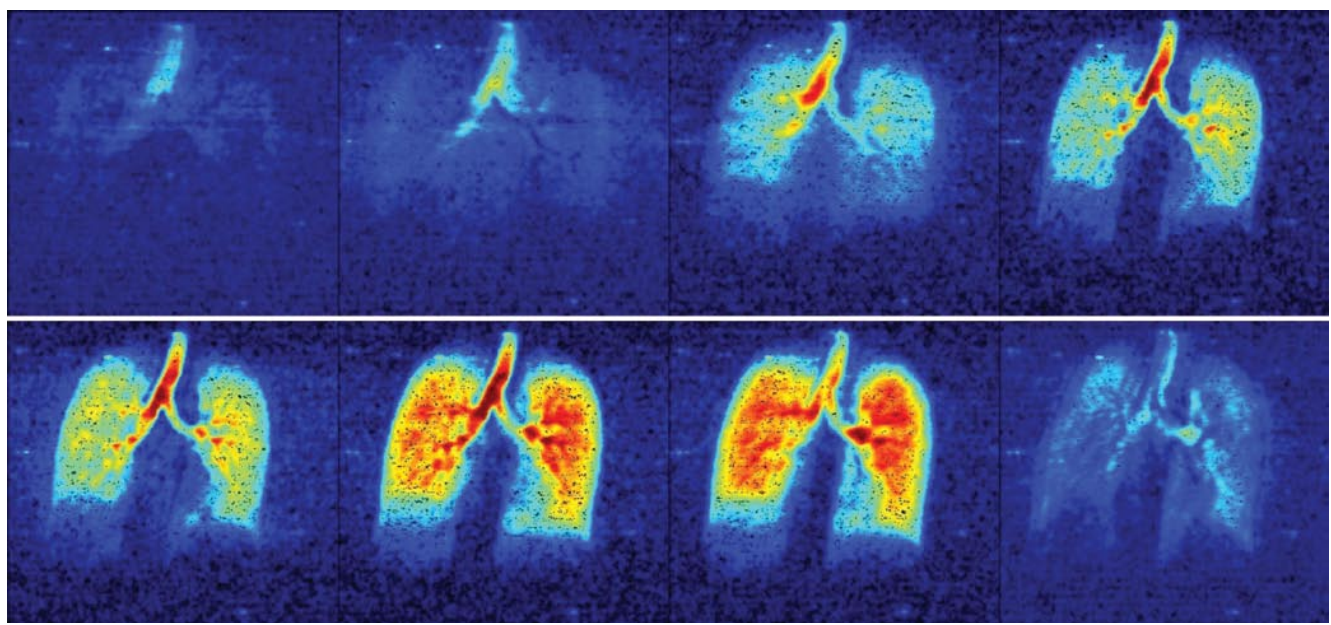

**Figure 5.6:** Magnetic resonance image of polarized $^{129}$Xe inhaled by a human. Each frame is a snapshot of the absorption of the polarized gas in the lung tissue during a normal respiration cycle. The $^{129}$Xe concentration is color coded with red indicating the highest concentration.

Heavy-ion radiotherapy was first developed at the Lawrence Berkeley National Laboratory in 1977. Now three synchrotron-based facilities are treating patients with carbon ion beams in Japan and Europe. In Japan, the Heavy Ion Medical Accelerator in Chiba and the Hyogo Ion Beam Medical Center provide treatments in a clinical setting. In Germany, a research therapy facility is in operation (GSI Darmstadt), and a fourth facility in a clinical setting is nearing completion (DKFZ Heidelberg, Germany). Additional heavy-ion therapy facilities have been proposed in Austria, Italy, and Germany.

# 6 Recommendations

This Plan is the fifth in a series that follows the tradition begun by NSAC nearly 30 years ago with the publication of the first Long Range Plan in December 1979. These Plans chronicle the enormous progress our field has made in its quest to understand the core of matter—the nucleons and the nuclei that they make.

Since the last Plan was published in 2002, nuclear scientists have made extraordinary discoveries, which are highlighted in the preceding chapters that discuss our science. But our community also has been frustrated with funding that has severely limited our ability to move forward in key areas. The last major construction project for nuclear science, RHIC, was launched over 20 years ago. The renewed emphasis on research in physical sciences by both the executive and legislative branches of the federal government comes at a very important time for nuclear science. New funding made available to our field will be used to revitalize the tools of our trade. The four recommendations of this Plan present a roadmap for new investment in facilities and the development of new experimental equipment that will allow us to continue to address the scientific challenges of our field into the future. These recommendations can be implemented in a staged approach over the course of the next decade. During this period, we must continue to operate effectively our existing flagship facilities to make optimal use of their capabilities. And just as researchers come from abroad to work in the United States, we must take advantage of the opportunities for forefront research that are being made available through investments in major new facilities in Europe and Asia. Implementing the four recommendations and sustaining a strong science program over the next decade fit within a budget profile consistent with the proposed federal reinvestment.

The four recommendations discussed below are given in priority order, starting with Recommendation I. Each of them represents a major advance in one of the subfields of our discipline—QCD and hadron structure, the phases of hadronic matter, the structure of nuclei and nuclear astrophysics, and fundamental symmetries and neutrinos. In addition, we discuss potential future directions and important initiatives that have been identified by the community. But the enterprise of nuclear science is much more than new facilities and initiatives. It is a coordinated effort of research, facility operations, education, stewardship, and research and development for the future. We highlight below some of the key aspects that make up the program.

> **RECOMMENDATION I**
>
> **We recommend completion of the 12 GeV CEBAF Upgrade at Jefferson Lab. The Upgrade will enable new insights into the structure of the nucleon, the transition between the hadronic and quark/gluon descriptions of nuclei, and the nature of confinement.**

A fundamental challenge for modern nuclear physics is to understand the structure and interactions of nucleons and nuclei in terms of quantum chromodynamics. Jefferson Lab's unique Continuous Electron Beam Accelerator Facility has given the United States leadership in addressing this challenge. Its first decade of research has already provided key insights into the structure of nucleons and the dynamics of finite nuclei.

Doubling the energy of the JLAB accelerator is equivalent to increasing the power of an optical microscope by a factor of two. The improved resolving power, together with the increased energy available for exciting the systems under study, will allow us to look more closely at the fundamental building blocks of nuclei. It will enable three-dimensional imaging of the nucleon, revealing hidden aspects of its internal dynamics. It will complete our understanding of the transition between the hadronic and quark/gluon descriptions of nuclei, and test definitively the existence of exotic hadrons, long-predicted by QCD as a consequence of the confinement of quarks and gluons. Through its studies of parity violation, it will provide low-energy probes of physics beyond the Standard Model complementing anticipated measurements at the highest accessible energy scales.

The upgrade of Jefferson Lab's accelerator to 12 GeV was a major recommendation in the 2002 Plan. While the technology needed to carry it out has been available for a number of years, funding limitations have delayed it. The Upgrade is currently scheduled to begin toward the end of 2008, and it is imperative that it proceed with sufficient funding to complete it as soon as possible. Investments in new experiments with the upgraded facility are being planned by both U.S. and foreign collaborations. Further delaying the start and the completion of the facility not only risks the loss of world leadership in this important area of physics but could also jeopardize planned investments by our colleagues abroad.

### RECOMMENDATION II

**We recommend construction of the Facility for Rare Isotope Beams, FRIB, a world-leading facility for the study of nuclear structure, reactions, and astrophysics. Experiments with the new isotopes produced at FRIB will lead to a comprehensive description of nuclei, elucidate the origin of the elements in the cosmos, provide an understanding of matter in the crust of neutron stars, and establish the scientific foundation for innovative applications of nuclear science to society.**

Nuclei are rich laboratories for the study of quantum phenomena, bridging the gap between quarks and gluons and the larger-scale worlds of nano- and microstructures. Nuclei are the power source of stars, their reactions are the agent for the chemical evolution of the universe, and their decays serve an increasing role in medicine. Their use in controlled nuclear fission and, in the more distant future, thermonuclear fusion shows increasing promise for meeting the world's low-carbon energy needs.

We now can see a path toward a more comprehensive understanding of nuclei, their interactions, and their role in the cosmos, based on the laboratory production of key rare isotopes that have previously existed only in the most violent explosions of stars or in neutron stars. Moreover, we have developed a theoretical roadmap that will use data from these rare-isotopes to link our understanding of the strong force to a description of complex nuclei. Rare isotopes are critical to this endeavor because they will exhibit new phenomena and amplify novel aspects of nuclear physics that have gone undetected in the limited span of isotopes thus far available. They will also play a key role, complementary to other approaches, in testing the fundamental symmetries of nature. Furthermore, access to a broad range of new isotopes and decay properties is a prerequisite for progress in many cross-disciplinary areas in basic sciences, medicine, national security, and other societal applications.

A recent report from the U.S. National Academies on the science of rare isotopes states: "[S]tudies of nuclei and nuclear astrophysics constitute a vital component of the nuclear science portfolio in the U.S.… The facility for rare-isotope beams [FRIB] envisaged for the United States would provide capabilities unmatched elsewhere that will directly address the key science of exotic nuclei… The committee concludes that a next generation, radioactive beam facility, of the type embodied in the U.S. FRIB concept represents a unique opportunity to explore the nature of nuclei under conditions that previously only existed in supernovae and to challenge our knowledge of nuclear structure by exploring new forms of nuclear matter."

To launch the field into this new era requires the immediate construction of FRIB with its ability to produce groundbreaking research. An NSAC subcommittee considered the options for FRIB that could fit within future budget projections and recommended a facility, based on a 200 MeV/nucleon, 400 kW superconducting heavy-ion driver linac, that will have outstanding capabilities for fast, stopped and reaccelerated beams and will be complementary to other facilities existing and planned, worldwide. As construction proceeds, it is essential to continue to break new ground in rare-isotope science and to continue to develop new methods and techniques through the effective utilization of the current user facilities at NSCL, HRIBF, and ATLAS to ensure that the nation is in the best position to realize the potential of the new facility.

### RECOMMENDATION III

**We recommend a targeted program of experiments to investigate neutrino properties and fundamental symmetries. These experiments aim to discover the nature of the neutrino, yet-unseen violations of time-reversal symmetry, and other key ingredients of the New Standard Model of fundamental interactions. Construction of a Deep Underground Science and Engineering Laboratory is vital to U.S. leadership in core aspects of this initiative.**

The discovery of flavor oscillations in solar, reactor, and atmospheric neutrino experiments—together with unexplained cosmological phenomena such as the dominance of matter over antimatter in the universe—call for a New Standard Model of fundamental interactions. Nuclear physicists are poised to discover the symmetries of this new model through searches for neutrinoless double beta decay and electric dipole moments, determination of neutrino properties and interactions, and precise measurements of electroweak phenomena. Among the fundamental questions that must be answered are: Is there a conserved symmetry for lepton number? What is the additional source of CP violation needed to account for the predominance of visible matter? What are the masses of the neutrinos? Are there additional interactions that were important in the early universe, such as those of supersymmetry? What is the origin of parity violation? How is gravitation incorporated into the model?

This initiative builds on the third recommendation of the 2002 Plan that called for the "immediate construction of the world's deepest underground science laboratory." Starting in 2003, NSF began developing plans to construct the multidisciplinary Deep Underground Science and Engineering Laboratory (DUSEL). Four sites for DUSEL were considered in a recent competition, and the Homestake gold mine was chosen. The next step is a full proposal for DUSEL at the Homestake mine. Following submission to NSF, the proposal will be considered by the National Science Board and if approved submitted to Congress for funding.

DUSEL and the experiments that will be carried out in it are vital components of the new initiative. Investments in complementary experiments at accelerator facilities are also integral to the initiative's discovery-oriented program. Particle-beam experiments will exploit new capabilities at the Fundamental Physics Neutron Beam Line at the Spallation Neutron Source at Oak Ridge National Laboratory and the 12 GeV CEBAF Upgrade, while other experiments will capitalize on existing capabilities at Brookhaven, Los Alamos, and other facilities.

> **RECOMMENDATION IV**
>
> **The experiments at the Relativistic Heavy Ion Collider have discovered a new state of matter at extreme temperature and density—a quark-gluon plasma that exhibits unexpected, almost perfect liquid dynamical behavior. We recommend implementation of the RHIC II luminosity upgrade, together with detector improvements, to determine the properties of this new state of matter.**

This recommendation builds on the outstanding success of the RHIC facility since it began operation in 2000. The striking discoveries of the first five years at RHIC compel us to carry out a broad, quantitative study of the fundamental properties of the quark-gluon plasma. This can be accomplished through significant increases in collider luminosity, detector upgrades, and advances in theory, which also create further discovery potential.

The RHIC II luminosity upgrade, using beam cooling, provides a 10-fold increase in collision rate, which enables measurements using uniquely sensitive probes of the plasma such as energetic jets and rare bound states of heavy quarks. Already measurements at RHIC using another cooling technique—stochastic cooling—show promise that beam cooling will lead to the projected luminosity increases. In order to utilize the higher counting rates, upgrades are needed for the two major RHIC detectors—PHENIX and STAR. These upgrades make important new types of measurements possible and extend significantly the physics reach of the experiments.

Achieving a quantitative understanding of the quark-gluon plasma also requires new investments in modeling of heavy-ion collisions, in analytic approaches, and in large-scale computing, aimed at quantifying theoretical uncertainties.

## FURTHER INTO THE FUTURE

The exchange of gluons between quarks mediates the color force that provides the internal binding in all of nuclear matter. Interactions among gluons determine the unique features of strong interactions: without gluons there are no protons, no neutrons, and no atomic nuclei. In fact, the mass of all visible matter in the universe arises predominantly from gluon self-interactions. Even though our understanding of QCD has advanced rapidly over the past decade, gluon properties in matter remain largely unexplored.

Recent theoretical breakthroughs and experimental results suggest that both nucleons and nuclei, when viewed at high energies, appear as dense systems of gluons, creating fields whose intensity may be the strongest allowed in nature. The emerging science of this universal gluonic matter drives the development of a next generation of high-luminosity

Electron-Ion Collider (EIC). The EIC's ability to collide high-energy electron beams with high-energy ion beams will provide access to those regions in the nucleon and nuclei where their structure is dominated by the glue. Moreover, the implementation of polarized beams in the EIC will give unprecedented access to the spatial and spin structure of gluons in the proton.

Thus an EIC with polarized beams has been embraced by the U.S. nuclear science community as embodying the vision for reaching the next QCD frontier. EIC would provide unique capabilities for the study of QCD well beyond those available at existing facilities worldwide and complementary to those planned for the next generation of accelerators in Europe and Asia. While significant progress has been made in developing concepts for an EIC, many open questions remain. Realization of an EIC will require advancements in accelerator science and technology, and detector research and development. The nuclear science community has recognized the importance of this future facility and makes the following recommendation.

- **We recommend the allocation of resources to develop accelerator and detector technology necessary to lay the foundation for a polarized Electron-Ion Collider. The EIC would explore the new QCD frontier of strong color fields in nuclei and precisely image the gluons in the proton.**

## INITIATIVES

Many new initiatives and facility upgrades were proposed by the nuclear science community as the components of this Plan were assembled. While all of them were well founded and merited serious consideration, funding constraints severely limited those that could be accommodated along with our four principle recommendations. Three initiatives were chosen for special consideration. They should be implemented as funding allows.

**Nuclear Theory**

Theory is a critical part of the nuclear science enterprise. Progress and new discoveries invariably come from the interplay between new experiments and new theoretical insights. Since 1990, for example, the Institute for Nuclear Theory has served as a catalyst for new developments by bringing together experimental and theoretical experts to work on problems at the forefront of nuclear physics. More generally, recent progress in nuclear theory has come from the development of new theoretical tools, such as effective field theory approaches, *ab initio* methods in determining the properties of nuclei up to the limits of their existence, novel applications from string theory for understanding RHIC phenomena, and lattice QCD approaches to the nuclear force—the latter being the inspiration for a new effort, launched by the U.S. nuclear science program in FY2006 to carry out lattice QCD calculations using the power of massively parallel computers. Research in lattice QCD as well as in computational nuclear structure and nuclear astrophysics is benefiting greatly from DOE's investment in its multidisciplinary program Scientific Discovery through Advanced Computing. Clearly, high-performance computing has become an indispensable tool for nuclear physics, allowing exploration of the internal structure and interactions of hadrons, the properties of hot and dense matter, the explosion mechanism of supernovae, the internal structure of nuclei, and the nuclear equation of state. Ensuring access to world-class computing resources is a prerequisite for nuclear theory to realize its potential. We strongly endorse the ongoing programs at DOE and NSF to provide resources for advanced scientific computing.

Looking to the future, meanwhile, a vigorous theory program is indispensable for realizing the full potential of the investments laid out in this Plan. Beyond continued strong support for a healthy base program, this requires additional investments targeted at theoretical problems of critical importance to the Long Range Plan goals. As an efficient approach to solve those problems, **we recommend the funding of finite-duration, multi-institutional topical collaborations initiated through a competitive peer-review process.** In addition to focusing efforts on important scientific problems, these initiatives are intended to bring together the best in the field, leverage resources of smaller research groups, and provide expanded opportunities for the next generation of nuclear theorists.

**Accelerator Research and Development**

Since the development of the first electrostatic accelerators by Van de Graaff and the first cyclotrons by Lawrence, accelerators have been essential to progress in nuclear science. Today, particle beams with energies ranging from less than 100 keV to nearly 40,000 GeV are used to explore the broad range of science discussed in this Plan. Superconducting radiofrequency technology for low-velocity ion beams was first demonstrated and used for nuclear physics at Argonne

National Laboratory. The large-scale development and utilization of superconducting radiofrequency acceleration technology has been pioneered in the United States by scientists at Jefferson Lab. The development of high-energy polarized proton beams at BNL has produced the world's first polarized proton collider at RHIC. Particle accelerators also are used for a wide range of applications that benefit society, e.g., in nuclear medicine and in the screening of cargo for dangerous materials using nuclear techniques.

Discoveries in nuclear science rely significantly on particle accelerators. Progress in accelerator science and technology is essential for the development of new capabilities that will enable future discoveries. Advances in accelerator science and technology and the ability to attract students to this field will be strongly enhanced by **targeted support of proposal-driven accelerator research and development supported by DOE and NSF**.

**Gamma-Ray Tracking**

During the past decade, much of the progress in our understanding of nuclear structure has been stimulated by measurements carried out with $4\pi$ gamma-ray detector arrays, such as Gammasphere in the United States, that feature high efficiency and excellent energy resolution. Gammasphere has proven ideal for experiments at low energies using stable beams and targets, but it quickly loses its ability to resolve closely spaced gamma-ray decays that occur when decaying ions have a significant recoil velocity. Early in this decade, a new detector technology that allows for tracking of gamma rays was being developed. With this, corrections for ion recoil velocity could be made, and thus the resolution and efficiency could be recovered even with large Doppler shifts in the gamma-ray energies. A $1\pi$ detector array, GRETINA, is now being built based on this new technology, and its first element has verified that the device will work as planned.

The implementation of a $4\pi$ gamma-ray tracking detector based on this new technology will revolutionize gamma-ray spectroscopy and provide sensitivity improvements of several orders of magnitude for studies in nuclear structure, nuclear astrophysics, and weak interactions. GRETA, a $4\pi$ spectrometer, is urgently needed to fully exploit the scientific opportunities at existing stable and radioactive beam facilities, as well as at FRIB. Thus **the construction of GRETA should begin upon successful completion of GRETINA. This gamma-ray energy tracking array will enable full exploitation of compelling science opportunities in nuclear structure, nuclear astrophysics, and weak interactions.**

## RESEARCH AND OPERATIONS

Continued progress in nuclear science depends critically on a healthy research program—both at our national laboratories and at universities—and the operation of the broad range of facilities needed to carry out that research. While the four recommendations in this Plan call for new investments for the future, **we must maintain a balance between funding those investments and providing support for the ongoing program**. This was a major consideration that led to the budget strategy, which is discussed in the next section, for implementing these recommendations.

The U.S. research program in nuclear science is broad based, and includes: individual faculty members, large university groups, and groups at national laboratories. Some of the larger groups operate their own accelerator facilities with an associated research infrastructure, while others, without accelerators, have significant research infrastructure in place. These groups play an essential role in the research enterprise. University researchers often take the lead in planning and executing the research programs of the national user facilities.

The integration of research and teaching at universities provides a natural environment for the education and training of the future scientific workforce. From the smallest to the largest, university laboratories and their faculty often provide the first exposure to nuclear science and its applications for the broader community of university students. Additionally the proximity to scholars in other disciplines provides an interface with other fields that enables and often facilitates interdisciplinary advances.

University-based accelerator facilities support a compelling and diverse portfolio of research in nuclear structure, nuclear reactions, nuclear astrophysics and fundamental symmetries. The infrastructure at universities, either in support of operations and research at local accelerator facilities or participation in development of detectors, instrumentation and equipment for collaborative research projects, enables university scientists to contribute significantly to new initiatives in the field. Support for the university facilities and research infrastructure is essential to enable them to remain significant partners in the nation's scientific endeavor.

An equally essential component of the U.S. nuclear science program is the operation of the major user facilities. Strategic investments by the United States in new and upgraded facilities have positioned the nation in a world leadership role in many areas of nuclear science. Exploiting the extraordinary opportunities for scientific discoveries made possible by these investments requires the effective operation of these facilities. Specifically, it is critical to sustain effective facility operations at CEBAF and RHIC in support of their world-leading research in QCD at levels that also allow facility upgrades to proceed in a timely way; to operate NSCL at optimal levels, taking advantage of its world-leading capabilities with fast rare-isotope beams; and to invest in ATLAS and HRIBF user facilities with their unique low-energy heavy-ion and rare-isotope beams in order to preserve scientific output and educate young scientists vital to meeting national needs.

## EDUCATION

Education and outreach are central to the mission of both DOE and NSF. They are the fundamental underpinnings that support the mandate of the agencies to advance the broad interests of society (e.g., in academia, medicine, energy, national security, industry, and government) and to help ensure U.S. competitiveness in the physical sciences and technology. Similarly, education and outreach are key components of any vision for the future of our field.

The United States is facing a potentially serious shortage of qualified workers in pure and applied nuclear science research, nuclear medicine, nuclear energy, and national security. In 2004, the Nuclear Science Advisory Committee recommended a significant increase in the number of new nuclear science Ph.D.s during the next decade, based on expected needs and the results from a comprehensive survey of the nuclear science workforce over the last decade. Increasing the number of Ph.D. nuclear scientists, especially U.S. citizens, requires introducing students to nuclear science and its research before they start graduate school, and requires increased participation from the full diversity of backgrounds. This needs to be addressed at the undergraduate level.

An effective program of nuclear science outreach is essential to ensure a broad, basic knowledge of nuclear science in U.S. society, enabling informed decisions by individuals and decision-making bodies on a wide range of important topics, including nuclear medicine, energy policy, homeland security, national defense and the importance and value of nuclear science research. At present, the public, and even scientists in other disciplines, are often uninformed or misinformed about nuclear science and its benefits. It is imperative for the health of the field that the excitement and value of nuclear science be effectively transmitted to our K–12 students, teachers, and the general public.

In order to increase the number and diversity of students interested in pursuing a graduate degree in nuclear science and to effect a change in the public perception of our field, the nuclear science community should **enhance existing programs and develop new ones that address the goals of increasing the visibility of nuclear science in undergraduate education and the involvement of undergraduates in research, and develop and disseminate materials and hands-on activities that demonstrate core nuclear science principles to a broad array of audiences.**

# 7 Resources

## INTRODUCTION

The spectacular successes of the U.S. nuclear science program in the past 30 years are directly attributable to investments made by the federal government with strategic guidance provided by NSAC, which have enabled the community to exploit new scientific opportunities. Through this process, new facilities identified as high priority for their potential science discoveries have been built, and older ones have closed. Over time there has been a consolidation of resources around major user facilities as the complexity of instrumentation has increased. Yet small facilities still thrive, often by tackling very difficult problems that are best addressed by a combination of more modest-scale instrumentation and long, complex measurements. This mixture of opportunities from small to large continues to serve our field well.

Over the past decade, nuclear science has been guided by two Long Range Plans—one published in 1996 and the other in 2002. The top priority of the field in both plans was to increase research funding and operations support to use existing facilities effectively. During this period, two new facilities came online—CEBAF at Jefferson Laboratory and RHIC at the Brookhaven National Laboratory—and one laboratory, the NSCL at MSU, carried out a major upgrade. Important new discoveries have come from the increased operations support of these flagship facilities, proving the wisdom of the 1996 and 2002 recommendations. But the emphasis on operations has come at a price to the field. Funding has been far below levels deemed necessary for a healthy nuclear science program in the past two plans. This funding shortfall has eroded the support for research at universities and national laboratories, and has not allowed major new initiatives essential for the vitality of the field.

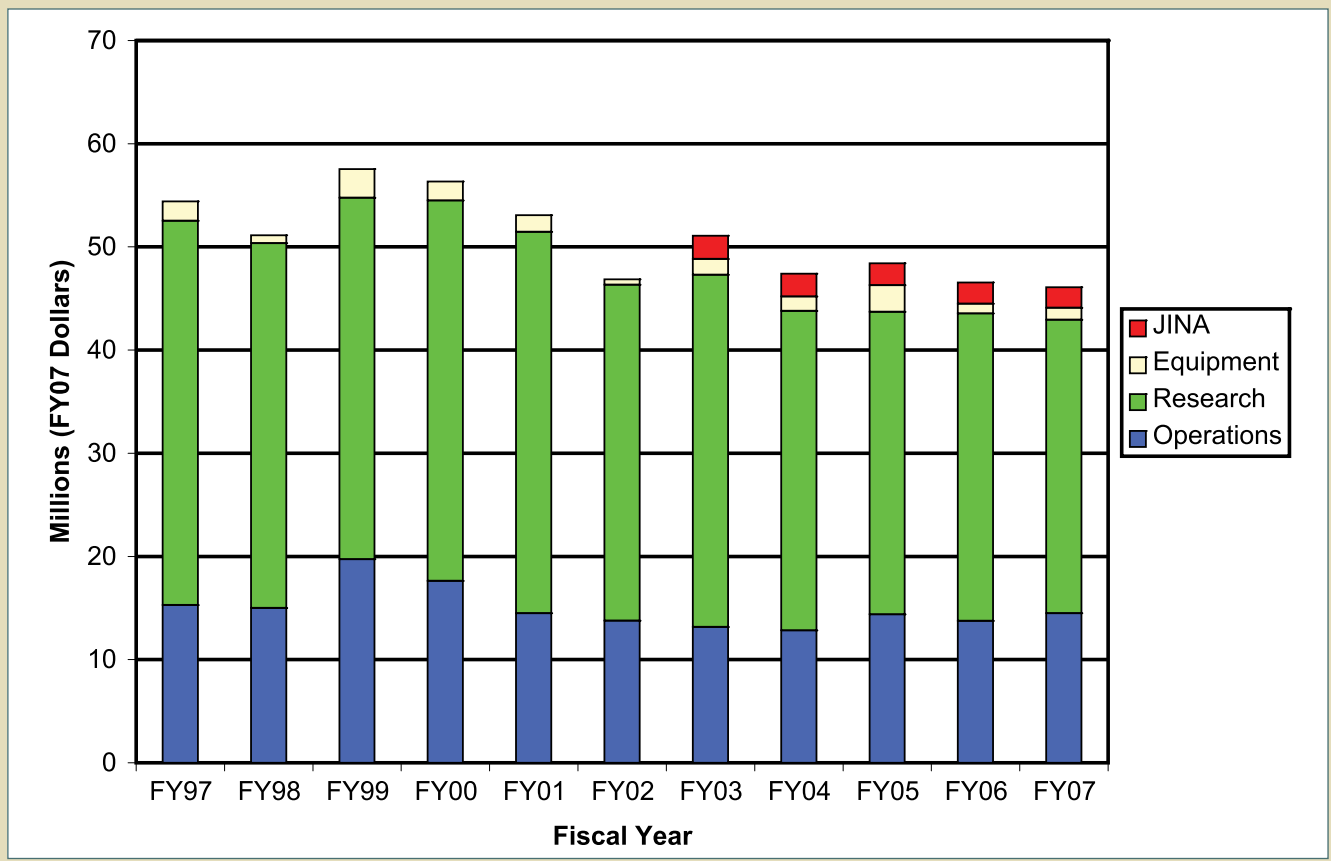

**Figure 7.1:** National Science Foundation funding history over the last decade inflation corrected to FY2007 dollars.

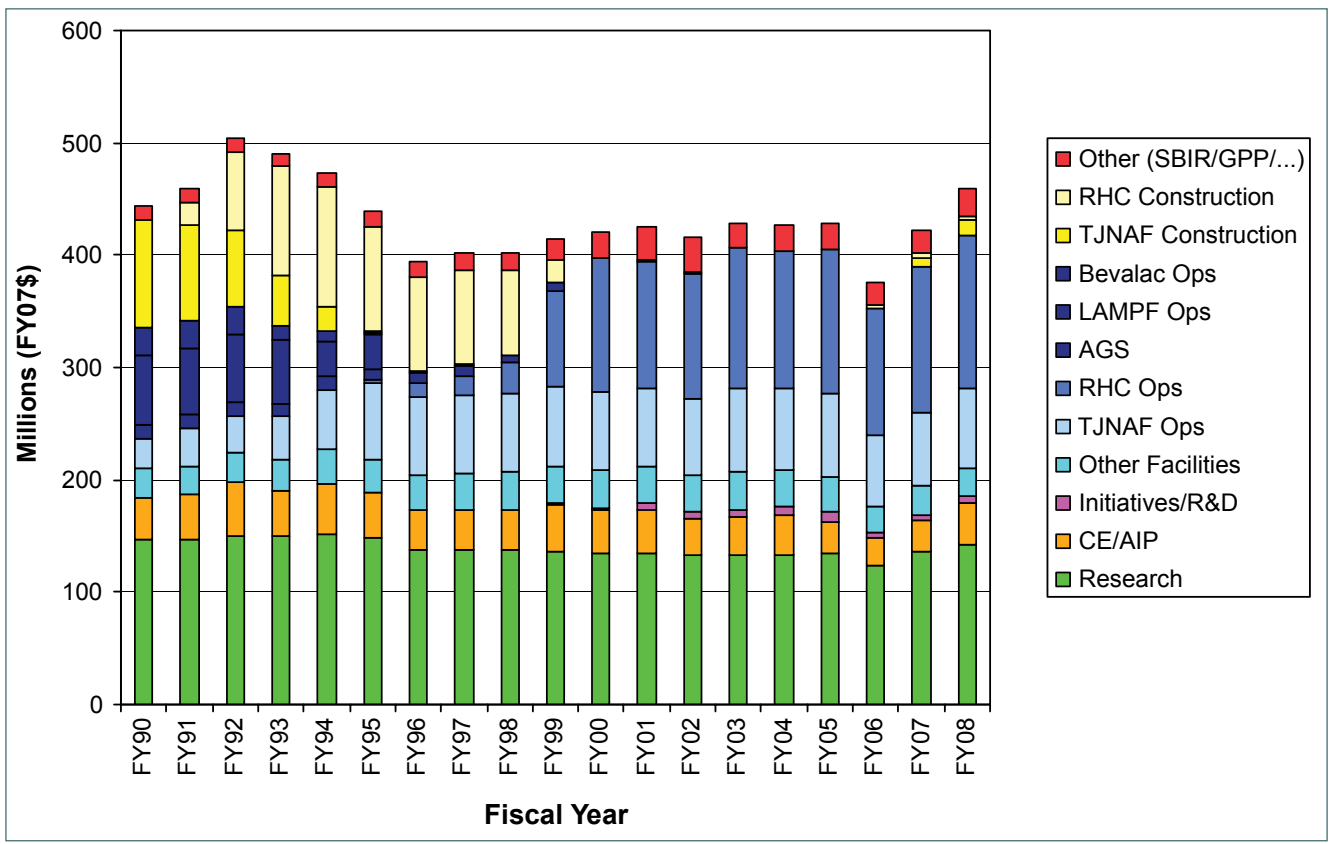

**Figure 7.2:** Department of Energy funding history in FY2007 dollars.

Federal funding for nuclear science comes primarily from DOE and NSF. Over the past decade, the budget for nuclear science research and operations at NSF has been close to a constant dollar level. This has had the unfortunate consequence of resulting in a significant drop in real support for the field after the effects of inflation are included, as shown in figure 7.1. As figure 7.2 indicates, DOE budgets for nuclear science over the last decade have been close to constant effort (with the exception of the disastrous FY2006 budget), effectively growing with inflation. This followed a period of declining budgets during the early 1990s as construction was completed at CEBAF. The major drop in funding in FY2006, coupled with poor out-year projections for improvement, forced our field, with guidance provided by NSAC, to reluctantly consider the possibility of closing one of its major facilities prematurely.

Nuclear science was not the only field with funding problems in FY2006. Indeed most areas in the physical sciences were strapped with declining budgets and out-year projections of constant effort, at best. The outlook changed dramatically, though, with the publication, in the fall of 2005, of the National Academy's report *Rising Above the Gathering Storm*. In contrast to many previous studies warning of problems engendered by stagnant funding for the physical sciences, and the dire consequences for future economic growth, the National Academy study stimulated an immediate response. Follow-up discussions with leaders in Congress and the nation's industries led President Bush to announce the American Competitiveness Initiative in January 2006. The administration requested in its FY2007 budget proposal that Congress double the budget for research in the physical sciences over the next decade. An increase of this scale has had bipartisan support in Congress for several years. This is best evidenced by letters to Appropriations Committee chairs that have been signed by a majority of members in the House and Senate calling for increased funding for the DOE's Office of Science and NSF. The FY2007 budget resolution, which was passed by Congress in March 2007, restored much of the budget losses of FY2006, and the budget request for FY2008 places the Office of Nuclear Physics (ONP) on the budget-doubling trajectory of the American Competitiveness Initiative.

As a consequence of the emphasis on supporting the base program (both research and facility operations), combined with tight budgets in the early years of this decade, no major construction in nuclear science has occurred in the United States since the completion of RHIC near the end of the last decade. In contrast, large new facilities for nuclear science are being built in Europe and Asia. This nearly 10-year period without major new construction is unprecedented in the history of U.S. nuclear science. The inevitable consequence is a program that is falling behind growing efforts in other parts of the world. If the United States is to maintain its world-leading position in nuclear science, it is imperative that we immediately begin making investments in new tools, while continuing to pursue scientific opportunities at our world-leading facilities.

## PRESENT PROGRAM

The nuclear science program today encompasses research in four related subfields: neutrinos and fundamental symmetries; structure of atomic nuclei and nuclear astrophysics; quantum chromodynamics (QCD) and the structure of hadrons; and the phases of QCD. Federal funding for the program in FY2007 was $423 million at DOE and $44 million at NSF. This represented a substantial increase for DOE over the FY2006 budget allocation, but after correction for inflation, it corresponds to nearly constant effort funding compared to the DOE FY2005 budget.

With the inherent interconnections between the four subfields, funding boundaries between them are diffuse. To provide an indication of the funding in each area, the approximate distribution of resources in the DOE and NSF FY2007 budgets, including operations support and theory, is shown in figures 7.3 and 7.4. In FY2007, operations support was provided for five user facilities, four funded by DOE—ATLAS at ANL, CEBAF at JLAB, HRIBF at ORNL, and RHIC at BNL—and one funded by NSF—NSCL at MSU. In addition, the two agencies supported low-energy accelerator-based programs at six university laboratories and one national laboratory.

In the mid 1990s, the DOE nuclear physics budget declined as construction was completed at CEBAF. Support for operations grew as RHIC was completed in the late 1990s, and reached about 60% of the DOE ONP budget. Operations support has stayed at approximately this same level for the past decade, even with the closure of two DOE national user facilities (Bates at MIT and the 88-Inch Cyclotron at LBL) following the last Long Range Plan. The combined research and operations funding at the ONP accounts for about 90% of the budget. NSF also closed one of its two user facilities in 2002. By FY2007, operations support at its remaining user facility was about 34% of the total NSF nuclear physics budget.

Early in this decade, NSF created Physics Frontier Centers to foster interdisciplinary research. The Joint Institute for Nuclear Astrophysics (JINA) was formed in the fall of 2002 as a Physics Frontier Center to stimulate closer interactions between astronomers and nuclear astrophysicists. The synergy between the members of JINA who represent the two fields has influenced both the requests for observational time and the nuclear physics experiments being carried out to understand the observations. Around the same time the Scientific Discovery through Advanced Computing (SciDAC) program began at DOE. This program brings together scientists and mathematicians, often with very different expertise, to work on important problems that require intensive computational effort. Problems in nuclear physics being attacked by SciDAC-funded teams include Lattice QCD calculations, simulations of stellar evolution and explosions, and nuclear structure and nuclear reaction calculations. This cross-disciplinary program has proven to be very valuable to the nuclear physics community.

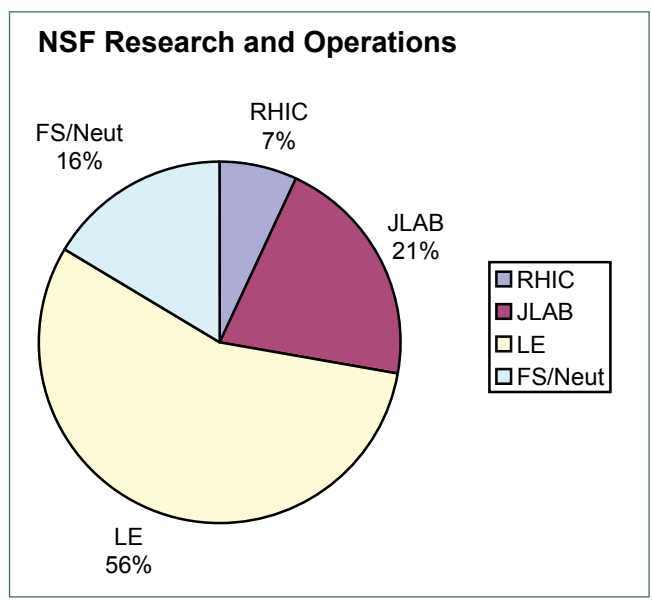

**Figure 7.3:** NSF budget breakdown for FY2006 by category—QCD and hadrons (JLAB), phases of QCD, fundamental symmetries and neutrinos (FS/Neut) and low-energy (LE) studies of nuclear structure and astrophysics.

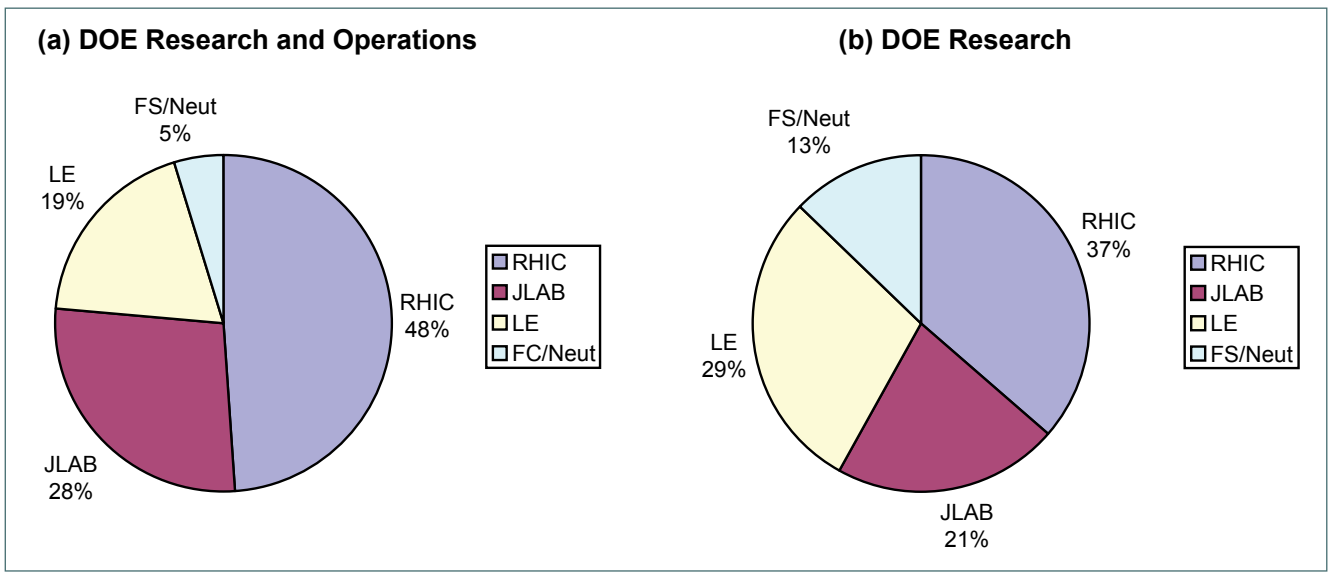

**Figure 7.4:** DOE budget breakdown for FY2007 by category—QCD and hadrons (JLAB), phases of QCD, fundamental symmetries and neutrinos (FS/Neut) and low-energy (LE) studies of nuclear structure and astrophysics—for research and operations (a) and just research (b).

## FUTURE REQUIREMENTS

The four recommendations in this Plan can be accommodated under a funding profile consistent with doubling the DOE ONP budget in actual year dollars over the next decade together with NSF funding for DUSEL, including some of the equipment for experiments to be carried out in DUSEL. The ONP budget assumed here is consistent with the request made in January 2006 to double the DOE Office of Science budget. The projected DOE ONP funding, including the major construction projects, is shown in figure 7.5. Within this budget profile, operations support would increase a bit slower than the assumed inflation rate through 2015 when it would begin to grow as new facilities come online. Research funding would have a very small but steady growth over inflation during the entire period. The majority of the growth in overall funding coming from the proposed doubling of the ONP budget would be invested in the construction of new tools that are essential to keep the field competitive.

Three major construction projects, corresponding to Recommendations I, II, and IV in this Plan, are included in the ONP profile. The first of these is the upgrade of CEBAF to 12 GeV. A conceptual design report for the upgrade was completed in 2006, and funding is included in the FY2008 president's budget request to complete project engineering and design. Construction funding is scheduled to begin in FY2009. The upgrade involves adding more superconducting accelerator cavities to the existing machine, upgrading the recirculating bending magnets to handle higher energies, adding a new experimental hall, and upgrading some of the instrumentation in the present experimental halls. Most of the construction would be finished by the end of 2013 and the Upgrade completed in 2015.

The second of the three projects is the construction of a next-generation facility for rare-isotope beams (FRIB). This has been a high priority for the nuclear science community since early in the decade. Initial plans were to construct a facility known as the Rare Isotope Accelerator (RIA). The central component of RIA was a superconducting heavy-ion linac that would accelerate uranium beams to 400 MeV/nucleon at a beam power of 400 kW. In 2006, based on departmental and national priorities, budgetary constraints, and current international efforts in rare-isotope research, DOE concluded that the construction of RIA, as originally conceived, would not be the most effective way to accomplish national nuclear physics research goals. Since then, a new concept has been developed for a 200 MeV/nucleon heavy-ion linac driver that would still produce a beam power of 400 kW for uranium, thanks to recent advances in ion-source technology. Research and development specific to the facility and conceptual design activities are planned to start in FY2009. Funding for construction would begin to grow in FY2013 as the construction funds needed to complete the CEBAF Upgrade drop.

The third project shown in the ONP profile is the upgrade of the RHIC accelerator/storage rings and its detectors. The upgrade (RHIC II) involves adding beam cooling to the existing rings and benefits from ongoing upgrades to the two major detectors—PHENIX and STAR. The result will be a significant increase in the luminosity the collider will deliver to detectors that can operate efficiently at the higher rates. The schedule for implementing the upgrade is driven by research and development on beam cooling, which is now underway.

The third recommendation of this Plan calls for a New Standard Model initiative. The new initiative encompasses a suite of important experiments. Construction for some of these is already underway at the new Fundamental Neutron Physics Beam Line at the ORNL Spallation Neutron Source. Others will be carried out at nuclear physics laboratories in the United States. Several critical experiments, such as new searches for neutrinoless double beta decay and a new detector for solar neutrinos, must be housed in an underground laboratory. Funding to carry out the New Standard Model initiative would be shared between DOE and NSF. If DUSEL is approved for construction, funds for an initial suite of experiments would be included in the cost of the project. Without DUSEL, other mechanisms would need to be found

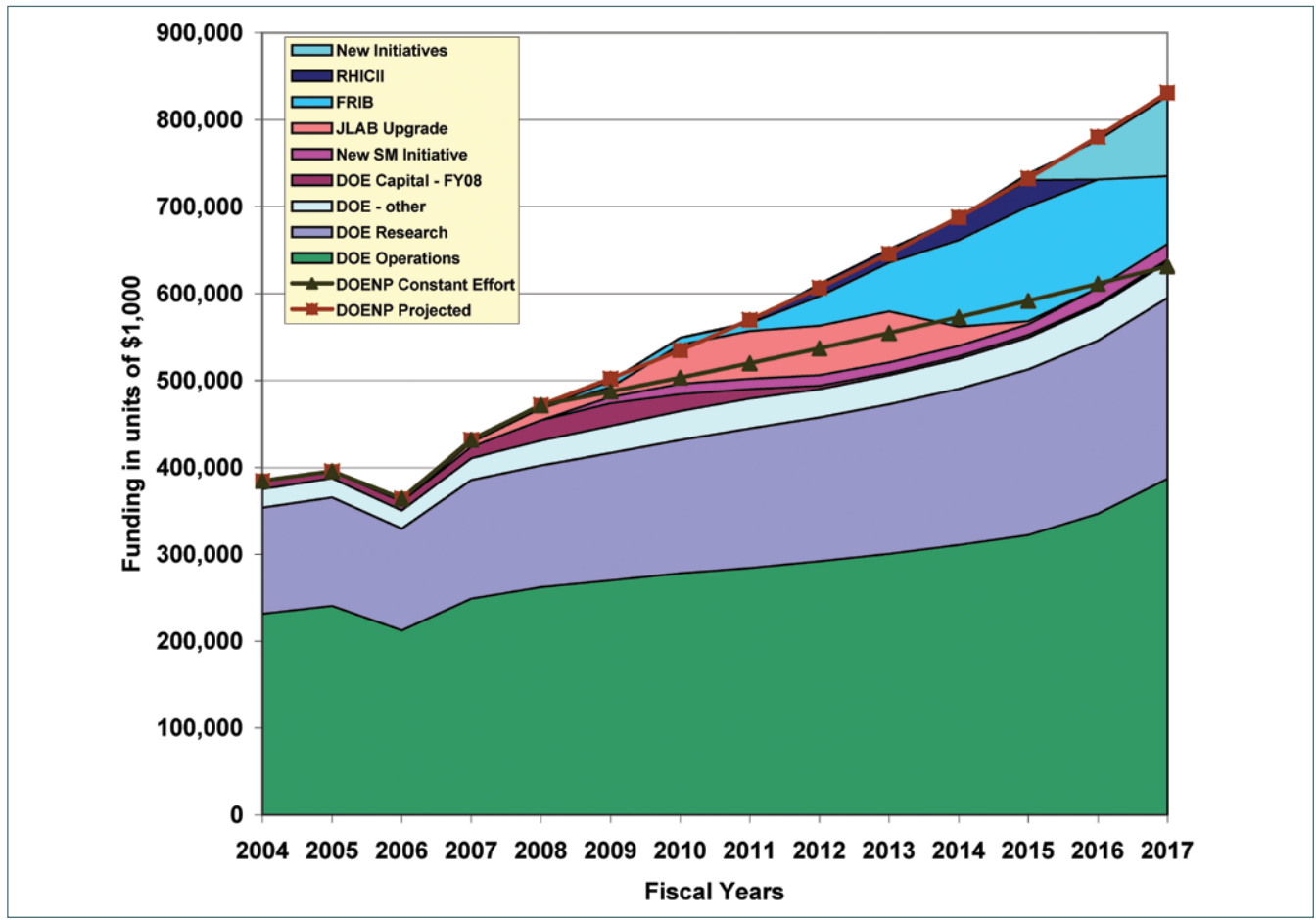

**Figure 7.5:** DOE budget projections (in as-spent dollars). The line that connects the triangles represents a constant effort budget.

to obtain the roughly $100 million in assumed NSF funding for this initiative.

The Long Range Plan as presented envisions a basic-research workforce in 2018 that is comparable in size to the present one. As detailed in the section on education, there is a need to increase the number of Ph.D.s produced per year in nuclear science by about 20% in order to meet the demand for a trained nuclear science workforce in the 21st century. This growth can be accommodated by the slow growth planned for research funding. However, with the emphasis on major construction in this Plan, it will not be possible under the present budget scenario to make a significant investment to rebuild the research infrastructure at universities that has been lost due to the slow erosion of research funding over the past two decades. Much of the training of the additional Ph.D.s will have to take place with involvement in experiments at major facilities. While the minimal increment in support for university infrastructure is regrettable, the community recognizes that investments in new and upgraded facilities must be the highest priority for the field at this time.

## THE EFFECT OF A CONSTANT EFFORT BUDGET

As part of the charge to develop this Plan, NSAC was asked to provide information on "what the impacts are and priorities should be, if funding available provides constant level of effort (FY2007 President's Budget Request) in the out-years (FY2008–2017)." Since starting work on this Plan, the President's Budget Request for FY2008 has been submitted to Congress. For purposes of discussion, the FY2008 request level has been used for projection, at constant effort, into the out-years. The funding profile under this constant effort scenario is included in figure 7.5.

It is clear that a constant-effort funding scenario falls far below the level needed to carry out the major projects recommended in this Plan. Maintaining the present level of operations and research funding would require that the work to complete the 12 GeV CEBAF Upgrade be extended well beyond FY2013. This would delay the start for construction of a new facility for rare-isotope beams by two to three years, as there would be no significant funding for its construction available before FY2015. FRIB would be further delayed due to the limited construction funding available each year and cost increases resulting from inflation. Following such a path would not lead to FRIB until well beyond the 10-year horizon of this Plan. With new projects now underway in Europe and Asia, a delay of this magnitude would severely cripple the U.S. program in a core component of our field. This approach also would provide no flexibility to carry out the upgrade at RHIC or to make an investment in major elements of the New Standard Model initiative through DOE, undercutting U.S. leadership in these areas of science as well.

If budgets fall far below those needed to implement this Plan, major decisions will need to be made for the future directions of U.S. nuclear science. The staged approach of upgrades and new facility construction that has been put forward already delays projects ready to be carried out sooner if funding were available. The U.S. nuclear science program will erode without significant new capital investments. At present, this need is most acute in research programs that require intense beams of rare isotopes—essential for advancing our understanding of both the physics of atomic nuclei and nuclear astrophysics. Maintaining U.S. leadership position in this vital subfield requires the generation of significant new capabilities for rare-isotope beams on a timely basis. If budgets were restricted to constant effort, proceeding with any of the new initiatives presented in this Plan would be possible only by reduced funding for operations and research, with clear adverse and potentially dire consequences for core components of the U.S. nuclear physics program. Since nuclear science, like all areas of basic research, evolves in time, it is impossible now to forecast what strategy would minimize damage to the field if future budgets dictated such stark choices.

## A FORWARD LOOK

We have witnessed many major new discoveries in nuclear science over the last decade that were the direct result of the construction and operation of new facilities and detectors during the 1980s and 1990s. We also have seen a growing use of exciting new technologies developed in nuclear science both in well-established areas of application, such as medicine, and in important new areas, such as homeland security. Continuing this growth and reaping the benefits it provides will require new investments. With these investments, the United States will maintain its present world-leading position in nuclear science. And we will continue to contribute to the economic growth, health, and security of our nation.

# 8 Appendix

# Charge Letter

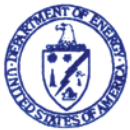 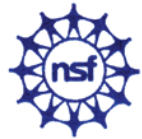

**U.S. Department of Energy
and the
National Science Foundation**

July 17, 2006

Professor Robert E. Tribble
Chair, DOE/NSF Nuclear Science Advisory Committee
Cyclotron Institute
Texas A&M University
College Station, TX 77843

Dear Professor Tribble:

This letter requests that the Department of Energy (DOE)/National Science Foundation (NSF) Nuclear Science Advisory Committee (NSAC) conduct a new study of the opportunities and priorities for United States nuclear physics research and recommend a long range plan that will provide a framework for coordinated advancement of the Nation's nuclear science research programs over the next decade.

The new NSAC Long Range Plan (LRP) should articulate the scope and the scientific challenges of nuclear physics today, what progress has been made since the last LRP and the impacts of these accomplishments both within and outside of the field. It should identify and prioritize the most compelling scientific opportunities for the U.S. program to pursue over the next decade and articulate their scientific impact. A national coordinated strategy for the use of existing and planned capabilities, both domestic and international, and the rationale for new investments should be articulated. To be most helpful, the plan should indicate what resources and funding levels would be required (including construction of new facilities) to maintain a world-leadership position in nuclear physics research, and what the impacts are and priorities should be, if the funding available provides constant level of effort (FY 2007 President's Budget Request) into the out-years (FY 2008-2017).

The recommendations and guidance in the NSAC 2002 LRP and subsequent reports have been utilized by the agencies as important input to their planning and programmatic decisions. Resources have been made available to the programs' major facilities and experiments that have allowed the U.S. program to be successful in delivering significant discoveries and advancements in nuclear physics over the last five years. This has occurred in the context of constrained funding that has resulted in a reduction in the number of DOE National User Facilities and limited the ability to pursue identified scientific opportunities. However, projected funding levels in the out-years would allow the agencies to begin to address the major project recommendations in the NSAC 2002 LRP. The projected funding for DOE is compatible with implementing the 12 GeV Upgrade of the Continuous Electron Beam Accelerator Facility, and starting construction of a rare isotope beam facility that is less costly than the proposed Rare

2Isotope Accelerator (RIA) facility early in the next decade. At NSF the process has been put in place for developing a deep underground laboratory project and bringing this project forward for a funding decision.

Since the submission of the NSAC 2002 LRP, increased emphasis has been placed within the federal government on international and interagency coordination of efforts in the fundamental sciences. The extent, benefits, impacts and opportunities of international coordination and collaborations afforded by current and planned major facilities and experiments in the U.S. and other countries, and of interagency coordination and collaboration in cross-cutting scientific opportunities identified in studies involving different scientific disciplines should be specifically addressed and articulated in the report. The scientific impacts of synergies with neighboring research disciplines and further opportunities for mutually beneficial interactions with outside disciplines, such as astrophysics, should be discussed.

An important dimension of your plan should be the role of nuclear physics in advancing the broad interests of society and ensuring the Nation's competitiveness in the physical sciences and technology. Education of young scientists is central to the mission of both agencies and integral to any vision of the future of the field. We ask NSAC to discuss the contribution of education in nuclear science to academia, medicine, security, industry, and government, and strategies to strengthen and improve the education process and to build a more diverse research community. Basic research plays a very important role in the economic competitiveness and security of our Nation. We ask that NSAC identify areas where nuclear physics is playing a role in meeting society's needs and how the program might enhance its contributions in maintaining the Nation's competitiveness in science and technology.

Activities across the federal government are being evaluated against established performance goals. In FY 2003, utilizing input from NSAC, the long-term goals for the DOE SC Nuclear Physics program and the metrics for evaluations of the program activities were established. It is timely during this long range planning exercise to gauge the progress towards these goals, and to recommend revised long-term goals and metrics for the DOE SC Nuclear Physics program, in the context of the new LRP, if appropriate. The findings and recommendations of this evaluation should be a separate report.

In the development of previous LRP's, the Division of Nuclear Physics of the American Physical Society (DNP/APS) was instrumental in obtaining broad community input by organizing town meetings of different nuclear physics sub-disciplines. The Division of Nuclear Chemistry and Technology of the American Chemical Society (DNC&T/ACS) was also involved. We encourage NSAC to exploit this method of obtaining widespread input again, and to further engage both the DNP/APS and DNC&T/ACS in laying out the broader issues of contributions of nuclear science research to society.

Charge Letter 167

Please submit an interim report containing the essential components of NSAC's recommendations to the DOE and the NSF by October 2007, and the final report by the end of calendar year 2007. The agencies very much appreciate NSAC's willingness to undertake this task. NSAC's previous long range plans have played a critical role in shaping the Nation's nuclear science research effort. Based on NSAC's laudable efforts in the past, we look forward to a new plan that can be used to chart a vital and forefront scientific program into the next decade.

Sincerely,

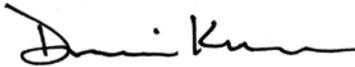

Dennis Kovar
Associate Director of the Office of Science
 for Nuclear Physics
Department of Energy

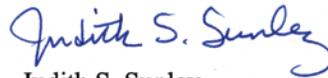

Judith S. Sunley
Acting Assistant Director
Mathematical and Physical Sciences
National Science Foundation

# NSAC Long Range Plan Working Group Meeting

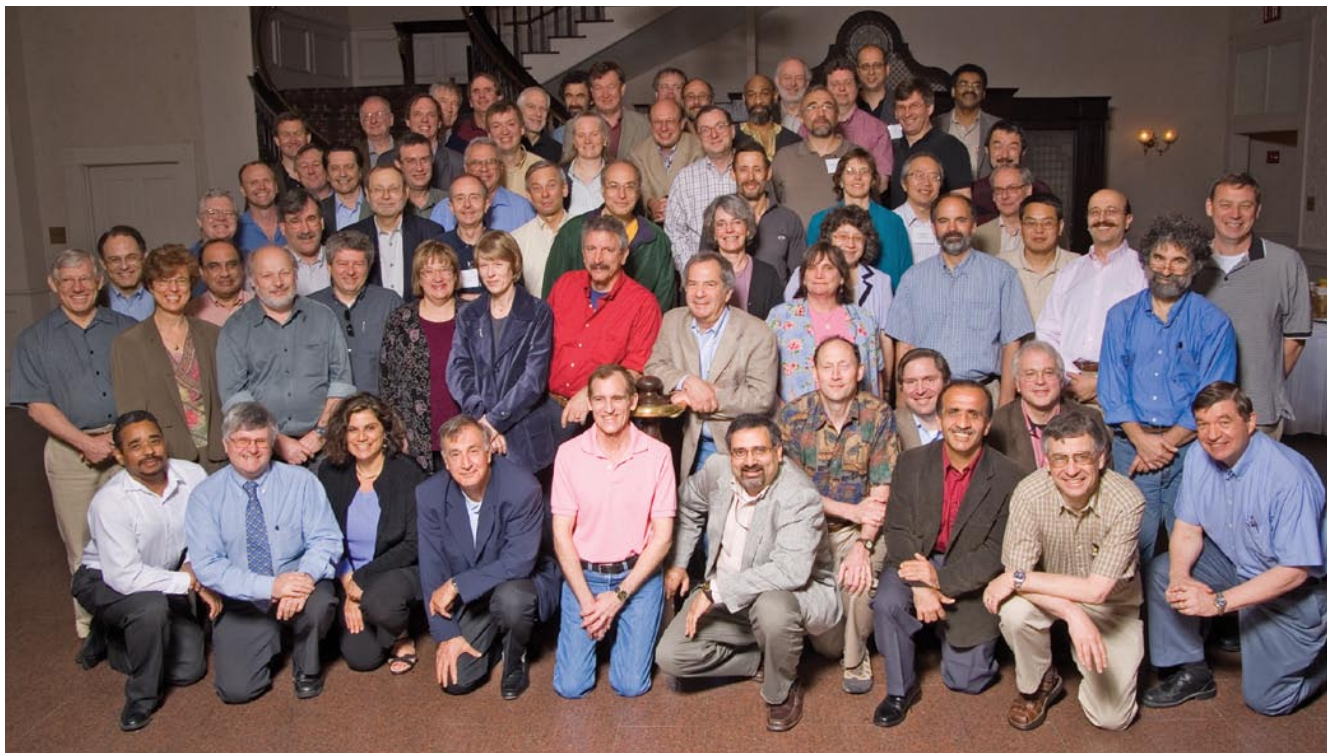

**Robert Tribble (chair),** Texas A&M University
**Elizabeth Beise,** University of Maryland
**Douglas Bryman,** TRIUMF
**Adam Burrows,** University of Arizona
**Lawrence Cardman,** University of Virginia/Jefferson Lab
**Richard F. Casten,** Yale University
**Gordon Cates,** University of Virginia
**Jolie A. Cizewski,** Rutgers University
**David J. Dean,** ORNL
**Abhay Deshpande,** SUNY–Stony Brook
**Charlotte Elster,** Ohio University
**Rolf Ent,** Jefferson Lab
**Bradley Filippone,** Caltech
**Stuart Freedman,** UC Berkeley/LBNL
**Thomas Glasmacher,** Michigan State University
**Timothy Hallman,** BNL
**Edward Hartouni,** LLNL
**Ulrich W. Heinz,** Ohio State University
**David Hertzog,** University of Illinois
**Roy Holt,** ANL
**Calvin Howell,** Duke University
**Barbara Jacak,** SUNY–Stony Brook
**Peter Jacobs,** LBNL
**Robert Janssens,** ANL
**Xiangdong Ji,** University of Maryland
**David Kaplan,** University of Washington

**Dmitri Kharzeev,** BNL
**Roy A. Lacey,** SUNY–Stony Brook
**David M. Lee,** LANL
**I-Yang Lee,** LBNL
**Naomi C. R. Makins,** University of Illinois
**Lia Merminga,** Jefferson Lab
**Curtis Meyer,** Carnegie Mellon University
**Zein-Eddine Meziani,** Temple University
**Richard Milner,** MIT
**Berndt Mueller,** Duke University
**Witold Nazarewicz,** University of Tennessee
**Heino Nitsche,** UC Berkeley/LBNL
**Margaret Norris (McMahan),** LBNL
**Michael Ramsey-Musolf,** Caltech/University of Wisconsin
**Winston Roberts,** Florida State University
**David Robertson,** University of Missouri
**Hamish Robertson,** University of Washington
**Thomas Roser,** BNL
**Guy Savard,** University of Chicago/ANL
**Hendrik Schatz,** Michigan State University
**Susan Seestrom,** LANL
**Bradley Sherrill,** Michigan State University
**James Symons,** LBNL

**Anthony W. Thomas,** Jefferson Lab
**Thomas S. Ullrich,** BNL
**Ubirajara van Kolck,** University of Arizona
**Steven Vigdor,** Indiana University/BNL
**Michael Wiescher,** Notre Dame
**John Wilkerson,** University of Washington
**Boleslaw Wyslouch,** MIT
**Sherry Yennello,** Texas A&M University
**Glenn Young,** ORNL
**William Zajc,** Columbia University

### International Guests
**Sidney Gales,** GANIL, France
**Walter Henning,** GSI, Germany
**Tohru Motobayashi,** RIKEN, Japan
**Naohito Saito,** JPARC, Japan
**Paul Schmor,** TRIUMF, Canada

### Agency Representatives
**Ani Aprahamian,** NSF
**Sidney Coon,** DOE
**Gene Henry,** DOE
**Bradley Keister,** NSF
**Dennis Kovar,** DOE
**Gulshan Rai,** DOE
**Brad Tippens,** DOE

## Long Range Plan Town Meetings

White Papers are available at http://www.sc.doe.gov/np/nsac/nsac.html

### NUCLEAR ASTROPHYSICS AND THE STUDY OF NUCLEI

**Chicago, IL, January 19–21, 2007**

**Organizing Committee**: A. Aprahamian, Univ. of Notre Dame; L. Cardman, Jefferson Lab; A. Champagne, Univ. of North Carolina, Chapel Hill; D. Dean, ORNL; G. Fuller, Univ. of California, San Diego; R. Grzywacz, Univ. of Tennessee; A. Hayes, LANL; R. Janssens (co-chair), ANL; I-Y Lee, LBNL; E. Ormand, LLNL; J. Piekarewicz, Florida State Univ.; H. Schatz (co-chair), Michigan State Univ.; M. Thoennessen, Michigan State Univ.; R. Wiringa, ANL; S. Yennello, Texas A&M Univ.

### NUCLEAR SCIENCE AND THE NEW STANDARD MODEL: FUNDAMENTAL SYMMETRIES AND NEUTRINOS IN THE NEXT DECADE

**Chicago, IL, January 19–21, 2007**

**Organizing Committee**: A. Balantekin, Univ. of Wisconsin; J. Beacom, Ohio State Univ.; Y. Efremenko, Univ. of Tennessee; S. Elliott, LANL; G. Drexlin, Insitut fur Kernphysik, Karlsruhe; B. Filippone, Caltech; G. Fuller, Univ. of California, San Diego; G. Greene, Univ. of Tennessee/ORNL; W. Haxton, Univ. of Washington; D. Hertzog, Univ. of Illinois; B. Holstein, Univ. of Massachusetts; P. Huffman, North Carolina State Univ.; J. Klein, Univ. of Texas; S. Klein, LBNL; K. Kumar, Univ. of Massachusetts; W. Louis, LANL; W. Marciano, BNL; G. McLaughlin, North Carolina State Univ.; J. Nico, NIST; A. Opper, George Washington Univ.; A. Poon, LBNL; M. Ramsey-Musolf (co-chair), Univ. of Wisconsin; H. Robertson (co-chair), Univ. of Washington; G. Savard, Univ. of Chicago/ANL; B. Vogelaar, Virginia Tech; S. Wilburn, LANL; J. Wilkerson, Univ. of Washington

### PHASES OF QCD

**Rutgers University, January 12–14, 2007**

**Organizing Committee**: P. Jacobs (co-chair), LBNL; D. Kharzeev, BNL; B. Müller (co-chair), Duke Univ.; J. Nagle, Univ. of Colorado; K. Rajagopal, MIT; S. Vigdor, Indiana Univ.

### HADRONIC PHYSICS

**Rutgers University, January 12–14, 2007**

**Organizing Committee**: L. Cardman, Jefferson Lab; A. Deshpande, SUNY–Stony Brook; S. Capstick, Florida State Univ.; J. Cizewski, Rutgers Univ.; X. Ji (co-chair), Univ. of Maryland; C. Keppel, Hampton Univ.; C. Meyer, Carnegie Mellon Univ.; Z. Meziani (co-chair), Temple Univ.; J. Negele, MIT; J. Peng, Univ. of Illinois; A. Thomas, Jefferson Lab

### A VISION FOR NUCLEAR SCIENCE EDUCATION AND OUTREACH FOR THE NEXT LONG RANGE PLAN

**Brookhaven National Laboratory, December 1–3, 2006**

**Organizing Committee**: J. Cizewski, Rutgers Univ.; T. Hallman, BNL; M. Norris (McMahan), LBNL

### NUCLEAR SCIENCE ENHANCING AMERICAN COMPETITIVENESS THROUGH BASIC RESEARCH

**Chicago, IL, January 19–20, 2007**

**Organizing Committee:** M. Chadwick, LANL; B. Gibson, LANL; T. Glasmacher, Michigan State Univ.; E. Hartouni (co-chair), LLNL; C. Howell (co-chair), Duke Univ.; D. McNabb, LLNL; D. Robertson, Univ. of Missouri

# Glossary

## A

| | |
|---|---|
| **ACTS** | Academies Creating Teacher Scientists (DOE Program) |
| **AGS** | Alternating Gradient Synchrotron (BNL) |
| **ALICE** | A Large Ion Collider Experiment (LHC Experiment) |
| **AMS** | Accelerator Mass Spectrometry |
| **ANL** | Argonne National Laboratory |
| **APS** | American Physical Society |
| **ASCR** | Office of Advanced Scientific Computing Research |
| **ATLAS** | Argonne Tandem Linac Accelerator System (ANL) |
| **ATLAS** | A Toroidal LHC Apparatus (LHC Experiment) |
| **ATTA** | Atom Trap Trace Analysis |
| **AdS/CFT** | Anti-de-Sitter/Conformal Field Theory |

## B

| | |
|---|---|
| **BASE** | Berkeley Accelerator Space Effects (LBNL Facility) |
| **BELLE** | Experiment to study the origin of CP Violation at KEK (Japan) |
| **BNL** | Brookhaven National Laboratory |
| **BRAHMS** | Broad Range Hadron Magnetic Spectrometers, RHIC Experiment (BNL) |
| **BigSOL** | The Superconducting Solenoid Rare-Isotope Beam Line (TAMU) |

## C

| | |
|---|---|
| **CARIBU** | CAlifornium Rare Ion Breeder Upgrade (ATLAS, ANL) |
| **CBM** | Compressed Baryonic Matter Experiment at FAIR (GSI) |
| **CDM** | Cold Dark Matter |
| **CEBAF** | Continuous Electron Beam Accelerator Facility (JLAB) |
| **CERN** | European Organization for Nuclear Research (Switzerland) |
| **SMC** | Spin Muon Collaboration, Nucleon Spin Structure Experiment at SPS (CERN) |
| **CGC** | Color Glass Condensate |
| **CKM** | Cabibbo-Kobayashi-Maskawa |
| **CLAIRE** | Center for Low Energy Astrophysics and Interdisciplinary Research (LBNL) |
| **CLAS** | CEBAF Large Acceptance Spectrometer (JLAB) |
| **CLEAN** | Cryogenic Low Energy Astrophysics with Noble Gases, Research and development Experiment (Yale) |
| **CLFV** | Charged Lepton Flavor Violation |
| **CMS** | Compact Muon Solenoid (LHC Experiment) |
| **CNO** | Carbon-Nitrogen-Oxygen |
| **COBE** | Cosmic Background Explorer (Satellite) |
| **COM** | Committee on Minorities of the APS |
| **COMPASS** | Common Muon Proton Apparatus for Structure and Spectroscopy (CERN/SPS Experiment) |
| **CP** | Charge-Parity |
| **CPT** | Charge-Parity-Time Reversal |
| **CSWP** | Committee on the Status of Women in Physics of the APS |
| **CT** | Computed Tomography |
| **CUORE** | Cryogenic Underground Observatory for Rare Events, double-beta decay experiment at Gran Sasso (Italy) |

## D

| | |
|---|---|
| **DANCE** | Detector for Advanced Neutron Capture Experiment at LANSCE facility (LANL) |
| **DESY** | Deutsches Elektronen-Synchrotron (Germany) |
| **DFT** | Density Functional Theory |
| **DHS** | Department of Homeland Security |
| **DIS** | Deep Inelastic Scattering |
| **DKFZ** | German Center for Cancer Research (Germany) |
| **DNP** | Division of Nuclear Physics (APS) |
| **DOD** | U.S. Department of Defense |
| **DOE** | U.S. Department of Energy |
| **DUSEL** | Deep Underground Science and Engineering Laboratory |
| **DVCS** | Deeply-Virtual Compton Scattering |

## E

| | |
|---|---|
| **EBIS** | Electron Beam Ion Source |
| **ECR** | Electron Cyclotron Resonance |
| **ECT*** | European Center for Theoretical Studies in Nuclear Theory |
| **EDM** | Electric Dipole Moment |
| **EFT** | Effective Field Theory |
| **EIC** | Electron-Ion Collider |
| **ELIC** | Electron Light Ion Collider (JLAB Concept of EIC) |

| | | | |
|---|---|---|---|
| **ELSA** | Experiment at Bonn University (Germany) | | |
| **EMC** | Electron Muon Collaboration, DIS Experiment at SPS (CERN) | | |

| | |
|---|---|
| **ELSA** | Experiment at Bonn University (Germany) |
| **EMC** | Electron Muon Collaboration, DIS Experiment at SPS (CERN) |
| **ENDF** | Evaluated Nuclear Data File |
| **EOS** | Equation-of-State |
| **EPR** | Electron Paramagnetic Resonance |
| **eRHIC** | Electron Relativistic Heavy Ion Collider (BNL Concept of EIC) |
| **ERL** | Energy Recovery Linac |
| **EXO** | Enriched Xenon Observatory, double-beta decay experiment |

### F

| | |
|---|---|
| **FAIR** | Facility for Antiproton and Ion Research at GSI (Germany) |
| **FDG** | 18F-labeled fluorodeoxyglucose (tracer used in PET) |
| **FEL** | Free-Electron Laser |
| **FMA** | Fragment Mass Analyzer (ANL) |
| **FNAL** | Fermi National Accelerator Laboratory |
| **FNPB** | Fundamental Neutron Physics Beamline (SNS at ORNL) |
| **FRIB** | Facility for Rare Isotope Beams |
| **FRS** | Fragment Recoil Separator |

### G

| | |
|---|---|
| **GANIL** | Grand Accelerateur National d'Ions Lourds, European Nuclear Structure Laboratory (France) |
| **GEM** | Gas Electron Multiplier |
| **GERDA** | Double-beta decay experiment at Gran Sasso (Italy) |
| **GFMC** | Greens Function Monte Carlo |
| **GMM** | Graphene-Magnet Multilayers |
| **GNEP** | Global Nuclear Energy Partnership |
| **GNO** | Gallium Neutrino Observatory, Solar Neutrino Experiment at Gran Sasso (Italy) |
| **GPD** | Generalized Parton Distribution |
| **GRETA** | $4\pi$ gamma-ray energy tracking detector |
| **GRETINA** | $1\pi$ gamma-ray energy tracking detector |
| **GSI** | Gesellschaft für Schwerionenforschung (Germany) |

### H

| | |
|---|---|
| **HEP** | High Energy Physics |
| **HERA** | Hadron-Electron Ring Accelerator at DESY (Germany) |
| **HERMES** | DIS Experiment for studies of the nucleon spin structure at HERA (DESY) |
| **HI$\gamma$S** | High-Intensity Gamma-Ray Source |
| **HIRA** | High-Resolution Silicon Array |
| **HKS** | Hyper-Nuclear Spectroscopy |
| **HPTL** | High Power Target Laboratory |
| **HRIBF** | Holifield Radioactive Ion Beam Facility (ORNL) |
| **HVP** | Hadronic Vacuum Polarization |

### I

| | |
|---|---|
| **IAEA** | International Atomic Energy Agency |
| **ICEY** | Conversion electron spectrometer (WNSL) |
| **ILL** | Institut Max von Laue-Langevin (France) |
| **IMB** | Irvine Michigan Brookhaven Detector (Lake Erie) |
| **INCITE** | Innovative and Novel Computational Impact on Theory and Experiment |
| **INFN** | Istituto Nazionale di Fisica Nucleare, National Institute of Nuclear Physics (Italy) |
| **INT** | Institute of Nuclear Theory (Seattle) |
| **INTEGRAL** | International Gamma-Ray Astrophysics Laboratory (ESA Satellite) |
| **IRIS** | Injector for Radioactive Ion Species (ORNL) |
| **ISAC** | Isotope Separator and Accelerator (TRIUMF) |
| **ISOL** | Isotope Separator on Line |
| **ISOLDE** | Radiactive Beam Facility (CERN) |
| **IUCF** | Indiana University Cyclotron Facility |

### J

| | |
|---|---|
| **J-PARC** | Proton Accelerator Research Complex (Japan) |
| **JHF** | Japanese Hadron Facility |
| **JINR** | Joint Institute for Nuclear Research (Russia) |
| **JINA** | Joint Institute for Nuclear Astrophysics (Notre Dame) |
| **JLAB** | Thomas Jefferson National Laboratory |
| **JSPS** | Japan Society for the Promotion of Science |
| **JUSTIPEN** | Japan-U.S. Theory Institute for Physics with Exotic Nuclei |

## K

| | |
|---|---|
| **KATRIN** | The Karlsruhe Tritium Neutrino Experiment |
| **KEK** | High Energy Accelerator Research Organization in Tsukuba (Japan) |

## L

| | |
|---|---|
| **LAMPF** | Los Alamos Meson Physics Facility (now LANSCE) (LANL) |
| **LANL** | Los Alamos National Laboratory |
| **LANSCE** | Los Alamos Neutron Science Center |
| **LBNL** | Lawrence Berkeley National Laboratory |
| **LEAR** | Low-Energy Anti-proton Ring (CERN) |
| **LEBAF** | Low Energy Beam Accelerator Facility |
| **LENA** | Laboratory for Experimental Nuclear Astrophysics (TUNL) |
| **LEP** | Large Electron and Positron Collider (CERN) |
| **LHC** | Large Hadron Collider (CERN) |
| **LINAC** | Linear Accelerator |
| **LQCD** | Lattice QCD |
| **LUNA** | Laboratory Underground for Nuclear Astrophysics, Gran Sasso (Italy) |

## M

| | |
|---|---|
| **MAMI** | Mainz Microtron Facility (Germany) |
| **MARS** | Momentum Achromat Recoil Separator (TAMU) |
| **MINERνA** | Neutrino Scattering Experiment (FNAL) |
| **MINOS** | Main Injector Neutrino Oscillation Search Experiment (FNAL) |
| **MIT** | Massachusetts Institute of Technology |
| **MONA** | Modular Large-Area Neutron Detector (NSCL) |
| **MRI** | Magnetic Resonance Imaging |
| **MSU** | Michigan State University |

## N

| | |
|---|---|
| **NCSM** | No Core Shell Model |
| **NERSC** | National Energy Research Scientific Computing Center (LBNL) |
| **NIA** | National Institute of Aerospace |
| **NIH** | National Institutes of Health |
| **NIST** | National Institute of Standards and Technology |
| **NMR** | Nuclear Magnetic Resonance |
| **NNDC** | National Nuclear Data Center |
| **NNLO** | Next-to-Next to Leading Order |
| **NNSA** | National Nuclear Security Administration |
| **NRC** | Nuclear Regulatory Commission |
| **NRDC** | Nuclear Reaction Data Centers Network |
| **NSAC** | Nuclear Science Advisory Committee |
| **NSCL** | National Superconducting Cyclotron Laboratory (Michigan State) |
| **NSDD** | Nuclear Structure and Decay Data |
| **NSF** | National Science Foundation |
| **NSL** | Nuclear Science Laboratory (Notre Dame) |

## O

| | |
|---|---|
| **ONP** | Office of Nuclear Physics of DOE |
| **ORELA** | Oak Ridge Electron Linear Accelerator |
| **ORLaND** | Oak Ridge Laboratory for Neutrino Detectors |
| **ORNL** | Oak Ridge National Laboratory |
| **ORRUBA** | Oak Ridge Rutgers University Barrel Array |

## P

| | |
|---|---|
| **PET** | Positron Emission Tomography |
| **PHENIX** | Pioneering High Energy Nuclear Interacting Experiment (RHIC, BNL) |
| **PHOBOS** | RHIC Experiment (BNL) |
| **PREX** | Lead Radius Experiment (JLAB) |
| **PSI** | Paul Scherrer Institut (Switzerland) |
| **PV** | Parity-Violating |
| **PVDIS** | Parity Violation in Deep Inelastic Scattering |

## Q

| | |
|---|---|
| **QCD** | Quantum Chromodynamics |
| **QCDOC** | QCD on a Chip (supercomputer developed by IBM and RIKEN-BNL) |
| **QED** | Quantum Electrodynamics |
| **QGP** | Quark-Gluon Plasma |

## R

| | |
|---|---|
| **RBRC** | RIKEN-BNL Research Center |
| **RCF** | RHIC Computing Facility (BNL) |
| **RCNP** | Research Center for Nuclear Physics (Japan) |
| **RESOLUT** | Resonator Solenoid with Upscale Transmission (Florida State) |
| **REU** | Research Experiences for Undergraduates |
| **RHIC** | Relativistic Heavy Ion Collider (Brookhaven) |
| **RIA** | Rare Isotope Accelerator |
| **RIB** | Rare Isotope Beam |
| **RIBF** | Rare Isotope Beam Facility (RIKEN) |
| **RIKEN** | Institute of Physical and Chemical Research (Japan) |
| **RIKEN-BNL** | RIKEN Research Center (RBRC) at BNL |
| **RISAC** | Rare Isotope Accelerator Assessment Committee |

## S

| | |
|---|---|
| **SASSYER** | Small Angle Separator System at Yale for Evaporation Residues (WSNL) |
| **SDSC** | San Diego Supercomputer Center |
| **SEE** | Single-Event Effects |
| **SENSIT** | Variation of Ultra-Low Field (ULF) NMR imaging |
| **SIS** | Heavy Ion Synchrotron at GSI (Germany) |
| **SLAC** | Stanford Linear Accelerator Center |
| **SM** | Standard Model |
| **SNO** | Sudbury Neutrino Observatory (Canada) |
| **SNS** | Spallation Neutron Source (Oak Ridge) |
| **SOL** | Standards of Learning |
| **SPIRAL** | Système de Production d'Ions Radioactifs en Ligne, RIB Facility (GANIL) |
| **SPS** | Super Proton Synchrotron (CERN) |
| **SQUID** | Superconducting Quantum Interference Device |
| **SRF** | Superconducting Radio-Frequency |
| **SSAA** | Stockpile Stewardship Academic Alliance |
| **STAR** | Solenoidal Tracker at RHIC (BNL) |
| **STEM** | Science, Technology, Engineering and Mathematics |
| **SciDAC** | Scientific Discovery through Advanced Computing |

## T

| | |
|---|---|
| **TAMU** | Texas A&M University |
| **TAPS** | Teacher Academy in Physical Science |
| **TPC** | Time Projection Chamber |
| **TRIUMF** | Tri-University Meson Facility (Canada) |
| **TUNL** | Triangle Universities Nuclear Laboratory (Duke) |

## U

| | |
|---|---|
| **UCN** | Ultra Cold Neutrons |
| **ULF** | Ultra Low Field |
| **USNDP** | U.S. Nuclear Data Program |
| **USQCD** | U.S. Lattice QCD consortium |

## V

| | |
|---|---|
| **VENUS** | Versatile ECR Ion Source for Nuclear Science (LBNL) |

## W

| | |
|---|---|
| **WECAN** | Women Encouraging Competitive Advancement in Nuclear science |
| **WMAP** | Wilkinson Microwave Anisotropy Probe (Satellite) |
| **WMD** | Weapons of Mass Destruction |
| **WNSL** | Wright Nuclear Structure Laboratory (Yale University) |
| **WPEC** | Working Party on International Nuclear Data Evaluation Co-operation |

**December 2007**